\documentclass[aps,rmp,reprint,amsmath,amssymb,graphicx,longbibliography]{revtex4-1}

\usepackage{graphicx}
\usepackage{bm}

\def\cl{{\cal L}}

\def\co{{\cal O}}

\def\svev#1{\left\langle #1\right\rangle}       % variable < >
\def\tr{{\rm tr}\,}
\def\Tr{{\rm Tr}\,}

\long \def \blockcomment #1\endcomment{}
\def\det{{\rm det}}

\def\tbeta{\tilde\beta}

\newcommand{\bee}{\begin{equation}}
\newcommand{\ee}{\end{equation}}
\newcommand{\beea}{\begin{eqnarray}}
\newcommand{\eea}{\end{eqnarray}}

\newcommand{\VEV}[1]{\left\langle #1\right\rangle}
\newcommand{\Dslash}{{D}\!\!\!\!/\,}

\begin{document}
%%%%%%%%%%%%%%%%%%%%%%%%%%%%%%%%%%%%%%%%%%%%%%%%%%%%%%%%%%%%%%%%%%%%%%

\title{Lattice tests of beyond  Standard Model dynamics}
\author{Thomas DeGrand}
%\email{thomas.degrand@colorado.edu}
\affiliation{Department of Physics,
University of Colorado, Boulder, CO 80309, USA}

\date{\today{}}

\begin{abstract}
Over the last few years lattice techniques have been used to investigate
candidate theories of new physics beyond the Standard
Model. This review gives a survey of  results from these studies.
Most of these investigations have been of systems of gauge fields and 
fermions that have slowly-running coupling constants.
A major portion of the review is a  critical discussion of work in this particular subfield, first
describing  the methods used, and then giving a compilation of results for specific models.
\end{abstract}

%\pacs{12.60-i 11.15.Ha 11.10.Hi  14.80.Bn, 14.80.Ec }
%\keywords{Suggested keywords}
\maketitle

\tableofcontents{}

%%%%%%%%%%%%%%%%%%%%%%%%%%%%%%%%%%%%%%%%%%%%%%%%%%%%%%%%%%%%%%%%%%%%%
\section{Introduction \label{sec:intro}}
%%%%%%%%%%%%%%%%%%%%%%%%%%%%%%%%%%%%%%%%%%%%%%%%%%%%%%%%%%%%%%%%%%%%%

\subsection{What does the title mean?}

With the recent discovery of the Higgs boson \cite{Aad:2012tfa,Chatrchyan:2012ufa},
 the Standard Model's particle spectrum seems to be  complete.
But, is the particle at 126 GeV really the Standard Model Higgs, a fundamental scalar field,
or is it something else? And,  what about experimental observations
that do not have a Standard Model explanation? Examples of such physics include
 neutrino masses and
oscillations, 
the origin of the matter-antimatter asymmetry of the Universe, the nature of
 dark matter, and of dark energy. For that matter,
why are the fundamental parameters of the Standard Model what they are? Why is the electron 
so light?
Attempts to answer these questions are what is generically meant by ``beyond Standard Model''
 physics.

``Lattice'' refers to lattice gauge theory, which is a collection of  analytic and numerical
techniques for studying
quantum field theories. In principle, 
 a lattice calculation starts with a Lagrangian and a cutoff and ends with a fully nonperturbative
prediction for some observable.
Lattice methods have
become a standard technique
to study nonperturbative
properties of the theory of the strong interactions, Quantum Chromodynamics or QCD.
Most of the information we have about the particle spectrum of baryons and mesons, and of many 
hadronic matrix elements
relevant to Standard Model tests, come from lattice calculations. 

And why put ``Lattice'' and ``beyond Standard Model'' in the same title? For almost forty years,
 phenomenologists have conjectured
that some beyond Standard Model physics might be nonperturbative. For example, the Higgs boson might 
not be fundamental; it could be
 a composite object held together by some new kind of strong force. About eight years
 ago, several physicists with lattice tool kits
 realized that the techniques they used for QCD might be applied to studies of
 candidate beyond Standard Model systems.
The field became very active. This review is an attempt to describe
 the systems that were studied, the
 techniques that were used, and the results that were obtained.

\subsection{A lattice perspective on issues relevant to beyond Standard Model physics}

Before going beyond the Standard Model, we should visit the Standard Model itself.
 (See \textcite{Logan:2014jla} for a more  pedagogic introduction.)
It has a product gauge symmetry $SU(3)\times SU(2)\times U(1)$ encoding respectively color
(the strong interactions), weak isospin and weak hypercharge: 
 the last two gauge symmetries break spontaneously, resulting in massive
 $W$ and $Z$ bosons and leaving the unbroken $U(1)$ symmetry of electromagnetism.
Quarks and leptons fall into three generations; the left and right-handed (negative
and positive helicity)
fermions have different electroweak couplings. 

In particular, the left-handed leptons and quarks form a doublet of weak isospin
\bee
E_L= \left(\begin{array}{c} \nu_e \\ e^- \end{array}\right)_L ;\qquad   Q_L =\left(\begin{array}{c} u \\ d \end{array}\right)_L
\ee
while the right-handed particles $e_R$, $u_R$, $d_R$ are singlets. This is for the first generation of fermions;
there are identical terms for the second and third generations.
The Lagrangian has three parts
\bee
\cl_{SM}= \cl_g + \cl_\Phi + \cl_m.
\ee
$\cl_g$  holds the kinetic terms for the fermions and gauge bosons
\beea
\cl_g &=& \sum_{j=1}^3 (\bar E^j_L (i \Dslash) E^j_L + \bar Q^j_L (i \Dslash) Q^j_L   \nonumber \\
& & + \bar e^j_R (i \Dslash) e^j_R 
+ \bar u^j_R (i \Dslash) u^j_R +  \bar d^j_R (i \Dslash) d^j_R    )          \nonumber \\
&  & - \frac{1}{4}F_{\mu\nu}^2 - \frac{1}{4}W_{\mu\nu}^2 - \frac{1}{4}B_{\mu\nu}^2 \nonumber \\
\label{eq:clg}
\eea
in terms of the field strengths of the gluons $F$, the $SU(2)$ weak fields $W$ and the $U(1)$ field $B$.
$D$ is the covariant derivative.
The index $j$ runs over the three generations of fermions. $\cl_g$ conceals three parameters, the three Standard Model gauge
couplings. These gauge invariant interactions
make up the part of the Standard Model that is most well tested: from $\cl_g$ follows all of electrodynamics,
asymptotic freedom, parity violation in the weak interactions, and much else.

Equation \ref{eq:clg} describes a set of massless gauge bosons. Electroweak symmetry is spontaneously broken, that is,
the symmetries of the Lagrangian are not respected by the vacuum. The mechanism for doing this is contained in 
$\cl_\Phi$. In the Standard Model this is achieved by  the Higgs field $\Phi$,
a single scalar field whose components form a complex doublet of weak $SU(2)$.
Three of the four components of $\Phi$, which would be Goldstone bosons, are ``eaten'' by the $SU(2)$ and $U(1)$
gauge fields to give the massive $W^+$, $W^-$ and $Z_0$. The fourth component becomes
the Higgs particle. The  potential is  arranged to
accomplish this. In the Standard Model, it is just
\bee
\cl_\Phi= |D_\mu \Phi|^2 - V(\Phi)
\ee
and
\bee
V(\Phi) = -\mu^2 \Phi^\dagger \Phi + \lambda(\Phi^\dagger \Phi)^2.
\ee
The sign of the quadratic term is taken by hand to be negative,
to insure spontaneous symmetry breaking. 
$V(\Phi)$ is characterized by two parameters $\mu$ and $\lambda$.
$\Phi$ develops a vacuum expectation value
\bee
\svev {\Phi} = \frac{1}{\sqrt{2}} \left(\begin{array}{c} 0 \\ v \end{array}\right)
\ee
where $v^2=\mu^2/\lambda$. The Higgs and Goldstone masses are found by considering small fluctuations
around the minimum,
\bee
\Phi= \frac{1}{\sqrt{2}} \left(\begin{array}{c} \phi_1+i\phi_2 \\ v+H + i\phi_3 \end{array}\right)
\ee
and the Higgs mass is 
$m_H^2=2\mu^2=2\lambda v^2$. 
 The measured masses of the $W$, $Z$, and the Higgs
(126 GeV) and the known value for the
$SU(2)$ and $U(1)$ couplings $g$ and $g'$ tell us that  $v=246$ GeV and
 $\lambda=0.13$, $\mu^2=  (89$ GeV$)^2$.

$\cl_m$ generates  the masses of quarks and leptons.
A mass term couples left-handed and right-handed fermions to each other. This coupling, if present,
would violate  gauge invariance, because left-handed and right-handed fermions transform differently under $SU(2)$.
 However, a trilinear coupling of the Higgs, a left-handed fermion, and a
right-handed fermion is consistent with gauge invariance. Thus the Standard Model's mass term is, schematically,
\beea
\cl_m= &
-\lambda_d^{ij} \bar Q_L^i \cdot \Phi d^j_R 
- \lambda_u^{ij} \epsilon^{ab} \bar Q_{La}^i \cdot \Phi_b^\dagger u^j_R  \nonumber \\
& -
\lambda_l^{ij}\bar E_L^i \cdot \Phi e_R^j + h.c.   \nonumber \\
\label{eq:clm}
\eea
When $\Phi$ gets its
vacuum expectation value $v$, this trilinear interaction generates  (generalized) mass terms for the quarks and leptons,
parametrized by the elements of the three complex matrices $\lambda^{ij}$. This is usually done in terms of the
Cabibbo - Kobayashi  - Maskawa matrix, but we will not need this in what follows.
$\cl_{\Phi}$ and $\cl_m$ are the parts of the Standard Model that are most often replaced
or augmented by some new beyond  Standard Model physics. Most often, 
$\cl_{\Phi}$ is replaced by some new mechanism to break electroweak symmetry, perhaps without a Higgs boson. Finding
something to replace $\cl_m$ which does not involve a Higgs field and also does not 
introduce unwanted (unobserved) new physics is often a serious issue.

The phrase ``beyond Standard Model'' has been around in the literature almost as long
as the phrase ``Standard Model.'' Why is that? There are several reasons.

First, there is known physics that is not part of the Standard Model: neutrino masses and mixings,
 dark matter, the origin of the Universe's matter - anti-matter asymmetry, the origin of inflation,
dark energy.

Second, the large number of parameters in the Standard Model seems excessive for a fundamental theory.
The Standard Model has many couplings: for three generations of particles, there are
 mass terms for the six quarks and three charged leptons
 and four Cabibbo-Kobayashi-Maskawa matrix elements responsible for mixing quark mass
 and weak interaction
eigenstates. Additionally, $\cl_\Phi$ holds
 two Higgs couplings ($\lambda$ and $\mu$), and  $\cl_g$ contains  three gauge couplings.
Could there not be something underneath, for which the Standard Model is just a low energy effective
field theory, and from which its couplings are derived?

Third, there are issues of principle with the Higgs sector. To begin, what is the origin of
electroweak symmetry breaking? In the Standard Model, the Higgs potential is simply postulated to have
a negative quadratic term, in order to
induce spontaneous symmetry breaking. This seems arbitrary.

The next  issue is the ``naturalness'' or ``hierarchy'' problem. Imagine that there is
 some new scale in 
nature, an ultraviolet (UV)  ``cutoff scale'' $\Lambda$
 for the Standard Model. If this scale is very high, why
 is the electroweak scale
so low compared to it? The problem is the instability of the Higgs mass against radiative corrections. There is
a quadratic dependence of the shift in the
  Higgs mass on  $\Lambda$. Any new physics scale $\Lambda$ induces a shift in the
 squared mass of the Higgs,  which is a value that is order $g^2 \Lambda^2$, in
size. ($g^2$ is a generic label for one of the Standard Model couplings.)

This effect can be seen in the Standard Model itself.
At one loop,
the shift in the mass term from its bare value (value at the cutoff) $\mu_0$  is
\bee
\mu^2 - \mu_0^2 = \frac{\lambda}{8\pi^2} \Lambda^2 - \frac{3 y_t^2}{8\pi^2} \Lambda^2 +
 \frac{3(3g^2+g'^2)}{16\pi^2}\Lambda^2.
\label{eq:hier}
\ee
The three terms on the right side of the equation come from of
 the Higgs self-interaction, the effect of top quarks, and the
 interaction of the Higgs
 with $W$ and $Z$ particles.
 $y_t$ is the top quark Yukawa coupling and $g$ and $g'$ are the $SU(2)$ and $U(1)$
 gauge couplings.  I neglected the other quarks, because their masses, and hence their Yukawa couplings,
  are so much smaller than the top quark's.
If $\Lambda$ is very large, $\mu_0$ must be delicately tuned to set $\mu$ to its low-energy 
value of 89 GeV.
This could be the case, but it also seems arbitrary.

Alternatively, imagine setting the Higgs mass to its known value, and
 setting the couplings to their observed values.
Solve Eq.~\ref{eq:hier}
 for $\Lambda$ assuming $\mu_0=0$. This gives $\Lambda \sim 5$ TeV. Is this a hint of a new physics scale?

Before the discovery of the Higgs, it was hoped that the Higgs mass itself would indicate a value
for the scale of new physics (which I will label $\Lambda$). The situation for a large Higgs mass is called the ``triviality bound,''
while too small a Higgs mass led to an ``instability bound.'' Either situation might have hinted at a low value for $\Lambda$.
Both bounds come from looking at the one-loop beta function for the Higgs self coupling.
Including only the Higgs self interaction and the top-quark Yukawa coupling
in the equation for the running coupling ($t=\log Q^2$) gives 
\bee
\frac{d\lambda}{dt} = \frac{3}{4\pi^2}[\lambda^2 + \lambda y_t^2 - y_t^4] .
\label{eq:hbf}
\ee

The ``triviality bound'' arises from the fact that the scale dependent Higgs self-interaction
becomes stronger as the momentum scale increases. At some point, it might become so strongly interacting
that perturbation theory would break down. If we neglect everything but the
self-interaction of the Higgs, we can integrate Eq.~\ref{eq:hbf} to find
\bee
\lambda(Q) = \frac{\lambda(Q_0)}{1-\frac{3}{4\pi^2}\lambda(Q_0)\log \frac{Q^2}{Q_0^2}} .
\label{eq:triv}
\ee
As the energy scale grows, so does $\lambda(Q)$. To prevent $1/\lambda(Q)$ from vanishing,
the Higgs mass must not become too large.
Replacing $\lambda(Q_0)$ by $m_H^2/2v^2$ in Eq.~\ref{eq:triv} gives
\bee
\frac{m_H^2}{v^2} < \frac{8\pi^2} {3 \log  \frac{\Lambda^2}{v^2} }.
\label{eq:upperH}
\ee

The other bound is the ``instability bound.'' Including only the top-quark Yukawa coupling
in Eq.~\ref{eq:hbf} gives
\bee
\frac{d\lambda}{dt} = Å -\frac{3}{4\pi^2}y_t^4 .
\ee
Then
\bee
\lambda(\Lambda) - \lambda(v) = -\frac{3}{4\pi^2} y_t^4 \log \frac{\Lambda^2}{v^2}.
\ee
To keep the vacuum stable, we need $\lambda(\Lambda)>0$.
Preventing this from happening gives a lower bound on the Higgs mass,
\bee
\frac{m_H^2}{v^2}> \frac{3}{2\pi^2} y_t^4 \log \frac{\Lambda^2}{v^2}.
\label{eq:lowerH}
\ee

These two bounds combine to give the ``Higgs chimney;'' if the Higgs mass were too small or too large, the
scale $\Lambda$ would become low, and the Standard Model would signal its own upper limit.
Unfortunately (or fortunately), 126 GeV is in the middle, and there seems to be no need for a nearby new physics
scale for stability. The Standard Model could simply be the low energy limit of delicately arranged dynamics at some high 
cutoff scale.

As it stands, the Standard Model is a renormalizable quantum field theory, which could be valid
all the way up to the Planck scale. Its low energy properties are independent of how it is cut off at
 arbitrarily short distance. It is a logical possibility that its couplings at the cutoff scale could
have been fine tuned.

But there is physics beyond the Standard Model (neutrino masses and so on). 
How can we combine the Standard Model with this new physics in some unified description?

The first possibility is that new physics is far away in energy. To deal with this
 situation, there
is another way to view the Standard Model:   it is an effective, low
 energy theory of Nature,
which arises from some as yet unknown dynamics. Choose units so that the Lagrange density has
 dimensions (energy)${}^4$
or $\Lambda^4$ where $\Lambda$ is a generic energy scale. Give all fields their engineering
dimensions to achieve this.  Write down the most general $SU(3)\times SU(2)\times U(1)$
symmetric Lagrangian with the field content of the Standard Model and coupling constants
 that are either dimensionless
or (only for $\mu^2$) of positive energy dimension. That Lagrangian is the Standard Model
 itself. Then, one can imagine adding new terms, which still involve only Standard Model fields
(since that is all there is) but with dimensionful couplings. Including these
terms (and symmetries), one  would write the electroweak Lagrangian as a set 
of terms $\cl_i$ where $i$
is the dimensionality of the operator: generically,
\bee
\cl = \cl_{SM} + \frac{1}{\Lambda} \cl_5 + \frac{1}{\Lambda^2} \cl_6 + \dots.
\label{eq:bseff}
\ee
The terms in this expansion have been cataloged.
 The enumeration was first given by \textcite{Buchmuller:1985jz}.
 There is only one dimension-5 term,
a Majorana  mass term for the neutrinos. Bounds on neutrino masses constrain its  $\Lambda$
to be larger than about $10^{13}$ GeV. At dimension six, there are eighty operators, each with its own $\Lambda$.
 Fortunately,
in any given process, only a few of them contribute, and so in principle, one
could consider some electroweak process, fit it to a combination of Standard Model
plus higher dimensional processes, and see if there is a signal for new physics.
 
 From the theory side, one could take some beyond Standard Model theory, 
 match it to the Lagrangian of Eq.~\ref{eq:bseff}  and either make predictions for its couplings or let
 data constrain the proposed new physics.
 For a useful discussion of this procedure, see
\textcite{Han:2004az}.
 This matching  might be done via a lattice calculation, if one imagined that new physics
was nonperturbative, lived at some very high scale, and was accessible to simulation.
This could range from something like a calculation of a decay constant in QCD (like $f_\pi$), 
to an analog of the kaon $B$ - parameter, to a full-blown
calculation of the expectation value of
some composite operator, plus its associated anomalous dimension.

Alternatively, new physics could involve
 particles that could actually be observed
in the near future in experiments. (This was the hope before the LHC turned on and
it is still the hope today.)
 These new degrees of freedom have to be realized
as explicit terms in the Lagrangian. If the new particle arose from nonperturbative dynamics, the appropriate
lattice calculation would be of its mass, like the spectroscopic ones done in a lattice QCD simulation.

In either case, the issue is, that the Standard Model by itself is a very accurate
description of many things; the new physics has to be carefully concealed. This is a strong
constraint on model building (or on the size of coefficients in the effective field theory
description). This problem has been well-analyzed in the particle physics literature.

These days, the branching ratios of the 126 GeV particle strongly constrain new physics.
(The Review of Particle Properties, \textcite{Agashe:2014kda} has the latest numbers.
To give a one-sentence summary, the couplings of the Higgs are broadly consistent with Standard Model
expectations, although experimental uncertainties are still large.)
Direct searches for new heavy particles also constrain new physics. (See \textcite{Halkiadakis:2014qda} for a recent summary.)
The situation involving the rest of the Standard Model can be described in a few sentences:
The most obvious manifestations of new physics in the low energy sector of the Standard Model
occur in the vacuum polarization of the $SU(2)\times U(1)$ gauge bosons
\cite{Peskin:1990zt,Peskin:1991sw,Altarelli:1990zd}. There are four
such amplitudes involving the physical photon, $Z$, their mixing, and the $W$, which are parametrized
by a set of $SU(2)\times U(1)$ quantities, conventionally called $\Pi_{QQ}$ (involving the photon),
$\Pi_{ii}$ for the $i$th component of weak isospin, and a mixing term $\Pi_{3Q}$.
All are functions of the squared momentum $q^2$ flowing through the gauge boson. At low energy, it is sensible to expand
these quantities in a power series in $q^2/M^2$, where $M$ sets the new physics scale.
Contributions to the electromagnetic current vanish at $q^2=0$ due to the usual Ward identity.
Then
\beea
\Pi_{QQ} &=& q^2\Pi'_{QQ}(0) + \dots \nonumber \\
\Pi_{3Q} &=& q^2\Pi'_{3Q}(0) + \dots \nonumber \\
\Pi_{11} &=& \Pi_{11}(0) + q^2\Pi'_{11}(0) + \dots \nonumber \\
\Pi_{33} &=& \Pi_{33}(0) + q^2\Pi'_{33}(0) + \dots \nonumber \\
\label{eq:vacpol}
\eea
There are six unknown on the right side of Eq.~\ref{eq:vacpol}. Three of them can be fixed by
experimental determinations of  the fine structure constant $\alpha$,
 the Fermi coupling $G_F$, and the Z-boson mass, leaving
three linear combinations, conventionally called $S$, $T$ and $U$, to be probes of new physics. They are
\beea
S &=& 16\pi[\Pi'_{33}(0)-\Pi'_{3Q}(0)] \nonumber \\
T &=& \frac{4\pi}{\sin^2 \theta_W \cos^2 \theta_W m_Z^2}[\Pi_{11}(0)-\Pi_{33}(0)] \nonumber \\
U &=& 16\pi[\Pi'_{33}(0)-\Pi'_{11}(0)]  . \nonumber \\
\label{eq:stu}
\eea

A decade ago, \textcite{Barbieri:2004qk,Han:2004az} combined precision electroweak data
with an effective field theory analysis of beyond Standard Model couplings, to
constrain the scale of new physics. Typical bounds, even then, were that $\Lambda$'s were in the few TeV range.
The recent analysis of \textcite{Ciuchini:2013pca} pushes the scale for many kinds
of new physics up to the 5-15 TeV range. Most of this new physics involves flavor structures
 which are different from the Standard Model's
$\cl_m$ (Eq.~\ref{eq:clm}). This is quite a different situation from the relatively low scales
 needed to address the hierarchy problem.

The absence of new physics for some distance above the Higgs mass itself constrains the
 kind of Lagrangians we can write down. We are forced to pause and think of symmetry reasons, for why 
particles could have masses far below the cutoff scale.
Unfortunately, the list is short:
\begin{itemize}
\item Gauge bosons remain massless due to gauge symmetry
\item Chiral symmetry protects the masses of (tree - level massless) fermions from additive renormalization
\item Goldstone bosons remain massless because their potentials obey a ``shift symmetry:''
Parameterizing the Higgs doublet in terms of the physical Higgs $H$, its vacuum expectation $v$, 
and the Goldstones $\theta_a$,
\bee
\left( \begin{array}{c} \phi_+ \\  \phi_0 \end{array}\right) = \exp(i \theta_a \tau_a) \left( \begin{array}{c} 0 \\ v+H\end{array}\right) .
\ee
An $SU(2)_L$ transformation just shifts the $\theta$'s by a constant. This implies that the Lagrangian can have no term
proportional to $\theta_a^2$.
\end{itemize}
Most New Physics constructions attempt to exploit these general observations. 
For example, supersymmetry protects the scalars by putting them in multiplets with chiral fermions.
In some higher dimensional completions of the Standard Model, the scalars are extra gauge degrees of freedom.
In some composite Higgs models, the Higgs begins life as a Goldstone boson.

Why do some people imagine that some of the new physics is nonperturbative?
 I do not have a good answer. It
is quite a contrast with the history
of the strong interactions: people have known that nuclear forces were strong from the
moment that they knew that nuclei were composite. After all, something had to overcome the protons'
Coulomb repulsion to hold the nucleus together. Here, there is no obvious need for a new strong force.

When I ask people why new physics is nonperturbative, the conversations all seem to come back to the 
hierarchy problem.
In principle, strong interactions could solve it.
Suppose that the Higgs is a bound state
of some new fermion and anti-fermion,
 with new gauge interactions that are asymptotically free.
(Clearly this is a special case of our first two special cases; we need to have gauge fields
 and massless fermions in our more-fundamental theory -- gauge fields to confine and massless
 fermions so that their
bound states are light.) Asymptotic freedom means that the
effective interaction between the fermions
 grows as one moves to lower energy scales and becomes strong at some
 scale $Q$, where it confines the
fermions into bound states.
This is precisely what happens in QCD. Perhaps it is more general.
The  quadratic dependence on $\Lambda$ of Eq.~\ref{eq:hier} is transformed into something smoother.
This result comes from the running of the coupling constant from a value $g^2(\Lambda)$ at the cutoff scale, down to scale $Q$:
\bee
\frac{1}{g^2(Q)} = \frac{1}{g^2(\Lambda)} + c \log \frac{Q}{\Lambda}.
\label{eq:AFsol}
\ee
As we run into the infrared,
the coupling grows.  Suppose $Q$ is the scale where confinement  and chiral symmetry
breaking is triggered
in some unknown way. Masses take their values around this scale. Setting the left hand side of
Eq.~\ref{eq:AFsol} to zero, and setting $Q=M_H$, we have
\bee
M_H \sim \Lambda \exp \left( -\frac{1}{cg^2(\Lambda)}\right).
\label{eq:AFrunning}
\ee
The fine tuning of the Higgs system is replaced
(hopefully) by some less-fine tuning; we only need to have a weakly-interacting
system at some high cutoff scale, and could take that scale to 
infinity while simultaneously tuning the bare coupling
to zero. All masses (except, of course, those of the Goldstone bosons)
 would have the same overall scale; all would be roughly around the scale
where the dynamics became strong. This is what happens in QCD.

Along the way, these  new states are ready for discovery at the Large Hadron Collider and hence
exciting for experimentalists.

Having raised the possibility of nonperturbative new physics,
we are back to the lattice.
It is easy to understand why one might want to apply lattice QCD methods to the study
of nonperturbative beyond Standard Model candidate models. For many candidates, the field content is similar
to QCD: gauge fields and fermions.  Asymptotic freedom is a QCD-like feature.
If a model were known to be confining  it would have a rich
spectrum of hadron-like states.
A chirally-broken model would have Goldstone bosons ready to be eaten by the W and Z.
The physical Higgs would be the analog of the sigma meson.
The pseudoscalar decay constant could be used to set the scale for all this new physics,
and might be tied to the scale of electroweak symmetry breaking.
From their experience with QCD, lattice practitioners might have all the tools to compute the
 masses, decay constants, and
other low energy constants associated with some specific new physics scenario, starting from the Lagrangian.

Perhaps, after this long general introduction, it is time to turn to specifics.
But I have to make one more introduction, to set the stage. Lattice models are
 typically built of gauge fields and fermions. Depending on the gauge group and fermion content,
there is a naive expectation for the vacuum structure of these systems, given by
the renormalization group. It is useful to pause and, in Sec.~\ref{sec:rg1},
  remind ourselves of this physics. Then,
Sec.~\ref{sec:catalog} is a ``review within a review,'' a set of thumbnail sketches of
the many beyond Standard Model systems that have been the targets of lattice investigation.
The range of topics in this section is so broad that it is almost impossible to describe coherently.
I then review lattice methodology, with attention to issues which arise in the context of beyond Standard Model candidates.
This is done in  Sec.~\ref{sec:lattice_meth}.
Most lattice work has involved systems with slowly running couplings.
The rest of the review treats this special case in detail. 
In Sec.~\ref{sec:vertical} I describe lattice methods used to study slowly-running systems.
The division of subject is by method, rather than by specific model. This allows me to illustrate
how the generic features of slowly-running systems reveal themselves to lattice probes.
Finally, in Sec.~\ref{sec:results} I describe the status of particular model systems.
A few tentative conclusions are presented in Sec.~\ref{sec:summary}.

I should finish the introduction with a few caveats: First, I cut off
the literature search on 1 April 2015, but this is an active area of research, and I expect that many things I say will
become obsolete. Next, the subject of beyond Standard Model physics is vast.
No one could cover it all in a review. I have tried to highlight places where there are lattice stories.
And finally, much of the lattice literature appears as short, unrefereed, and often preliminary contributions to the annual International
Symposium on Lattice Field Theory series of meetings. I have tried to avoid referring to these articles when
a longer, refereed publication is available.

%%%%%%%%%%%%%%%%%%%%%%%%%%%%%%%%%%%%%%%%%%%%%%%%%%%%%%%%%%%
\section{A pause for context: formulas from the renormalization group \label{sec:rg1}}
%%%%%%%%%%%%%%%%%%%%%%%%%%%%%%%%%%%%%%%%%%%%%%%%%%%%%%%%%%%%%%%%%%%%%
Imagine that we have an $SU(N_c)$ gauge theory with $N_f$ flavors of massless Dirac fermions in representation $R$.
 The gauge coupling is scale dependent.
At two loops, the beta function  is  \cite{Caswell:1974gg,Jones:1974mm}
\bee
\beta(g^2)=\frac{dg^2}{d\log \mu^2}=-\frac{b_1}{16\pi^2}g^4-\frac{b_2}{(16\pi^2)^2}g^6+\cdots,
\label{2loopbeta}
\ee
where
\beea
b_1&=&  \frac{11}{3}\, C_2(G) - \frac{4}{3}\,N_f T(R) \label{2loopbeta1}\\
b_2&=& \frac{34}{3}\, [C_2(G)]^2
  -N_f T(R) \left[\frac{20}{3}\, C_2(G) %\right.\nonumber\\&&
  + 4 C_2(R) \right] .
\label{eq:2loopbeta2}
\eea
Here $C_2(R)$ is the value of the quadratic Casimir operator in representation $R$
($G$ denotes the adjoint representation, so $C_2(G)=N_c$), while $T(R)$ is the
 conventional trace normalization. $\mu$ is a momentum scale. 
 
When the number of fermionic degrees of freedom, basically parametrized by $N_f T(R)$, is small,  both $b_1$
and $b_2$ are positive (in my conventions). The beta function has a zero, a fixed point,
called the Gaussian fixed point,
at $g^2=0$. The fixed point is infrared unstable; the coupling increases  as $\mu$ decreases, ``under flow into the infrared.''
It is thought that in this case the coupling increases without bound under flow to the infrared,
and it is further presumed that this implies that the vacuum is confining and chirally broken.
Examples of such systems are QCD and its near relatives.

When the number of fermionic degrees of freedom is sufficiently large, $b_1$ changes sign. The scale dependent coupling
falls to zero in the infrared.
The Gaussian fixed point becomes infrared stable. At long distances the system is believed to
be non-interacting, or ``trivial,'' similar to (for example) $\phi^4$ theory in dimension $D \ge 4$.

At an intermediate number of fermionic degrees of freedom, it could happen that
  $b_1>0$ and $b_2<0$. The system would have
an infrared attractive fixed point (IRFP) where the beta function vanishes,
 $\beta(g_f^2)=0$. This is often called a ``\textcite{Banks:1981nn}  fixed point.''
 Under a change of scale from the ultraviolet to the infrared,
the gauge coupling would flow into the fixed point and remain there.
 In the particle physics literature this is often referred to as
 ``conformal''  or ``infrared conformal'' behavior.
We speak of the ``conformal window'' as the values of  $N_f$, for a given representation, for
which the system is neither
confining nor trivial. 

Inside the conformal window, all correlation functions show a power law behavior
at long distances. There are no intrinsic mass gaps, hence no particles.
Chiral symmetry is unbroken. This is the analog of the familiar case of a statistical
system at a second order critical point. This is certainly nothing like we see in electroweak Nature.
Thus, candidate beyond Standard Model theories must, generally,  not be inside the conformal window.

Of course, unless the zero of the beta function is at a very small value of $g^2$, it is
 unlikely that 
the first two terms in the beta function would show it. Why should the higher order terms be small?
The smaller the number of fermion degrees of freedom, the larger the value of $g^2$ becomes,
at a place where the beta function vanishes, and the more uncontrolled would be a 
perturbative calculation. And the beta function is only scheme independent through two loops.
To make sense of the story we are trying to tell requires recasting it in the more general language
of the renormalization group, 
outside the narrow statements of perturbation theory, in terms of relevant and irrelevant operators.
 The investigation of the system
would need a better set of tools,
 perhaps associated with a lattice calculation.

Let us return to that point later. For the time being, just carry the thought: a system might
 have a quickly running coupling constant, or a slowly running one.

%%%%%%%%%%%%%%%%%%%%%%%%%%%%%%%%%%%%%%%%%%%%%%%%%%%%%%%%%%%%%%%%%%%%%
\section{The landscape of models with lattice investigations\label{sec:catalog}}
%%%%%%%%%%%%%%%%%%%%%%%%%%%%%%%%%%%%%%%%%%%%%%%%%%%%%%%%%%%%%%%%%%%%%

Let me briefly summarize the particular scenarios for beyond Standard Model physics
 that have either seen,
or might see, lattice studies.
I have rewritten this section multiple times, trying to describe them in some
 kind of coherent order.
I do not think I have succeeded in doing this. But I think the problem is that to ask for
 coherence is impossible.
There are many unrelated possibilities for physics beyond the Standard Model.
Instead, what I will do is start with QCD, and then move increasingly farther away from it.

\subsection{QCD}
A large fraction of lattice QCD literature has beyond Standard Model physics as its back story.
The rate for any hadronic weak interaction process -- or for some process driven by new physics --
typically involves a hadronic matrix element of some operator. These matrix elements are computed on the lattice.
This subject is  huge; for example,
\textcite{Aoki:2013ldr} is a 179 page  review of it.
Most of these tests are associated with the flavor structure of the Standard Model,
either checking Eq.~\ref{eq:clm} or looking for modifications to it.
 The hope, of course, is that the Standard Model rate will
show some disagreement with low energy experiment, so revealing the need  for new physics.

Some lattice QCD calculations make direct contact with Higgs physics.
\textcite{Lepage:2014fla} recently emphasized the importance of good quality
 measurements of the strong coupling
constant and of the charm and bottom quark masses on precision measurements of the Higgs width.

\subsection{Slightly beyond QCD}

A small amount of lattice work has been devoted to systems that are believed
to be like QCD.  These are systems that are almost certainly
 confining and chirally broken. Examples of these systems are $SU(N_c)$ gauge theories
with $N_c>3$ and a small number of fermionic degrees of freedom.
 Usually the physics issues discussed in the literature of these systems
  are related to QCD rather than beyond  Standard Model
dynamics. For example, most of the qualitative knowledge about QCD we have comes
from the large-$N_c$ expansion of  \textcite{'tHooft:1973jz}.
This knowledge can be -- and is being -- tested by lattice simulation.

\begin{figure}
\begin{center}
\includegraphics[width=\columnwidth]{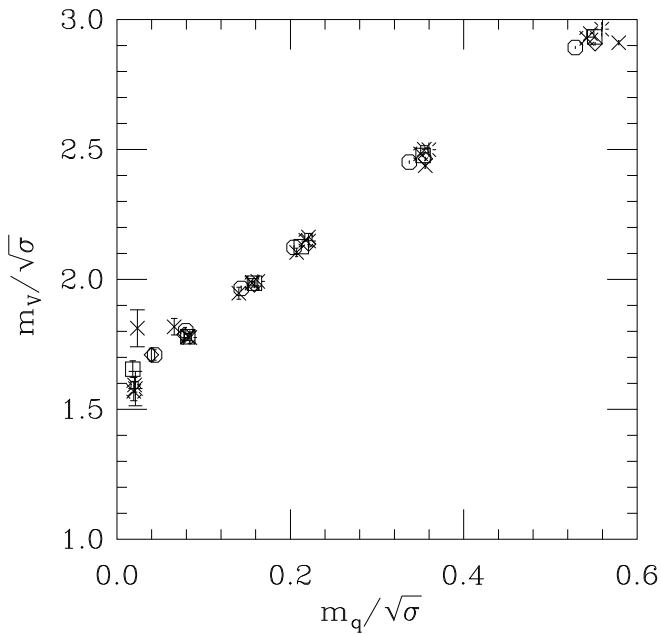}
\caption{The vector meson spectrum versus quark mass for different $N_c$,  using data 
from \protect{\textcite{Bali:2013kia}}. The data points
are
crosses for $N_c=2$,
diamonds for $N_c=3$,
octagons for $N_c=4$,
squares for $N_c=5$,
fancy crosses for $N_c=6$,
fancy squares for $N_c=7$,
and
bursts for $N_c=17$,
  }
\label{fig:vectorbali}
\end{center}
\end{figure}

\textcite{Lucini:2012gg} gave a recent review of work on large-$N_c$ QCD.
Studies of physical systems with a particle content similar to QCD include
the familiar large-$N_c$
 limit of `t Hooft (where the fermions are in the fundamental representation and $N_f$ is held fixed and small),
   or variants such as
 $SU(N_c)$ gauge theories coupled to a small number of fermions, not in the
 fundamental representation.
The situation with these systems is quite simple to state: large $N_c$ scaling works quite well.
A nice example of a comparison, from \textcite{Bali:2013kia}, is shown in Fig.~\ref{fig:vectorbali}.
This is a plot of the vector meson mass versus the quark mass, both scaled
in units of the square root of the string tension. They have many more examples.
Large $N_c$ scaling predicts that
meson masses show little dependence on $N_c$. Decay constants scale as $\sqrt{N_c}$
(for  fundamental representation fermions; the scaling is as $N_c$ for
two-index representation fermions).
At least for $N_c=3$, and small $N_f=2-3$,
 the $N_f$ dependence of masses and matrix elements is small, according to \textcite{Aoki:2013ldr}.

These studies have a role in beyond Standard Model applications. Often, one sees large-$N_c$ arguments
quoted in general discussions of composite Higgs systems.
For example, one might have a new physics scenario with $SU(N_c)$ gauge fields.
One might be interested in the ratio of scale of the masses of excitations  to the
size of chiral symmetry breaking. This might be  parametrized by, say, the pseudoscalar decay constant $f_\pi$,
which, in turn, might be related to some electroweak parameter, such as the Higgs vacuum expectation value.
Often, phenomenological papers
just take the known (QCD, real world) numbers and scale them appropriately.
This is, of course, just an assumption.
These days, one could do a simulation to get the ratio directly. Then, absent some direct experimental
measurement, a comparison to large-$N_c$ counting
is an appropriate way to put the lattice number into some bigger context.

\subsection{Models with slowly running couplings and the Higgs as a bound state\label{sec:tc}}
The oldest scenarios in which the Higgs is not a fundamental scalar date back to
\textcite{Susskind:1978ms,Weinberg:1979bn}. The physics idea for generating gauge boson masses is elegant:
 Imagine having  some new gauge fields coupled to massless fermions, whose
dynamics is confining and breaks  chiral symmetry. (The original model assumed a doublet of quarks, like
in QCD, so the pattern of chiral symmetry breaking was $SU(2)\times SU(2) \rightarrow SU(2)$.)
 There will be a set of massless Goldstone bosons.
The spontaneously broken symmetry is associated with axial vector currents
\bee
J^{\mu 5 a} = \bar q \gamma^\mu \gamma^5 \tau^a q
\ee
and the matrix element of the axial vector current can be parametrized as
\bee
\langle 0 | J^{\mu 5 a} |\pi^b\rangle = i f_\pi p^\mu \delta^{ab}
\ee
where in QCD $f_\pi$ is the pseudoscalar decay constant.

The spontaneous symmetry breaking mechanism that gives mass to the $W$ and $Z$ when they
``eat'' a Goldstone boson  does not require that the eaten Goldstone must be fundamental.
Any coupling ($L$ for left-handed, now)
\bee
\cl_{int}= g J^{\mu L a} A_\mu^a
\ee
will induce a mixing of the gauge boson with the Goldstone through the generation of
a term in the vacuum polarization tensor
\bee
\Pi^{ab}_{\mu\nu} = (\frac{g f_\pi}{2})^2[g_{\mu\nu} - \frac{p_\mu p_\nu}{p^2} ]\delta^{ab} .
\ee
This is a mass for the vector meson, $m_W= g f_\pi/2$. The vacuum expectation 
value of the usual Higgs field $v$ in the formula for the gauge boson's mass
is replaced by the pseudoscalar decay constant, $f_\pi$. To use this dynamics to generate
the Standard Model result, we then need to assume that the new dynamics
naturally generates a scale $f_\pi=246$ GeV. More complicated models would scale this equality by
an order - unity numerical value.  Such a model is referred to, generically, as ``Technicolor.''
The new fermions are called ``techniquarks.''
This subject has an enormous literature. For a review of it, see \textcite{Hill:2002ap}.

The Higgs also generates fermion masses.  This is generally awkward to achieve with technicolor.
In technicolor models there are just bound states of techniquarks.
To generate masses for the quarks and leptons,  they must be coupled somehow to the techniquarks.
This is commonly done by introducing a new level of dynamics, at some much higher scale, called 
 Extended Technicolor (ETC) interactions \cite{Dimopoulos:1979es,Eichten:1979ah}.
If the ETC gauge bosons are very heavy, they induce four fermion interactions
(here, between two capital letter techniquarks and two lower case ordinary Standard Model fermions)
\bee
\cl_4 = (g_E \bar u_L \gamma_\mu U_L)\frac{-i}{m_{ETC}^2}(g_E \bar U_R \gamma_\mu u_R).
\ee
Now let the ETC fermions condense. Replacing $U_L \bar U_R$ by its vacuum expectation value $\Sigma$,
we generate a fermion mass
\bee
m_u = \frac{g_E^2}{m_{ETC}^2} \Sigma .
\label{eq:tc1}
\ee
One can make a guess at the scale $m_{ETC}$ using the known quark masses. It must be on
the order of  1 - 100 TeV.

This mechanism
 has a number of phenomenological issues. The first one is that the same interactions that
couple two quarks to two techniquarks should also couple four quarks together. This
is a problem, because such interactions must be very weak; they give rise to flavor changing
neutral currents, which are too large to be consistent with observation
 if the quarks are to acquire their observed masses.
A  potential resolution of this problem is called ``walking technicolor''
\cite{Holdom:1981rm,Holdom:1984sk,Yamawaki:1985zg,Appelquist:1986tr,Appelquist:1987fc}.
  To describe it, we have to
 rewrite Eq.~\ref{eq:tc1} more carefully.

Fermion masses arise from physics at the ETC scale. Labeling this scale as  $\Lambda_{ETC}$,
a fermion mass is
\bee
m = \frac{ \svev{U_L \bar U_R}_{ETC}}{\Lambda^2_{ETC}}.
\label{eq:metc}
\ee
The flavor changing neutral current term is also ETC scale physics,
\bee
\cl_{FCNC} \sim \frac{ (\bar s \gamma_5 d)(\bar s\gamma_5 d)}{\Lambda^2_{ETC}} .
\ee
In Eq.~\ref{eq:metc} the condensate is scale dependent.
 $\svev{U_L \bar U_R}_{TC}$ ought to be  a typical electroweak size, say about $v^3$.
 Its value at the ETC scale is related
to its value at the TC scale by renormalization group running,
\bee
\svev{U_L \bar U_R}_{ETC} =
\svev{U_L \bar U_R}_{TC} \exp \left( \int_{\Lambda_{TC}}^{\Lambda_{ETC}} \gamma_m(g_{TC}(\mu)) \frac{d\mu}{\mu} \right)
\label{eq:rung}
\ee
where $\gamma_m$ is the anomalous dimension of the technifermion mass operator.
If the gauge coupling runs very slowly as the energy scale drops from the high ETC scale to the low TC (or electroweak) scale, 
then $\gamma_m$ does not change much either, and
the soft running expected for a typical QCD-like theory is replaced by 
a power law.
We have
\bee
\svev{U_L \bar U_R}_{ETC} = \left( \frac{\Lambda_{ETC} }{\Lambda_{TC}} \right)^{\gamma_m}
\svev{U_L \bar U_R}_{TC}.
\ee
Slow running is, of course, ``walking.''
Finally, if  $\gamma_m$ is large at the values of $g^2$'s that run slowly,
 one might be able
to have one's cake (generate phenomenologically viable fermion masses) and eat
 it, too (make $\Lambda_{ETC}$
large enough to suppress flavor-changing neutral currents).

So many ``if's''. But
 the situation for the lattice simulator is pretty well laid out: Does
a candidate theory exhibit walking? Is it confining and chirally broken? If so, is
its mass anomalous dimension large? If the answer to all these questions is Yes,
 the perhaps it is a viable technicolor candidate.
What is its spectrum and what are its low energy constants?

Technicolor candidates would lie in the confining phase, but very close to the conformal window.
To search for them, a first task  might be to try to map out the boundary between
confining and chirally broken theories, and ones in the conformal window. The relevant
parameters are of course the number of colors and the number of flavors of fermions and their representations.
Two-loop perturbation theory might be suspect.
Higher order terms for the beta and gamma functions have been computed, in $\overline{MS}$ scheme.
\textcite{Ryttov:2010iz,Pica:2010xq}
have used these results to explore the location and properties of the IRFP.
My impression of these results is that when the fixed point coupling becomes strong, perturbative
predictions for the location of a fixed point and of the value of the critical exponents at the fixed point
are  not 
particularly stable.

\begin{figure}
\begin{center}
\includegraphics[width=0.5\textwidth,clip]{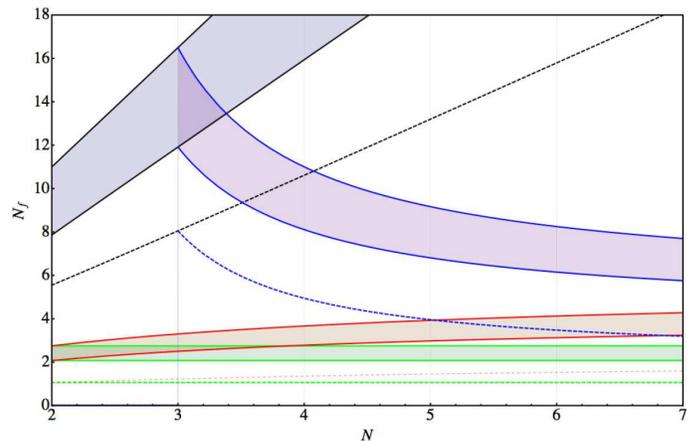}
\end{center}
\caption{
Conjectured phase diagram (due to {\protect{\textcite{Dietrich:2006cm}}})
 for non-supersymmetric theories with fermions
in the: i) fundamental representation (blue), ii) two-index
antisymmetric representation (purple), iii) two-index symmetric
representation (red), iv) adjoint representation (green) as a
function of the number of flavors $N_f$ and the number of colors $N$. The
shaded areas depict the corresponding conformal windows from their calculation.
The dashed curve represents the change of sign in the second
coefficient of the beta function.  }
\label{fig:ds}
\end{figure}

\textcite{Dietrich:2006cm} combined one-loop running with expectations from solving 
Schwinger-Dyson relations, to make a map of the $N_c - N_f$ plane for various representations of fermions.
 Figure \ref{fig:ds} shows  their prediction for a phase diagram. This figure has served
as the target for many lattice calculations.

%%%%%%%%%%%%%%%%%%%%%%%%%%%%%%%%%%%%%%%%%%%%%%%%%%%%%%%%%%%%%%%%%%%%%
\begin{figure}
\begin{center}
\includegraphics[width=0.4\textwidth,clip]{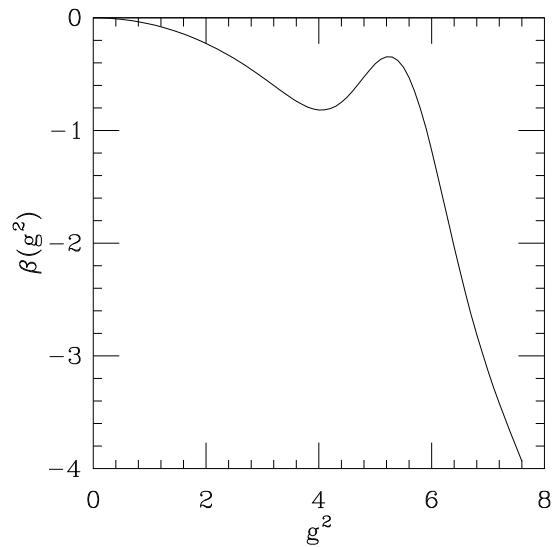}
\end{center}
\caption{Artist's conception of a beta function for a walking theory -- a negative beta function that
approaches the origin, then turns away.
\label{fig:walkingdream}}
\end{figure}

A cartoon of the expected coupling constant evolution of a walking theory
is shown in Fig.~\ref{fig:walkingdream}.
The beta function starts out negative, then bends toward zero.
 Walking occurs at the coupling where the beta function is smallest. At the inflection point, something
must make the beta function bend over steeply.  What could that be?
One possibility  \cite{Appelquist:1996dq,Miransky:1996pd}  is that chiral symmetry breaking occurs at a
 coupling near the cusp. Some
of the fermions condense into colorless pions and decouple from the gauge bosons, reducing the
effective number of fermionic degrees of freedom and letting the coupling grow.
Self-consistent (Schwinger-Dyson) calculations support this scenario, but can they be trusted?

One can imagine theories whose beta function looks like Fig.~\ref{fig:walkingdream}.
Several toy models of walking theories have  been proposed
\cite{Nogradi:2012dj,Aoki:2014yra}. Another way to turn the beta function over might
involve introducing extra  external scales. (This is different from the technicolor scenario,
where the scale of the turning appears dynamically.)
The simplest possibility is  a system with  many flavors of massive fermions.
At momentum scales that are much greater than the
 fermion masses, the fermions behave as if they are massless. 
But as the energy scale falls below the fermion masses,  they decouple from the gauge fields.
The effective number of fermions in the beta function changes. One might
set the number of flavors large enough,
 that at  sufficiently high energy, the coupling might be arranged
to show an IR flow toward a fixed point. As the energy scale drops, the fermions decouple, the coupling runs 
differently (faster) and the true long distance behavior would cross over to some strongly coupled 
theory -- almost certainly confining and chirally broken. Perhaps parameters could be tuned to
produce walking. 
(See \textcite{Brower:2014ita} for a recent  study of this.)

(This description might be too poetic. Recall \cite{Rodrigo:1993hc} that
in $\overline{MS}$ schemes, where coupling constants are mass-independent, 
one has to  treat theories with different $N_f$'s as effective field theories and match  the running
couplings at a scale $\mu$ equal to the fermion mass. With an $n$th order beta function, this has to be done
at order $O(n-1)$. The coupling constant steps discontinuously at thresholds, rather than showing a smooth
behavior like Fig.~\ref{fig:walkingdream}.)

A more serious issue with Fig.~\ref{fig:walkingdream} is that it suggests that the physics
 of strong coupling is described by a single coupling constant. This may not be the case.
In QCD, for example, the coupling constant monotonically strengthens as the momentum scale falls.
At long distances, the coupling constant loses its utility as a useful quantity, in the sense that
one cannot use it to parametrize calculations of interesting observables.

Technicolor has additional issues. Most of them are difficult to quantify because the
 dynamics of technicolor is strongly interacting.
 
 The range of quark masses is wide, from a few MeV for the
up and down quarks, to 173 GeV for the top quark. Complicated constructions seem to be needed to
generate all these masses.

The technipions might be eaten by the W and Z, but where are
the other particles, the technirho and beyond? One typically imagines that the scale of non-chiral
physics is about $4\pi f$. That value is in the range of LHC searches, and, so far, they have not been seen.
(\textcite{Halkiadakis:2014qda}  has limits.)

 Technicolor has issues with precision 
electroweak measurements. Usually, the $S-$ parameter is too large.
However, is it really possible to compute the technicolor vacuum polarization
contribution to the gauge bosons in a reliable way? The literature often falls back on
analog QCD calculations, suitably rescaled. [See, for example, the discussion in
\textcite{Peskin:1991sw,Contino:2010rs}.]

In technicolor models, electroweak symmetry breaking does not involve a Higgs boson. If technicolor dynamics were
sufficiently QCD-like, one would expect to see a scalar state in the spectrum, in analogy with the
situation in QCD.  There, the scalar state is the $f_0(500)$,
 a light ($M\sim 400-550$ MeV) broad ($\Gamma\sim 400-700$ MeV)
resonance. The observed Higgs is narrow, so this is an issue for technicolor phenomenology.
But, if the candidate theory is not very QCD-like,  can QCD analogies be trusted?
All of these questions could be addressed by lattice simulations.

I think that technicolor is the only beyond Standard Model scenario that
 has enough of a lattice literature
 to justify a detailed review. 
 The subject turned out to be filled with  surprises.
 After completing this survey section, I will return 
to a detailed discussion of results for these systems.

\subsection{Composite Higgs: the Higgs as a pseudo - Nambu - Goldstone boson\label{sec:partialcomp}}
In Sec.~\ref{sec:tc} electroweak symmetry breaking  occurs when the techniquarks
 form a condensate which
transforms non-trivially under $SU(2)\times U(1)$. In such models the condensate scale is
the weak scale, around 246 GeV.
Another alternative is to arrange that the new physics generates a condensate,
 but the condensate preserves
$SU(2)\times U(1)$. Then, electroweak symmetry breaking could occur at a scale
which is much lower than the condensate scale.  (From our point of view, the new physics is at a higher
scale than the electroweak symmetry breaking scale.)
 A scalar excitation present at the
high scale would develop a vacuum expectation value, and some of its degrees of
freedom would eaten by the $W$ and $Z$.

The earliest discussions of this approach go back to
\textcite{Kaplan:1983fs,Kaplan:1983sm,Georgi:1984ef,Banks:1984gj,Georgi:1984af,Dugan:1984hq,Georgi:1985hf}.
 The idea was to make the Higgs a Goldstone boson corresponding
to a spontaneously broken global symmetry of a new strongly interacting sector.
Gauge and Yukawa interactions of the Higgs explicitly violate the global symmetry and generate
a potential, including a mass term, for the Higgs.

Let us approach the issue in a top-down way: We have some new dynamics at a scale $\Lambda$.
It encodes a global symmetry $G$ which is spontaneously broken to a subgroup $H$,
and the Goldstones are described by a $G/H$ nonlinear sigma model.
Some of them are destined to become the real Higgs doublet.
At this point,  all of the components are massless.
They have  non-renormalizable interactions parametrized by
a scale $f$, where $4\pi f$ is presumed to roughly equal $\Lambda$
(the usual connection between chiral and non-chiral dynamics).

Electroweak interactions are introduced
 by gauging an $SU(2)\times U(1)$ subgroup of $G$.
Electroweak gauge interactions, and the interactions of Standard Model fermions with
the Goldstones, explicitly break the shift symmetry.
They generate a potential for the Goldstones which has a nontrivial minimum.
The task of the model builder is to do this without re-introducing quadratic divergences along the lines of
 Eq.~\ref{eq:hier}.
The literature refers to these models as  ``composite Higgs models.''

Avoiding a mass shift like  Eq.~\ref{eq:hier} is a nontrivial task. One mechanism which
could succeed in principle was the
idea of collective symmetry breaking \cite{ArkaniHamed:2002qy,ArkaniHamed:2002qx}.
The resulting systems are called ``little Higgs'' systems. In these models,
electroweak interactions are introduced
by gauging a subgroup which is a direct product of several
factors, $G_1\times G_2 \times \dots$ in such a way that each $G_i$ commutes with a subgroup of $G$
that acts non-linearly on the Higgs. This means that if any one $G_i$ is gauged, the
unbroken global symmetry insures that the Higgs remains massless. Only when the full
product of $G_i$'s is  gauged does the Higgs cease to be a Goldstone boson.
The consequence of this dynamics is that the induced mass of the Higgs is proportional to a product
of all the gauge coupling constants corresponding to the different $G_i$ factors.
The terms in Eq.~\ref{eq:hier} are not of this form and so they are absent.

Most of the literature of composite Higgs models confines itself to the low energy effective theory
of the would-be Goldstones. Explicit examples of such actions may be found,
for example, in \textcite{Giudice:2007fh,Contino:2010mh,Azatov:2012bz,Contino:2013kra,Buchalla:2014eca}.
General surveys, such as \textcite{Bellazzini:2014yua}, organize their discussion in terms
of the ratio $\xi$  of two dimensionful parameters, $f$, as described above,
the scale of the nonlinear sigma model, and the Higgs vacuum expectation value
$v=246$ GeV, $\xi=v^2/f^2$.

In partial compositeness scenarios, the top quark and the Higgs share a common dynamics.
One such scenario is due to \textcite{Kaplan:1991dc}:
the top quark couples linearly to strong-sector baryons, which, in effect, allows
 it to couple to the composite Higgs as well.
This scenario is used to generate fermion masses. Here the
interaction involves the coupling of a Standard Model fermion $\psi$ to a composite operator $O$,
\bee
{\cal L}= \lambda \bar \psi O + h.c.
\label{eq:partialcomp}
\ee
In the original \textcite{Kaplan:1991dc} version of this idea, the
composite operator $O$ is a three-quark technibaryon so $\cal L$ is a four-fermi interaction,
whose origin is (perhaps)  some ETC theory at a yet higher scale. Diagonalizing the 
resulting mass matrix gives states which are linear superpositions
of the fermion and the technibaryon -- hence the phrase ``partial compositeness.''
The mixing of a Standard Model fermion with a composite
 is also used to generate part of the effective potential for the Goldstones.

This is a huge field. However, unlike the systems described in
 Sec.~\ref{sec:tc}, it has a tiny lattice
literature. Why that is so I will come back to, below. What it means is that this section
has a different orientation from the rest of the review. There, the  story
is ``Here is some
physics; here is what lattice simulations showed,'' and the intended audience is (mostly)
physicists who
did not do the simulations. For this section, my goal is to try to convince lattice
physicists that there
are interesting issues which can be addressed on the lattice.

A complete technical analysis of the issues facing a lattice calculation remains to be written.
Here is my attempt at an overview.

First, the Standard Model gauge group must be a subgroup of the unbroken group $H$, otherwise
the Standard Model gauge fields would develop masses on the scale of $g \Lambda$.
Next, phenomenology needs to know the
couplings in an effective Lagrangian. Given a specific choice
of an ultraviolet completion of a composite Higgs model, the situation might be exactly like QCD:
Introduce $\Sigma$, the nonlinearly-realized field, $\Sigma \sim \exp(-i \tau^a \pi^a/f)$
 (for generic generators $\tau^a$, Goldstone
fields $\pi^a$ and decay constant $f$).
The goal of a lattice calculation would be to start with the ultraviolet completion and
compute the
effective potential of the nonlinear sigma model, $V_{eff}(\Sigma)$.
It is necessary to specify the electroweak quantum numbers of the fields in $\Sigma$.
For a viable model, four of them have to self-assemble into a complex $SU(2)$ doublet, the Higgs.
Typically there will be members of other $SU(2)$ multiplets. The members of other nonsinglet
irreducible representations should not condense.
(For lattice QCD practitioners, it is standard to assume that the vacuum can be rotated into the identity in flavor space.)

Generically,  $V_{eff}(\Sigma)$ receives contributions from the Standard Model gauge bosons and fermions.
The gauge boson part comes  from the part of the
lowest order chiral Lagrangian which is quadratic
in the gauge fields, from
\bee
{\cal L} = \frac{f^2}{4} \Tr |D_\mu \Sigma|^2
\ee
where $D_\mu$ is a covariant derivative. The Lagrangian includes
  a $g^2 W^2 \pi^2$ vertex, which in turn, generates a quadratically
divergent contribution to the potential. Typically, though, the energy scale is $f$, not $\Lambda$:
\bee
V_{eff} =  c\frac{g^2 f^2}{16\pi^2} \pi^2 + \dots.
\ee

The constant $c$ is calculable on the lattice.
The procedure is similar to that  for the $\pi^+-\pi^0$ mass difference in QCD,
via the Das, Guralnik, Mathur, Low, Young sum rule \cite{Das:1967it}. It uses the
 the difference of the vector and axial current correlators
\beea
\Pi_{\mu\nu}(q) &=& \int {d^4 q} \exp(iqx) \langle J_\mu^L(x) J_\nu^R(0)\rangle \nonumber \\
   &\equiv & (q^2 \delta_{\mu\nu} - q_\mu q_\nu)\Pi_T^{LR}(q^2) + q_\mu q_\nu\Pi_L^{LR}(q^2) . \nonumber \\
\label{eq:continPi}
\eea
$J_\mu^L$ and $J_\mu^R$ are the left and right currents $\bar \psi \gamma_\mu(1\pm \gamma_5)\psi$,
so the object in the integral is the appropriate difference.
The coefficient $c$ is proportional to the integral
\bee
c \sim \int_0^\infty dq^2 q^2 \Pi_T^{LR}(q^2) .
\ee
Several lattice groups \cite{Shintani:2008qe,Boyle:2009xi,Appelquist:2010xv}
 have published calculations of the  $\pi^+-\pi^0$ mass difference using this
observable. \textcite{Contino:2010rs}
 gives a pretty explicit description of what to do for a composite Higgs model.

The sign of $c$ can be inferred  in advance, without the need for a lattice calculation.
 This is the phenomenon of vacuum alignment, first described by
 \textcite{Peskin:1980gc,Preskill:1980mz}, and related to
 Witten's inequality \cite{Witten:1983ut}. The contribution of gauge bosons is positive,
so the gauge symmetry remains unbroken. Something else must break it.
But sometimes, one can use the calculation
as an estimate for the masses given by electroweak symmetry breaking to  the uneaten, now-pseudo Goldstones.

Typically, the negative term in $V_{eff}$ comes from the fermions.
Models vary in details, but many involve partial compositeness: the Standard Model fermions mix with
new physics baryons, which in turn can couple to the Higgs.

In  published models, the Yukawa couplings are numbers and so the derived value of $v$ depends on them.
I am not sure if their actual values are accessible to a lattice calculation, or not.
However, they are running couplings, and their anomalous dimensions
are related to those of the technibaryon operators.
For example, \textcite{Contino:2010rs} rewrites Eq.~\ref{eq:partialcomp} as
\bee
 {\cal L} = \sum_n \lambda \svev{0| O|\chi_n} \bar q \chi_n + h.c.
\ee
introducing a tower of composite fermions $\chi_n$.
 To the lattice practitioner $\svev{0| O|\chi_n}$  is just
a baryon creation amplitude. Lattice techniques could be adapted to find its anomalous dimension.
There is a recent discussion by \textcite{Golterman:2015zwa}
 of lattice issues involved in computing partial compositeness observables. The subject needs more theoretical analysis.

Interesting ultraviolet completions require QCD - like theories with different
numbers of colors, or quarks in non-fundamental representations, or both.
The Littlest Higgs model \cite{ArkaniHamed:2002qy} relies on the non-linear
sigma model $SU(5)/SO(5)$.  A possible ultraviolet completion
is any confining gauge theory with five Majorana fermions in some real
representation.  The most economical way to realize this scenario
is an $SU(4)$ gauge theory, where the two-index antisymmetric representation (AS2) is real.
The $SU(5)/SO(5)$ sigma model is also central to the more recent
composite-Higgs models of
\textcite{Vecchi:2013bja,Ferretti:2014qta,Ferretti:2013kya}.
In particular, \textcite{Ferretti:2013kya} makes the case why the
$SU(4)$ theory with AS2 fermions is the most attractive candidate
within this approach, whereas  \textcite{Ferretti:2014qta}
elaborates on the phenomenology of this composite-Higgs model.
The models of \textcite{Ferretti:2014qta,Ferretti:2013kya}
require fermions in the fundamental representation in addition
to the AS2 ones, in order to give the top quark a mass via partial compositeness.

Another ultraviolet completion is \textcite{Barnard:2013zea},
with an $Sp(2N)$ gauge group and two representations of fermions. The global symmetry breaking pattern
is $SO(6)/SO(5)$.

The pattern of chiral
symmetry breaking can be different from QCD.
When the fermions in the ultraviolet completion are Dirac fermions in a complex representation,
 parity and charge conjugation are
good symmetries, and the Goldstone bosons associated with chiral symmetry breaking are all
 pseudoscalars.
The Higgs is a scalar, so there is apparently no way it can be a Goldstone boson. However,
 the fermions associated with the new dynamics
could belong to a real or to a pseudoreal representation. Then there is no a-priori distinction
 between a scalar bilinear or a pseudoscalar one.
The quantum numbers will be determined after the fact when the Standard Model quantum numbers of
 the appropriate fields are assigned.
For consistency, the condensate must be a scalar.

The situation
was first described by 
 \textcite{Peskin:1980gc,Preskill:1980mz,Kosower:1984aw}.
 When the fermions make up a complex representation of the gauge group, the expected
pattern of chiral symmetry breaking is $SU(N_f)\times SU(N_f)\rightarrow SU(N_f)$. With $N_f$ Dirac fermions (or $2N_f$ Majoranas)
in a real representation of the gauge group, the symmetry breaking pattern is $SU(2N_f)\rightarrow SO(2N_f)$.
With a pseudoreal fermion representation, it is $SU(2N_f)\rightarrow Sp(2N_f)$.

There is already a small lattice literature on these systems: see \textcite{Damgaard:2001fg} and its citations.
These early papers observed the pattern of chiral symmetry breaking through  regularities in the spectrum of Dirac eigenvalues.
 \textcite{Lewis:2011zb,Hietanen:2014xca} recently studied the spectroscopy
of $SU(2)$ gauge fields and $N_f=2$ fundamentals, a pseudoreal representation.
 \textcite{DeGrand:2015lna} did similar work for $SU(4)$ with
$N_f=2$ two-index antisymmetric (real representation) fermions.
There is more to do. Direct calculations of the $V_{eff}(\Sigma)$ are an obvious target for future work.

Finally, why is the lattice literature for this subject so small? I can think of several reasons.

First,  lattice simulations are performed discretizing ultraviolet complete Lagrangians.
Most of the literature of composite Higgs is concerned with its low energy effective theory.
Until recently, there were few examples of ultraviolet completions.
For example, the two (very complete)
 review articles of \textcite{Perelstein:2005ka,Bellazzini:2014yua} total about
eighty pages of print, but their combined discussion of ultraviolet completions is only about three pages
long.

Second, many of the published
ultraviolet completions are difficult venues for lattice simulations:
they involve theories in more than four dimensions, or supersymmetry, or both.

Third, the ultraviolet completions typically involve gauge groups with $N_c \ne 3$, or
fermions in higher dimensional representations, or Weyl or Majorana fermions rather than Dirac fermions.
New code must be written.
This should not be a barrier, but historically, it has been.

Fourth, some of the key calculations require lattice fermions with good chiral properties,
at least for the valence quarks. An example is $\Pi_T^{LR}(q^2)$.
The matching factors converting lattice to continuum regularization
 for the vector and axial vector currents are different unless
the lattice action can support a Ward identity pinning them together.

And last, particularly for some versions of the partial compositeness scenario, one needs to simulate
several representations of fermions at once. 
(Of course, there are interesting physics questions for these systems
on their own.)

So far, there is not enough lattice
work in this area to justify a review. Perhaps in a few years there will be.

\subsection{Composite dark matter}
Not much is known about dark matter other than it exists, that it is long - lived,
that its density is about a quarter
of the mass density needed to close the Universe, and that it is dark, lacking electromagnetic interactions.
In some cases dark matter candidates naturally arise in other models of beyond  Standard Model physics:
for example, in supersymmetric extensions of the Standard Model that have R-parity as a
 symmetry, the lightest
supersymmetric partner is a dark matter candidate. But there are also many models
for dark matter with no direct extension to other physics issues.
There is a small speculative literature arguing that dark matter could be strongly interacting,
a sort of hidden version of QCD, coupling somehow weakly to Standard Model particles.
Early references include \textcite{Nussinov:1985xr,Barr:1990ca} and the recent lattice study by
\textcite{Detmold:2014qqa} of one candidate system
lists about thirty phenomenological papers.
Not surprisingly, there are
 lattice studies of composite dark matter models.
The literature I know of includes studies of $SU(2)$ gauge theories coupled to $N_f=2$ flavors
of fundamental fermions
\cite{Lewis:2011zb,Hietanen:2013fya,Hietanen:2014xca,Detmold:2014qqa,Detmold:2014kba}
and
$SU(4)$ gauge theory with quenched  fundamental representation fermions \cite{Appelquist:2014jch,Appelquist:2015yfa}.
Most of the work is about the spectroscopy of these systems, mostly their baryon spectroscopy because
one is interested in knowing what
is likely to be the most stable particle. There is also some discussion about matrix
 elements appropriate
for dark matter detection. [For  examples of such a calculation, see \textcite{Appelquist:2013ms,Appelquist:2015zfa}.]

%%%%%%%%%%%%%%%%%%%%%%%%%%%%%%%%%%%%%%%%%%%%%%%%%%%%%%%%%%%%%%%%%%%%%
\begin{figure}
\begin{center}
\includegraphics[width=\columnwidth,clip]{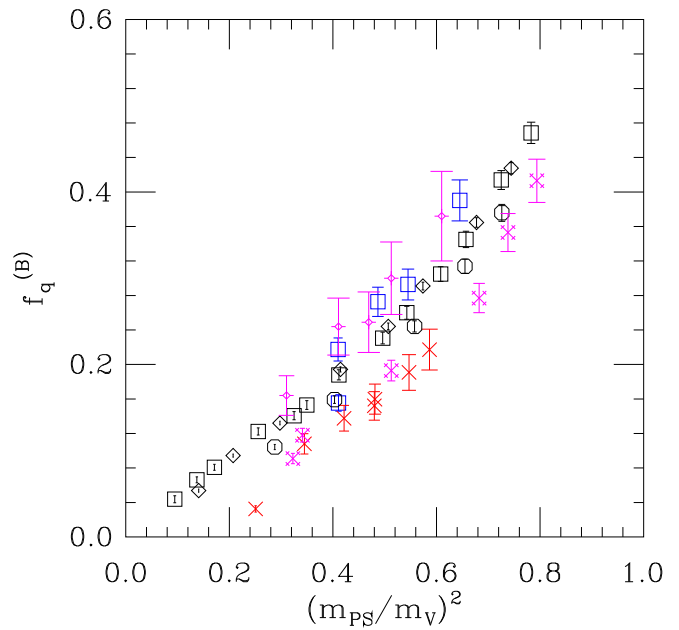}
\end{center}
\caption{
The quantity $f_q^{(B)}$ defined in Eq.~(\ref{eq:barysigma}), plotted vs the squared pseudoscalar to vector meson mass
ratio $(m_{PS}/m_V)^2$.
 Data shown include quenched fundamental SU$(3)$, SU$(5)$ and SU$(7)$ (black squares, diamonds, octagons),
 dynamical SU$(3)$ (blue squares), and dynamical SU$(4)$ AS2 (red crosses).
 This data is from \protect{\textcite{DeGrand:2015lna}}.
Also plotted in purple©
 are results from \protect{\textcite{Appelquist:2014jch}} for quenched fundamental SU$(4)$, for bare gauge
coupling $\beta = 11.5$ (fancy diamonds) and $\beta = 12.0$ (fancy crosses).
\label{fig:barysigma}}
\end{figure}
%%%%%%%%%%%%%%%%%%%%%%%%%%%%%%%%%%%%%%%%%%%%%%%%%%%%%%%%%%%%%%%%%%%%%

In the models which have been studied to date, the dominant nuclear interaction with a
 dark matter particle is through Higgs exchange.
The interesting quantity is the matrix element between a nucleon $a$ through its constituent quarks $Q$
 and the dark matter baryon $B$ through its constituent fermions $q$. Schematically, this quantity is
proportional to
\bee
{M_a}=\frac{ y_Q y_q}{m_{Higgs}^2} \sum_q\svev{B|\bar q q|B}\sum_Q\svev{a|\bar Q Q |a}.
\ee
The factors $y_Q$ and $y_q$ are Yukawa couplings. The expectation values
 in the expression are the QCD sigma term and its dark matter analog.
There are several lattice calculations of this quantity, typically given in terms of 
\bee
\label{eq:barysigma}
f_q^{(B)} \equiv \frac{m_q}{M_B} \frac{\partial M_B}{\partial m_q} =
\frac{m_q}{M_B} \langle B | \bar q q | B \rangle.
\ee
A recent compilation  \cite{DeGrand:2015lna}  is shown in Fig.~\ref{fig:barysigma}.
(Not shown is data by \textcite{Detmold:2014kba}  which is for an $SU(2)$ system and is presented over 
a tiny range of fermion masses, with similar results.)
It appears that $f_q^{(B)}$ is reasonably independent of the underlying dynamics.

 I believe that  dark matter phenomenology does not demand technicolor - like dynamics (a slowly
 running coupling constant) and so to the lattice practitioner, these systems 
are QCD - like and are reasonably
 easy to study. The issue, of course, is motivation for any particular  model in the absence of an experimental
signal.

\subsection{Dilatonic Higgs\label{sec:dilaton}}
Another possibility to generate a light Higgs is to somehow tune the ultraviolet theory so that
its couplings are close to some critical value, where its correlation length diverges.
A diverging correlation length is the same thing as a very light particle, which would be
a candidate to replace the Higgs. Of course, it also brings along new physics at some higher scale.

The ``homework example'' for these systems is the mean field behavior of
an $O(N)$ spin model with a potential
 $V(\phi)=a_2 \phi^T \cdot \phi + a_4 (\phi^T \cdot \phi)^2$,
With $a_2>0$, the $O(N)$ symmetry is unbroken and all fields have a squared mass $\sim a_2$.
The symmetry is spontaneously
broken for $a_2<0$, the Goldstone bosons are massless
and the Higgs has a squared mass $m_H^2\sim -a_2$.
 At criticality, where $a_2=0$, the Higgs mass also vanishes. That is the 
state we are interested in. 

(Of course, in a better treatment, all masses vanish
at criticality where the system experiences scaling behavior. But close to criticality,
there should be a light state.)

 I used the $O(N)$ model rather than a $Z(2)$ model, where a massless state also appears at criticality, 
in order to make the point in the symmetry-broken phase there are Goldstone bosons; 
the scalar channel will have
a two-particle branch cut in addition to a Higgs pole. A numerical simulation
 will have to disentangle
the branch cut from the desired signal.

Similar behavior is expected in $N_f=2$ QCD with massless quarks. Precisely at the 
critical temperature, the system should exhibit scaling, with power law decay for all correlation functions.
Slightly away from criticality, this branch cut behavior should dissolve into a set of resonances, one of 
which, an isoscalar scalar meson, will be very light. 

I have not found any  definitive study of such a state in the finite temperature QCD literature.
These calculations are technically quite demanding. There are two (related) problems.
The first one is that the state has the quantum numbers of the vacuum. A mass $M$ is  determined by fitting 
a correlation function of a source and sink operator separated by a distance $t$ to the functional form
\bee
\svev{O(t)O(0)} \sim A + B\exp(-Mt)+\dots
\ee
The constant term $A$ 
is only present when the states created by $O$ have vacuum quantum numbers, otherwise it vanishes.
When it is nonzero it dominates the mass-dependent term when  $t$ becomes large.
The second issue is that the calculation involves disconnected diagrams. Think of the 
state as a $\bar q q$ pair. The correlator has a contribution where the source pair annihilates
into gluons, which then reconvert at the sink. These correlators are intrinsically noisy.
There is, however, one related observation.
\textcite{Cheng:2010fe} have published measurements of the isotriplet scalar screening mass in finite temperature QCD.
It shows a dip near the transition temperature, while always remaining greater than the pseudoscalar mass.

The particle physics literature refers to these states as ``dilatons.''
A dilaton is a pseudo-Nambu Goldstone boson associated with scale symmetry
breaking. 
The divergence of the dilatation current
is the trace of the energy-momentum tensor.  This trace is anomalous in a massless gauge theory and its size is
proportional to the beta function.  The Goldstone boson which comes from spontaneously broken
dilatation symmetry (i.e. the scale put in by renormalization) has a mass
proportional to the anomaly, and thus to the beta function.  A walking theory has a
small beta function, hence a light dilaton.  When the fermions acquire masses
they couple to the dilaton in proportion to their masses, so the scalar
couples to fermions like a Higgs. 

Such states have a long citation trail [two of many early papers are
\textcite{Gildener:1976ih,Yamawaki:1985zg}] and the idea continues to appear as a beyond Standard
Model possibility. 
They have a somewhat fraught phenomenology. The issue is that our world
 is not conformal. If the world of very high energy is conformal,
there must be a crossover to its behavior, and it is quite difficult to keep such a Higgs from
moving up in mass to the scale where the crossover begins. [Compare the discussion
 in \textcite{Bellazzini:2012vz,Tavares:2013dga}.]

In the lattice literature, the words  ``dilatonic Higgs'' seem to me to be a shorthand for having
a confining and chirally broken system in which there is a scalar particle which is parametrically
lighter than the rest of the spectrum (apart from the Goldstones, of course).
The first question is to determine whether the system is confining and chirally broken, or not.
Such light states could appear in the symmetry-restored phase (as in the $O(N)$ example). They
could also appear in a system which is conformal in the zero fermion mass limit. There,
all masses fall to zero as $m_H = A_H m_q^\gamma$, but the $A_H$'s can be different in different channels.
Then, where does the a $0^{++}$ state fit into the spectrum,
either at nonzero fermion mass, or as the fermion mass is taken to zero?
The value at nonzero fermion mass is what is actually measured in a simulation. The zero mass case
is an extrapolation. Finite simulation volume is another issue when light particles are involved.
It had better be the case that the Higgs candidate is much lighter than everything except the
would-be Goldstone modes. Even then, there are other issues: an important one
 is the ratio of $f_\pi$ to other mass values. If $f_\pi$ is set by electroweak physics, the
other states must be out of reach of where the LHC has already scanned, or the model is not viable.
 And if the light state is going to replace the Higgs,
its branching ratios had better be close to Standard Model values.
 I will return to the discussion of lattice
 results for these states in Sec.~\ref{sec:results}.

\subsection{Fundamental scalars on the lattice }
Lattice studies of strongly coupled scalar fields have a long history, going back into the 1980's.
There was a literature about self-interacting scalar fields, of scalar fields interacting
with gauge fields, and of scalars interacting with fermions.
A major area of research in that era, which extended up to the discovery of the Higgs, was
constructing upper and lower bounds on the Higgs mass. My discussion of the issues, 
around Eqs.~\ref{eq:upperH}-\ref{eq:lowerH}, was quite naive. When the Higgs gets close to 
its upper bound, or to its lower bound, its interactions (either with itself, or with the top quark)
 become strong. A perturbative story is suspect. Of course, people were hopeful -- 
perhaps the Higgs would not be found, or it might have been pushed to a mass value where new physics
could be nearby. They wanted to make nonperturbative bounds, to get a better indication of where new physics might be.
Two papers which studied this, from just before the  Higgs discovery, are
\textcite{Fodor:2007fn,Gerhold:2010bh}. 

Now that we have the Higgs, the story might be different: suppose there are heavier generations
of fermions. Does the observed Higgs mass constrain their masses?
\textcite{Bulava:2013ep} say Yes, and that the maximum allowed mass of a fourth generation quark is about 300 GeV.

In the late 90's, several lattice groups studied the scenario of electroweak baryogenesis \cite{Kuzmin:1985mm}.
If the electroweak sector had a strongly first order transition, the metastability associated
with the transition would lead to thermal non-equilibrium. This is one of the
necessary Sakharov conditions for baryogenesis. The other conditions (baryon number violation, C and CP
nonconservation) also exist in the Standard Model, so in principle, the 
generation of the baryon asymmetry in the early Universe
could arise from electroweak interactions. A series of beautiful lattice calculations
showed that the transition was a crossover for Higgs masses above 72 GeV. Even at the time,
this was already inconsistent
with experiment, ruling out the scenario.
A recent conference proceedings \cite{Laine:2013raa}  has references to the original literature.
Knowing the Higgs mass allows one to refine the calculations, and perhaps constrain other models
of baryogenesis. See \textcite{D'Onofrio:2014kta}.

Finally, there is a small recent literature of lattice simulations of the  gauge-Higgs sector of the Standard Model itself.
The formal issue is that the Standard Model is a gauge theory. Observables must be gauge invariant,
gauge invariant observables are represented by composite operators, and composite operators
can have very different spectral properties than simple ones. Take QCD as an example.
\textcite{Maas:2012tj,Maas:2013aia,Maas:2014pba}     have studied these issues. My interpretation of their results is that
the weakly coupled Standard Model is still what we think it is, even on the lattice,
but that it could have been different.

\subsection{Lattice-regulated supersymmetry}
Phenomenological  supersymmetric extensions of the Standard Model are, of course, completely perturbative.
No lattice calculations are needed  to make predictions. But
there is also a small literature devoted to lattice-regulated supersymmetry.
These are simulations of ${\cal N}=1$ and ${\cal N}=4$ supersymmetric Yang - Mills theory
 in space - time dimension $D=4$ and various models in $D=2$. These papers are not
about phenomenology, per se. Rather, the questions are along the line of
``does the lattice system exhibit features of supersymmetry?''

People want to put
 supersymmetric theories on the lattice because
many of the nonperturbative features which 
appear in ordinary (non-supersymmetric) theories, such as spontaneous chiral symmetry breaking,
confinement, magnetic monopole condensation, strong coupling - to - weak coupling duality
(to name a few) were first studied in a supersymmetric context. It would be useful to have 
a nonperturbative formulation of these specific systems, which checks these calculations.

Of course, one has somehow to evade the problem that supersymmetry is an extension of the
usual Poincar\'e algebra and is broken completely by naive discretization.
However, this is a problem that has been mostly solved.
A good place to begin a literature search is with the
review article by \textcite{Catterall:2009it}, and with citations to it.

${\cal N}=1$ probably has the greater literature. This is a  system of adjoint Majorana fermions
coupled to gauge fields. It is simulated with chiral lattice fermions, such as domain wall fermions,
 using the rational Hybrid Monte Carlo
algorithm (see
Eq.~\ref{eq:rhmc}, below). 
 The supersymmetric limit is the limit of vanishing fermion mass.
Some representative papers include \textcite{Fleming:2000fa,Giedt:2008xm,Endres:2009yp,Kim:2011fw,Bergner:2013nwa}.

${\cal N}=4$ is much trickier. The issue is, not surprisingly, the scalars. An intricate
construction allows one to simulate a theory with a single scalar supercharge. The other
fifteen supercharges of  ${\cal N}=4$ are broken by the lattice discretization.
It is believed that the situation is like the loss of rotational invariance in a usual lattice 
system: the breaking of the symmetry is due to irrelevant operators. This means that  these
supersymmetries are recovered in the continuum limit. Exactly how to do that in an efficient way
 is at present a research problem. A recent paper, \textcite{Catterall:2014vka}, discusses this issue.
 It has citations to earlier literature.
Alternative formulations of lattice supersymmetry include
 \textcite{Honda:2013nfa, Hanada:2013rga,Honda:2011qk,Ishiki:2009sg,Ishii:2008ib,Ishiki:2008te}.

\subsection{Gauge bosons and matter in space-time  dimensions $D>4$}
Higher dimensional extensions of the Standard Model have an enormous and rich 
continuum literature.
Lattice studies, however, are very sparse. The fundamental issue is that gauge couplings
in $D>4$ are dimensionful, and hence the systems are non-renormalizable.
In fact, the extra-dimensional gauge theory has to be understood only as a low energy
excitation of some more fundamental theory. At its cutoff scale ($\Lambda$
in energy, or for us on the lattice, the lattice spacing $a$)
the effective description breaks down and details of the underlying theory become important.
Typically, the systems of interest have compact extra dimensions. Calling their
scale $L$, the effective description only makes sense if the compactification length $L$
is large compared to the cutoff, or $L\Lambda \gg 1$.

Most of the work I know about is in $D=5$, with $SU(N)$ gauge fields and small $N$ (mostly $N=2$).
The fifth dimension is compact, sometimes orbifolded 
\cite{Irges:2006hg,Knechtli:2014ioa,Irges:2013rya,Irges:2012mp}, sometimes not \cite{deForcrand:2010be}.

Many of the simulations introduce one lattice spacing for the four large dimensions and
a different lattice spacing $a_5$ for the fifth dimension. One cannot take both cutoffs to zero;
power divergences appear that cannot be absorbed into a finite number of counterterms.
But one can tune one of the dimensions to zero, holding the others fixed. Then one can explore
the phase diagram of the system, looking for critical points or lines. At these places,
the correlation length $\xi$ diverges,  in units of the lattice spacing $a$,
 $a/\xi \rightarrow 0$. In that sense, the lattice spacing is removed,
and a  four-dimensional theory is left. Slightly off the critical line, there is a four 
dimensional theory, but with extra irrelevant operators.

Like lattice supersymmetry,
the question here seems to me to be more ``Can I make it work?'' rather than ``What can I do with it?''

%%%%%%%%%%%%%%%%%%%%%%%%%%%%%%%%%%%%%%%%%%%%%%%%%%%%%%%%%%%%%%%%%%%%%
\section{Lattice methodology\label{sec:lattice_meth}}
%%%%%%%%%%%%%%%%%%%%%%%%%%%%%%%%%%%%%%%%%%%%%%%%%%%%%%%%%%%%%%%%%%%%%

%%%%%%%%%%%%%%%%%%%%%%%%%%%%%%%%%%%%%%%%%%%%%%%%%%%%%%%%%%%%%%%%%%%%%
\subsection{A lightning introduction to lattice calculations \label{sec:latrev}}
%%%%%%%%%%%%%%%%%%%%%%%%%%%%%%%%%%%%%%%%%%%%%%%%%%%%%%%%%%%%%%%%%%%%%

Before we  go on,  we have to recall how lattice calculations are
performed. Good textbooks, for example \textcite{DeGrand:2006zz,Gattringer:2010zz}, provide a detailed
introduction to the subject. What follows is a synopsis, the bare minimum the reader who does
not do lattice simulations of beyond Standard Model systems needs to know to have a context for the results.

 Imagine that we are interested in studying some  quantum field theory
with lattice techniques. 
We discretize the system, that is, we replace space and time by a grid of points. We then
define field variables that live on the links or sites of the lattice,
and construct an action that couples them together.
 We do
this in some way that preserves as many symmetries as possible.
Preserving gauge symmetry is vital to maintain current conservation, so nearly all lattice
calculations use gauge invariant actions and integration measures. Space-time symmetries and chiral
symmetries may be more problematic to enforce, so let us defer a discussion of them for a while. The lattice theory is then
 an effective field theory
 defined with an UV cutoff, the lattice spacing $a$. One can think of this
cutoff as being roughly equivalent to
an ultraviolet momentum cutoff $\Lambda \sim 1/a$.

 The lattice path integral is used as a
 probability measure to generate configurations of the field variables. For example, the
functional integral (or partition function) for a lattice bosonic
field $\phi_n$ has the form
\bee
   Z = \int [d\phi]\, \exp[-S(\phi)]
\label{eq:MCintro.partfunc}
\ee
where $S(\phi)$ is some lattice action and $[d\phi] = \prod_n d\phi_n$ is
 an integration over the values of
the field on each lattice site $n$.  Any physical observable ${\cal
O}$ can be expressed as a function of the field $\phi$.  Its formal
expectation value is
\bee
  \VEV{\cal O} = \left.\int [d\phi]\, {\cal O}(\phi) \exp[-S(\phi)]\right/
              \int [d\phi]\, \exp[-S(\phi)].
\label{eq:textbook}
\ee
This is just the average of the observable with respect to the measure
\bee
  P(\phi) \propto \exp[-S(\phi)].
\label{eq:MCintro.equil}
\ee
 Thus, the average
value of some observable is an ensemble average over the configurations of field variables.
In a lattice calculation, the
 generation of configurations is done numerically, by some stochastic algorithm.
Monte Carlo methods generate a sequence of $N$ random field
configurations $\phi^{(k)}$ with a probability distribution given by Eq.~\ref{eq:MCintro.equil}.
The expectation value of the observable is then just the simple average of
the observable over the ensemble of  configurations:
\bee
    \VEV{\cal O} = \frac{1}{N}\sum_{k=1}^N {\cal O}(\phi^{(k)}).
\label{eq:obsavg}
\ee
The uncertainty in the observation typically scales like $1/\sqrt{N}$.  Lattice correlation
functions
are then  compared to some theoretical model to extract the values
of desired observables.

Generic correlation functions measured in a  lattice simulation in a finite
simulation volume usually show an exponential falloff with distance, characterized
by a correlation length $\xi$. A particle mass $m$ is of course just the inverse of
the correlation length.

All lattice calculations are performed with the cutoff present. It is clear 
 that the cutoff is unphysical; we want to remove it from the calculation
and present cutoff-independent results. Lattice people talk about taking
 the momentum cutoff $\Lambda$ to infinity,
 or the lattice spacing $a$ to zero,  while fixing some fiducial mass scale. This
 is a shorthand for the requirement that the correlation length measured in units of the 
cutoff, $\xi/a$, must diverge.
 The correlation length will, of course, be a function
of the bare parameters that characterize the simulation.
Making the correlation length diverge is done by tuning
the bare parameters of the theory.

A fiducial scale is needed to set against the correlation length.
In lattice simulations, this scale is almost universally taken,
as a first step, to be the lattice spacing  $a$ itself.
When this  is done, pure numbers come out of the simulation; all predictions
of dimensionful quantities (like masses) appear 
 with an appropriate power of the cutoff (that is, a
calculation produces the product $a \times m$).
Almost all real lattice Monte Carlo
predictions are of dimensionless ratios of dimensionful quantities, like mass ratios.
Lattice people like to say
that one prediction of a mass determines the lattice spacing, when the value of that mass
 is fixed
by experiment. This is just the statement that $a= ma/m_{expt}$. One then uses this $a$ to make
predictions in energy units for other masses or dimensionful quantities.

Recall the usual definition that a running coupling is (infrared) relevant, marginal, 
or irrelevant with respect to
changes of scale, depending on whether it grows, remains almost unchanged, or
shrinks, as it is evaluated at longer and longer distance scales. That
a coupling is relevant or not can be empirically determined: can it be  varied, so
that  the correlation length grows?
If so, it is probably relevant. The  increase in the correlation length occurs as the relevant bare coupling
is tuned toward its critical value. Most lattice simulations are of theories with one or two relevant couplings. They also have
many irrelevant ones, typically arising when the continuum theory is transferred to the lattice.

While the correlation length is finite, the fact that the lattice action is an effective
field theory becomes important: one's answers ought to -- and generally do -- depend on the value of cutoff. One would
 observe this
in  measured mass ratios, as a function of the bare parameters in the simulation.

Most lattice simulations are done for asymptotically free theories. Their one or
 two relevant couplings are the gauge coupling $g$ and fermion masses $m$.
 The system has a critical 
surface in the space of all couplings
 that encloses a Gaussian fixed point at $g=0$ and $m=0$.  Tuning the two relevant couplings to zero
causes the correlation length, measured in units of $a$,
 to diverge.

 Much of the lattice language
for understanding cutoff effects implicitly makes use of the fact that one tunes $g$ and $m$ to zero
to remove them. Focus on the gauge coupling for a moment.
The advantage of having an asymptotically free theory is that when the bare coupling is taken smaller
and smaller,
the short distance behavior of the theory becomes increasingly perturbative and hence increasingly controlled.
In particular, field dimensions approach their engineering dimensions.
This allows us to parametrize the dependence of an observable on the cutoff scale.
  It is nearly given by naive
dimensional analysis.
In an asymptotically free theory, if the lattice spacing were small enough, a typical mass ratio would behave as
\beea
[a m_1 (a)]/[a m_2 (a)] &=& m_1(0)/m_2(0) + {\cal O}(m_1a) + \nonumber\\
  & & {\cal O}[(m_1 a)^2] +\dots
\label{eq:SCALING}
\eea
(modulo powers of $\log(m_1a)$).
The leading term does not depend on the value of the UV cutoff.  That
is our cutoff-independent prediction.  Everything else is an artifact
of the calculation. 

This is important because it gives control over the calculation.
Away from weak coupling, scaling dimensions of operators may be different
 from their engineering dimensions.
Corrections to scaling may not scale with their expected power laws. It may not be possible to identify
relevant versus irrelevant operators. Worse, the system may happen to
lie in the basin of attraction of other fixed points, or may be susceptible
to non-universal lattice-artifact phase transitions which depend on the particular choice of discretization.

Running of the gauge coupling to zero in the UV is generally only observed qualitatively in ``spectral''
calculations (of masses or matrix elements), through the observation that the correlation length increases
(``$a$ goes to zero'') as the bare gauge coupling is decreased.
 This was not the case in the earliest days of lattice
simulations, where people attempted to relate a mass to a bare lattice gauge coupling
along the lines of Eq.~\ref{eq:AFrunning}. We now know that lattice perturbation theory is much dirtier than
its continuum counterpart, and corrections to this naive behavior are large due to lattice artifacts.
 Instead of this, almost all lattice data is extrapolated to the continuum
with an analog of Eq.~\ref{eq:SCALING}.
Nowadays completely separate calculations of non-spectral observables are used to make quantitative 
statements about running couplings.

It is much easier to see that the mass is a relevant coupling; masses of all multiquark bound states vary strongly
as the bare lattice mass is tuned, and only become small as it is tuned to zero.

All lattice gauge theories replace the gauge fields $A_\mu(x)$ by ``link variables''
connecting adjacent sites. The link variables are group elements 
\bee
U_\mu(x) = \exp ig aA_\mu(x).
\ee
The gauge field functional integration measure
 is a product of integrals for each link variable  over the Haar measure of the gauge
group.
All lattice actions are traces over products of the $U$'s around some closed path. In the
so-called Wilson or plaquette action, this path is the minimal four-link one around a unit square.
There are many other possibilities, of course. All these actions, and all fermionic actions, differ from
the expected continuum action of fermions coupled to gauge fields by the addition of
extra irrelevant operators, so simulations with any of these actions done sufficiently close
to the Gaussian fixed point are expected to produce cutoff - independent predictions
of the continuum theory. In particular, space-time symmetries are broken by the lattice
discretization, but the operators which break them are irrelevant ones, and these
symmetries (such as rotational invariance)  are expected to be restored in the naive continuum limit.

\subsection{What systems can be studied on the lattice?}
Technical issues associated with putting fermions on the lattice strongly affect how
easy it is to simulate any particular theory.

Briefly, there are three generic kinds of lattice fermions. To summarize a (long) textbook discussion,
the constraint is the Nielsen-Ninomiya theorem, which says (loosely; this actually not precisely correct)
 that one cannot write down a well-behaved lattice
fermion action that is simultaneously chiral and undoubled.  ``Doubling'' is a shorthand way to say that
the lattice system has extra,
usually unwanted, fermionic degrees of freedom. These states are the doublers. 
The three kinds of fermions  are
\begin{itemize}
\item
Wilson fermions and their variants (clover or twisted mass fermions): a four-component spinor
sits on each site of the lattice. Their actions contain terms which, while
 formally irrelevant, explicitly
break chiral symmetry. The benefit of this breaking is that the lattice theory has the same number of
fermionic degrees of freedom as its continuum analog.
\item
Staggered fermions maintain some chiral symmetry, but at the cost of introducing doublers.  ``A 
single staggered fermion corresponds to four degenerate flavors in the naive continuum limit,'' we say.
\item Domain wall and overlap fermions, which
live in five dimensions (domain wall fermions), or are the four dimensional effective field theories
of five-dimensional fermions (overlap fermions), remain undoubled and replace the continuum definition of
 chirality by
a more complicated one, called the Ginsparg-Wilson relation. They are theoretically beautiful,
exactly encoding  Ward identities associated with chiral symmetry.
 From a practical point of view
these fermions are quite expensive to simulate.
\end{itemize}
A specific fermion action will lie in one of these classses, but beyond that, it will have a variety
of different lattice terms, typically different ways of discretizing the derivative operator.

All lattice simulations I know of are of vector theories.
Direct simulation of chiral gauge theories, like the Standard Model itself, is
quite difficult.
\textcite{Luscher:2000hn} gives a fairly complete overview of the subject.
To even begin, by imagining an ultraviolet regulator for a chiral gauge theory, the theory
must be anomaly free. But the consequence is that any consistent regulator that preserves
gauge invariance must refer to the fermion representation. This is hard to do;
a simple lattice cutoff will not suffice. People who want to study chiral gauge theories on the lattice
typically feel that they
are forced to use regulators that break gauge symmetry, and then attempt to tune their bare parameters to a critical point
which will produce a chiral gauge theory when the correlation length diverges.
\textcite{Golterman:2000hr,Golterman:2004qv} describe approaches along these lines.

The lattice introduces additional issues. The doublers which appear in an action with chiral symmetry
 turn out to have the opposite chirality to their partners; at the end, there will be
equal numbers of left- and right- handed fermions. Domain wall or overlap fermions
 allow one to go farther, and \textcite{Luscher:2000hn} describes all-orders perturbative
constructions of chiral gauge theories. I do not know of any numerical studies
of these systems, though.

The next issue is that 
$P(\phi)$ (see Eq.~\ref{eq:MCintro.equil}) has to
have a probability interpretation, in order to perform importance sampling.
One cannot directly simulate dynamical fermions, because
Grassmann variables are not numbers. One must formally integrate
them out, leaving behind a functional integral for bosons. Being classical, this
can be studied numerically.
To see how this is done, 
consider a system of gauge fields and a single species of fermion. Its partition function is
\bee
   Z = \int [dU] [d\bar \psi ] [d \psi]\, \exp[-S_G(U)-\bar \psi M(U)\psi]
\ee
where  $M = D + m$.
After integrating out the fermionic Grassmann variables, the partition
function  becomes
\bee
   Z = \int [dU]\, \exp[-S_G(U)] \det M(U) .
\label{eq:fermionmethod.fullQCD}
\ee
The determinant is nonlocal, so computing its change under a change in the
gauge field is very expensive.  The standard way to deal with this
is to simulate the determinant by introducing a set of scalar
``pseudofermion'' fields $\Phi$. This is done via the formal identity
\bee
  \det M(U) = \int [d\Phi^* d\Phi]\exp\, [ -\Phi^* M^{-1}\Phi].
\label{eq:pf1}
\ee

Expanding $\Phi$ in terms of eigenmodes $\psi_j$ of $M$
and the corresponding eigenvalues $\lambda_j$
\bee
 \Phi^* M^{-1}\Phi = \sum_j \svev{\Phi |  \psi_j} \frac{1}{\lambda_j}\svev{ \psi_j | \Phi}
\ee
exposes a cascade of problems, all arising from the fact that the
eigenvalues of lattice Dirac operators are complex and their real
parts may not be positive-definite. Individual terms in the exponential can be
complex or carry a net negative sign. Then the exponential in
Eq.~(\ref{eq:pf1}) cannot be interpreted as a conventional probability
measure.

There are often ways to avoid this. With Wilson fermions, one can show, using lattice
 symmetries of the action,
 that simulations of
pairs of degenerate mass fermions (i.e., even $N_f$) give a positive-definite determinant.
(Basically,  $D^\dagger = \gamma_5 D \gamma_5$, so $(\det D)^2 = \det D^\dagger D$.)
Staggered fermions naturally come in multiples of four flavors, and the four flavor
 combination has a positive
determinant.

Often, one wants to have a different fermion content than what is possible in these favorable situations.
 Odd numbers of flavors require caution.
For example, in QCD, one might want to simulate a degenerate up and down quark pair, and a heavier strange quark.
One replaces the strange quark's determinant by
\bee
\det M(U) \rightarrow (\det |M(U)|^2) ^{1/2} .
\label{eq:rhmc0}
\ee
This can be simulated with the RHMC (``rational Hybrid Monte Carlo'') algorithm, with
a pseudofermion action
\bee
\det H(U)^p \rightarrow \int [d\Phi^* d\Phi]\exp\, [ -\Phi^* \sum_j \frac{c_j}{H(U) + d_j} \Phi].
\label{eq:rhmc}
\ee
 The determinant could
try to change sign during the simulation. That would invalidate Eq.~\ref{eq:rhmc0}.
This might not be noticed, nor treated properly, by its approximation, Eq.~\ref{eq:rhmc}.

There are related issues with staggered fermions, going from the doubled 
number of degrees of freedom
that staggered fermions naturally encode, to the desired counting for a
 single continuum flavor.
One must make the replacement
\bee
 \det M(U) = \det M_{stagg}^{1/4}(U)
\label{eq:quarter}
\ee
to simulate a single continuum flavor. There is a long controversy in the QCD literature about how to correctly deal
with this replacement. I believe that the situation is well understood for chirally broken theories
simulated in the vicinity of the Gaussian fixed point. (The conference proceedings by \textcite{Sharpe:2006re}
is an excellent overview.) Briefly, at nonzero lattice spacing, the action associated with Eq.~\ref{eq:quarter} is nonlocal.
Rooted staggered fermions cannot be described by a local theory corresponding to a single Dirac fermion.
Associated with this nonlocality, there are all kinds of artifacts, such as negative norm states. However,
when chiral symmetry is broken,
a low energy theory can be construct which correctly describes the Goldstone sector of the rooted theory. This theory
has a set of low energy constants which include those of the continuum theory, plus additional ones. Continuum predictions can be 
made -- and are made -- using this more complicated chiral perturbation theory.

 Simulations of QCD at nonzero chemical potential are
difficult because the fermionic determinant is complex.

Finally, some vocabulary. To label the bare gauge coupling $g$ of an $SU(N_c)$ gauge theory,
lattice people work with the quantity $\beta=C/g^2$, where $C$ is a constant. For
the plaquette action, $C=2N_c$. The bare quark mass $m_0$ in 
simulations with Wilson or clover fermions
is usually replaced by a hopping parameter $\kappa = 1/2(4+am_0)^{-1}$, and people almost
always quote $\kappa$ rather than $am_0$.

Patterns of chiral symmetry breaking (``vacuum alignment'') for different numbers of colors
and fermionic representation were first described by
 \textcite{Peskin:1980gc,Preskill:1980mz,Kosower:1984aw} and were listed in Sec.~\ref{sec:partialcomp} above.
\textcite{Golterman:2014lea,Golterman:2014yha} describe the complications of lattice artifacts
for this physics.

\subsection{Lattice issues for beyond Standard Model calculations with slowly running couplings}

The situation for a lattice practitioner faced with a proposed nonperturbative
extension of the Standard Model is, at first sight, similar to
the situation of lattice QCD: Given an ultraviolet complete action that might
encode some nonperturbative low energy physics, the way to proceed is as follows:
\begin{enumerate}
\item Write down a lattice discretization and simulate it
\item From the simulation, determine the vacuum structure of the system:
does it have a mass gap in the infinite volume limit, is it
confining, is it chirally broken, is it something else?
\item If the system has a mass gap, compute the spectrum and perhaps appropriately interesting matrix elements
\item Alternatively, use the expectation value of some operator to define a
running coupling constant (typically, the scale at which the coupling is measured is set by
the size of the simulation volume) and see how it runs
\item From the results of (3) or (4), evaluate the possibility that the action
might be a viable scenario for beyond Standard Model physics
\end{enumerate}
Most lattice studies of beyond Standard Model dynamics involved systems with slowly running gauge couplings.
As a result, getting beyond item (2) proved to be very difficult. The issue was that all the techniques lattice people
had at their disposal were designed for QCD, where the coupling constant runs quickly.  Several years later,
I think there is a reasonable consensus between different groups about the answer
to point (2) for most of the systems that have been studied. However, agreement is
not universal and one can find controversy throughout the literature of the subject.

This is quite different from the situation in modern lattice QCD simulation.
There, the disagreements are about very specific points, such the particular value
of some mass or matrix element. In fact, the flow chart I gave for beyond Standard Model candidate
theories already differs from its QCD analog. Lattice QCD simulations never really had
to deal with item (2): the vacuum structure of QCD was, broadly speaking, noncontroversial
before the first simulations were carried out. 
Before QCD, experiment showed that strongly interacting matter was composite and chirally broken.
After the discovery of asymptotic freedom and 
before lattice gauge theory was invented, the question was,  are asymptotic freedom
and confinement related? 
Confinement was the most important phenomenological feature of the \textcite{Wilson:1974sk}  formulation
of lattice gauge theory. He showed that essentially all lattice gauge theories are confining
 in their strong coupling limit. The important question  then became, does confinement
persist in the continuum limit?  The earliest numerical simulations of lattice gauge theories
by \textcite{Creutz:1980zw,Creutz:1980wj} showed the coexistence of confinement and asymptotic freedom
in a single phase for a non-Abelian gauge theory.

Early analytic lattice
 work \cite{Blairon:1980pk,Greensite:1980hy,Svetitsky:1980pa,Weinstein:1980jk,KlubergStern:1982eh}
argued strongly that
chiral symmetry was broken in the strong coupled limit of lattice QCD, and again,
questions of interest were about the value of quantities such as the condensate
or the pion decay constant in the continuum limit, not about whether chiral symmetry was actually broken.
Lattice QCD very quickly became a subject about numbers, not about qualitative behavior.
And so it has remained. Not knowing the answer ahead of time made lattice studies of beyond Standard
Model
candidates very different from lattice QCD.

The origin of the difficulty in analyzing systems with slowly running couplings
 is most easily seen from the formula for the
 one-loop beta function:
with a scale change of $s$, the inverse coupling changes by an amount
\bee
\frac{1}{g^2(s)} - \frac{1}{g^2(1)} = \frac{b_1}{8\pi^2} \log s .
\label{eq:naiverun}
\ee
For the $SU(3)$ gauge group with $N_f$ flavors of fundamental representation Dirac fermions,
$b_1=11-2/3N_f$.  Consider ordinary QCD, with $N_f=3$, for which $b_1=9$. We know empirically that
a scale change of about $s=10$ causes the system to go from weakly coupled to strongly coupled:
this can be seen from the potential between heavy quarks, which is Coulombic at short
distance (0.1 fm) but confining at long distance (1 fm). A single lattice simulation with
a lattice spacing of say 0.05 fm and a size of 20-40 lattice spacings can capture both ends of this behavior,
so that the system can be perturbative at the shortest distance and nonperturbative at the longest distance.
Simulations involve the action at the cutoff scale, and if the system is weakly interacting at the cutoff 
scale, we know what we are doing.

Now consider the case of $N_f=12$, where
$b_1=3$.  With one loop running, we would need a scale change of $s=1000$
to make the coupling constant change by the same amount as the $N_f=3$ system changed with $s=10$.
 Such a scale factor cannot be accommodated on any
single lattice size which is capable of simulation today or in the foreseeable future.

At this point, a reader  objects: You are telling a one-loop story. You are a lattice person working
in strong coupling. Why should I believe a one-loop story?

The answer is: Yes,  the story could be wrong. But either it is wrong in a favorable way, or
an unfavorable way. In a favorable way, physics evolves more rapidly with scale than $b_1$ 
suggests (this happens
in  $N_f=3$ QCD).
This is easy to see in a simulation; you do not need to know about the story. But the physical systems
I am thinking about are candidates for walking technicolor. Recall  Fig.~\ref{fig:walkingdream}.
To the left of the inflection point, the coupling runs more slowly than the one loop formula. 
$b_1$ is effectively smaller.
The one-loop result for how the coupling changes with scale is too optimistic. Instead of $s=1000$, 
suppose the beta function is half the size of $b_1$. Then
you need $s=10^6$ to see the same change in the coupling.

There are many equivalent ways to state the consequences of having a slowly running coupling in a finite volume lattice
simulation:
\begin{itemize}
\item In such a theory, if the coupling constant is small at short distances (that is, at the cutoff
scale) in any simulation,
it remains small at long distance. Then, how can nonperturbative physics appear?
\item If the coupling constant is large at long distances, it must be large at the shortest distance
(at the cutoff scale) on the lattice. Then, how closely does the lattice theory resemble its continuum analog?
\item The coupling constant effectively does not run with scale in any practical simulation volume
\item If a simulation does show a potential which has both a Coulomb term and a linear confining piece,
it must also be characterized by a quickly running coupling constant, over the range of scales present in the simulation.
\end{itemize}

Nearly all lattice systems with many fermion degrees of freedom show this generic behavior.

How to deal with lattice artifacts in QCD is 
under reasonable control. However, that is because continuum QCD
 is qualitatively well understood.
Making sense of the ``quarter root trick'' (Eq.~\ref{eq:quarter} for staggered fermions)
 is done in the context of
chiral perturbation theory. But, suppose that one is simulating a theory which might not be
chirally broken? One might not be able to distinguish a lattice artifact from real physics.
In particular, little is known about the universality properties of a rooted theory. Its global symmetries
are simply different from those of an unrooted system.

Since most lattice studies involve slow running, we should think a bit more about what to expect.
We can continue to do this using perturbation theory.
The two loop beta function can be integrated exactly to find a relation between scale and coupling.
Defining $\overline b_1 = b_1/(16\pi^2)$ and $\overline b_2=b_2/(16\pi^2)^2$, it is
\bee
\overline b_1 \log \frac{\mu^2}{\mu_0^2} = \frac{1}{g^2(\mu)} - \frac{1}{g^2(\mu_0)}
-\frac{\overline b_2}{\overline b_1} \log \left(\frac{\overline b_1 + \overline b_2 g^2(\mu)}
{\overline b_1 + \overline b_2 g^2(\mu_0)} \right)  .
\label{eq:2looprun}
\ee

When the coefficients  $b_1$ and $b_2$ have opposite signs, there is a fixed point, at
$g_f^2= -{\overline b_1}/{\overline b_2}$. Equation \ref{eq:2looprun} takes the compact form
\bee
\overline b_1 \log \frac{\mu^2}{\mu_0^2} = \frac{1}{g^2(\mu)} - \frac{1}{g^2(\mu_0)}
-\frac{1}{g_f^2} \log \left (\frac{g^2(\mu)-g_f^2}{g^2(\mu_0)-g_f^2}  \right).
\label{eq:2looprunfp}
\ee

Now we can examine some useful limits. If $g^2(\mu)$ and $g^2(\mu_0)$ are both small, the logarithm is small
compared to the first terms and we have the familiar one loop running formula. However,
when $g^2(\mu)-g_f^2$ is small, the logarithm is the dominant term, and the coupling evolves
in a different (but equally familiar) way:
\bee
g^2(\mu)-g_f^2 = (g^2(\mu_0)-g_f^2) \left( \frac{\mu}{\mu_0}   \right)^{\overline b_1 g_f^2}
\label{eq:usualrun}
\ee
The beta function has a linear zero: $\beta(g^2)\sim -\overline b_1 (g^2-g_f^2)$.
At ever smaller $\mu$, the coupling runs into the fixed point. This is an infrared attractive
 fixed point.
${\overline b_1}g_f^2$ is an example of a critical exponent. In this case we will label it $y_g$.

\begin{figure}
\begin{center}
\includegraphics[width=\columnwidth,clip]{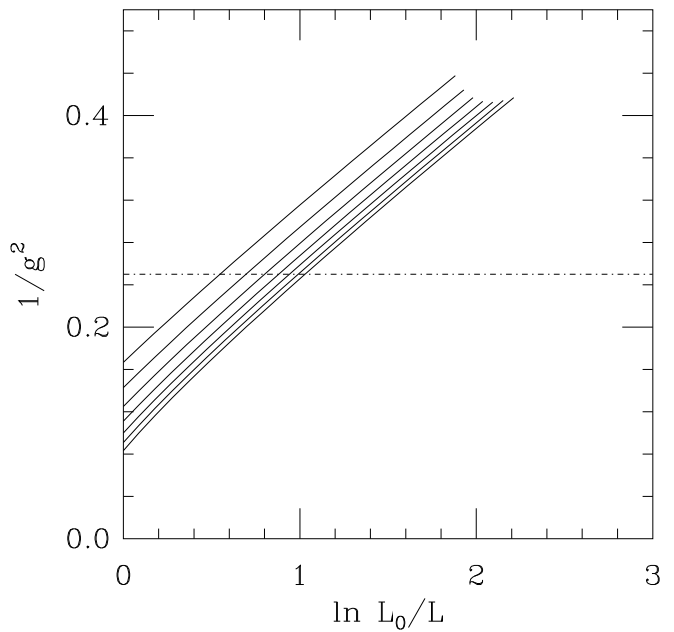}
\end{center}
\caption{Running coupling constant from Eq.~\protect{\ref{eq:2looprun}} for a set of initial
couplings: $N_c=3$, $N_f=3$. The dotted line is a line of constant $1/g^2$.
\label{fig:bfnpert3}}
\end{figure}

\begin{figure}
\begin{center}
\includegraphics[width=\columnwidth,clip]{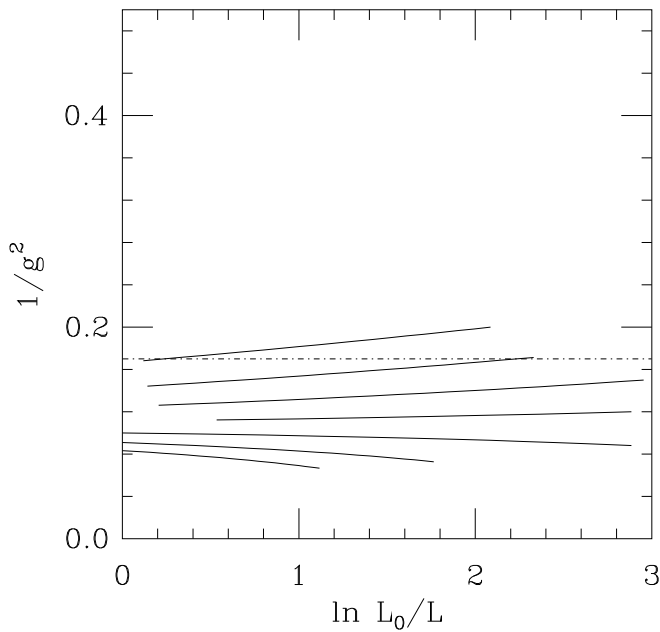}
\end{center}
\caption{Running coupling constant from Eq.~\protect{\ref{eq:2looprun}} for a set of initial
couplings: $N_c=3$, $N_f=12$. The dotted line is a line of constant $1/g^2$.
\label{fig:bfnpert12}}
\end{figure}

Setting $\mu/\mu_0 = L_0/L$, we can define a coupling measured at a distance scale $L$. This will
be useful to anticipate definitions of running couplings used in lattice simulations.
We contrast the running of the coupling constant in two cases
in Figs.~\ref{fig:bfnpert3}-\ref{fig:bfnpert12}.
 The first picture shows the case of $N_c=3$ and $N_f=3$ fundamental flavors; the
second shows the case
for $N_c=3$ and $N_f=12$. The initial $g^2(\mu_0)$ is taken to be equally spaced values 1,2,3, $\dots$.
The fixed point coupling for $N_f=12$ is at $g^2=9.47$.
We will come back to the dotted lines in Sec.~\ref{sec:running}.

%%%%%%%%%%%%%%%%%%%%%%%%%%%%%%%%%%%%%%%%%%%%%%%%%%%%%%%%%%%%%%%%%%%%%

%%%%%%%%%%%%%%%%%%%%%%%%%%%%%%%%%%%%%%%%%%%%%%%%%%%%%%%%%%%%%%%%%%%%%

This analysis is incomplete, because it leaves out the fermion mass. Inside the conformal window,
the fermion mass is a relevant coupling. 
In fact, in most simple systems, it is the only relevant coupling, and to make the correlation
 length diverge, it must be fine-tuned to zero. Its evolution equation has a linear zero, as
does the 
renormalization group equation for $g^2$ in the vicinity of $g_f^2$. This is the usual textbook
 situation for second order criticality,
where the
correlation length  diverges algebraically,
\bee
\frac{\xi}{a} \sim \left(a m_q \right)^{-\frac{1}{y_m}}.
\label{eq:corrlen}
\ee
(\textcite{Cardy:1996xt} is a good reference.)
Systems with this behavior are often called ``infrared conformal'' in the particle physics literature.
The quantity $y_m$ is the leading relevant exponent for the system, in statistical physics language.
This exponent is related to the anomalous dimension $\gamma_m$ of the mass operator $\bar\psi\psi$,
and determines the running of the mass parameter according to
\bee
\mu \frac{dm(\mu)}{d\mu} = -\gamma_m(g^2) m(\mu).
\ee
The relation is $y_m = 1+\gamma_m(g^2)$.

For future reference, in lowest order in perturbation theory,
\bee
\gamma_m= \frac{6 C_2(R)}{16\pi^2} g^2.
\label{eq:gmpt}
\ee
The actual exponent is $\gamma_m(g_f^2)$. In perturbation theory, it grows as the bottom of the conformal
 window is approached from above (say, by decreasing the number of flavors).

Inside the conformal window, all couplings other than $m_q$ are irrelevant. Note that the gauge coupling
(more properly, the distance of the gauge coupling
from its fixed point value)  is one of these couplings. Taking the continuum limit has nothing to do 
with tuning the bare gauge coupling, other than setting it within the basin of attraction of $g_f^2$. 
(The literature is occasionally confused about this point.)
In the two loop example, that happens naturally,
for any value of $g^2$. In most cases, $0<g^2<g_f^2$ is in the basin of attraction of $g_f^2$.
So is a region $g^2 >g_f^2$. This is a curious region, a bit like QED, because the coupling
constant
becomes larger at shorter distances. But again, tuning $m_q$ to zero is how we take the continuum
limit of $a/\xi \rightarrow 0$. We expect that values of $g^2$ which remain in the basin of
attraction of $g_f^2$ cannot become too great, because lattice theories generically confine when the
gauge coupling becomes large. Thus, there should be  a strongest
coupling, a boundary of the basin of attraction. If the boundary were 
marked by another second order transition,
it would be characterized by a UV attractive fixed point, at some $g_{UV}$.
More complicated possibilities can be imagined \cite{Kaplan:2009kr}.

(The absence of a scale in the conformal window should not be confused with the presence of a massless
state in the spectrum.
In the confining phase, when chiral symmetry is broken, there is an infinite correlation length,
the inverse pion mass. But in other channels there is a mass gap,
 and there are other physical scales, such as $f_\pi$.)

The irrelevance of the gauge coupling has the consequence that 
the location of an infrared attractive fixed point is not
physical. Contrast this behavior to that of a relevant coupling, which marks a real
 qualitative change in long-distance physics: $m_q=0$ for the fermion mass.
 A change in the renormalization scheme
can shift the location of $g_f^2$. This means that in the scaling limit ($\xi/a \gg 1$),
changes in $\xi$ as the bare  $g^2$ is varied can only be order unity corrections. 
(For a lattice QCD practitioner used to simulating clover fermions, the situation
is similar to what one would find when tuning the clover coefficient.)
There is no good reason for
$\xi/a$  to increase (or decrease, either) versus increasing $g^2$.
The scaling limit is the limit of vanishing quark mass, or more generally, of the limit
that all relevant couplings are taken to their critical values.

One complicating issue in this discussion is that while the
 gauge coupling is irrelevant, the critical exponent associated with the gauge coupling is often close to zero.
(Dimensional analysis ``predicts'' $y_m \sim 1$, $y_g \sim 0$.)
This has an unfortunate practical consequence which I already mentioned:  
In finite volume simulations, the gauge coupling  will evolve so slowly
with scale toward its fixed-point value that the system is effectively conformal, regardless
of the actual value of the cutoff-scale gauge coupling.
One is likely to observe a leading exponent $y_m$ that shows a 
slow, smooth dependence
on bare gauge coupling. This may be very hard to analyze.

Finally, while I have discussed the relevant $m_q$ and irrelevant $g^2$, we cannot forget
all the other couplings. To choose a lattice action is to implicitly fix the initial values of
all the irrelevant couplings, too. But these couplings also run.  While they run to zero in the
long distance limit, it might be that
over the range of length scales accessible in a simulation in finite volume, these couplings could
exhibit significant running, which could contaminate results.
(And remember, far away from the Gaussian fixed point, one may not know what is relevant and what is not.
The flow may even find another fixed point.)
This is a source of systematic error.
It is also an important practical issue which arises when one wants to compare results from two different simulations
which are performed with different lattice actions.

And of course, one may not know a priori that one is dealing with a conformal system.

Once the fermion mass becomes large, we expect that the fermions decouple 
from long - distance dynamics. The most likely scenario in that case is that the system
becomes confining, since the fermions no longer screen the gauge fields. 
One would expect to see a linear potential re - emerge. Probably the lightest
excitations would be gluonic in nature, glueballs. It is unknown how much of 
spontaneous chiral symmetry breaking would remain. Of course, 
explicit chiral symmetry breaking by the fermion masses would also be present.

More direct applications of the statistical literature to physics in the conformal window
can be found in 
\textcite{DeGrand:2009mt}
and
\textcite{DelDebbio:2010ze,DelDebbio:2010jy,DelDebbio:2013qta}.

%%%%%%%%%%%%%%%%%%%%%%%%%%%%%%%%%%%%%%%%%%%%%%%%%%%%%%%%%%%%%%%%%%%%%
\section{Lattice results for systems with slowly running couplings -- by method\label{sec:vertical}}
%%%%%%%%%%%%%%%%%%%%%%%%%%%%%%%%%%%%%%%%%%%%%%%%%%%%%%%%%%%%%%%%%%%%%

\subsection{Spectroscopy and related observables}

Spectroscopy can, in principle, distinguish between systems which are confining and chirally broken,
and systems which are nearly conformal. So, let us imagine
doing a simulation. Recall that, at any nonzero $m_q$, the system is ``ordinary,'' not conformal, with
a mass gap, regardless of what happens at
$m_q=0$. It will have a discrete spectrum.

Collect spectroscopic data (probably one might begin with a similar set of bare couplings, perhaps
one bare gauge coupling and several fermion masses).
Is the spectrum of excitations QCD-like? That is, 
as the fermion mass is made smaller, does the pseudoscalar state become much lighter
than the vector state?
Does the pseudoscalar mass extrapolate to zero with the fermion mass, like $m_\pi^2 \propto m_q$? 
Do other masses extrapolate to nonzero values at $m_q=0$?
Is the vector meson mass different from the axial vector mass?
Is the static potential linear at long distance?

If the answer to these questions is Yes, then probably the system is
 confined and chirally broken and,
given the discussion in the last section, it probably also has
a quickly running coupling constant.

(The question ``Is the vector meson mass different from the axial vector mass?'' refers to the fact that
in a system in which chiral symmetry is unbroken, opposite parity states are degenerate, being related to each other by chiral rotations.)

Now simulate at weaker bare gauge coupling. Does it seem that  the correlation length
 grows, while
the good features seen so far appear to maintain themselves?
Is it possible to move to ever weaker coupling without encountering
 a discontinuity, a phase transition between
the strong coupling phase and some new phase? If there is a transition,
 is it induced by the size
 of the lattice? If the answer to these questions is Yes, then confinement
 and chiral symmetry breaking probably coexist with asymptotic freedom.

In a system inside the conformal window, spectroscopy would be qualitatively different.
All masses would track towards zero as the fermion mass were made smaller. The system
 would not
exhibit spontaneous chiral symmetry breaking; the vector and pseudoscalar mesons would
 not behave particularly
differently. In a system with only explicit chiral symmetry breaking,
the spectrum is  parity - doubled, in the $m_q \rightarrow 0$ limit,
so that one would observe approximate equality of the vector meson and
 axial vector meson
mass, and of the and scalar and pseudoscalar masses. 
If there was a nonzero string tension at nonzero fermion mass, one would expect that it would 
go to zero
with the fermion mass.

\begin{figure}
\begin{center}
\includegraphics[width=\columnwidth,clip]{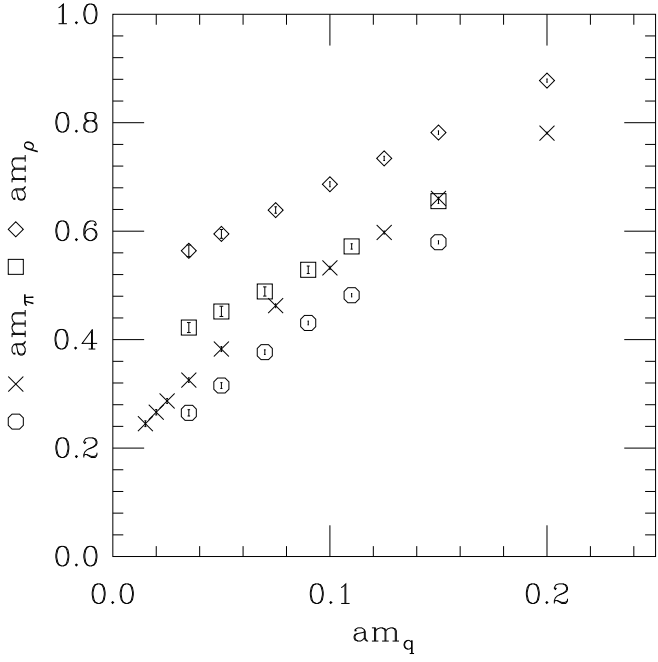}
\end{center}
\caption{Pseudoscalar and vector meson masses, in lattice units, from quenched $SU(3)$ gauge theory.
Octagons and squares are data from a weaker gauge coupling
 simulation; crosses and diamonds from stronger coupling.
 [This is raw data from \protect{\textcite{DeGrand:2003in}}.]
\label{fig:confined}}
\end{figure}

We can illustrate these differences with an example. Figure \ref{fig:confined} shows the
pseudoscalar mass and the vector meson mass, as a function of the
quark mass, from a typical calculation in quenched $SU(3)$, at 
each of  two different gauge couplings. 
All of the parameters are given in units of 
the lattice spacing $a$. The separation of the pseudoscalar and vector meson at small quark mass is apparent.
The data sets labeled by octagons and squares are collected at a smaller gauge coupling than the data sets
shown by crosses and diamonds. It is clear that if one uses the zero fermion mass limit of the vector mass
 to define the lattice spacing,
then weaker coupling corresponds to smaller lattice spacing.

\begin{figure}
\begin{center}
\includegraphics[width=\columnwidth,clip]{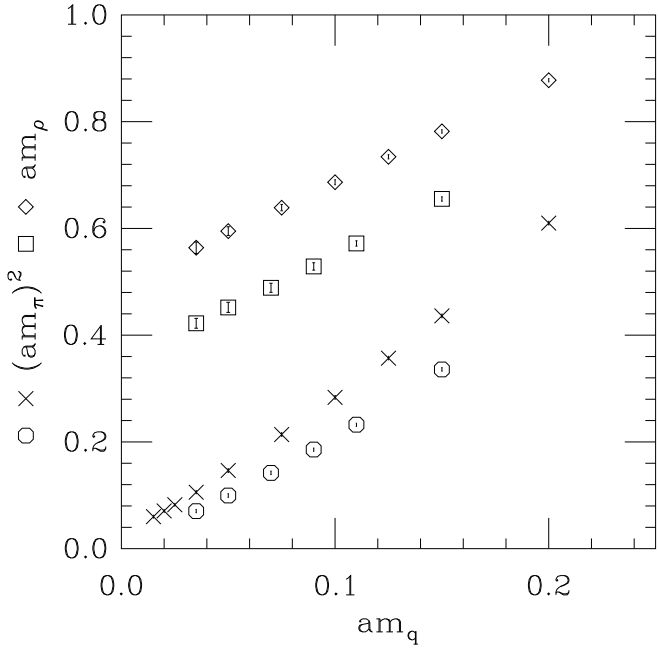}
\end{center}
\caption{Squared peudoscalar and vector meson masses, in lattice units, from quenched $SU(3)$ gauge theory.
Octagons and squares are data from a weaker gauge coupling
 simulation; crosses and diamonds from stronger coupling. The data is identical to 
what is shown in Fig.~\protect{\ref{fig:confined}}.]
\label{fig:confined2}}
\end{figure}

Figure \ref{fig:confined2} is a presentation of spectroscopy which is more common in the QCD literature.
The data is identical to Fig.~\ref{fig:confined}, but I am plotting
the squared pseudoscalar mass. Its approximate linearity is the qualitative signal that
chiral symmetry is broken, the Gell-Mann, Oakes, Renner formula in action.

Contrast this case with that of a data set from a simulation of $SU(2)$ gauge 
theory with $N_f=2$ adjoint representation fermions.
I have plotted data from \textcite{Bursa:2011ru}. Many other collaborations including
\textcite{Catterall:2007yx,Catterall:2009sb,Hietanen:2008mr} have similar results.
 The particular lattice system that was simulated had a strongly coupled phase which is
chirally broken and a weakly coupled phase which is almost certainly
conformal in the zero quark mass limit. Data from the strong coupling phase
qualitatively resembles that in Fig.~\ref{fig:confined}. But in weak coupling, shown in the figure,
the pseudoscalar and vector masses never separate, and while there is some dependence of
the bound state mass 
on bare gauge coupling, there is strong dependence on the fermion mass. 

\begin{figure}
\begin{center}
\includegraphics[width=\columnwidth,clip]{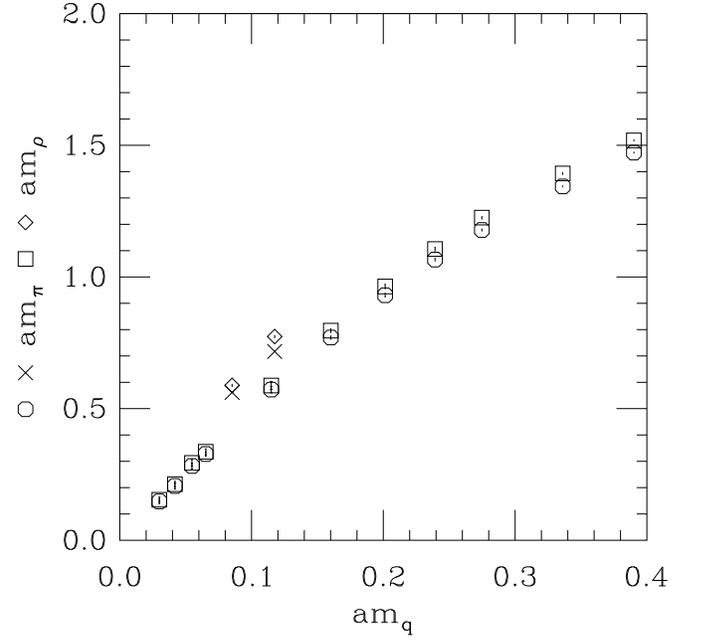}
\end{center}
\caption{Pseudoscalar and vector meson masses, in lattice units, from  $SU(2)$ gauge theory
coupled to $N_f=2$ adjoint fermions.
Octagons and squares are data from a weaker gauge coupling
 simulation; crosses and diamonds from stronger coupling.
 This is data from \protect{\textcite{Bursa:2011ru}}, a $12^3\times 24$  volume lattice at weaker coupling
and a $24^3\times 64$ volume at stronger coupling.
\label{fig:conformal}}
\end{figure}

Unfortunately, a data set may not be so clean-cut.

Recall, that first, at nonzero $m_q$, even a would - be conformal
system is ``ordinary,'' with a mass gap, regardless of what happens at
$m_q=0$.  In heavy quark systems the pseudoscalar and vector
 states are nearly degenerate.  If the fermion mass is too large, it may be impossible to distinguish
 systems which are trending conformal in the zero mass limit from confining ones.

The Gell-Mann, Oakes, Renner  dependence of pseudoscalar mass on fermion mass,
$m_\pi^2 \propto m_q$, which is a signal  of chiral symmetry breaking,
 is only the leading behavior in a chiral expansion.
Higher orders might be important.
With large $N_f$, one-loop chiral logs can be huge. 
For example, the non-analytic correction to the pseudoscalar decay constant  $f$
(for symmetry breaking pattern $SU(N_f)\times SU(N_f)\rightarrow SU(N_f)$) is
\bee
\frac{\delta f_{PS}(m)}{f} = \frac{N_f}{2} \left(\frac{m}{4\pi f}\right)^2 \ln \frac{m^2}{\Lambda^2} .
\ee
Such terms might overwhelm any naive (analytic) expansion of lattice data to the zero mass limit.

Finite volume affects spectroscopy in many ways. Imagine first that we do have a confining, 
chirally broken system in infinite volume.
When the volume is large, in the sense that $M_H L$ for any hadron mass $M_H$ and system size $L$,
is large, the dominant effect of finite volume is from pion loops. Instead of returning to 
the point of emission,
 the pions scatter ``around the world'' or ``off image charges.''
The typical situation is that finite volume effects go as $\exp(-m_\pi L)$.
When the observable in question has a chiral logarithm in its expansion, as in the case of $f_{PS}$
above, the coefficient of the logarithm is also the coefficient of the finite volume correction.
 If that coefficient is large,
there will be large finite volume corrections.

Next, suppose that  $m_\pi L <<1$ but $m_H L>>1$ for all other states, 
and $f_\pi L>>1$ as well.
This is called the ``epsilon regime.'' Symmetries cannot break in finite volume,
so the condensate $\Sigma(V)$ will fall from its infinite volume value, $\Sigma$,
 to zero with the quark mass. The relevant dimensionless variable is
$\zeta =m_q \Sigma V$ with $V=L^4$. If one is certain one is in this situation, one can use 
measurements of the condensate
and of correlators in various channels to extract chiral observables.
For example, the finite-volume condensate is
\bee
\Sigma(V) = \Sigma f(\zeta) \sim m_q\Sigma^2 V +\dots
\label{eq:fvs}
\ee
But suppose one is not certain? One might interpret the vanishing of $\Sigma(V)$ as evidence for infrared conformality.

 In between
the epsilon regime and the large volume ``p-regime'' there is another regime, the ``delta-regime,''
where the pseudoscalar correlator shows a rotor spectrum. Naive chiral behavior is once again absent.

And notice the repeated use of the words ``extrapolate to $m_q=0$.'' Even in QCD, that is 
a nontrivial task.

 Typically, a finite simulation volume can induce
 phase transitions in a lattice system.
 For an asymmetric box, with
 $N_t < N_x$, the short time direction implies a finite temperature,
 $T= 1/(aN_t)$.  Typically, simulations of confining systems in these asymmetric volumes will 
show a phase transition from a strong-coupling confined phase to a weak-coupling deconfined and 
chirally restored phase. The weak coupling phase is (usually) analytically connected to the Gaussian fixed point,
which is (usually) where we might want to tune, to take the continuum limit.
So, is the continuum limit confined?

 One way to test this is to
 vary $N_t$ and see if the transition moves in bare parameter space.
If it moves to weaker coupling, in a way that $T$ remains roughly constant in
physical units, the transition is, most likely, a finite temperature transition, and the zero temperature phase
is likely to remain confined, while still
analytically connected to the Gaussian fixed point. If the transition remains fixed in bare coupling
as the lattice volume is varied, and the system  is deconfined on the weak coupling side,
the transition is a ``bulk transition.'' The
 weak coupling phase is analytically
connected to the Gaussian fixed point and the system has a
 continuum limit that is deconfined. So there is a test: does the deconfinement
 transition move appropriately with $N_t$?
 
 The problem with this test is that even a bulk transition moves a bit
when the volume is small. And there is another problem: how can one tell
 that the motion is consistent with a finite 
temperature transition, anyway? Often, one imagines analyzing a formula like Eq.~\ref{eq:AFrunning},
where $M_H$ is $T_c$ and $g^2(\Lambda)$ is the bare lattice gauge coupling. But typically,
one is simulating at strong coupling, and 
asymptotic freedom does not work well as a descriptor of physics at strong coupling.
One really has to compute the value of some other observable $M$ with the dimensions of a mass,
typically in a zero temperature simulation
at the same bare parameters, and
look for variation of $T/M$ as $N_t$ is varied. It is Eq.~\ref{eq:SCALING} all over again.
 (This is how the deconfinement or chiral restoration
 crossover temperature in QCD
is determined.)
This is rapidly becoming an expensive proposition.

Inside the conformal window, and in infinite volume, tuning
 the  fermion mass $m_q$ to zero causes the correlation length to diverge algebraically,
 as in Eq.~\ref{eq:corrlen}. One might hope to use this functional dependence
 as a diagnostic.
However, no simulation is ever done in infinite volume. The system size $L$ is
 also a relevant parameter since
the correlation length only diverges in the $1/L\rightarrow 0$ limit.
When the correlation length measured in a system of size $L$
(call it $\xi_L$) becomes comparable to $L$, $\xi_L$ will saturate at $L$ even as $m_q$
vanishes. Equivalently, bound state masses  will become independent of the fermion mass when it is small.
This is what non-Goldstone excitations are expected to do in a confining system.
Again, the finite volume can induce confusion between  ``confining'' and ``infrared conformal'' behavior.
Simulations with the same bare parameters, but at several volumes,
are needed to sort out this behavior. We will return to a detailed description of the necessary analysis
in Sec.~\ref{sec:anomalous}.

Regardless of whether the data looks confining or looks conformal, for a definitive answer,
one needs simulations at many values of the bare coupling. This is quite similar to the case
in precision QCD, where one has to extrapolate to $a=0$ to produce a cutoff - free number. But
it is expensive. In QCD, when one is doing something new, one might attempt
to simulate at one value of $a$ (or perhaps one value of the bare gauge coupling and 
several quark masses) and to present results with the claim that the lattice spacing is small enough,
and the lattice volume is large enough, that 
at the level of accuracy, only small quantitative changes in the numbers are expected.
But this is for a system whose gross behavior is reasonably well understood.
When the properties of the system are unknown,
 making inferences based on data from one bare gauge coupling is risky.

\subsection{Running coupling constants from observables\label{sec:running}}

A  running coupling constant is determined in a lattice simulation by measuring some observable
that has a perturbative expansion (that gives the coupling), at some convenient length scale,
such that variation of the observable with the length scale gives the running.

In QCD, the physical problem whose solution is desired is ``What is $\alpha_s$ at the Z-pole?''
This is a needed ingredient in precision tests of the Standard Model. In the  lattice calculations
 reported in the Review of Particle Properties \cite{Agashe:2014kda},
 the length scale is taken from some
low energy observable computed on the lattice, and the running to the $Z$ uses perturbation theory,
rather than treating the running as something to be determined.
However, there are lattice techniques that (at least for QCD) are capable of doing the running
completely nonperturbatively. These are the ones that have been adapted to beyond Standard Model systems.

Recall that generic beyond Standard Model candidates have two potentially relevant couplings, 
a fermion mass and the gauge coupling.
Coupling constant evolution takes place in (at least) a two dimensional space. This is difficult
 to deal with in a simulation,
and so lattice studies typically begin by setting one of the relevant couplings to zero. This coupling is usually
the fermion mass, and the theorist considers the gauge coupling in isolation.
 This is (often) not an easy place to do simulations.
At a minimum, one must use boundary conditions for which the massless Dirac operator is invertible.

 With an action with
 good enough chiral properties (staggered, domain wall, or overlap fermions) the massless limit is simply achieved,
 by setting the bare 
fermion mass to zero. With Wilson fermions, one must tune the bare mass so that a derived mass is zero.
The fermion mass
whose vanishing signals
the chiral limit is the  so-called Axial Ward Identity (AWI) fermion mass, defined through the ratio
of correlators
\begin{equation}
am_q=\frac12\,\frac{\partial_4\svev{A^b_4(t)\co^b(0)}}{\svev{P^b(t)\co^b(0)}}.
\label{eq:AWI}
\end{equation}
Here $A^b_4(t)=\bar\psi\gamma_5\gamma_4\tau^b\psi$ is the time component of the
local axial vector current with flavor $b$, taken at zero spatial momentum on
the time slice $t$; $P^b(t)$ is the local pseudoscalar density.
The operator $\co^b(0)$ is a source.  Any calculation that must be done at $am_q=0$ is carried out along a
 line called the critical kappa line,
$\kappa_c(\beta$) vs $\beta$.

However the coupling is determined, it must be analyzed. To contrast the issues encountered
while studying slowly running versus quickly running couplings, it is instructive to return to
Figures \ref{fig:bfnpert3}-\ref{fig:bfnpert12}. The first figure shows the situation for a quickly running
 coupling; the second figure is the 
situation with twelve fundamentals, and slow running.
The two problems in data analysis are to determine the shape of the $1/g^2(L)$ versus $\ln L_0/L$
curves, and to show that the determination is free of cutoff effects.

The key to doing this is to take $L$ as the independent variable. $g^2(L)$ is the coupling defined at
scale $L$. Interpret the $L_0$ in $\ln L_0/L$ as the cutoff. In a lattice calculation, $L_0$ is the 
lattice spacing $a$.
$L_0/L = 1/N$ the number of lattice points the simulation length is divided into.
Do a simulation at some value of the bare parameters. This gives a point in the figure. Now change
$L_0$, holding the bare simulation parameters fixed. In a perfect world, we might imagine doing this infinitesimally.
The running coupling will shift along the solid lines in the figure.

Next, change $L$ to some new value $L'$, and change the bare parameters.
 Tune them, until $g^2(L')=g^2(L)$.
These couplings are connected by the dotted line in Fig.~\ref{fig:bfnpert3}.
 Now the question is,
how does the slope of the line change as $L$ is changed. Or,
what are the slopes of the two curves?  Are they different? And if they are different,
can they be used to extrapolate to  tiny $L_0$?

 In Fig.~\ref{fig:bfnpert3}, the slopes are
 the same, by construction. But the picture also shows that
 when the beta function is large, it is easy to shift the bare coupling by
a large amount and match renormalized couplings at a dense set of $L$'s.
In a real simulation, the line is replaced by a set of points at (integer) $L$'s. The slope
is typically replaced by measurements at two $L$ values related by a common scale change:
$L=6-12$, $L=8-16$, and so on. One would then have a set of measurements of the slope
as a function of $g^2(L)$, at many values of $L_0$. One could then proceed to an 
extrapolation to the continuum limit.

This shifting and matching is reasonably straightforward to perform when the coupling
runs quickly, as in  Fig.~\ref{fig:bfnpert3}. However, look at Fig.~\ref{fig:bfnpert12}.
We sit, say, at $\ln L_0/L=2$ and tune bare parameters so that
 $1/g^2(L)=0.17$ (the right edge of the dotted line).
 Now we change
the bare parameters by some amount and try to reacquire the
 same value of $g^2$ at some other $L$.
When the coupling runs slowly, matching $g^2(L')$ to the fiducial
 $g^2(L)$ by shifting the bare
parameters requires an enormous
change in $L$; the change diverges as we move to the fixed point. Computer resources 
are finite,
and at some point one can no longer support the necessary $L'/L$ ratio
 in a set of simulations.

One can imagine shrinking the ratio by reducing the size of the shift in couplings.
(That is, the lines in Fig.~\ref{fig:bfnpert12} are spaced $\Delta 1/g_0^2=0.2$ apart;
 reduce the shift
to 0.1 and try again.) Now the problem is statistics.
In my experience (which is limited to the Schr\"odinger functional, to be described below),
the uncertainty in a $1/g^2(L)$ measurement is not too dependent on the slope of the line,
or even on $N_c$ or $N_f$, so the intrinsic {\it fractional} uncertainty on the slope,
 from the
difference of two $1/g^2(L)$'s, scales inversely with the slope. This is not a favorable
result for a slowly running theory, for if the slope cannot be measured,
 the change in the slope also cannot be measured.
Clearly, a less noisy coupling will allow one to take a
 smaller interval of $L$ by reducing
$\Delta(1/g^2)$, but as one approaches the critical coupling, the slope of the line
will vanish regardless of definition.

Finally, whatever method is used to measure a running coupling constant, it is important to
check it by collecting data in weak coupling, to validate the method against an analytic result.
The goal is to see one-loop or two
loop running. I, personally, do not know how to evaluate results I see in the literature which do not
have such anchor points.

Two methods dominate in lattice calculations of running coupling constants. The 
Schr\"odinger functional is the older
of the two. More recent calculations tend to use variations on a method called ``Wilson flow.''

\subsubsection{ Schr\"odinger functional}
The Schr\"odinger functional (SF)~\cite{Luscher:1992an,Luscher:1993gh,Sint:1995ch,Jansen:1998mx,DellaMorte:2004bc}
 is an implementation of the background field method that is especially suited for  lattice
 calculations.
It is done by performing simulations in a finite volume of linear
dimension $L$, while imposing fixed boundary conditions on the gauge field (at Euclidean times $t=0$ and $t=L$).
 The usual partition function $Z = \Tr \exp(- LH)$ is replaced by
the ``Schr\'odinger functional'' $Z(\phi_b,\phi_a) = \svev{\phi_b| \exp(-LH) | \phi_a}$.
This fixing involves a free parameter $\eta$, so call the Schr\'odinger functional $Z(\eta)$.
A coupling constant is defined through the variation of the effective action $\Gamma$ (which in turn is defined as
$\Gamma= - \ln Z(\eta)$).
The classical field that minimizes the Yang--Mills action
subject to these boundary conditions is a background color-electric field.
By construction the only distance scale that characterizes the background field is $L$, so
$\Gamma$ gives the running coupling via
\bee
\label{eq:Gamma}
\Gamma = g(L)^{-2} S_{\textrm{YM}}^{\textrm{cl}} ,
\ee
where
$S_{\textrm{YM}}^{\textrm{cl}}$
is the classical action of the background field.
When $\Gamma$ is calculated non-perturbatively, Eq.~\ref{eq:Gamma} gives a non-perturbative
definition of the running coupling at scale $L$.
In a simulation, the coupling constant is determined through differentiation,
\begin{eqnarray}
  \left.\frac{\partial \Gamma}{\partial\eta} \right|_{\eta=0}
  &=&
  \left.\svev{\frac{\partial S_{YM}}{\partial\eta}
  -\frac{N_f}{2} \tr \left( \frac{1}{D_F D_F^\dagger}\;
        \frac{\partial (D_F^\dagger D_F)}{\partial\eta}\;
            \right)}\right|_{\eta=0}  \label{eq:sfa}\\
&\equiv & \frac{K}{g^2(L)} .
            \label{deta}
\end{eqnarray}
$D_F$ is the lattice Dirac operator.
The constant $K$ is chosen to match to a perturbative evaluation of Eq.~\ref{eq:sfa}.
In words, the expectation value $\svev{\dots}$ gives $g^2(L)$.

By calculating the inverse running coupling
on lattices of size $L$ and $sL$, we obtain the discrete beta function (DBF)
\bee
B(u,s) = \frac1{g^2(sL)}-\frac1{g^2(L)}, \qquad u \equiv \frac1{g^2(L)}\ .
\label{DBF}
\ee
It is necessary to deal with lattice artifacts in $B(u,s)$. This is often done by
comparing data from systems at fixed aspect ratio $s$, for example,
$L=6$ and 12, 8 and 16, 12 and 24.

With the definition of the beta function for the inverse coupling in terms of the usual beta function
\bee
  \tbeta(1/g^2) \equiv \frac{d(1/g^2)}{d\ln L}
  = 2\beta(g^2)/g^4 = 2u^2 \beta(1/u) ,
\label{invbeta}
\ee
the discrete beta function is
\bee
\ln s = \int_{L}^{sL} \frac{dL'}{L'} =
\int_{u}^{u+B(u,s)}\frac{du'}{\tbeta(u')}\ .
\label{DBF2}
\ee

%%%%%%%%%%%%%%%%%%%%%%%%%%%%%%%%%%%%%%%%%%%%%%%%%%%%%%%%%%%%%%%%%%%%%
\begin{figure}
\begin{center}
\includegraphics[width=\columnwidth,clip]{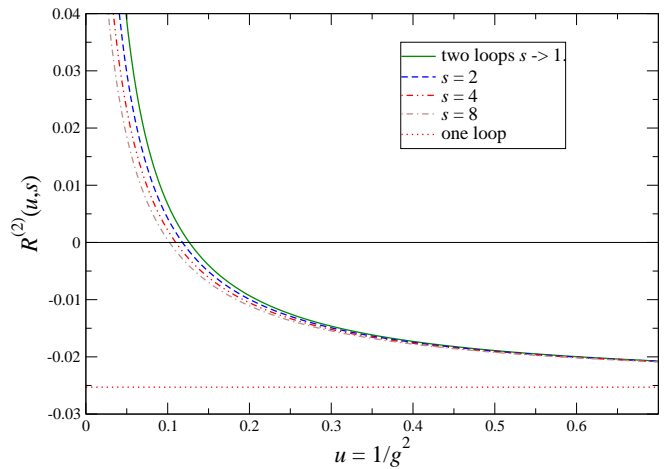}
\end{center}
\caption{Rescaled discrete beta function for $SU(2)$ gauge theory with $N_f=2$ adjoints,
calculated in two loops for various scaling factors $s$.
Also shown are the one- and two-loop beta functions;
the rescaled DBF approaches the two-loop beta function when $s\to1$.
Top to bottom, the curves are in the order shown in the legend.
The figure is taken from \protect{\textcite{DeGrand:2011qd}}.
\label{fig:bfn}}
\end{figure}

The literature is often careful to distinguish between the DBF
and the usual beta function. 
For a quickly running system like QCD, it is necessary to do this.
But in a slowly running system
 the DBF's we can measure are, to high accuracy,
just proportional to the beta function itself. This occurs
because the coupling runs slowly and because the values of $s$ accessible
 in a simulation are small.
In that case
the \textit{rescaled} DBF, defined as
\bee
R(u,s)=\frac{B(u,s)}{\ln s}\ ,
\label{RDBF}
\ee
will be approximately equal to the beta function $\tbeta(u)$.
The situation for $SU(2)$ with $N_f=2$ adjoints  is illustrated in Fig.~\ref{fig:bfn}.
The figure shows the two-loop result,
\begin{eqnarray}
R^{(2)}(u,s) &=& -\frac{2b_1}{16\pi^2}- \frac{b_2}{16\pi^2b_1}
\nonumber\\
&&\times\frac{\ln\left[1+(2b_1/16
\pi^2) u^{-1} \ln s\right]}{\ln s}\ ,
\label{2loopggen}
\end{eqnarray}
for the rescaled DBF for scale factor $s=2$, 4, 8, compared to the
one-loop and two-loop beta functions.
The rescaled DBF for $s=2$ is hardly distinguishable from the beta function.

There are two lessons to be drawn from Fig.~\ref{fig:bfn}.
If the actual DBF resembles the two-loop result,
we can combine the rescaled DBF's for many scale factors $s$ onto a single plot
to give a good approximation to the actual beta function.
Furthermore, since any value of $s\alt2$ is as good as another,
we can combine the couplings for all lattice volumes studied to extract the
beta function via a fit. Most of the scaling violations
will be at the smallest $a/L$, so we can simply look at the largest $L$ data points.
With slow running, one is really asking whether the slope of the $1/g^2(L)$ versus $\ln L$
line varies with $L$.

\begin{figure}
\begin{center}
\includegraphics[width=\columnwidth,clip]{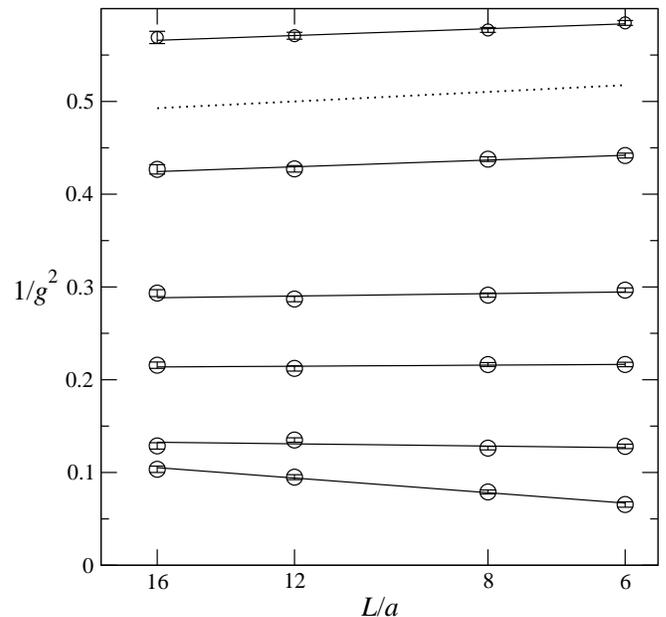}
\end{center}
\caption{SF coupling $1/g^2$ \textit{vs.}~$L/a$ (plotted on a logarithmic scale),
 for $SU(2)$ gauge theory coupled to $N_f=2$ adjoint fermions, from {\protect{\textcite{DeGrand:2011qd}}}.
Data are at lattice gauge coupling (from the top) $\beta=3.0$, 2.5, 2.0, 1.75,  1.5, and~1.4.
The lines through the data points are fits to the
data at each $\beta$ of the form $1/g^2(L)=a+b\ln (L/a)$.
The dotted line has the slope $2b_1/(16\pi^2)$ as given by the lowest-order beta function,
Eq.~(\protect{\ref{eq:2loopbeta2}}).
\label{fig:1g2}}
\end{figure}

An example of a plot of $1/g^2(L)$ versus $\ln L$ is shown in  Fig.~\ref{fig:1g2}.  It is
for the case of
 $SU(2)$ gauge theory coupled to $N_f=2$ adjoint fermions, from \textcite{DeGrand:2011qd}.
The slope changes sign. This is the clearest example of
IRFP behavior from a Schr\"odinger functional analysis, that I know. The picture can also
be used to illustrate various ways of dealing with lattice artifacts: different methods amount to computing
the slope of each line by taking different mixes of $L$ values. For example, one could
 compare the slope from $L$'s of fixed ratio, or from the whole line, or by dropping
 data points at smaller $L$'s.

There are studies of alternative choices of boundary conditions of the Schr\"odinger functional,
with the idea of finding a set with reduced lattice artifacts \cite{Karavirta:2012qd,Sint:2012ae,Sint:2011gv}.
Typically, this is done using perturbation theory.
The issue with using them for slowly-running systems near the bottom of the conformal window
is that the place where one really wants to simulate (typically, looking for
a zero of a beta function) is at strong coupling. There, perturbation theory is unreliable.
 Choosing a functional form to extrapolate to zero cutoff that includes 
lattice artifacts is, at best,
phenomenology.

\subsubsection{ ``Flow''}
The new alternative goes by names such as ``gradient flow'' or ``Wilson flow.''  It is a smoothing
method for gauge fields achieved by diffusion in a fictitious (fifth dimensional) time $t$.
 In the continuum version, a
smooth gauge field $B_{t,\mu}$ is defined in terms of the original gauge field $A_\mu$ through
an iterative process
\beea
\partial_t B_{t,\mu} &=& D_{t,\mu}B_{t,\mu\nu}  \nonumber \\
B_{t,\mu\nu} &=& \partial_\mu B_{t,\nu} - \partial_\nu B_{t,\mu} + [B_{t,\mu},B_{t,\nu}] , \nonumber \\
\label{eq:flow}
\eea
where the smoothed field begins as the original one,
\bee
B_{0,\mu}(x)=A_\mu(x).
\ee
Correlators of the flow field can be used to define a coupling constant \cite{Fodor:2012td}. For example,
 one possibility, due to    \textcite{Luscher:2010iy}, is
\bee
\svev{E(t)} = \frac{1}{4}\svev{G_{\mu\nu}(t)G_{\mu\nu}(t)} = N_c \frac{g^2}{t^2} + O(g^4).
\ee
This can be used to define a renormalized coupling at a scale $t$,
\bee
g^2_{flow}(t) = \frac{t^2 \svev{E(t)}}{N_c}.
\label{eq:flowg2}
\ee
 Simulations in a box of size $L$ set the overall scale,
and the second scale, $t$, is taken to be a fixed fraction of $L$.
The method has many variations. For example, the spatial averaging term in the diffusion equation could be 
identical to, or different from, the discretized gradient term in the action which is simulated.
Flow can be combined with the Schr\"odinger functional \cite{Fritzsch:2013je}, or 
can be used by itself to define a coupling constant.

\begin{figure}
\begin{center}
\includegraphics[width=\columnwidth,clip]{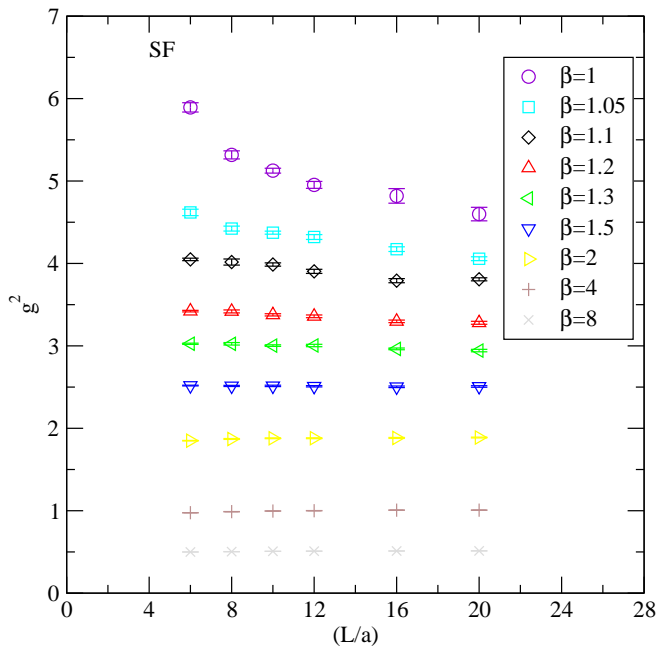}
\end{center}
\caption{Continuum extrapolation of the Schr\"odinger functional
step scaling function for $SU(2)$ with $N_f=2$ adjoints, from \protect{\textcite{Rantaharju:2013bva}}.
\label{fig:SU2sf}}
\end{figure}

\begin{figure}
\begin{center}
\includegraphics[width=\columnwidth,clip]{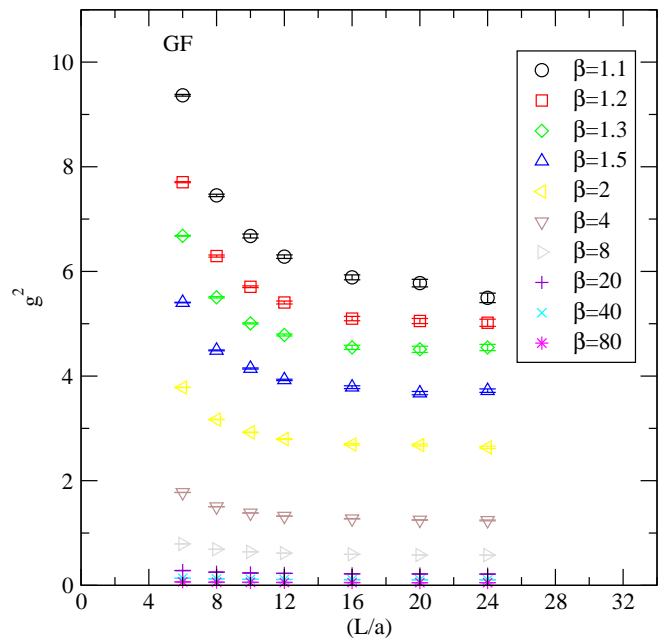}
\end{center}
\caption{Continuum extrapolation of the gradient flow
step scaling function for $SU(2)$ with $N_f=2$ adjoints, from \protect{\textcite{Rantaharju:2013bva}}.
\label{fig:SU2gf}}
\end{figure}

People who have used it report that they can compute a coupling constant with much smaller errors
than a Schr\"odinger functional calculation would give with equivalent statistics. Since the choice of $t$
defines its own coupling, it is possible to collect simultaneous data for different definitions of couplings
$g^2_{flow}(t)$ and select the best one (by some criterion) later.
Discretization errors must still be removed along the lines previously described.
Recently, \textcite{Rantaharju:2013bva} compared the Schr\"odinger functional coupling
 to gradient flow in $SU(2)$ with $N_f=2$ adjoints.
 Here there was an issue
with the simplest version of a flow running coupling: discretization errors were observed to
 be larger than for the
Schr\"odinger functional.
 I reproduce his figures in Figs.~\ref{fig:SU2sf}-\ref{fig:SU2gf}.

These pictures are only the beginning of a presently ongoing
research area, studying how to suppress lattice artifacts in measurements with flow.
Tree level improvement is described by \textcite{Fodor:2014cpa} and
 (in a preliminary version) by \textcite{Ramos:2014kka}.
However (as for the Sch\"rodinger functional) the theoretical analysis assumes 
closeness to free field behavior.
Fixed points for interesting slowly running systems occur in strong coupling
 (if at all), and dealing with lattice artifacts
in strong coupling will, I think, always be phenomenological.

%%%%%%%%%%%%%%%%%%%%%%%%%%%%%%%%%%%%%%%%%%%%%%%%%%%%%%%%%%%%%%%%%%%%%
%%%%%%%%%%%%%%%%%%%%%%%%%%%%%%%%%%%%%%%%%%%%%%%%%%%%%%%%%%%%%%%%%%%%%

\subsubsection{ Monte Carlo Renormalization Group}

Another approach, called  ``Monte Carlo Renormalization Group'' (MCRG),
 is an implementation of the real space renormalization group.
Take a system defined with a momentum space cutoff $\Lambda$ (or a lattice spacing $a$) and
some set of dynamical variables $U$.  Introduce some averaging algorithm which replaces the fine
 grained $U$'s with some coarse grained $V$'s.
Now
define a system with a smaller $\Lambda'$ or a bigger lattice spacing, by
integrating out the $U$'s, to give a partition function expressed in terms of the coarse-grained variables and their
action $S'(V)$:
\beea
Z &=& \int DU e^{- S(U)} \nonumber \\
&=& \int dV T(U,V) \int dU  e^{- S(U)} \nonumber \\
&\equiv &       \int dV     e^{- S'(V)}              .   \nonumber \\    \label{eq:mcrg}
\eea
Repeat this procedure.
The change in the action is encoded in a set of transformation rules for the coupling constants
in the action $\{K_1,K_2,K_3\dots\} \rightarrow \{K'_1,K'_2,K'_3\dots\}$.
In a lattice system, the range of the averaging of $U$'s to $V$ gives the scale change $s$,
so that we can speak of the coupling constants running over a (discrete) scale $s$.
How the couplings run depends on the choice of blocking kernel $T(U,V)$;
this is the analog of a renormalization scheme.
 As the system is repeatedly blocked,
the (infrared) irrelevant couplings die away leaving the marginal and relevant ones. These couplings
will approach a unique renormalized trajectory emanating from the critical surface.
Different bare couplings begin at different places but end upon the renormalized trajectory.

The issue now is, how to measure the couplings. In the ``two-lattice matching MCRG method''
this is done indirectly, through observables. The idea is that if observables are measured in two different
lattice simulations, and if all the observables have identical expectation values, than 
the systems are identical, so their coupling constants are matched.
Now imagine two systems with different $K$'s. Take one system 
and perform $n$ blocking steps so that the cutoff is reduced by a factor $s^n$. Measure many observables.
Next, suppose that a second system, with its own set of $K$'s, is blocked, and suppose that after $n-1$ steps
its observables coincide exactly with those of the first system (and remains identical under 
further blocking).
 We would say that when the bare couplings flow from 
 $\{K_1,K_2,K_3\dots\} \rightarrow \{K'_1,K'_2,K'_3\dots\}$, long distance physics is unchanged
under a scale factor $s$. This is a renormalization group equation for the bare parameters.

\begin{figure}
\includegraphics[width=\columnwidth,clip]{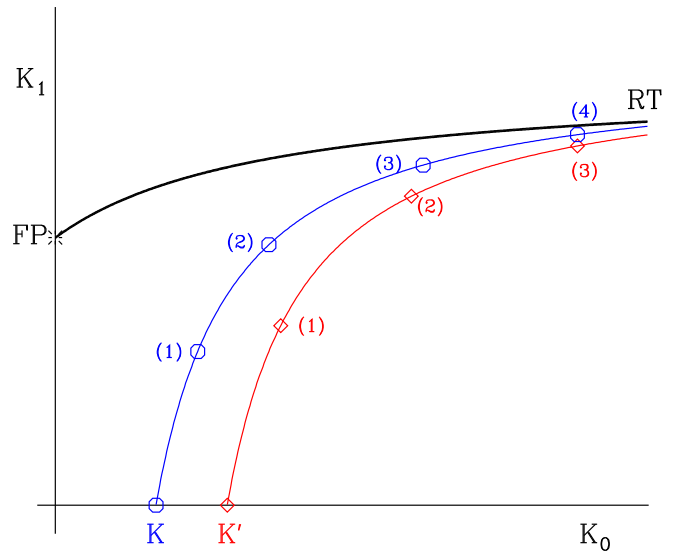}
\caption{Sketch (from \protect{\textcite{Hasenfratz:2009ea}}) of the RG flow around a fixed point (FP)
 with one relevant operator. The coupling pair $(K,K')$
indicates matched couplings whose correlation length differ by a factor of $s$.
The line labeled ``RT'' is the renormalized trajectory.
 \label{fig:RGflow}}
\end{figure}

So, the rubric is:
\begin{enumerate}
\item Generate a first configuration ensemble of size $L^{d}$ with action $S(K)$.
Block each configuration $n$ times and measure a set of expectation
values on the resulting $(L/s^{n})^{d}$ set.
\item Generate a second ensemble of configurations of size $(L/s)^{d}$ with action $S(K')$.
 Block each configuration $n-1$
times and measure the same expectation values on the resulting $(L/s^{n})^{d}$
set. Compare the results with that obtained in step 1. and tune the
coupling $K'$ such that the expectation values agree. A cartoon is shown in Fig.~\ref{fig:RGflow}.
\end{enumerate}

The method has many good features: one can use smallish lattices and measurements of local operators
usually can be done accurately. It has some not so good features: the location of the fixed point,
the renormalized trajectory, and the number of steps needed to reach the renormalized trajectory
all depend on the choice of action and of blocking kernel.  Of course, it is  advantageous to be
able to tune $T(U,V)$.  The analysis  is much easier when there is only
one relevant variable (for example, in pure gauge theory, the gauge coupling) than 
when there might be more than one (typically, the mass and perhaps the gauge coupling).

Early references to these methods for spin models  are \textcite{Swendsen:1979gn,Swendsen:1984vu} and for QCD, 
\textcite{Bowler:1984hv,Hasenfratz:1984bx,Hasenfratz:1984hx}.
The use of these methods for slowly running theories was revived by 
\textcite{Hasenfratz:2009ea,Hasenfratz:2010fi,Hasenfratz:2011xn}. Most of her work was 
on $SU(3)$ with 8 and 12 flavors of fundamentals. Results will be discussed below.

%%%%%%%%%%%%%%%%%%%%%%%%%%%%%%%%

A number of other possibilities for renormalized couplings have been proposed;
 none has a long citation trail.
One worth mentioning is a technique \cite{deDivitiis:1993hj} that defines a coupling through the
correlation of Polyakov loops, measured over distances that are a fixed fraction of the lattice size.
This has been used by \textcite{Lin:2012iw}
for many-flavor studies in $SU(3)$.

\subsection{Computing the mass anomalous dimension $\gamma_m$ \label{sec:anomalous}}

\subsubsection{ Schr\"odinger functional}

The Schr\"odinger functional gives $\gamma_m$ through the
 volume dependence of the renormalization factor $Z_P$ of the
isovector pseudoscalar density $P^a=\bar\psi\gamma_5(\tau^a/2)\psi$.
(The pseudoscalar density is related by a chiral rotation to $\bar\psi\psi$,
which is the object of interest.) It is computed from two correlators
via~\cite{Sint:1998iq,Capitani:1998mq,DellaMorte:2005kg,Bursa:2009we}
\bee
Z_P = \frac {c \sqrt{f_1}}{f_P(L/2)}.
\label{eq:ZP}
\ee
$f_P$ is the propagator from the $t=0$ boundary to a point pseudoscalar
operator at time $x_0$,
\beea
  f_P(x_0)&=&-\frac{1}{3}\sum_a \int d^3y\, d^3z\,  \left\langle
  \overline{\psi}(x_0)\gamma_5\frac{\tau^a}{2}\psi(x_0)\right.\nonumber\\
  &&\times\left.\overline{\zeta}(y)\gamma_5\frac{\tau^a}{2}\zeta(z)
  \right\rangle.
  \label{eq:fPdef}
\eea
It is conventional to take $x_0=L/2$. In the expression,  $\zeta$ and $\bar\zeta$ are
gauge-invariant wall sources at $t=a$, i.~e., one lattice layer away from
the $t=0$ boundary.  The $f_1$ factor is the boundary-to-boundary correlator,
which cancels the normalization of the wall source. Explicitly, it is
\beea
  f_1&=&-\frac{1}{3L^6}\sum_a \int d^3u\, d^3v\, d^3y\, d^3z\,
  \left\langle
  \overline{\zeta}^\prime(u)\gamma_5 \frac{\tau^a}{2}{\zeta}^\prime(v)
  \right.\nonumber\\
  &&\times\left.\overline{\zeta}(y)\gamma_5\frac{\tau^a}{2}\zeta(z)
  \right\rangle,
  \label{eq:f1def}
\eea
and $\zeta'$ and $\bar\zeta'$ are wall sources at $t=L-a$.

\begin{figure}
\begin{center}
\includegraphics[width=\columnwidth,clip]{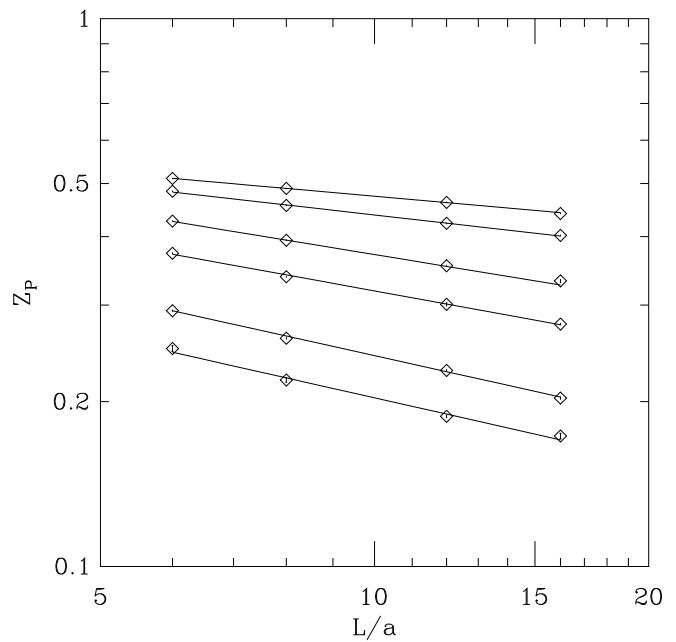}
\end{center}
\caption{Pseudoscalar renormalization constant $Z_P$, from \protect{\textcite{DeGrand:2011qd}}.
From the top, data are from $\beta=3.0$, 2.5, 2.0, 1.75, 1.5 and~1.4.
Lines are fits to
$\ln Z_P(L) = -\gamma_m\ln (L/a) + \text{const}$ for each $\beta$.
\label{fig:zp}}
\end{figure}

The (continuum) mass step scaling
function~\cite{Sint:1998iq,Capitani:1998mq,DellaMorte:2005kg,Bursa:2009we} is
\bee
  \label{eq:sigma_p}
  \sigma_P(v,s) = \left. {\frac{Z_P(sL)}{Z_P(L)}}
  \right|_{g^2(L)=v}.
\ee
It is related to the mass anomalous dimension via
\bee
\label{eq:sigPgamma}
  \sigma_P(v,s) = \exp\left[-\int_1^s \frac {dt}{t}\,
  \gamma_m\left(g^2(tL)\right)\right] .
\ee
When the SF coupling $g^2(L)$ runs  slowly,
Eq.~\ref{eq:sigPgamma} is well approximated by
\bee
\sigma_P(g^2,s) = s^{ - \gamma_m(g^2)}.
  \label{eq:gamma}
\ee
We can therefore combine many $sL$ values collected at the same bare parameter
values into one fit function giving $\gamma_m$,
\bee
\ln Z_P(L)=-\gamma_m \ln L +\text{const}.
  \label{eq:lgamma}
\ee
An example of data for $Z_P$ is shown in Fig.~\ref{fig:zp}.
%%%%%%%%%%%%%%%%%%%%%%%%%%%%%%%%%%%%%%%%%%%%%%%%%%%%%%%%%%%%%%%%%%%%%
%%%%%%%%%%%%%%%%%%%%%%%%%%%%%%%%%%%%%%%%%%%%%%%%%%%%%%%%%%%%%%%%%%%%%

As in the case of the running coupling, the question is whether the slope of the line changes
with $L$, and what its value is at large $L$. This can be done either by comparing
the slope from pairs of points at fixed $s$, or of the whole line. Again, there are many possibilities.

\subsubsection{Finite size scaling}

\begin{figure}
\begin{center}
\includegraphics[width=\columnwidth,clip]{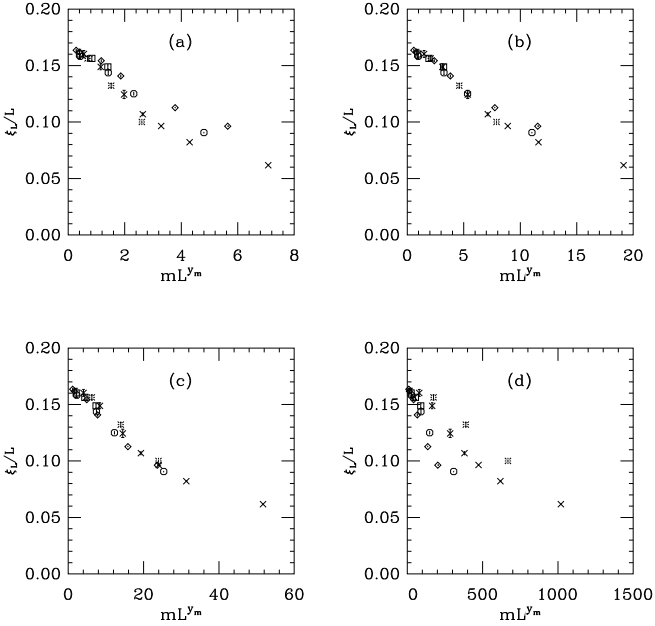}
\end{center}
\caption{
Curve collapse in $SU(3)$ gauge theory with $N_f=2$ symmetric-representation fermions, from 
{\protect{\textcite{DeGrand:2009hu}}}.
Plots of $\xi_L/L$ vs $m_q L^{y_m}$ at $\beta=5.2$ for four choices of $y_m$:
(a) $y_m=1.0$, (b) $y_m=1.4$, (c) $y_m=1.8$ (d) $y_m=3.0$.
Plotting symbols are for different simulation volumes,
diamonds, $12^3\times 6$ ($L=6$);
octagons, $12^3\times 8$ ($L=8$);
squares, $16^3\times 8$ ($L=8$);
crosses, $12^4$ ($L=12$);
bursts, $16^4$ ($L=16$). Curve collapse seems to be best in panels (b) and (c).
\label{fig:alltest}}
\end{figure}

Recall that the correlation length  $\xi$ of an infrared conformal system would diverge
 as the fermion mass $m_q$   were taken to zero, but the finite system
size $L$ prevents it.
If the only large length scales in the problem are $\xi$ and $L$,
then  observables can only involve the scales $\xi$ and $L$, and their ratio.
This ``finite size scaling'' argument says that the correlation length in finite volume $\xi_L$
must scale as
\bee
\xi_L = L F(\xi/L)
\label{eq:fss1}
\ee
where $F(x)$ is some unknown function of $\xi/L$. A somewhat more useful version of this
relation invokes Eq.~\ref{eq:corrlen}, to say 
\bee
\xi_L = L f(L^{y_m} m_q)   .
\label{eq:fss2}
\ee
This expression can be used to find the exponent $y_m$. One
 can plot $\xi_L/L$ vs $L^{y_m} m_q$ for many $L$'s, and vary $y_m$. Under this variation,
 data from different $L$'s will march across the $x$ axis
at different rates. The exponent can be determined by tuning $y_m$ to collapse
the data onto a single curve. An example of such an analysis is shown in Fig.~\ref{fig:alltest}.
It is for $SU(3)$ gauge theory and $N_f=2$ symmetric representation fermions,  by \textcite{DeGrand:2009hu}.

Often, it is unknown whether the system under investigation is infrared conformal, or not.
A comparison  of its data with Eq.~\ref{eq:fss2} is used to decide the question. This could
be misleading: a coupling which runs so slowly that it scarcely changes over the range
of available $L$'s would induce effectively conformal behavior.

Many finite size scaling analyses of lattice data replace curve collapse with a fit to some
functional form for $F(x)$. The shape of $F$ is known for extreme values of its argument.
For example, in Eq.~\ref{eq:fss1}, $F(x) \sim x$ for small $x$ and $F(x) \sim 1$ for large $x$.
Fitting to a curve allows one to quote a goodness - of - fit parameter (such as a chi-squared) along with the fit
 value of $y_m$.
 The problem with this is that,
generally, the complete functional form of $F(x)$ or $f(x)$ is unknown.
A poor fit could occur because the guessed functional dependence of $f(x)$ was incorrect.
Sometimes, one can fit the scaling functions to high quality Monte Carlo data from one model
which is a member of its universality class, and use those fits to test whether other systems
lie in that class. An example of this analysis is that of \textcite{Engels:2011km}, who fit scaling functions
of the three-dimensional $O(4)$ spin model with the aim of  making comparisons with $N_f=2$ QCD
near its chiral transition. One should also keep in mind
that different quantities have their own scaling functions. A fit to lattice data for (say) the 
pseudoscalar mass, the vector mass, and $f_\pi$ would have to use three different scaling functions,
one for each quantity.

In contrast, it is difficult to assign a goodness-of-fit parameter to curve collapse.

An issue with this analysis  bedevils many of the systems which have been studied:
the gauge coupling $g_0$ runs very slowly. This means that its
exponent $y_g \sim 0$. An analysis that left it out would produce a leading critical exponent
$y_m$ that appeared to drift with bare gauge coupling.
If the marginal coupling is included in the scaling analysis, 
Eq.~\ref{eq:fss2} is modified to
\bee
\xi_L   = L f_H\left(x, g_0 m^{\omega}\right),
\ee
where $\omega \equiv -y_g / y_m  $.
The scaling function $f_H\left(x, g_0 m^{\omega}\right)$ is analytic even at the fixed 
point, and can be expanded as
\bee
  \label{eq:expansion}
  \xi_L = L F_H(x)\left\{1 + g_0 m^{\omega} G_H(x) + O\left(g_0^2 m^{2\omega}\right)\right\}.
\ee
The first term is the usual  expression while the second  accounts for the leading corrections to scaling.

The first group to go beyond 
 Eq.~\ref{eq:fss2} was  \textcite{Cheng:2013xha}. These authors studied the system with $N_c=3$ and  
$N_f=12$ fundamental fermions. They fit (with $1/\xi_L = M_H$)
\bee
  \label{eq:corrected}
  \frac{L M_H}{1 + c_G g_0 m^{\omega}} = F_H(x).
\ee
They did fits to several dimensionful parameters (pseudoscalar and vector masses, $f_\pi$) over a wide range of
volumes and fermion masses. Weaknesses of the calculation are that first, the authors assumed some functional form
for the scaling function (to be fair, I do not see how to do curve collapse in a multidimensional space)
and second, the confidence levels associated with the chi-squareds of a number of the fits are poor.
Nevertheless, I find it quite impressive. $y_m$  is nearly independent of bare gauge coupling over a wide range.
Including the non-leading exponent renders all previous studies obsolete.

\begin{figure}
\begin{center}
\includegraphics[width=\columnwidth,clip]{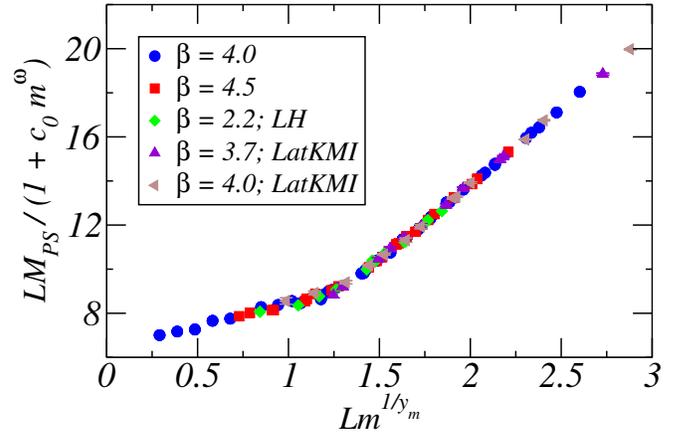}
\end{center}
\caption{\protect{\textcite{Cheng:2013xha}}'s best curve collapse fit, combining their data
 and that of of the LH and LatKMI collaborations \protect{\cite{Fodor:2011tu,Aoki:2012eq}}.
\label{fig:annapion}}
\end{figure}

\begin{figure}
\begin{center}
\includegraphics[width=\columnwidth,clip]{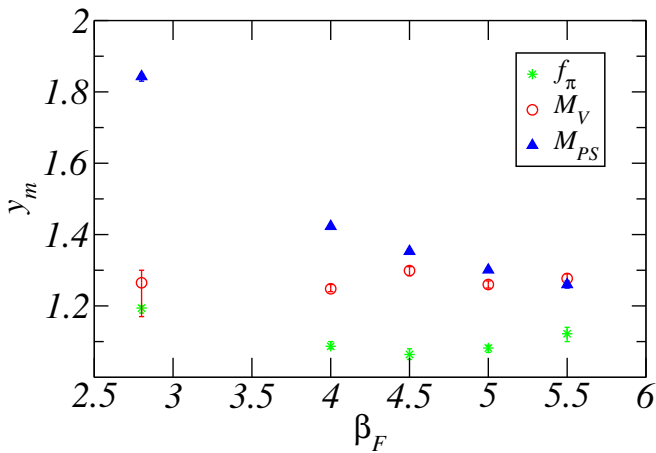}
\end{center}
\caption{\protect{Exponent $y_m$  from \textcite{Cheng:2013xha}} for the pseudoscalar mass, vector mass,
and pseudoscalar decay constant, not including the non-leading coupling. The abscissa is the bare gauge coupling.
\label{fig:annaymc0}}
\end{figure}

\begin{figure}
\begin{center}
\includegraphics[width=\columnwidth,clip]{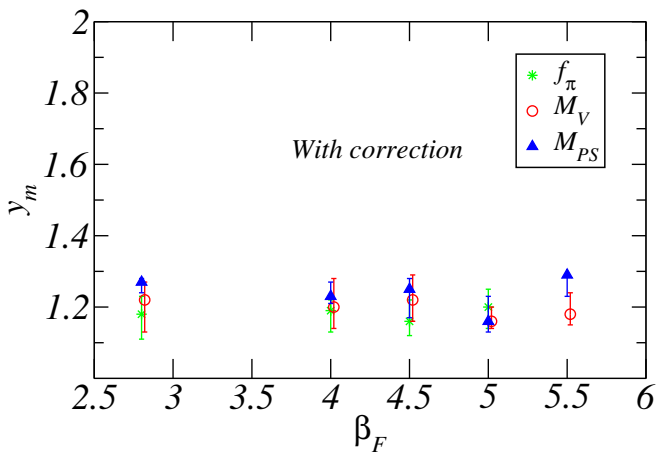}
\end{center}
\caption{\protect{Exponent $y_m$  from \textcite{Cheng:2013xha}} for the pseudoscalar mass, vector mass,
and pseudoscalar decay constant,  including the non-leading coupling.
\label{fig:annaym}}
\end{figure}

Figure \ref{fig:annapion} shows the best curve collapse fit for the pseudoscalar mass from these authors.
It uses their data plus results from two other collaborations,
 with many $L$'s and many $\beta$'s. Compare the $y_m$'s with and without the correction,
Figs.~\ref{fig:annaymc0}-\ref{fig:annaym}.

\subsubsection{Mass anomalous dimension from Dirac eigenvalues}
Next, there are
 a set of related methods extracting $y_m$ from the spectral density of eigenvalues $\lambda$
of the Dirac operator. The physics seems simple:
The Banks-Casher relation \cite{Banks:1979yr} connects the condensate $\Sigma$ and the
 density of eigenvalues $\lambda$
 of the Dirac operator
$\rho(\lambda)$. At nonzero mass it is
\bee
\Sigma(m_q) = - \int \rho(\lambda) d \lambda \frac{2m_q}{\lambda^2+m_q^2}.
\ee
If the massless theory is conformal, and if
the condensate $\Sigma(m_q)$ scales as $ m_q^{\alpha}$ for small mass, then $\rho(\lambda)$ also scales
as $\lambda^{\alpha}$.

A finite-size scaling argument \cite{Akemann:1997wi}
 relates the scaling for the density $\rho$ to the scaling of
the value of individual eigenvalues. If we consider the average value of the $i$th eigenvalue
of the Dirac operator in a box of volume $V=L^D$, and  if  $\rho(\lambda)\sim \lambda^\alpha$,
then we expect
\bee
\langle \lambda_i \rangle \sim \left(\frac{1}{L}\right)^p
\label{eq:eigscale}
\ee
where
\bee
p=\frac{D}{1+\alpha}.
\label{eq:scaling}
\ee
For the case of a theory with an IRFP, $p$ is equal to $y_m$, the leading
exponent. Thus $\rho(\lambda) \sim \lambda^{D/y_m -1}$.

In QCD, or in other chirally broken theories, $\alpha=0$ and $p=D$. Here the story is rich,
and involves an interplay of confinement, chiral symmetry breaking and random matrix
theory. [See \cite{Damgaard:1998xy,Osborn:1998qb}.]
Even the probability distribution of individual eigenvalues can be used to determine the condensate.
It is a universal function of the product $\lambda \Sigma V$ with $V=L^D$. Results obtained build on ones like Eq.~\ref{eq:fvs}.

Most of the beyond Standard Model literature uses the integrated spectral density or mode number.
This technique is adapted from its original QCD venue, following the discussion in \textcite{Giusti:2008vb}.
Their approach has been applied to near-conformal theories by a number of authors. 
The most-cited  beyond Standard Model study is \textcite{Patella:2012da},
who studied the integrated spectral density for $SU(2)$ gauge theory coupled to $N_f=2$ adjoint fermions.

He split up the eigenvalues into three classes:
\begin{itemize}
\item Very small ones, which are sensitive to the simulation volume
\item Intermediate ones which show the desired power law scaling behavior
\item Large ones which (for an asymptotically free system) go over to free field $\rho(\lambda) \sim \lambda^3$
behavior
\end{itemize}
He integrated over the intermediate eigenvalues to find an exponent.

An issue with using this method is that the exponent depends on the range of eigenvalues used to
 measure it.
The authors of \textcite{Cheng:2013eu} combine the intermediate and large eigenvalues to construct
a ``scale dependent mass anomalous dimension,'' whose scale in energy space is given by $\lambda$ itself,
and whose extrapolation to small $\lambda$ gives the actual $y_m$ (assuming, of course, that the
 system studied is truly conformal).
 They are able to compare and contrast a confining theory ($SU(3)$ with $N_f=4$ fundamentals)
 with a slowly running one ($SU(3)$ with $N_f=12$ fundamentals),
which they identify as conformal. Their prediction for $y_m$ will be quoted below.

When I read these papers, I cannot help thinking: are the smallest eigenvalues, which are the ones
 most sensitive to the volume,
not also the ones that are sensitive to the longest distance physics? And if so, is there not some kind of
finite size scaling or curve collapse story that can be told about them? No such story exists in the 
literature, as far as I know.

There is another issue with the use of eigenvalues,
which appears when one thinks about what is actually being measured.

Briefly, the spectral density of the massless Dirac operator
\bee
\rho(\lambda) = \frac{1}{V} \svev{\sum_k \delta(\lambda-\lambda_k)}
\ee
is the discontinuity across the imaginary mass axis of the resolvant,
\bee
\rho(\lambda)= \frac{1}{2\pi} \lim_{\epsilon \rightarrow 0} 
\Sigma_{val}(i\lambda+\epsilon)-\Sigma_{val}(i\lambda-\epsilon)
\ee
where
\bee
\Sigma_{val}(m_v)= \frac{1}{V}\sum_k \svev{\frac{1}{m_v+i\lambda_k}}
\ee
is the expectation value  $\svev{\psi(0)\bar \psi(0)}$ for a fermion of mass $m_v$.
The resolvant cannot be computed in ordinary field theory. The ordinary partition function is simply
not a generator for it. We need a generator, and that can be found, but in a partially quenched
version of our theory, where the valence fermions have a different mass from the dynamical fermions
(and the system has additional bosonic degrees of freedom to remove the valence fermions from the
partition function). Only this extended field theory can probe the spectral density. 

So, can a partially quenched theory tell us things about an unquenched theory?
For a chirally broken and confining theory like QCD, it can, and partial quenching is one of the standard
techniques for computing low energy constants. But outside of this framework,
I know of no precise statement of the connection.
The end result is  that if
chiral symmetry is unbroken, the physics of the measured spectral
density may not
be quite what people think it is.

%%%%%%%%%%%%%%%%%%%%%%%%%%%%%%%%%%%%%%%%%%%%%%%%%%%%%%%%%%%%%%%%%%%%%
\section{Lattice results for systems with slowly running couplings -- by system\label{sec:results}}
%%%%%%%%%%%%%%%%%%%%%%%%%%%%%%%%%%%%%%%%%%%%%%%%%%%%%%%%%%%%%%%%%%%%%

Now we begin a survey of lattice calculations, separated by specific model properties.

\subsection{Early studies (before about 2007)}
There is a long history of lattice studies of systems with many fermionic degrees of freedom.
Most of the early ones involved thermodynamics. The question was, did the deconfinement
temperature $T_c$ scale appropriately (remain a constant ratio with respect to any other massive observable)
as the lattice spacing was taken away? The data were ambiguous.
An early review, \textcite{Fleming:2008gy}, contains citations to this work.
There were also a number of simulations of Wilson fermions with many flavors of fundamentals
 by Iwasaki and collaborators
[two papers are \textcite{Iwasaki:1991mr,Iwasaki:2003de}].
These studies also searched for the loss of confinement as the number of flavors
increased.  (They were actually interested in seeing whether a deconfined phase persisted all the way to
$\beta=0$.) Many of the features of later simulations with Wilson fermions are
 first present in these studies.

\textcite{Damgaard:1997ut} studied $SU(3)$ gauge theory coupled to sixteen fundamental flavors, and observed that while
the system had a strong coupling phase, its weak coupling phase was chirally restored. They argued that they 
could define a running coupling from the string tension, and that its beta function was positive in the weak coupling phase.
 Looking back, this was
the first appearance of the interior of the conformal window in a simulation. It was followed by
 \textcite{Heller:1997vh} --  the first  Schr\"odinger functional  measurement of a running coupling in a many-flavor system.
Heller also observed a positive beta function. This paper
 was the inspiration for the later, beyond Standard Model Schr\"odinger functional work.

The earliest numerical simulation of a system with an explicit place in beyond Standard Model phenomenology was by
\textcite{Catterall:2007yx}, who studied what they called ``minimal walking technicolor''
 ($SU(2)$ with $N_f=2$ adjoint representation fermions). ``Minimal'' refers to the particle content:
with $N_f=2$ there are three Goldstones to be eaten by the $W$ and $Z$ leaving no technipions behind.
``Walking,'' of course, because the beta function is small in one loop, and the system might be confining
according to the analysis of \textcite{Sannino:2004qp,Dietrich:2006cm}.
They carried out spectroscopic measurements and observed what was at the time very peculiar behavior, that I have
already described above --  
recall Fig.~\ref{fig:conformal}. This was later 
 recognized as the spectroscopy of a near conformal system in finite volume.

The field then became very active. To go on, we should separate the discussion of 
different physical systems into their own sections.

\subsection{Studies of $N_c=3$ and many flavors of fundamental fermions}
I do not think these systems were ever taken seriously as true technicolor candidates.
They have  have too many ($N_f^2-1)$
Goldstone bosons. Electroweak symmetry breaking only eats three of them,
 leaving $N_f^2-4$ technipions to be observed in experiment, or somehow explained away.
But, all lattice QCD people have computer programs to simulate $SU(3)$ gauge fields
and it is easy to modify the code to do many flavors of fermions. With 
staggered fermions, multiples
of 4 are easy, with Wilson fermions, multiples of two. The motivation was just to see
whether walking actually occurred, or not.
For these reasons,  I believe it is still the most-studied lattice beyond Standard Model sector,
both in number of papers  written and in computer hours consumed.

\begin{figure}
\begin{center}
\includegraphics[width=\columnwidth,clip]{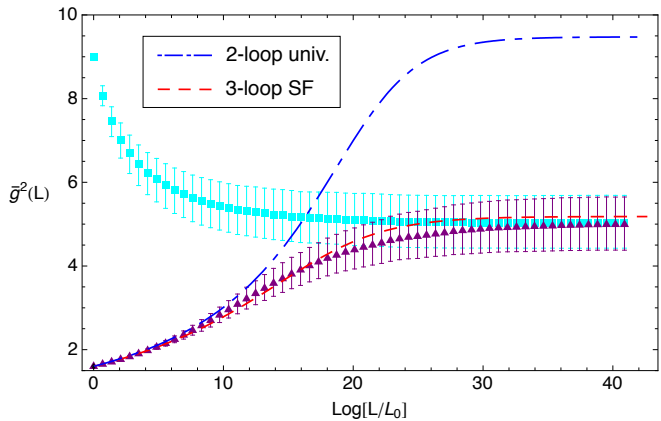}
\end{center}
\caption{Continuum running for $SU(3)$ gauge theory coupled to $N_f=12$ fundamentals, from 
\protect{\textcite{Appelquist:2009ty}}.
Results shown for running from below the infrared fixed point (purple triangles) are based
 on $\overline{g}^2(L_0) \equiv 1.6$.  Also shown is continuum backwards running from above the fixed point
 (light blue squares), based on $\overline{g}^2(L_0) \equiv 9.0$.
\label{fig:yale}}
\end{figure}

The earliest studies in this area were the
 large scale Schr\"odinger functional simulations of $N_f=8$ and 12 by
\textcite{Appelquist:2009ty,Appelquist:2007hu}. They claimed to observe an IRFP for $N_f=12$, while
the beta function for $N_f=8$ was everywhere negative. Thus, the boundary for the conformal window
was claimed to  be somewhere between 8 and 12 flavors.
 Their results for $N_f=12$, as they presented them, are shown in 
Fig.~\ref{fig:yale}.

\begin{figure}
\begin{center}
\includegraphics[width=\columnwidth,clip]{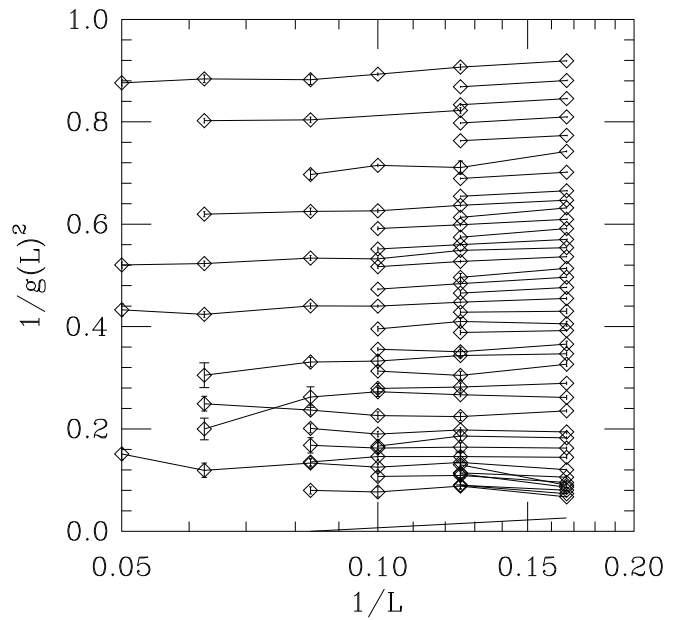}
\end{center}
\caption{Raw lattice data for $SU(3)$ gauge theory coupled to $N_f=12$ fundamentals, from
\protect{\textcite{Appelquist:2009ty}}. Lines connect data with the same bare couplings. The line at the bottom
is the slope expected from one-loop running.
\label{fig:yale12}}
\end{figure}

This figure uses heavily processed lattice data. It comes from a many-parameter fit to 
all of their data at many bare
couplings and many volumes, of the form
\bee
\label{eq:interpolating_function}
\frac{1}{\overline{g}^2(\beta, L/a)} = \frac{\beta}{6} \left[ 1 - \sum_{i=1}^{n} c_{i,L/a} \left(\frac{6}{\beta}\right)^{i} \right].
\ee
I cannot evaluate results from such global fits. Fortunately, these authors published
 their data, and it is possible to look at
it directly. This is shown in Fig.~\ref{fig:yale12}. Lines connect
data collected at the same bare gauge couplings. I have drawn a line whose slope is
 the one-loop beta function result.
As the bare gauge coupling moves from weak to strong coupling, the slope of the lines 
in Fig.~\ref{fig:yale12}
flattens slightly. Does it change sign? They said Yes, but the existence
 of a large literature  about this system indicates that others looked at the figure and said Maybe.

\begin{figure}
\begin{center}
\includegraphics[width=\columnwidth,clip]{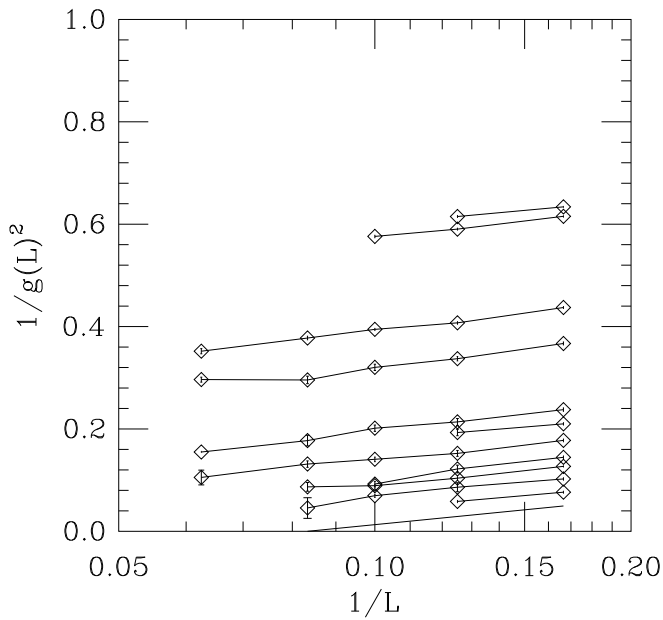}
\end{center}
\caption{Raw lattice data for $SU(3)$ gauge theory coupled to $N_f=8$ fundamentals, from
\protect{\textcite{Appelquist:2009ty}}. Lines connect data with the same bare couplings. The line at the bottom
is the slope expected from one-loop running.
\label{fig:yale8}}
\end{figure}

The situation with eight flavors seemed to be much more clear cut: the beta function was
 everywhere negative.
This can be seen in a plot of $1/g^2(L)$ versus $\ln L$, Fig.~\ref{fig:yale8}. In fact, all 
the lines show nearly the same slope.
This is not surprising from a perturbative viewpoint; $b_2$ (recall Eq.~\ref{eq:2loopbeta2}) is nearly zero.

\subsubsection{$N_f=12$}

The largest set of lattice results concerns $N_f=12$.

Shortly after \textcite{Appelquist:2009ty} appeared, \textcite{Fodor:2009wk} carried out 
spectroscopic studies with $N_f=4$, 8 and 9 flavors
and argued that even the larger $N_f$ systems were chirally broken. In a conference proceedings
followed by a journal article, \textcite{Fodor:2011tu}  claimed that $N_f=12$ was also confining.
Their lattice data was taken at many volumes and fermion masses, but one bare gauge coupling.
\textcite{Appelquist:2011dp,DeGrand:2011cu} did finite size scaling studies to the data sets of  \textcite{Fodor:2011tu}  and
 concluded that they were
consistent with infrared conformality.
 A later large volume study  by \textcite{Aoki:2012eq} contributed data sets at two
bare couplings and  concluded that the data favored infrared conformality.
Finally,  \textcite{Cheng:2013xha} performed a finite size scaling analysis, 
with the leading irrelevant operator,
on all these data sets, and claimed that all the data were consistent with infrared conformality,
strongly affected by a nonleading exponent. Figures were shown above, Figs.~\ref{fig:annapion}-\ref{fig:annaym}.

Besides \textcite{Appelquist:2009ty}, several groups claimed to observe an IRFP.
\textcite{Lin:2012iw} computed the  renormalized coupling from  twisted Polyakov loops
\cite{deDivitiis:1993hj} and claimed this.
\textcite{Hasenfratz:2011xn} used MCRG to observe a positive beta function (in my conventions)
for the bare step scaling function, in strong coupling. This is evidence for an IRFP since the beta function 
is negative in weak coupling.
And most recently,
\textcite{Cheng:2014jba} observed a fixed point using a gradient flow definition of a running coupling.

To summarize: All groups except \textcite{Fodor:2011tu} observe behavior consistent with infrared
 conformality
for  $N_f=12$. The mass anomalous dimension  $\gamma_m$
is found to be small by all who report a measurement. 
There are enough studies of this system that a table is useful. See Tab.~\ref{tab:nf12gamma}.
 I think that the evidence in favor of
infrared conformality is overwhelming.

\begin{table}
\begin{tabular}{c c c c}
\hline
Reference  &  method & result & $\gamma_m$ \\
\hline
\textcite{Appelquist:2009ty} &  SF & I* & \\
\textcite{Fodor:2011tu} & spectra & C  & \\
\textcite{Appelquist:2011dp} & FSS (fit) & I & 0.40(1) \\
\textcite{DeGrand:2011cu} & FSS (cc)  &  I & 0.35(23) \\
\textcite{Hasenfratz:2011xn} & MCRG & I* & \\
\textcite{Aoki:2012eq}   & FSS (cc) & I & 0.4-0.5 \\
\textcite{Lin:2012iw}    &  other & I* & \\
\textcite{Cheng:2013eu}  &   spectral & I & 0.32(3) \\
 \textcite{Cheng:2013xha}&  FSS (fit)  & I & 0.235(15) \\
\textcite{Cheng:2014jba} & flow & I* &   \\
\textcite{Lombardo:2014pda} & FSS (fit) & I & 0.235(46)) \\
 \hline
 \end{tabular}
\caption{Claims for the phase structure  for $SU(3)$, $N_f=12$ fundamentals. The ``result''
column is keyed with a C if the authors claim to observe a confined, chirally broken system,
I if infrared conformal behavior is claimed or assumed. A ``*'' indicates that an IRFP was observed.
Under ``method,''
``FSS'' refers to finite size scaling with ``fit'' for a fit to a known scaling function and ``cc''
for curve collapse. ``Spectral'' refers to use of the spectral density of Dirac eigenvalues.
``Spectra'' refers to spectroscopy.
SF is Schr\"odinger functional. ``Flow'' refers to some variant of Wilson flow.
Predictions for $\gamma_m$ are given where available.
\label{tab:nf12gamma}}
\end{table}

\textcite{Aoki:2013zsa} measure the mass of a scalar resonance in $N_f=12$. They find it is slightly lighter
than the pseudoscalar mass at the nonzero fermion masses where they simulate.
This would be the light state consistent with incipient criticality described in Sec.~\ref{sec:dilaton}.

This is probably a good place to talk about lattice artifacts in strong coupling.
 Recall the situation with these slowly - running theories:
if the gauge coupling is strong at long distance, than it is also strong at the cutoff scale.
 This is an invitation for lattice artifacts
to appear. Universality should be lost. This is not just a problem of principle. These days,
essentially all lattice groups simulate with different lattice actions.
In the small lattice spacing limit, all these actions differ by irrelevant operators, 
and they all should give identical predictions. But
 in strong coupling,
one group might see something which another group does not,  just because their actions are different.

 However, there are some general features that can
 be described.
I know the situation for Wilson type fermions the best. Recall that the bare Wilson fermion mass 
is additively renormalized.  Any calculation that must be done at $am_q=0$, such as a
Schr\"odinger functional calculation, is carried out along the
 the critical kappa line,
$\kappa_c(\beta$) vs $\beta$. The generic Wilson fermion artifact is that when the number of fermionic degrees
 of freedom is large enough, at strong coupling
the $\kappa_c$ line vanishes: there is a line of discontinuity in which the AWI quark 
mass jumps abruptly from  positive to negative.
This was seen first in  \textcite{Iwasaki:1991mr,Iwasaki:2003de}, and nearly every paper with many Wilson fermion
degrees of freedom reports it.
\textcite{Nagai:2009ip} is a particularly complete example: the authors studied $SU(2)$ and some $SU(3)$ gauge theories coupled
to many flavors of fundamental fermions, at $\beta=0$. A first order transition appears at around $N_f=6$
for $SU(2)$.

What is annoying about this transition is that the interesting region for slowly running 
theories is at strong coupling,
but if there is no place where the fermion mass vanishes, one cannot do lattice studies.
 In particular, running coupling
studies typically chase a running coupling into strong coupling, watch it run ever more slowly,
 and then the transition appears just before
(or just after) a zero of the beta function is about to occur.

The precise location of the transition is not universal, and it is possible to design (empirically)
 actions for which the
transition is pushed to stronger coupling. Shamir, Svititsky, and I found it quite useful to do this.

As far as I can tell, there are no other transitions generically observed 
on the weak coupling side of this transition, so
at least the weak coupling phase of a Wilson fermion simulation seems to be 
analytically connected to the Gaussian fixed point
at $g^2=0$ or infinite $\beta$.

Staggered fermions seem to be more complicated, but 
maybe that is just because I have no personal experience with them.  \textcite{Cheng:2011ic} has a collection 
of earlier references and a description of their new strong coupling
 phase. It is bracketed by jumps in the condensate.
It is a phase where lattice translational symmetry is broken: the condensate $\svev{ \bar \psi \psi}$
is different on even and odd lattice sites. The phase form a pocket extending from small (zero?)
 quark mass to some maximum value, over a range of strong values of $\beta$. 
This is seen both for $N_f=12$ and 8
fundamentals. The phase is confining but apparently chirally restored.  (Such continuum language may not be appropriate for
a strong coupling phase.)
Other groups 
\cite{Deuzeman:2012ee,Fodor:2012uu,Jin:2013hpa}  have reported similar structure.

Of course, groups are careful to avoid such phases when they see them. But that may not be good enough.
One is really interested in physics in the basin of attraction of either the Gaussian fixed point 
or of an IRFP. A nearby transition may affect what one is seeing, as much as the IRFP or the Gaussian fixed point.
This was probably an issue for Wilson fermion Schr\"odinger functional studies, which were looking
for a fixed point very close to a strong coupling transition.

\subsubsection{$N_f=10$}
\begin{figure}
\begin{center}
\includegraphics[width=0.9\columnwidth]{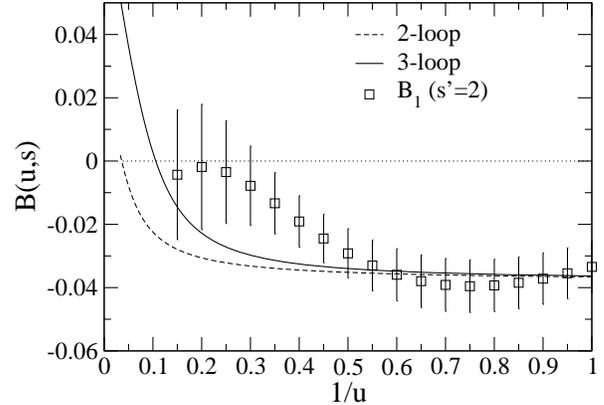}
\caption{$N_f=10$ beta function for $u=g^2$ from \protect{\textcite{Hayakawa:2010yn}},
along with perturbative expectations.
\label{fig:nf10bfn}
}
\end{center}
\end{figure}

\textcite{Hayakawa:2010yn} computed a running coupling in a Sch\"rodinger functional simulation.
Their $s=2$ discrete beta function is shown in Fig.~\ref{fig:nf10bfn}.
They certainly observe slower running than the perturbative result. Is there a zero?
I am afraid to say Yes, although they have no such fear. This result is significant with respect to
$N_f=12$, because if $N_f=10$ is infrared conformal, it is hard to see how
$N_f=12$ could not be.

\subsubsection{$N_f=8$}
$N_f=8$ is quite curious:
The Schr\"odinger functional beta function of \textcite{Appelquist:2009ty} is negative. 
As Fig.~\ref{fig:yale8} shows, the beta function basically runs at its 
one-loop value over the entire observed range.

Early work by \textcite{Deuzeman:2008sc}
claimed to see a thermal transition that moved to weaker bare coupling as the lattice size increased.
So far, so good, for confinement and chiral symmetry breaking.  But recently \textcite{Aoki:2013xza} studied $N_f=8$.
Most of their data is from three volumes, but one bare gauge coupling.
They
described observing behavior at small fermion mass consistent with chiral breaking
(nonzero pseudoscalar decay constant, nonzero vector meson mass, zero pseudoscalar mass all 
 in the chiral limit). At the same time they found
behavior at large fermion mass consistent with power law scaling and a large $\gamma_m\sim 1$.
This seems strange; if a data set is going to be infrared conformal, it will be most infrared
conformal at the
smallest fermion mass,   subject to the caveat that finite volume effects are largest there.

\begin{figure}
\begin{center}
\includegraphics[width=\columnwidth,clip]{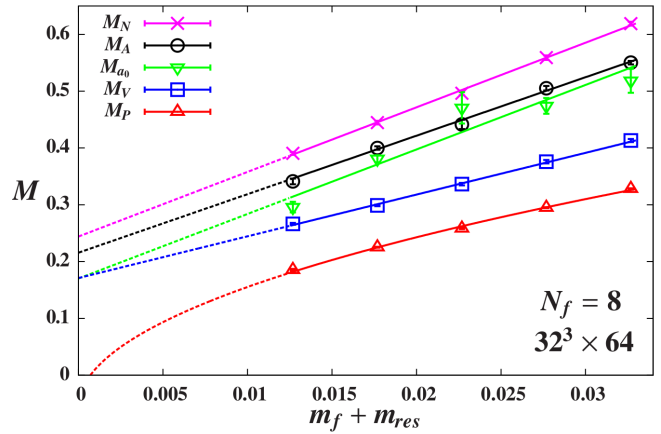}
\end{center}
\caption{$N_f=8$ spectroscopy from \protect{\textcite{Appelquist:2014zsa}} with fits motivated by
chiral symmetry breaking.
\label{fig:nf8chiral}}
\end{figure}

\begin{figure}
\begin{center}
\includegraphics[width=\columnwidth,clip]{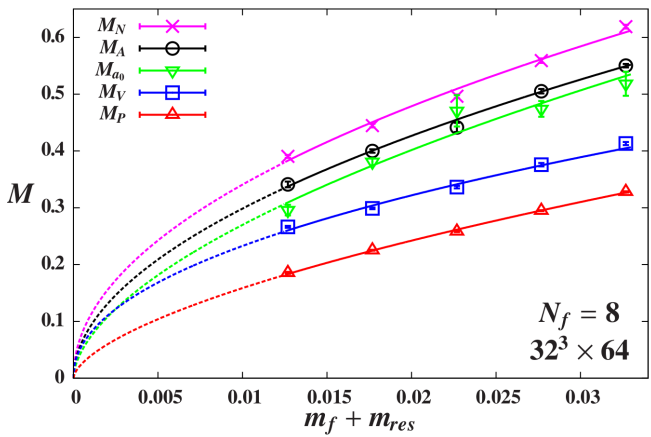}
\end{center}
\caption{$N_f=8$ spectroscopy from \protect{\textcite{Appelquist:2014zsa}} with fits 
to power laws, Eq.~\protect{\ref{eq:corrlen}}.
\label{fig:nf8power}}
\end{figure}

\textcite{Appelquist:2014zsa} also have data at one gauge coupling, two large volumes, 
and several quark masses. They see separation between the pseudoscalar and vector masses
and lack of parity doubling in the vector and axial vector channels, all increasing
at their smallest fermion masses. However, simple power law fits (like Eq.~\ref{eq:corrlen})
also reproduce the data with good quality. (Their two volumes did not have the overlapping region needed
for a real finite size scaling analysis.) Compare Figs.~\ref{fig:nf8chiral} and \ref{fig:nf8power}.

%%%%%%%%%%%%%%%%%%%%%%%%%%%%%%%%%%%%%%%%%%%%%%%%%%%%%%%%%%%%%%%%%%%%%
%%%%%%%%%%%%%%%%%%%%%%%%%%%%%%%%%%%%%%%%%%%%%%%%%%%%%%%%%%%%%%%%%%%%%

The Dirac eigenvalue study of \textcite{Cheng:2013eu} reported a large $\gamma_m \sim 1$
from the integrated spectral density. (Recall Eq.~\ref{eq:scaling}.) They did not observe
good quality chiral behavior a la Banks-Casher.

Finally, two recent groups, \textcite{Hasenfratz:2014rna,Fodor:2015baa}, report 
calculations of a gradient flow running coupling. The beta function
is everywhere negative, smaller than \cite{Hasenfratz:2014rna} or consistent with \cite{Fodor:2015baa} its 
small two loop value.

I think that lattice calculations of running couplings provide strong evidence that
$N_f=8$ is not inside the conformal window. I am not sure what can be done with spectroscopy to
support this claim.
Simulations at several bare gauge couplings, along with data at enough volumes for
a real finite size scaling analysis, might help.
But it might just be that the coupling is  running so slowly, that it can never grow across any
imaginable simulation volume; then the  range of accessible volumes makes the system
 effectively conformal.
One would have to simulate  deep in strong coupling to see signals of confinement
 or chiral symmetry breaking.
 But then, the system would  be strongly
 interacting at its shortest lattice distances. Where would be a connection to asymptotic freedom?

\textcite{Aoki:2014oha} report a light isoscalar scalar state, whose mass is roughly equal to that
of the pseudoscalar at the nonzero fermion masses where they collected data. They argue that its
mass extrapolates to a nonzero value at zero fermion mass, and thus it is a candidate for a dilatonic Higgs.
I am not prepared to believe this claim since where they have data, the mass of the scalar
 is degenerate with the mass of 
the pseudoscalar state.

\subsubsection{$N_f\le 6$}
With $N_f=6$ and below, we are back on more familiar ground. These systems are
 confining and chirally broken.
At $N_t=6$, there is a finite temperature transition that moves with lattice size in
 a reasonable way \cite{Miura:2011mc}.
The LSD collaboration has done several studies comparing observables with some relation
to electroweak physics at $N_f=2$ and 6.
$N_f=6$ has a running, not a walking, coupling, but
the collaboration was hoping to see trends that might become stronger
closer to the edge of the conformal window.
 Their data is from one $\beta$ value per $N_f$, with
 several fermion masses.

\begin{figure}
\begin{center}
\includegraphics[width=\columnwidth]{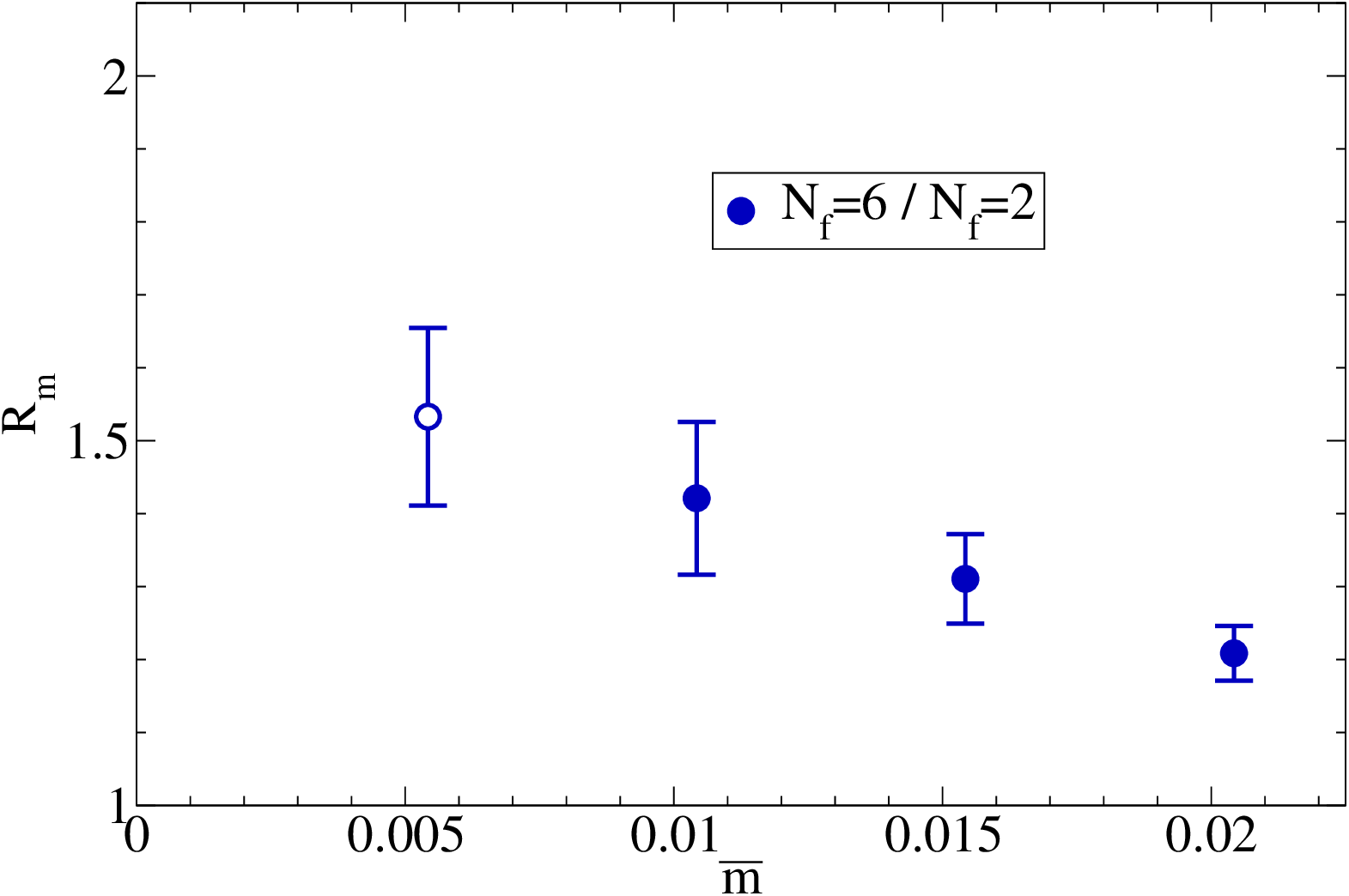}
\caption{
The ratio $\svev{\bar \psi \psi}/F^3$ of  $N_f=6$ to $N_f=2$,
 $R_m \equiv [M_m^{2}/2mF_m]_{6f}/[M_m^{2}/2mF_m]_{2f}$, 
versus average fermion mass $\overline{m}$, from \protect{\textcite{Appelquist:2009ka}}.
\label{fig:ratio}
}
\end{center}
\end{figure}

\textcite{Appelquist:2009ka} shows that  the ratio of the condensate to the cube of the pseudoscalar decay constant $F$,
$\svev{\bar \psi \psi}/F^3$,  increases
with increasing $N_f$. (They actually compute $\svev{\bar \psi \psi}$
from $M_\pi^2 F^2/(2m_q )$.)  Their plot is shown in Fig.~\ref{fig:ratio}.
 The larger condensate allows for larger quark masses
while keeping flavor changing neutral currents small. (Recall the discussion around
Eq.~\ref{eq:metc}.) They tell us that their $N_f=6$ data sets are not
at small enough quark mass and big enough volume to be reliably deep
 into the chiral limit, so their results are tentative.

Next, \textcite{Appelquist:2010xv}   compute  the S-parameter.
It is proportional to the limiting value of $d(q^2\Pi_T^{LR}(q^2))/dq^2$ (recall
Eq.~\ref{eq:continPi}) at small $q^2$,
after Goldstone boson effects are subtracted.

\begin{figure}
\begin{center}
\includegraphics[width=\columnwidth]{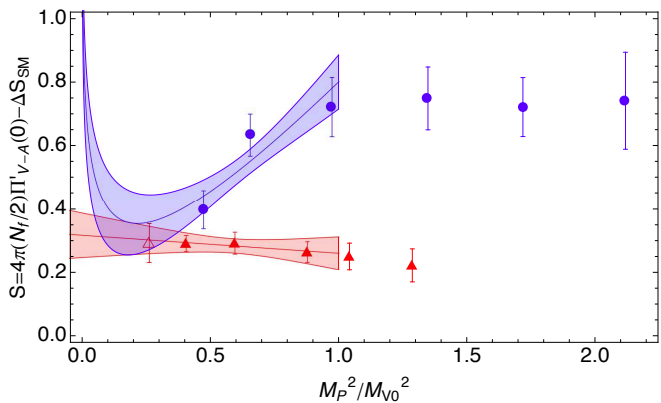}
  \caption{$S$ parameter for $N_f = 2$ (red diamonds and band) and $N_f = 6$ (blue circles and band),
from \protect{\textcite{Appelquist:2010xv}}.
 For each of the solid points, $M_{P}L >4$.
\label{fig:lsdspar}
}
\end{center}
\end{figure}

At heavier quark masses, their S parameter scales roughly linearly with $N_f$
(or more simply, the $N_f=6$ value is three times the $N_f=2$ value). This is expected
behavior, just counting degrees of freedom. However, 
at their smallest $N_f=6$ fermion mass 
their $S$ parameter plunges to become nearly equal to the $N_f=2$ value. This is shown in 
Fig.~\ref{fig:lsdspar}.

This is only one point, but they argue it is a real effect, with the following cause:
If the correlator can be saturated by a sum of resonances, it can be written as
\bee
\Pi^{LR}_T(q^2) = \sum_V\frac{f_V^2M_V^2}{q^2+M_V^2} - \sum_A \frac{f_A^2M_A^2}{q^2+M_A^2} - \frac{f_\pi^2}{q^2}
\label{eq:SR}
\ee
and the $S$-parameter is dominated by the difference of
vector and axial vector contributions of this expression.
The masses of the lightest vector and axial vector mesons are relatively easy to extract from 
lattice data. The authors observed that these states became more degenerate at $N_f=6$
than they were at $N_f=2$ and at the same time the S parameter decreased.
 They took the decrease
to be a favorable generic result for a technicolor solution to beyond Standard Model physics.

Their third calculation is of  $WW$ scattering parameters \cite{Appelquist:2012sm}.
This is done using the Goldstone equivalence theorem; longitudinal $W$ scattering
 amplitudes
can be computed in terms of Goldstone boson scattering amplitudes. The 
scattering amplitude (more precisely, the low energy scattering phase shift) can be
computed from the shift in energy of the two-pion state in finite volume.
This is not an easy calculation even in QCD, and LSD only had one volume.
They could measure a scattering length for the maximal isospin channel in
 $N_f=2$ and 6. It is consistent with
 the lowest order chiral perturbative result.

\subsection{$N_c=2$ and many fundamental flavors}

The cost of a simulation scales as $N_c^3$, so these systems are cheaper than $N_c=3$.
 This means that, in principle,
 one can study a larger range of volumes for an equivalent use of resources. However,
 they are less studied than $N_c=3$. 

A Schr\'odinger functional analysis by 
\textcite{Karavirta:2011zg} claims that $N_f=10$ has an IRFP and $N_f=4$ has a negative beta function.
A conference proceedings by
\textcite{Ohki:2010sr}  argues that $N_f=8$ has an IRFP.
\textcite{Rantaharju:2014ila} recently presented a conference proceedings with preliminary results of a gradient flow coupling
for $N_f=8$. They observe perturbative running in weak coupling with no direct evidence for a fixed point.

$N_f=6$ is the most controversial point. 
The two-loop beta function has a zero deep in strong coupling.
\textcite{Bursa:2010xn} claimed slow running,  but could not tell if there is a fixed point.
\textcite{Hayakawa:2013yfa} claimed to see an IRFP, with
a small $\gamma_m$, though with large errors ($0.26 \le \gamma_m^* \le 0.74$).
\textcite{Karavirta:2011zg} have inconclusive results for $N_f=6$.
Their $\gamma_m$ for $N_f=6$ ranges from 0.1-0.25 over observed $g^2$ range,
smaller than the one loop perturbative value. 
The largest statistics study to date of $N_f=6$ is the Schr\"odinger functional study of 
 \textcite{Appelquist:2013pqa}. They found no evidence for an IRFP.

\subsection{Fermions in higher dimensional representations}

An alternative way to achieve slow running is to bundle the many fermion degrees of freedom into a small
 number of higher dimensional representations. This could be phenomenologically attractive: with $N_f=2$ 
there are no un-eaten
Goldstones to become technipions. On the other hand, this could be phenomenologically unattractive:
technifermions are in different color representations from
 Standard Model fermions, so they cannot be members of the same multiplet.

The most studied of these systems is
$SU(2)$ with $N_f=2$ adjoints (``minimal walking technicolor'').
Every technique I have mentioned -- and probably more -- has been applied to this system.
I have already shown examples of its spectroscopy, by \textcite{Hietanen:2008mr}.
Interesting Schr\"odinger functional studies include \textcite{Hietanen:2009az,Bursa:2009we,DeGrand:2011qd}.
 It is the clearest example of an IRFP system.

\begin{figure}
\begin{center}
\includegraphics[width=\columnwidth,clip]{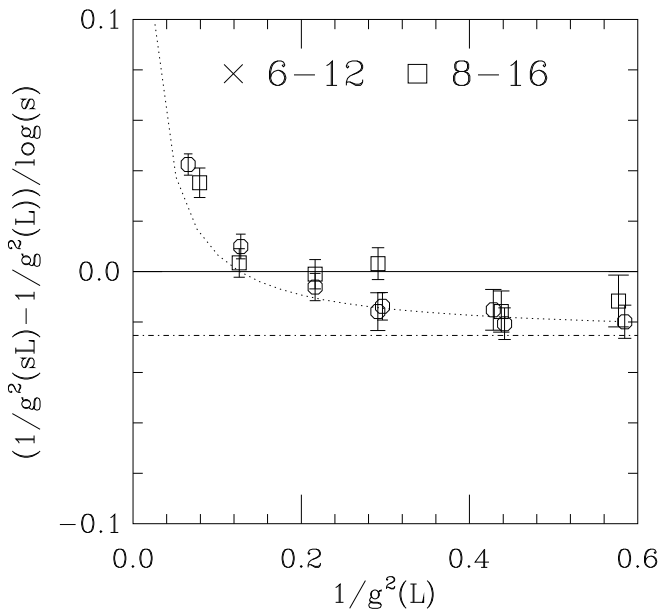}
\end{center}
\caption{Beta function for $SU(2)$ with $N_f=2$ adjoints, from  \protect{\textcite{DeGrand:2011qd}}.
This is based on comparisons of two $L$ values related by a scale factor $s=2$.
The straight line is the one-loop beta function, and the curved line is the two loop beta function.
\label{fig:betasu2}}
\end{figure}

\begin{figure}
\begin{center}
\includegraphics[width=\columnwidth,clip]{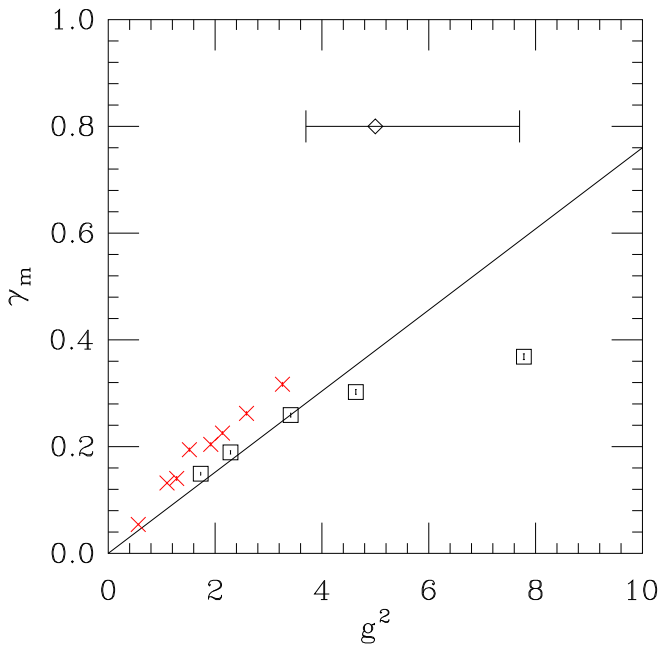}
\end{center}
\caption{Mass anomalous dimension $\gamma_m(g^2)$ from  \protect{\textcite{DeGrand:2011qd}}
The horizontal bar at the top marks our location (with its uncertainty) for the critical coupling.
The crosses are the data of \protect{\textcite{Bursa:2009we}},
analyzed with the same  fit.
The diagonal line is the lowest order perturbative result.
\label{fig:gammamg2}}
\end{figure}

 I will pause and show a few more pictures from \textcite{DeGrand:2011qd}.
Figure \ref{fig:1g2} and Fig.~\ref{fig:zp}
showed raw lattice data for the running coupling and Sch\"rodinger functional $Z_P(L)$.
These data can be turned into plots of the beta function and coupling-dependent mass anomalous dimension.
These are shown in Figs.~\ref{fig:betasu2} and \ref{fig:gammamg2}.

 Many groups have observed that $\gamma_m$ is small. Some numbers are given in Table~\ref{tab:su2agamma}.

\begin{table}
\begin{tabular}{c c c}
\hline
Reference  &  method &  $\gamma_m$ \\
\hline
\textcite{DelDebbio:2010hx} & scaling &  0.22(6))  \\
 \textcite{DeGrand:2011qd} & SF & 0.31(6) \\
\textcite{Patella:2012da} & spectral & 0.371(20) \\
\textcite{Giedt:2012rj} & FSS & 0.50(26) \\
\textcite{DelDebbio:2013hha} & spectral &  0.38(2)  \\
\hline
 \end{tabular}
\caption{Mass anomalous dimension $\gamma_m$ in $SU(2)$ with $N_f=2$ adjoint fermions, from
 publications with reasonably small uncertainties.
 Under ``method,''
FSS refers to finite size scaling, ``scaling'' to a fit to Eq~\protect{\ref{eq:corrlen}}.
 ``Spectral'' refers to use of the spectral density of Dirac eigenvalues.
SF is Schr\"odinger functional.
\label{tab:su2agamma}}
\end{table}

%%%%%%%%%%%%%%%%%%%%%%%%%%%%%%%%%%%%%%%%%%%%%%%%%%%%%%%%%%%%%%%%%%%%%
%%%%%%%%%%%%%%%%%%%%%%%%%%%%%%%%%%%%%%%%%%%%%%%%%%%%%%%%%%%%%%%%%%%%%

%%%%%%%%%%%%%%%%%%%%%%%%%%%%%%%%%%%%%%%%%%%%%%%%%%%%%%%%%%%%%%%%%%%%%
%%%%%%%%%%%%%%%%%%%%%%%%%%%%%%%%%%%%%%%%%%%%%%%%%%%%%%%%%%%%%%%%%%%%%

Recently  \textcite{Athenodorou:2014eua} report
that $SU(2)$ with one adjoint
flavor is near conformal with a mass anomalous dimension near unity.
 This comes from 
finite size scaling of the spectrum and the integrated spectral density.
 Perturbatively, this system is like $SU(3)$ with eight fundamentals; $b_1$ is small and $b_2$ is even smaller.

The other work in this area I know of is mostly by me and my collaborators, all with Wilson fermions, mostly
using Sch\"rodinger functional:
\begin{itemize}
\item $SU(3)$ with $N_f=2$ two-index symmetric (S2) representation fermions,
 \cite{DeGrand:2008kx,DeGrand:2010na,DeGrand:2012yq},  also with spectroscopy \cite{DeGrand:2008kx},
and finite size scaling \cite{DeGrand:2009hu}
\item $SU(4)$ with $N_f=2$ S2 representation fermions \cite{DeGrand:2012qa}
\item $SU(3)$ with $N_f=2$ adjoints \cite{DeGrand:2013uha}
\item $SU(4)$ with $N_f=6$ AS2 fermions  \cite{DeGrand:2013uha}
\end{itemize}
 We could not tell if the beta function
had a zero for any of these systems -- it either followed, or ran more slowly than, the two loop formula,
deep into strong coupling. At this point we lost control of the calculation: either we hit
the Wilson fermion transition where zero quark mass was lost, or the calculation simply 
became too expensive. Spectral data for $SU(3)$ S2  at several bare gauge couplings shows curve collapse consistent
with near conformal behavior distorted by the finite volume.
(That was shown in Fig.~\ref{fig:alltest}.)
All these systems are claimed to have small mass anomalous dimension
at values of the coupling constant where the observed beta function is small.
This region dominates the integral in the formula for the running condensate, Eq.~\ref{eq:rung},
so most of the evolution is at small $\gamma_m$. This renders these systems
uninteresting for technicolor, we said, even if the beta function were to become large and
 negative at even stronger coupling.

\begin{figure}
\begin{center}
\includegraphics[width=0.4\textwidth,clip]{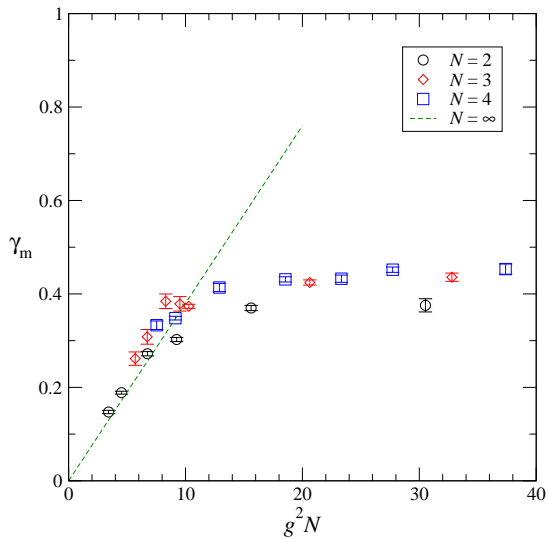}
\end{center}
\caption{ $\gamma_m$  from simulations with two flavors of symmetric representation fermions,
displayed as a function of $\lambda=g^2N$ where $g^2$ is the Schr\"odinger functional coupling.
Data is from \protect{\textcite{DeGrand:2012qa}}.
The line is the lowest order large-$N$ prediction.
\label{fig:gammamlargeN}}
\end{figure}

\begin{figure}
\begin{center}
\includegraphics[width=0.4\textwidth,clip]{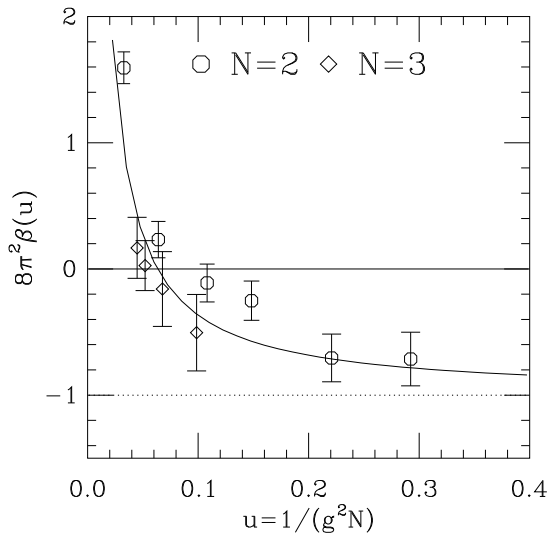}
\end{center}
\caption{ The beta function from  from simulations with two flavors of adjoint fermions,
Data is from \protect{\textcite{DeGrand:2011qd,DeGrand:2013uha}}.
The lines are the beta functions at one and two loops.
\label{fig:betalargeN}}
\end{figure}

Large-$N_c$ scaling is a nice way to present these systems. Figure 
\ref{fig:gammamlargeN} shows $\gamma_m$ from
our S2 studies and Fig.~\ref{fig:betalargeN} shows the beta function with two adjoint flavors.

%%%%%%%%%%%%%%%%%%%%%%%%%%%%%%%%%%%%%%%%%%%%%%%%%%%%%%%%%%%%%%%%%%%%%
%%%%%%%%%%%%%%%%%%%%%%%%%%%%%%%%%%%%%%%%%%%%%%%%%%%%%%%%%%%%%%%%%%%%%

There is a controversy about  $SU(3)$ with $N_f=2$ S2 fermions:
\textcite{Fodor:2012ty}
claim that the system is confining and chirally broken. Their results have, so far, mostly only been
presented in a long series of conference proceedings. It is difficult to evaluate
such works in progress, so
 my description of their results might be incomplete. Let me try:

Their calculations use staggered fermions with $N_f=2$ flavors achieved by rooting
 the fermion determinant.
The bulk of their published simulations are almost all at one gauge coupling, 
deep in strong coupling,
although data at four couplings is said to exist. Data are collected at
many bare fermion masses and lattice volumes, and the pseudoscalar mass
and chiral condensate are presented after an extrapolation to infinite volume using
chiral perturbation theory. The chiral condensate extrapolates to a nonzero value in the zero mass limit.
 The vector and axial vector mesons do not appear to be degenerate
  (so that the parity doubling which would indicate chiral restoration is absent).
  The pseudoscalar and vector masses 
appear to separate; looking at the figures in \textcite{Fodor:2012ty}, at the lightest fermion mass, the
pseudoscalar's mass is about half the vector's. 
They observe $m_\rho/f_\pi \sim 7$ in the chiral limit.
Fits of different (infinite-volume extrapolated) quantities to the naive Eq.~\ref{eq:corrlen} 
done in \textcite{Fodor:2012ty} do not give 
$y_m$'s    which are consistent with each other. Finally, \textcite{Fodor:2014pqa,Fodor:2015vwa}
claim evidence for a light isoscalar scalar state. When I compare  figures in their two papers,
it appears to be lighter than their pseudoscalar mass at the lightest recorded quark mass.

Observable related to the potential and a running coupling present contrasting pictures.
 \textcite{Fodor:2012uw} shows a plot of the static force
versus distance $r$. At $r/a=4-5$, it looks Coulombic to the eye, and by $r/a=10$ the force is constant.
This says that the potential changes qualitatively over a scale factor of distance of about two.
With their vector meson masses $am_\rho$ to give a dimensionless number, the crossover is at a distance
where $m_\rho r \sim 2$. In QCD, the inflection point is  at about 
$r \sim 0.3$ fm, so $m_\rho r \sim 1$. This comparison plus the spectroscopy reported in the previous
paragraph argues for a QCD-like system with a rapidly varying coupling constant.

However, \textcite{Fodor:2015vwa} presents a calculation of a ``flow'' coupling constant. They show a figure
overlaying their result on the one and two-loop perturbative beta function. It shows a coupling
which increases over the observed range, without a fixed point. The coupling
appears to be running at much lower rate  than that of the one-loop beta function, and at their largest coupling,
it runs more slowly than the two-loop formula. (Recall that the one loop $b_1=13/3$ for this system, as opposed to
$b_1=9$ for $SU(3)$ with three fundamental
flavors or $b_1=3$ for twelve fundamentals.) 
I do not know how to reconcile the results of this paragraph with those
of the previous one.
There is insufficient published data from these authors to allow further conclusions to be drawn.

Over the last few years, \textcite{Kogut:2010cz,Kogut:2011ty,Kogut:2014kla}
have investigated this system, and the $N_f=3$ S2 system, using the motion of  finite temperature
phase transition as a potential indicator of confining versus infrared  conformal behavior. Their results
are ambiguous: with small lattices in the temporal direction ($N_t=4$ and 6)
 the transitions, which are located at strong coupling, move quickly,
while at larger $N_t$ the transition continues to move, but more slowly.
``However, further simulations at larger $N_t(s)$ are needed,'' write \textcite{Kogut:2014kla}.

\subsection{An attempt to sum up}

I think that  nearly all lattice results from systems with many fermion degrees of freedom
show behavior which is consistent with expectations from the one-loop beta function, as described around
Eq.~\ref{eq:naiverun}. (The one exception is  the work by one group on $SU(3)$ with $N_f=2$ S2 fermions, noted immediately above.)
Nearly all systems studied have both spectroscopic data and coupling constant measurements,
and more-or-less power law behavior for spectra (Eqs.~\ref{eq:corrlen} and \ref{eq:fss1}) is correlated with
the presence of slowly running couplings.
What is still unknown is the precise boundary of the conformal window. This is not surprising: 
slow running is not that different from no running.

%%%%%%%%%%%%%%%%%%%%%%%%%%%%%%%%%%%%%%%

The  plot of \textcite{Dietrich:2006cm}, Fig.~\ref{fig:ds}, has been frequently mentioned.
Let us try to build our own picture, using lattice data. Figure \ref{fig:ds}
 shows the boundary of the conformal window 
for various representations as lines, as if one could imagine systems with fractional flavor number. The
situation for small $N_f$ and $N_c$ is more discrete, of course.
 How to present things?
The relevant variable is something like the number of fermionic degrees of freedom, versus the number
of gluonic ones.
Looking at the two-loop beta function, a reasonable choice is $N_f T(R)/N_c$. (Of course, $b_2$ includes
$C_2(R)$, but it basically scales as $N_c$, and it has a small coefficient compared to 20/3.)

\begin{figure}
\begin{center}
\includegraphics[width=\columnwidth,clip]{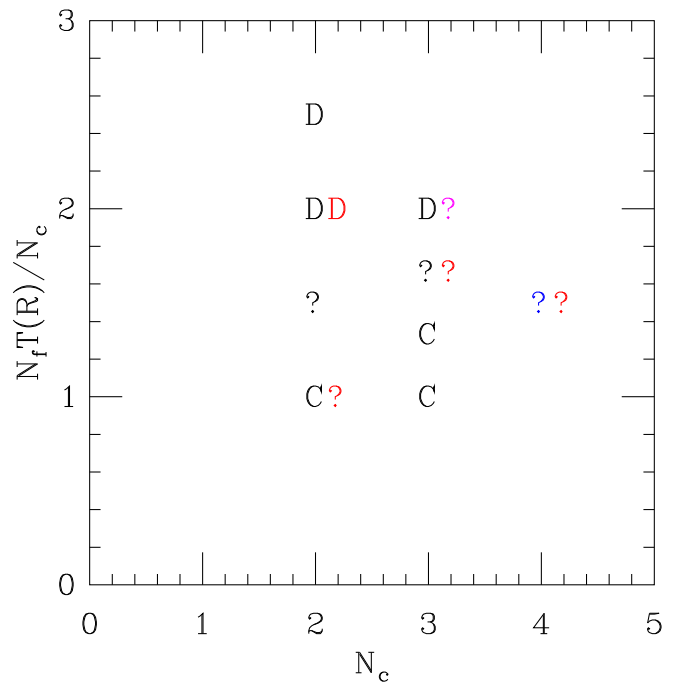}
\end{center}
\caption{ (Color online) My attempt at a synthesis of lattice results for the vacuum structure of
various systems. The labels are ``C''  for confined and chirally broken,  ``D'' for 
deconfined, chirally restored, and probably conformal in the massless limit, 
and ``?''for unknown.
Colors are black  for fundamental representation, red  for S2, blue
 for AS2 and purple  for
adjoint $N\ne 2$ (close in color to red because they are the same for $SU(2)$).
\label{fig:everything}}
\end{figure}

Figure \ref{fig:everything} is my best guess at the status. 
 The labels are ``C''  for confined and chirally broken,  ``D''  for 
deconfined, chirally restored, and probably conformal in the massless limit, and ``?'' 
 for unknown. Colors are black  for fundamental representation, red  for S2, blue
 for AS2 and purple for
adjoint $N\ne 2$.
If you are viewing this figure in black and white, the rightmost symbols for $N_c=2$ and 3 are S2 (equivalent to adjoint
for $N_c=2$) and the leftmost ones are for fundamentals. The two $N_c=4$ entries are S2 and AS2.

%%%%%%%%%%%%%%%%%%%%%%%%%%%%%%%%%%%%%%%%%%%%%%%%%%%%%%%%%%%%%%%%%%%%%

I assigned the following systems question marks: for $N_c=2$, $N_f=6$ fundamentals and $N_f=1$ adjoints;
for $N_c=3$, $N_f=10$ fundamentals. I listed all of the higher-representation systems with $N_c=3$ and 4 as
``unknown.''
Yes, after seven years of work, there are still
question marks. But, people did not study systems where they knew the answer.
 These are all difficult systems.

And did anyone ever publish a plot with the `technicolor dream'' beta function,
 Fig.~\ref{fig:walkingdream}?
Not in a simulation of a four dimensional  system of gauge fields and fermions.

\begin{figure}
\begin{center}
\includegraphics[width=0.4\textwidth,clip]{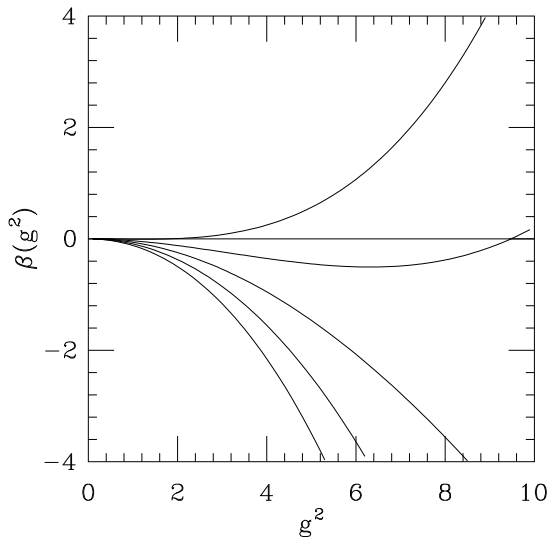}
\end{center}
\caption{Two-loop beta function for $SU(3)$ with (from the bottom up) $N_f=3$, 6, 9, 12, and 15
 flavors of fundamental representation fermions. ($N_f=9$ has a zero at $g^2=66$, far to the right off the plot.)
\label{fig:beta3691215}}
\end{figure}

To say once more why slow running was difficult: A dip in the beta function was not the issue, 
the problem was the extremely
small value of the one loop beta function as the number of fermion degrees of freedom increased.
 Contrast Fig.~\ref{fig:walkingdream}  with real two-loop running,
 Fig.~\ref{fig:beta3691215}.
 
 I think that generally, when they began studying systems
with slowly running couplings,
people did not appreciate how different they were from QCD. Much of the context QCD studies used to 
evaluate results was absent. For example, dealing with rooted staggered fermions
 requires knowledge that the system is chirally broken, plus access to the Gaussian fixed point,
the ability to make the system weakly interacting at short distance while maintaining strong
interactions at long distance.

Several approaches worked poorly. 
I do not think that simulations at finite temperature have been too useful.
Large scale spectroscopic simulations at a single
value of the bare gauge coupling generally proved inadequate for determining whether a system was
confining and chirally broken, or infrared conformal. 
Simulations at many volumes were useful. Simulations at one large volume, hoping to approximate infinite volume,
 were less so. Remember
Eq.~\ref{eq:naiverun}.
When the coupling constant runs slowly, no volume is large enough.
I think that the case of $N_f=12$ fundamentals clearly illustrates this conclusion. Fig.~\ref{fig:annapion}
is a smoking gun for infrared conformality, and the authors needed data from many bare parameters and volumes
 to build it.

%%%%%%%%%%%%%%%%%%%%%%%%%%%%%%%%%%%%%%%%%%%%%%%%%%%%%%%%%%%%%%%%%%%%%

The situation for the
mass anomalous dimension is harder to summarize. The best-determined numbers are for
$SU(2)$ with $N_f=2$ adjoints, and $SU(3)$ with $N_f=12$ fundamentals. In both cases $\gamma_m$ is small.
Only for the systems $SU(3)$ with $N_f=8$ fundamentals, $SU(2)$ with $N_f=6$ fundamentals,
and $SU(2)$ with $N_f=1$ adjoint are there claims of large $\gamma_m$'s, and the claims are presented
very cautiously.
All other systems appear to have small mass anomalous dimensions. Apparently, a large mass anomalous 
dimension can only occur for a theory which is right on the sill of the conformal window (if at all).

%%%%%%%%%%%%%%%%%%%%%%%%%%%%%%%%%%%%%%%%%%%%%%%%%%%%%%%%%%%%%%%%%%%%%%%%%%%%%%%%%%%%%%%%%%%%%%%%%%%
\section{Conclusions \label{sec:summary}}
%%%%%%%%%%%%%%%%%%%%%%%%%%%%%%%%%%%%%%%%%%%%%%%%%%%%%%%%%%%%%%%%%%%%%
At the time I am writing, there is no evidence for any particular beyond Standard Model scenario
in data. The Higgs exists with near Standard Model couplings, and evidence for beyond Standard Model
physics (neutrino masses and mixing, dark matter, and so on) remains a set of disconnected
observations.
As an outsider in the  beyond Standard Model field, it seems to me that the dominant theoretical 
motivation for new physics
is not so much that there is physics beyond the Standard Model, but that
there is a theoretical issue with the Standard Model itself -- the hierarchy problem.
 This takes us back to 
a nonperturbative resolution of the hierarchy problem as a possibly attractive choice, Eq.~\ref{eq:AFrunning}
and beyond.

 This is a niche for the lattice. But it requires  an ultraviolet completion, before any
lattice calculation can be envisioned.
The lattice is not about symmetries, it is about low energy constants. Phenomenologists who have a favorite
beyond Standard Model scenario and want lattice people to study it have to give them
a concrete ultraviolet completion.

 Most lattice work focused on one particular corner of beyond Standard Model dynamics,
 technicolor, and
on one small area of technicolor, mostly $SU(3)$ and many fundamentals. This certainly seems peculiar,
given the wide set of continuum beyond Standard Model possibilities in the literature.
Why did this happen? I am not sure. It might be because lattice simulations have to begin with
some ultraviolet completion, and
because  the framework of candidate ultraviolet completions
was naturally present in the technicolor literature, in a lattice-friendly way: non-Abelian gauge 
theories with fermions
in four dimensions.

Studies of near-conformal systems tell us that when $N_c$ is small,
 there are actually only a small number of confining and chirally broken systems. Most of them are not appropriate
for beyond Standard Model physics associated with the Higgs: the coupling constants probably run
 too fast for technicolor, and the flavor symmetry groups
are often too small for composite Higgs scenarios. However, some of them are
composite Higgs candidates, and some of them could be composite Dark Matter candidates.
Most of them are also interesting as analog systems for QCD. Little is known about their
mass (and other)  anomalous dimensions. All of them could be explored with today's available
software and computer power. In particular, much of the technology for computing QCD matrix elements
can be straightforwardly applied to these systems.
 This could be an interesting thing to do. It would check large $N_c$ counting
of matrix elements, and the variation of observables on $N_f$ could be probed. Recall how, in
Sec.~\ref{sec:tc}, I said  that if technicolor was like QCD, it would be ruled out by experiment,
but that technicolor might not be QCD-like? A larger version of this question is to ask how much like QCD
are theories which are nearby it in the space of $N_c$, $N_f$ and fermion representation.

Of course, no theoretical calculation by itself is going to reveal the existence of some particular
beyond Standard Model scenario. Without new experimental data, all theory can do is suggest possibilities.
In the long term, whether or not the words ``Lattice'' and ``beyond Standard Model''  
should -- or will -- appear  again in the same title
is a question only experiment can decide.
%%%%%%%%%%%%%%%%%%%%%%%%%%%%%%%%%%%%%%%%%%%%%%%%%%%%%%%%%%%%%%%%%%%%%
\begin{acknowledgments}
%%%%%%%%%%%%%%%%%%%%%%%%%%%%%%%%%%%%%%%%%%%%%%%%%%%%%%%%%%%%%%%%%%%%%%
I am grateful for  conversations  and correspondence about the material in this review with
S.~Catterall,
P.~Damgaard,
C.~DeTar,
M.~Golterman,
A.~Hasenfratz,
W.~Jay,
F.~Knechtli,
Y.~Liu,
E.~Neil,
D.~Schaich,
Y.~Shamir,
R.~Shrock,
and
B.~Svetitsky.
I appreciate the comments of Jay, Shamir, and Svetitsky on a draft of this review.
This work was supported in part by the U.~S. Department of Energy.
%%%%%%%%%%%%%%%%%%%%%%%%%%%%%%%%%%%%%%%%%%%%%%%%%%%%%%%%%%%%%%%%%%%%%
\end{acknowledgments}
%%%%%%%%%%%%%%%%%%%%%%%%%%%%%%%%%%%%%%%%%%%%%%%%%%%%%%%%%%%%%%%%%%%%%%

%%%%%%%%%%%%%%%%%%%%%%%%%%%%%%%%%%%%%%%%%%%%%%%%%%%%%%%%%%%%%%%%%%%%%
\bibliography{review}

%merlin.mbs 2010-03-15 4.21a (PWD, AO, DPC)
%Control: key (0)
%Control: author (3) reversed first dotless
%Control: editor formatted (0) differently from author
%Control: production of article title (0) allowed
%Control: page (1) range
%Control: year (0) verbatim
%Control: production of eprint (0) enabled
\begin{thebibliography}{234}%
\makeatletter
\providecommand \@ifxundefined [1]{%
 \@ifx{#1\undefined}
}%
\providecommand \@ifnum [1]{%
 \ifnum #1\expandafter \@firstoftwo
 \else \expandafter \@secondoftwo
 \fi
}%
\providecommand \@ifx [1]{%
 \ifx #1\expandafter \@firstoftwo
 \else \expandafter \@secondoftwo
 \fi
}%
\providecommand \natexlab [1]{#1}%
\providecommand \enquote  [1]{``#1''}%
\providecommand \bibnamefont  [1]{#1}%
\providecommand \bibfnamefont [1]{#1}%
\providecommand \citenamefont [1]{#1}%
\providecommand \href@noop [0]{\@secondoftwo}%
\providecommand \href [0]{\begingroup \@sanitize@url \@href}%
\providecommand \@href[1]{\@@startlink{#1}\@@href}%
\providecommand \@@href[1]{\endgroup#1\@@endlink}%
\providecommand \@sanitize@url [0]{\catcode `\\12\catcode `\$12\catcode
  `\&12\catcode `\#12\catcode `\^12\catcode `\_12\catcode `\%12\relax}%
\providecommand \@@startlink[1]{}%
\providecommand \@@endlink[0]{}%
\providecommand \url  [0]{\begingroup\@sanitize@url \@url }%
\providecommand \@url [1]{\endgroup\@href {#1}{\urlprefix }}%
\providecommand \urlprefix  [0]{URL }%
\providecommand \Eprint [0]{\href }%
\@ifxundefined \urlstyle {%
  \providecommand \doi  [0]{\begingroup \@sanitize@url \@doi}%
  \providecommand \@doi [1]{\endgroup \@@startlink {\doibase
  #1}doi:\discretionary {}{}{}#1\@@endlink }%
}{%
  \providecommand \doi  [0]{doi:\discretionary{}{}{}\begingroup
  \urlstyle{rm}\Url }%
}%
\providecommand \doibase [0]{http://dx.doi.org/}%
\providecommand \Doi [0]{\begingroup \@sanitize@url \@Doi }%
\providecommand \@Doi  [1]{\endgroup\@@startlink{\doibase#1}\@@Doi}%
\providecommand \@@Doi [1]{#1\@@endlink}%
\providecommand \selectlanguage [0]{\@gobble}%
\providecommand \bibinfo  [0]{\@secondoftwo}%
\providecommand \bibfield  [0]{\@secondoftwo}%
\providecommand \translation [1]{[#1]}%
\providecommand \BibitemOpen [0]{}%
\providecommand \bibitemStop [0]{}%
\providecommand \bibitemNoStop [0]{.\EOS\space}%
\providecommand \EOS [0]{\spacefactor3000\relax}%
\providecommand \BibitemShut  [1]{\csname bibitem#1\endcsname}%
%</preamble>
\bibitem [{\citenamefont {Aad}\ \emph {et~al.}(2012)\citenamefont {Aad} \emph
  {et~al.}}]{Aad:2012tfa}%
  \BibitemOpen
  \bibfield  {author} {\bibinfo {author} {\bibnamefont {Aad} \bibfnamefont
  {Georges}},  \emph {et~al.} (\bibinfo {collaboration} {ATLAS Collaboration})}
  (\bibinfo {year} {2012}),\ \bibfield  {title} {\enquote {\bibinfo {title}
  {{Observation of a new particle in the search for the Standard Model Higgs
  boson with the ATLAS detector at the LHC}},}\ }\Doi
  {10.1016/j.physletb.2012.08.020} {\bibfield  {journal} {\bibinfo  {journal}
  {Phys.~Lett.},\ }\textbf {\bibinfo {volume} {B716}},\ \bibinfo {pages}
  {1--29}},\ \Eprint {http://arxiv.org/abs/1207.7214} {arXiv:1207.7214
  [hep-ex]} \BibitemShut {NoStop}%
%%CITATION = ARXIV:1207.7214;%%
\bibitem [{\citenamefont {Akemann}\ \emph {et~al.}(1998)\citenamefont
  {Akemann}, \citenamefont {Damgaard}, \citenamefont {Magnea},\ and\
  \citenamefont {Nishigaki}}]{Akemann:1997wi}%
  \BibitemOpen
  \bibfield  {author} {\bibinfo {author} {\bibnamefont {Akemann} \bibfnamefont
  {G}}, \bibinfo {author} {\bibfnamefont {P.H.}\ \bibnamefont {Damgaard}},
  \bibinfo {author} {\bibfnamefont {Ulrika}\ \bibnamefont {Magnea}}, \ and\
  \bibinfo {author} {\bibfnamefont {S.M.}\ \bibnamefont {Nishigaki}}} (\bibinfo
  {year} {1998}),\ \bibfield  {title} {\enquote {\bibinfo {title}
  {{Multicritical microscopic spectral correlators of Hermitian and complex
  matrices}},}\ }\Doi {10.1016/S0550-3213(98)00143-6} {\bibfield  {journal}
  {\bibinfo  {journal} {Nucl.~Phys.},\ }\textbf {\bibinfo {volume} {B519}},\
  \bibinfo {pages} {682--714}},\ \Eprint {http://arxiv.org/abs/hep-th/9712006}
  {arXiv:hep-th/9712006 [hep-th]} \BibitemShut {NoStop}%
%%CITATION = HEP-TH/9712006;%%
\bibitem [{\citenamefont {Altarelli}\ and\ \citenamefont
  {Barbieri}(1991)}]{Altarelli:1990zd}%
  \BibitemOpen
  \bibfield  {author} {\bibinfo {author} {\bibnamefont {Altarelli}
  \bibfnamefont {Guido}}, \ and\ \bibinfo {author} {\bibfnamefont {Riccardo}\
  \bibnamefont {Barbieri}}} (\bibinfo {year} {1991}),\ \bibfield  {title}
  {\enquote {\bibinfo {title} {{Vacuum polarization effects of new physics on
  electroweak processes}},}\ }\Doi {10.1016/0370-2693(91)91378-9} {\bibfield
  {journal} {\bibinfo  {journal} {Phys. Lett.},\ }\textbf {\bibinfo {volume}
  {B253}},\ \bibinfo {pages} {161--167}}\BibitemShut {NoStop}%
%%CITATION = PHLTA,B253,161;%%
\bibitem [{\citenamefont {Aoki}\ \emph {et~al.}(2014a)\citenamefont {Aoki},
  \citenamefont {Aoki}, \citenamefont {Bernard}, \citenamefont {Blum},
  \citenamefont {Colangelo} \emph {et~al.}}]{Aoki:2013ldr}%
  \BibitemOpen
  \bibfield  {author} {\bibinfo {author} {\bibnamefont {Aoki} \bibfnamefont
  {Sinya}}, \bibinfo {author} {\bibfnamefont {Yasumichi}\ \bibnamefont {Aoki}},
  \bibinfo {author} {\bibfnamefont {Claude}\ \bibnamefont {Bernard}}, \bibinfo
  {author} {\bibfnamefont {Tom}\ \bibnamefont {Blum}}, \bibinfo {author}
  {\bibfnamefont {Gilberto}\ \bibnamefont {Colangelo}},  \emph {et~al.}}
  (\bibinfo {year} {2014a}),\ \bibfield  {title} {\enquote {\bibinfo {title}
  {{Review of lattice results concerning low-energy particle physics}},}\ }\Doi
  {10.1140/epjc/s10052-014-2890-7} {\bibfield  {journal} {\bibinfo  {journal}
  {Eur.~Phys.~J.},\ }\textbf {\bibinfo {volume} {C74}}~(\bibinfo {number}
  {9}),\ \bibinfo {pages} {2890}},\ \Eprint {http://arxiv.org/abs/1310.8555}
  {arXiv:1310.8555 [hep-lat]} \BibitemShut {NoStop}%
%%CITATION = ARXIV:1310.8555;%%
\bibitem [{\citenamefont {Aoki}\ \emph {et~al.}(2014b)\citenamefont {Aoki},
  \citenamefont {Balog},\ and\ \citenamefont {Weisz}}]{Aoki:2014yra}%
  \BibitemOpen
  \bibfield  {author} {\bibinfo {author} {\bibnamefont {Aoki} \bibfnamefont
  {Sinya}}, \bibinfo {author} {\bibfnamefont {Janos}\ \bibnamefont {Balog}}, \
  and\ \bibinfo {author} {\bibfnamefont {Peter}\ \bibnamefont {Weisz}}}
  (\bibinfo {year} {2014b}),\ \bibfield  {title} {\enquote {\bibinfo {title}
  {{Walking in the 3-dimensional large $N$ scalar model}},}\ }\Doi
  {10.1007/JHEP09(2014)167} {\bibfield  {journal} {\bibinfo  {journal} {JHEP},\
  }\textbf {\bibinfo {volume} {1409}},\ \bibinfo {pages} {167}},\ \Eprint
  {http://arxiv.org/abs/1407.7079} {arXiv:1407.7079 [hep-lat]} \BibitemShut
  {NoStop}%
%%CITATION = ARXIV:1407.7079;%%
\bibitem [{\citenamefont {Aoki}\ \emph {et~al.}(2012)\citenamefont {Aoki},
  \citenamefont {Aoyama}, \citenamefont {Kurachi}, \citenamefont {Maskawa},
  \citenamefont {Nagai} \emph {et~al.}}]{Aoki:2012eq}%
  \BibitemOpen
  \bibfield  {author} {\bibinfo {author} {\bibnamefont {Aoki} \bibfnamefont
  {Yasumichi}}, \bibinfo {author} {\bibfnamefont {Tatsumi}\ \bibnamefont
  {Aoyama}}, \bibinfo {author} {\bibfnamefont {Masafumi}\ \bibnamefont
  {Kurachi}}, \bibinfo {author} {\bibfnamefont {Toshihide}\ \bibnamefont
  {Maskawa}}, \bibinfo {author} {\bibfnamefont {Kei-ichi}\ \bibnamefont
  {Nagai}},  \emph {et~al.}} (\bibinfo {year} {2012}),\ \bibfield  {title}
  {\enquote {\bibinfo {title} {{Lattice study of conformality in twelve-flavor
  QCD}},}\ }\Doi {10.1103/PhysRevD.86.059903, 10.1103/PhysRevD.86.054506}
  {\bibfield  {journal} {\bibinfo  {journal} {Phys.~Rev.},\ }\textbf {\bibinfo
  {volume} {D86}},\ \bibinfo {pages} {054506}},\ \Eprint
  {http://arxiv.org/abs/1207.3060} {arXiv:1207.3060 [hep-lat]} \BibitemShut
  {NoStop}%
%%CITATION = ARXIV:1207.3060;%%
\bibitem [{\citenamefont {Aoki}\ \emph {et~al.}(2013a)\citenamefont {Aoki}
  \emph {et~al.}}]{Aoki:2013zsa}%
  \BibitemOpen
  \bibfield  {author} {\bibinfo {author} {\bibnamefont {Aoki} \bibfnamefont
  {Yasumichi}},  \emph {et~al.} (\bibinfo {collaboration} {LatKMI
  Collaboration})} (\bibinfo {year} {2013a}),\ \bibfield  {title} {\enquote
  {\bibinfo {title} {{Light composite scalar in twelve-flavor QCD on the
  lattice}},}\ }\Doi {10.1103/PhysRevLett.111.162001} {\bibfield  {journal}
  {\bibinfo  {journal} {Phys.~Rev.~Lett.},\ }\textbf {\bibinfo {volume}
  {111}}~(\bibinfo {number} {16}),\ \bibinfo {pages} {162001}},\ \Eprint
  {http://arxiv.org/abs/1305.6006} {arXiv:1305.6006 [hep-lat]} \BibitemShut
  {NoStop}%
%%CITATION = ARXIV:1305.6006;%%
\bibitem [{\citenamefont {Aoki}\ \emph {et~al.}(2013b)\citenamefont {Aoki}
  \emph {et~al.}}]{Aoki:2013xza}%
  \BibitemOpen
  \bibfield  {author} {\bibinfo {author} {\bibnamefont {Aoki} \bibfnamefont
  {Yasumichi}},  \emph {et~al.} (\bibinfo {collaboration} {LatKMI
  Collaboration})} (\bibinfo {year} {2013b}),\ \bibfield  {title} {\enquote
  {\bibinfo {title} {{Walking signals in $N_f=8$ QCD on the lattice}},}\ }\Doi
  {10.1103/PhysRevD.87.094511} {\bibfield  {journal} {\bibinfo  {journal}
  {Phys.~Rev.},\ }\textbf {\bibinfo {volume} {D87}}~(\bibinfo {number} {9}),\
  \bibinfo {pages} {094511}},\ \Eprint {http://arxiv.org/abs/1302.6859}
  {arXiv:1302.6859 [hep-lat]} \BibitemShut {NoStop}%
%%CITATION = ARXIV:1302.6859;%%
\bibitem [{\citenamefont {Aoki}\ \emph {et~al.}(2014)\citenamefont {Aoki} \emph
  {et~al.}}]{Aoki:2014oha}%
  \BibitemOpen
  \bibfield  {author} {\bibinfo {author} {\bibnamefont {Aoki} \bibfnamefont
  {Yasumichi}},  \emph {et~al.} (\bibinfo {collaboration} {the LatKMI
  Collaboration})} (\bibinfo {year} {2014}),\ \bibfield  {title} {\enquote
  {\bibinfo {title} {{Light composite scalar in eight-flavor QCD on the
  lattice}},}\ }\Doi {10.1103/PhysRevD.89.111502} {\bibfield  {journal}
  {\bibinfo  {journal} {Phys.~Rev.},\ }\textbf {\bibinfo {volume} {D89}},\
  \bibinfo {pages} {111502}},\ \Eprint {http://arxiv.org/abs/1403.5000}
  {arXiv:1403.5000 [hep-lat]} \BibitemShut {NoStop}%
%%CITATION = ARXIV:1403.5000;%%
\bibitem [{\citenamefont {Appelquist}\ \emph {et~al.}(2014c)\citenamefont
  {Appelquist}, \citenamefont {Brower}, \citenamefont {Buchoff}, \citenamefont
  {Cheng}, \citenamefont {Fleming} \emph {et~al.}}]{Appelquist:2013pqa}%
  \BibitemOpen
  \bibfield  {author} {\bibinfo {author} {\bibnamefont {Appelquist}
  \bibfnamefont {T}}, \bibinfo {author} {\bibfnamefont {R.C.}\ \bibnamefont
  {Brower}}, \bibinfo {author} {\bibfnamefont {M.I.}\ \bibnamefont {Buchoff}},
  \bibinfo {author} {\bibfnamefont {M.}~\bibnamefont {Cheng}}, \bibinfo
  {author} {\bibfnamefont {G.T.}\ \bibnamefont {Fleming}},  \emph {et~al.}}
  (\bibinfo {year} {2014c}),\ \bibfield  {title} {\enquote {\bibinfo {title}
  {{Two-Color Theory with Novel Infrared Behavior}},}\ }\Doi
  {10.1103/PhysRevLett.112.111601} {\bibfield  {journal} {\bibinfo  {journal}
  {Phys.~Rev.~Lett.},\ }\textbf {\bibinfo {volume} {112}},\ \bibinfo {pages}
  {111601}},\ \Eprint {http://arxiv.org/abs/1311.4889} {arXiv:1311.4889
  [hep-ph]} \BibitemShut {NoStop}%
%%CITATION = ARXIV:1311.4889;%%
\bibitem [{\citenamefont {Appelquist}\ \emph {et~al.}(2011a)\citenamefont
  {Appelquist}, \citenamefont {Fleming}, \citenamefont {Lin}, \citenamefont
  {Neil},\ and\ \citenamefont {Schaich}}]{Appelquist:2011dp}%
  \BibitemOpen
  \bibfield  {author} {\bibinfo {author} {\bibnamefont {Appelquist}
  \bibfnamefont {T}}, \bibinfo {author} {\bibfnamefont {G.T.}\ \bibnamefont
  {Fleming}}, \bibinfo {author} {\bibfnamefont {M.F.}\ \bibnamefont {Lin}},
  \bibinfo {author} {\bibfnamefont {E.T.}\ \bibnamefont {Neil}}, \ and\
  \bibinfo {author} {\bibfnamefont {D.A.}\ \bibnamefont {Schaich}}} (\bibinfo
  {year} {2011a}),\ \bibfield  {title} {\enquote {\bibinfo {title} {{Lattice
  Simulations and Infrared Conformality}},}\ }\Doi {10.1103/PhysRevD.84.054501}
  {\bibfield  {journal} {\bibinfo  {journal} {Phys.~Rev.},\ }\textbf {\bibinfo
  {volume} {D84}},\ \bibinfo {pages} {054501}},\ \Eprint
  {http://arxiv.org/abs/1106.2148} {arXiv:1106.2148 [hep-lat]} \BibitemShut
  {NoStop}%
%%CITATION = ARXIV:1106.2148;%%
\bibitem [{\citenamefont {Appelquist}\ \emph {et~al.}(2013)\citenamefont
  {Appelquist} \emph {et~al.}}]{Appelquist:2013ms}%
  \BibitemOpen
  \bibfield  {author} {\bibinfo {author} {\bibnamefont {Appelquist}
  \bibfnamefont {T}},  \emph {et~al.} (\bibinfo {collaboration} {Lattice Strong
  Dynamics (LSD) Collaboration})} (\bibinfo {year} {2013}),\ \bibfield  {title}
  {\enquote {\bibinfo {title} {{Lattice calculation of composite dark matter
  form factors}},}\ }\Doi {10.1103/PhysRevD.88.014502} {\bibfield  {journal}
  {\bibinfo  {journal} {Phys.~Rev.},\ }\textbf {\bibinfo {volume}
  {D88}}~(\bibinfo {number} {1}),\ \bibinfo {pages} {014502}},\ \Eprint
  {http://arxiv.org/abs/1301.1693} {arXiv:1301.1693 [hep-ph]} \BibitemShut
  {NoStop}%
%%CITATION = ARXIV:1301.1693;%%
\bibitem [{\citenamefont {Appelquist}\ \emph {et~al.}(2014a)\citenamefont
  {Appelquist} \emph {et~al.}}]{Appelquist:2014zsa}%
  \BibitemOpen
  \bibfield  {author} {\bibinfo {author} {\bibnamefont {Appelquist}
  \bibfnamefont {T}},  \emph {et~al.} (\bibinfo {collaboration} {LSD
  Collaboration})} (\bibinfo {year} {2014a}),\ \bibfield  {title} {\enquote
  {\bibinfo {title} {{Lattice simulations with eight flavors of domain wall
  fermions in SU(3) gauge theory}},}\ }\Doi {10.1103/PhysRevD.90.114502}
  {\bibfield  {journal} {\bibinfo  {journal} {Phys.~Rev.},\ }\textbf {\bibinfo
  {volume} {D90}}~(\bibinfo {number} {11}),\ \bibinfo {pages} {114502}},\
  \Eprint {http://arxiv.org/abs/1405.4752} {arXiv:1405.4752 [hep-lat]}
  \BibitemShut {NoStop}%
%%CITATION = ARXIV:1405.4752;%%
\bibitem [{\citenamefont {Appelquist}\ \emph {et~al.}(2014b)\citenamefont
  {Appelquist} \emph {et~al.}}]{Appelquist:2014jch}%
  \BibitemOpen
  \bibfield  {author} {\bibinfo {author} {\bibnamefont {Appelquist}
  \bibfnamefont {T}},  \emph {et~al.} (\bibinfo {collaboration} {Lattice Strong
  Dynamics (LSD) Collaboration})} (\bibinfo {year} {2014b}),\ \bibfield
  {title} {\enquote {\bibinfo {title} {{Composite bosonic baryon dark matter on
  the lattice: SU(4) baryon spectrum and the effective Higgs interaction}},}\
  }\Doi {10.1103/PhysRevD.89.094508} {\bibfield  {journal} {\bibinfo  {journal}
  {Phys.~Rev.},\ }\textbf {\bibinfo {volume} {D89}},\ \bibinfo {pages}
  {094508}},\ \Eprint {http://arxiv.org/abs/1402.6656} {arXiv:1402.6656
  [hep-lat]} \BibitemShut {NoStop}%
%%CITATION = ARXIV:1402.6656;%%
\bibitem [{\citenamefont {Appelquist}\ \emph {et~al.}(2012)\citenamefont
  {Appelquist}, \citenamefont {Babich}, \citenamefont {Brower}, \citenamefont
  {Buchoff}, \citenamefont {Cheng} \emph {et~al.}}]{Appelquist:2012sm}%
  \BibitemOpen
  \bibfield  {author} {\bibinfo {author} {\bibnamefont {Appelquist}
  \bibfnamefont {Thomas}}, \bibinfo {author} {\bibfnamefont {Ron}\ \bibnamefont
  {Babich}}, \bibinfo {author} {\bibfnamefont {Richard~C.}\ \bibnamefont
  {Brower}}, \bibinfo {author} {\bibfnamefont {Michael~I.}\ \bibnamefont
  {Buchoff}}, \bibinfo {author} {\bibfnamefont {Michael}\ \bibnamefont
  {Cheng}},  \emph {et~al.}} (\bibinfo {year} {2012}),\ \bibfield  {title}
  {\enquote {\bibinfo {title} {{WW Scattering Parameters via Pseudoscalar Phase
  Shifts}},}\ }\Doi {10.1103/PhysRevD.85.074505} {\bibfield  {journal}
  {\bibinfo  {journal} {Phys.~Rev.},\ }\textbf {\bibinfo {volume} {D85}},\
  \bibinfo {pages} {074505}},\ \Eprint {http://arxiv.org/abs/1201.3977}
  {arXiv:1201.3977 [hep-lat]} \BibitemShut {NoStop}%
%%CITATION = ARXIV:1201.3977;%%
\bibitem [{\citenamefont {Appelquist}\ \emph {et~al.}(2015b)\citenamefont
  {Appelquist}, \citenamefont {Berkowitz}, \citenamefont {Brower},
  \citenamefont {Buchoff}, \citenamefont {Fleming} \emph
  {et~al.}}]{Appelquist:2015zfa}%
  \BibitemOpen
  \bibfield  {author} {\bibinfo {author} {\bibnamefont {Appelquist}
  \bibfnamefont {Thomas}}, \bibinfo {author} {\bibfnamefont {Evan}\
  \bibnamefont {Berkowitz}}, \bibinfo {author} {\bibfnamefont {Richard~C.}\
  \bibnamefont {Brower}}, \bibinfo {author} {\bibfnamefont {Michael~I.}\
  \bibnamefont {Buchoff}}, \bibinfo {author} {\bibfnamefont {George~T.}\
  \bibnamefont {Fleming}},  \emph {et~al.}} (\bibinfo {year} {2015b}),\
  \bibfield  {title} {\enquote {\bibinfo {title} {{Direct Detection of Stealth
  Dark Matter through Electromagnetic Polarizability}},}\ }\href@noop {}
  {}\Eprint {http://arxiv.org/abs/1503.04205} {arXiv:1503.04205 [hep-ph]}
  \BibitemShut {NoStop}%
%%CITATION = ARXIV:1503.04205;%%
\bibitem [{\citenamefont {Appelquist}\ \emph {et~al.}(2015a)\citenamefont
  {Appelquist}, \citenamefont {Brower}, \citenamefont {Buchoff}, \citenamefont
  {Fleming}, \citenamefont {Jin} \emph {et~al.}}]{Appelquist:2015yfa}%
  \BibitemOpen
  \bibfield  {author} {\bibinfo {author} {\bibnamefont {Appelquist}
  \bibfnamefont {Thomas}}, \bibinfo {author} {\bibfnamefont {Richard~C.}\
  \bibnamefont {Brower}}, \bibinfo {author} {\bibfnamefont {Michael~I.}\
  \bibnamefont {Buchoff}}, \bibinfo {author} {\bibfnamefont {George~T.}\
  \bibnamefont {Fleming}}, \bibinfo {author} {\bibfnamefont {Xiao-Yong}\
  \bibnamefont {Jin}},  \emph {et~al.}} (\bibinfo {year} {2015a}),\ \bibfield
  {title} {\enquote {\bibinfo {title} {{Stealth Dark Matter: Dark scalar
  baryons through the Higgs portal}},}\ }\href@noop {} {}\Eprint
  {http://arxiv.org/abs/1503.04203} {arXiv:1503.04203 [hep-ph]} \BibitemShut
  {NoStop}%
%%CITATION = ARXIV:1503.04203;%%
\bibitem [{\citenamefont {Appelquist}\ \emph {et~al.}(2008)\citenamefont
  {Appelquist}, \citenamefont {Fleming},\ and\ \citenamefont
  {Neil}}]{Appelquist:2007hu}%
  \BibitemOpen
  \bibfield  {author} {\bibinfo {author} {\bibnamefont {Appelquist}
  \bibfnamefont {Thomas}}, \bibinfo {author} {\bibfnamefont {George~T.}\
  \bibnamefont {Fleming}}, \ and\ \bibinfo {author} {\bibfnamefont {Ethan~T.}\
  \bibnamefont {Neil}}} (\bibinfo {year} {2008}),\ \bibfield  {title} {\enquote
  {\bibinfo {title} {{Lattice study of the conformal window in QCD-like
  theories}},}\ }\Doi {10.1103/PhysRevLett.100.171607} {\bibfield  {journal}
  {\bibinfo  {journal} {Phys.~Rev.~Lett.},\ }\textbf {\bibinfo {volume}
  {100}},\ \bibinfo {pages} {171607}},\ \Eprint
  {http://arxiv.org/abs/0712.0609} {arXiv:0712.0609 [hep-ph]} \BibitemShut
  {NoStop}%
%%CITATION = ARXIV:0712.0609;%%
\bibitem [{\citenamefont {Appelquist}\ \emph {et~al.}(2009)\citenamefont
  {Appelquist}, \citenamefont {Fleming},\ and\ \citenamefont
  {Neil}}]{Appelquist:2009ty}%
  \BibitemOpen
  \bibfield  {author} {\bibinfo {author} {\bibnamefont {Appelquist}
  \bibfnamefont {Thomas}}, \bibinfo {author} {\bibfnamefont {George~T.}\
  \bibnamefont {Fleming}}, \ and\ \bibinfo {author} {\bibfnamefont {Ethan~T.}\
  \bibnamefont {Neil}}} (\bibinfo {year} {2009}),\ \bibfield  {title} {\enquote
  {\bibinfo {title} {{Lattice Study of Conformal Behavior in SU(3) Yang-Mills
  Theories}},}\ }\Doi {10.1103/PhysRevD.79.076010} {\bibfield  {journal}
  {\bibinfo  {journal} {Phys.~Rev.},\ }\textbf {\bibinfo {volume} {D79}},\
  \bibinfo {pages} {076010}},\ \Eprint {http://arxiv.org/abs/0901.3766}
  {arXiv:0901.3766 [hep-ph]} \BibitemShut {NoStop}%
%%CITATION = ARXIV:0901.3766;%%
\bibitem [{\citenamefont {Appelquist}\ \emph {et~al.}(1996)\citenamefont
  {Appelquist}, \citenamefont {Terning},\ and\ \citenamefont
  {Wijewardhana}}]{Appelquist:1996dq}%
  \BibitemOpen
  \bibfield  {author} {\bibinfo {author} {\bibnamefont {Appelquist}
  \bibfnamefont {Thomas}}, \bibinfo {author} {\bibfnamefont {John}\
  \bibnamefont {Terning}}, \ and\ \bibinfo {author} {\bibfnamefont {L.C.R.}\
  \bibnamefont {Wijewardhana}}} (\bibinfo {year} {1996}),\ \bibfield  {title}
  {\enquote {\bibinfo {title} {{The Zero temperature chiral phase transition in
  SU(N) gauge theories}},}\ }\Doi {10.1103/PhysRevLett.77.1214} {\bibfield
  {journal} {\bibinfo  {journal} {Phys.~Rev.~Lett.},\ }\textbf {\bibinfo
  {volume} {77}},\ \bibinfo {pages} {1214--1217}},\ \Eprint
  {http://arxiv.org/abs/hep-ph/9602385} {arXiv:hep-ph/9602385 [hep-ph]}
  \BibitemShut {NoStop}%
%%CITATION = HEP-PH/9602385;%%
\bibitem [{\citenamefont {Appelquist}\ and\ \citenamefont
  {Wijewardhana}(1987a)}]{Appelquist:1986tr}%
  \BibitemOpen
  \bibfield  {author} {\bibinfo {author} {\bibnamefont {Appelquist}
  \bibfnamefont {Thomas}}, \ and\ \bibinfo {author} {\bibfnamefont {L.C.R.}\
  \bibnamefont {Wijewardhana}}} (\bibinfo {year} {1987a}),\ \bibfield  {title}
  {\enquote {\bibinfo {title} {{Chiral Hierarchies and Chiral Perturbations in
  Technicolor}},}\ }\Doi {10.1103/PhysRevD.35.774} {\bibfield  {journal}
  {\bibinfo  {journal} {Phys.~Rev.},\ }\textbf {\bibinfo {volume} {D35}},\
  \bibinfo {pages} {774}}\BibitemShut {NoStop}%
%%CITATION = PHRVA,D35,774;%%
\bibitem [{\citenamefont {Appelquist}\ and\ \citenamefont
  {Wijewardhana}(1987b)}]{Appelquist:1987fc}%
  \BibitemOpen
  \bibfield  {author} {\bibinfo {author} {\bibnamefont {Appelquist}
  \bibfnamefont {Thomas}}, \ and\ \bibinfo {author} {\bibfnamefont {L.C.R.}\
  \bibnamefont {Wijewardhana}}} (\bibinfo {year} {1987b}),\ \bibfield  {title}
  {\enquote {\bibinfo {title} {{Chiral Hierarchies from Slowly Running
  Couplings in Technicolor Theories}},}\ }\Doi {10.1103/PhysRevD.36.568}
  {\bibfield  {journal} {\bibinfo  {journal} {Phys.~Rev.},\ }\textbf {\bibinfo
  {volume} {D36}},\ \bibinfo {pages} {568}}\BibitemShut {NoStop}%
%%CITATION = PHRVA,D36,568;%%
\bibitem [{\citenamefont {Appelquist}\ \emph {et~al.}(2010)\citenamefont
  {Appelquist} \emph {et~al.}}]{Appelquist:2009ka}%
  \BibitemOpen
  \bibfield  {author} {\bibinfo {author} {\bibnamefont {Appelquist}
  \bibfnamefont {Thomas}},  \emph {et~al.} (\bibinfo {collaboration} {LSD
  Collaboration})} (\bibinfo {year} {2010}),\ \bibfield  {title} {\enquote
  {\bibinfo {title} {{Toward TeV Conformality}},}\ }\Doi
  {10.1103/PhysRevLett.104.071601} {\bibfield  {journal} {\bibinfo  {journal}
  {Phys.~Rev.~Lett.},\ }\textbf {\bibinfo {volume} {104}},\ \bibinfo {pages}
  {071601}},\ \Eprint {http://arxiv.org/abs/0910.2224} {arXiv:0910.2224
  [hep-ph]} \BibitemShut {NoStop}%
%%CITATION = ARXIV:0910.2224;%%
\bibitem [{\citenamefont {Appelquist}\ \emph {et~al.}(2011b)\citenamefont
  {Appelquist} \emph {et~al.}}]{Appelquist:2010xv}%
  \BibitemOpen
  \bibfield  {author} {\bibinfo {author} {\bibnamefont {Appelquist}
  \bibfnamefont {Thomas}},  \emph {et~al.} (\bibinfo {collaboration} {LSD
  Collaboration})} (\bibinfo {year} {2011b}),\ \bibfield  {title} {\enquote
  {\bibinfo {title} {{Parity Doubling and the S Parameter Below the Conformal
  Window}},}\ }\Doi {10.1103/PhysRevLett.106.231601} {\bibfield  {journal}
  {\bibinfo  {journal} {Phys.~Rev.~Lett.},\ }\textbf {\bibinfo {volume}
  {106}},\ \bibinfo {pages} {231601}},\ \Eprint
  {http://arxiv.org/abs/1009.5967} {arXiv:1009.5967 [hep-ph]} \BibitemShut
  {NoStop}%
%%CITATION = ARXIV:1009.5967;%%
\bibitem [{\citenamefont {Arkani-Hamed}\ \emph {et~al.}(2002a)\citenamefont
  {Arkani-Hamed}, \citenamefont {Cohen}, \citenamefont {Katz},\ and\
  \citenamefont {Nelson}}]{ArkaniHamed:2002qy}%
  \BibitemOpen
  \bibfield  {author} {\bibinfo {author} {\bibnamefont {Arkani-Hamed}
  \bibfnamefont {N}}, \bibinfo {author} {\bibfnamefont {A.G.}\ \bibnamefont
  {Cohen}}, \bibinfo {author} {\bibfnamefont {E.}~\bibnamefont {Katz}}, \ and\
  \bibinfo {author} {\bibfnamefont {A.E.}\ \bibnamefont {Nelson}}} (\bibinfo
  {year} {2002a}),\ \bibfield  {title} {\enquote {\bibinfo {title} {{The
  Littlest Higgs}},}\ }\Doi {10.1088/1126-6708/2002/07/034} {\bibfield
  {journal} {\bibinfo  {journal} {JHEP},\ }\textbf {\bibinfo {volume} {0207}},\
  \bibinfo {pages} {034}},\ \Eprint {http://arxiv.org/abs/hep-ph/0206021}
  {arXiv:hep-ph/0206021 [hep-ph]} \BibitemShut {NoStop}%
%%CITATION = HEP-PH/0206021;%%
\bibitem [{\citenamefont {Arkani-Hamed}\ \emph {et~al.}(2002b)\citenamefont
  {Arkani-Hamed}, \citenamefont {Cohen}, \citenamefont {Katz}, \citenamefont
  {Nelson}, \citenamefont {Gregoire} \emph {et~al.}}]{ArkaniHamed:2002qx}%
  \BibitemOpen
  \bibfield  {author} {\bibinfo {author} {\bibnamefont {Arkani-Hamed}
  \bibfnamefont {N}}, \bibinfo {author} {\bibfnamefont {A.G.}\ \bibnamefont
  {Cohen}}, \bibinfo {author} {\bibfnamefont {E.}~\bibnamefont {Katz}},
  \bibinfo {author} {\bibfnamefont {A.E.}\ \bibnamefont {Nelson}}, \bibinfo
  {author} {\bibfnamefont {T.}~\bibnamefont {Gregoire}},  \emph {et~al.}}
  (\bibinfo {year} {2002b}),\ \bibfield  {title} {\enquote {\bibinfo {title}
  {{The Minimal moose for a little Higgs}},}\ }\Doi
  {10.1088/1126-6708/2002/08/021} {\bibfield  {journal} {\bibinfo  {journal}
  {JHEP},\ }\textbf {\bibinfo {volume} {0208}},\ \bibinfo {pages} {021}},\
  \Eprint {http://arxiv.org/abs/hep-ph/0206020} {arXiv:hep-ph/0206020 [hep-ph]}
  \BibitemShut {NoStop}%
%%CITATION = HEP-PH/0206020;%%
\bibitem [{\citenamefont {Athenodorou}\ \emph {et~al.}(2015)\citenamefont
  {Athenodorou}, \citenamefont {Bennett}, \citenamefont {Bergner},\ and\
  \citenamefont {Lucini}}]{Athenodorou:2014eua}%
  \BibitemOpen
  \bibfield  {author} {\bibinfo {author} {\bibnamefont {Athenodorou}
  \bibfnamefont {Andreas}}, \bibinfo {author} {\bibfnamefont {Ed}~\bibnamefont
  {Bennett}}, \bibinfo {author} {\bibfnamefont {Georg}\ \bibnamefont
  {Bergner}}, \ and\ \bibinfo {author} {\bibfnamefont {Biagio}\ \bibnamefont
  {Lucini}}} (\bibinfo {year} {2015}),\ \bibfield  {title} {\enquote {\bibinfo
  {title} {{Infrared regime of SU(2) with one adjoint Dirac flavor}},}\ }\Doi
  {10.1103/PhysRevD.91.114508} {\bibfield  {journal} {\bibinfo  {journal}
  {Phys. Rev.},\ }\textbf {\bibinfo {volume} {D91}}~(\bibinfo {number} {11}),\
  \bibinfo {pages} {114508}},\ \Eprint {http://arxiv.org/abs/1412.5994}
  {arXiv:1412.5994 [hep-lat]} \BibitemShut {NoStop}%
%%CITATION = ARXIV:1412.5994;%%
\bibitem [{\citenamefont {Azatov}\ \emph {et~al.}(2012)\citenamefont {Azatov},
  \citenamefont {Contino},\ and\ \citenamefont {Galloway}}]{Azatov:2012bz}%
  \BibitemOpen
  \bibfield  {author} {\bibinfo {author} {\bibnamefont {Azatov} \bibfnamefont
  {Aleksandr}}, \bibinfo {author} {\bibfnamefont {Roberto}\ \bibnamefont
  {Contino}}, \ and\ \bibinfo {author} {\bibfnamefont {Jamison}\ \bibnamefont
  {Galloway}}} (\bibinfo {year} {2012}),\ \bibfield  {title} {\enquote
  {\bibinfo {title} {{Model-Independent Bounds on a Light Higgs}},}\ }\Doi
  {10.1007/JHEP04(2012)127, 10.1007/JHEP04(2013)140} {\bibfield  {journal}
  {\bibinfo  {journal} {JHEP},\ }\textbf {\bibinfo {volume} {1204}},\ \bibinfo
  {pages} {127}},\ \Eprint {http://arxiv.org/abs/1202.3415} {arXiv:1202.3415
  [hep-ph]} \BibitemShut {NoStop}%
%%CITATION = ARXIV:1202.3415;%%
\bibitem [{\citenamefont {Bali}\ \emph {et~al.}(2013)\citenamefont {Bali},
  \citenamefont {Bursa}, \citenamefont {Castagnini}, \citenamefont {Collins},
  \citenamefont {Del~Debbio} \emph {et~al.}}]{Bali:2013kia}%
  \BibitemOpen
  \bibfield  {author} {\bibinfo {author} {\bibnamefont {Bali} \bibfnamefont
  {Gunnar~S}}, \bibinfo {author} {\bibfnamefont {Francis}\ \bibnamefont
  {Bursa}}, \bibinfo {author} {\bibfnamefont {Luca}\ \bibnamefont
  {Castagnini}}, \bibinfo {author} {\bibfnamefont {Sara}\ \bibnamefont
  {Collins}}, \bibinfo {author} {\bibfnamefont {Luigi}\ \bibnamefont
  {Del~Debbio}},  \emph {et~al.}} (\bibinfo {year} {2013}),\ \bibfield  {title}
  {\enquote {\bibinfo {title} {{Mesons in large-N QCD}},}\ }\Doi
  {10.1007/JHEP06(2013)071} {\bibfield  {journal} {\bibinfo  {journal} {JHEP},\
  }\textbf {\bibinfo {volume} {1306}},\ \bibinfo {pages} {071}},\ \Eprint
  {http://arxiv.org/abs/1304.4437} {arXiv:1304.4437 [hep-lat]} \BibitemShut
  {NoStop}%
%%CITATION = ARXIV:1304.4437;%%
\bibitem [{\citenamefont {Banks}(1984)}]{Banks:1984gj}%
  \BibitemOpen
  \bibfield  {author} {\bibinfo {author} {\bibnamefont {Banks} \bibfnamefont
  {Tom}}} (\bibinfo {year} {1984}),\ \bibfield  {title} {\enquote {\bibinfo
  {title} {{Constraints on SU(2) x U(1) breaking by vacuum misalignment}},}\
  }\Doi {10.1016/0550-3213(84)90389-4} {\bibfield  {journal} {\bibinfo
  {journal} {Nucl.~Phys.},\ }\textbf {\bibinfo {volume} {B243}},\ \bibinfo
  {pages} {125}}\BibitemShut {NoStop}%
%%CITATION = NUPHA,B243,125;%%
\bibitem [{\citenamefont {Banks}\ and\ \citenamefont
  {Casher}(1980)}]{Banks:1979yr}%
  \BibitemOpen
  \bibfield  {author} {\bibinfo {author} {\bibnamefont {Banks} \bibfnamefont
  {Tom}}, \ and\ \bibinfo {author} {\bibfnamefont {A.}~\bibnamefont {Casher}}}
  (\bibinfo {year} {1980}),\ \bibfield  {title} {\enquote {\bibinfo {title}
  {{Chiral Symmetry Breaking in Confining Theories}},}\ }\Doi
  {10.1016/0550-3213(80)90255-2} {\bibfield  {journal} {\bibinfo  {journal}
  {Nucl.~Phys.},\ }\textbf {\bibinfo {volume} {B169}},\ \bibinfo {pages}
  {103}}\BibitemShut {NoStop}%
%%CITATION = NUPHA,B169,103;%%
\bibitem [{\citenamefont {Banks}\ and\ \citenamefont
  {Zaks}(1982)}]{Banks:1981nn}%
  \BibitemOpen
  \bibfield  {author} {\bibinfo {author} {\bibnamefont {Banks} \bibfnamefont
  {Tom}}, \ and\ \bibinfo {author} {\bibfnamefont {A.}~\bibnamefont {Zaks}}}
  (\bibinfo {year} {1982}),\ \bibfield  {title} {\enquote {\bibinfo {title}
  {{On the Phase Structure of Vector-Like Gauge Theories with Massless
  Fermions}},}\ }\Doi {10.1016/0550-3213(82)90035-9} {\bibfield  {journal}
  {\bibinfo  {journal} {Nucl.~Phys.},\ }\textbf {\bibinfo {volume} {B196}},\
  \bibinfo {pages} {189}}\BibitemShut {NoStop}%
%%CITATION = NUPHA,B196,189;%%
\bibitem [{\citenamefont {Barbieri}\ \emph {et~al.}(2004)\citenamefont
  {Barbieri}, \citenamefont {Pomarol}, \citenamefont {Rattazzi},\ and\
  \citenamefont {Strumia}}]{Barbieri:2004qk}%
  \BibitemOpen
  \bibfield  {author} {\bibinfo {author} {\bibnamefont {Barbieri} \bibfnamefont
  {Riccardo}}, \bibinfo {author} {\bibfnamefont {Alex}\ \bibnamefont
  {Pomarol}}, \bibinfo {author} {\bibfnamefont {Riccardo}\ \bibnamefont
  {Rattazzi}}, \ and\ \bibinfo {author} {\bibfnamefont {Alessandro}\
  \bibnamefont {Strumia}}} (\bibinfo {year} {2004}),\ \bibfield  {title}
  {\enquote {\bibinfo {title} {{Electroweak symmetry breaking after LEP-1 and
  LEP-2}},}\ }\Doi {10.1016/j.nuclphysb.2004.10.014} {\bibfield  {journal}
  {\bibinfo  {journal} {Nucl.~Phys.},\ }\textbf {\bibinfo {volume} {B703}},\
  \bibinfo {pages} {127--146}},\ \Eprint {http://arxiv.org/abs/hep-ph/0405040}
  {arXiv:hep-ph/0405040 [hep-ph]} \BibitemShut {NoStop}%
%%CITATION = HEP-PH/0405040;%%
\bibitem [{\citenamefont {Barnard}\ \emph {et~al.}(2014)\citenamefont
  {Barnard}, \citenamefont {Gherghetta},\ and\ \citenamefont
  {Ray}}]{Barnard:2013zea}%
  \BibitemOpen
  \bibfield  {author} {\bibinfo {author} {\bibnamefont {Barnard} \bibfnamefont
  {James}}, \bibinfo {author} {\bibfnamefont {Tony}\ \bibnamefont
  {Gherghetta}}, \ and\ \bibinfo {author} {\bibfnamefont {Tirtha~Sankar}\
  \bibnamefont {Ray}}} (\bibinfo {year} {2014}),\ \bibfield  {title} {\enquote
  {\bibinfo {title} {{UV descriptions of composite Higgs models without
  elementary scalars}},}\ }\Doi {10.1007/JHEP02(2014)002} {\bibfield  {journal}
  {\bibinfo  {journal} {JHEP},\ }\textbf {\bibinfo {volume} {02}},\ \bibinfo
  {pages} {002}},\ \Eprint {http://arxiv.org/abs/1311.6562} {arXiv:1311.6562
  [hep-ph]} \BibitemShut {NoStop}%
%%CITATION = ARXIV:1311.6562;%%
\bibitem [{\citenamefont {Barr}\ \emph {et~al.}(1990)\citenamefont {Barr},
  \citenamefont {Chivukula},\ and\ \citenamefont {Farhi}}]{Barr:1990ca}%
  \BibitemOpen
  \bibfield  {author} {\bibinfo {author} {\bibnamefont {Barr} \bibfnamefont
  {Stephen~M}}, \bibinfo {author} {\bibfnamefont {R.~Sekhar}\ \bibnamefont
  {Chivukula}}, \ and\ \bibinfo {author} {\bibfnamefont {Edward}\ \bibnamefont
  {Farhi}}} (\bibinfo {year} {1990}),\ \bibfield  {title} {\enquote {\bibinfo
  {title} {{Electroweak Fermion Number Violation and the Production of Stable
  Particles in the Early Universe}},}\ }\Doi {10.1016/0370-2693(90)91661-T}
  {\bibfield  {journal} {\bibinfo  {journal} {Phys.~Lett.},\ }\textbf {\bibinfo
  {volume} {B241}},\ \bibinfo {pages} {387--391}}\BibitemShut {NoStop}%
%%CITATION = PHLTA,B241,387;%%
\bibitem [{\citenamefont {Bellazzini}\ \emph {et~al.}(2013)\citenamefont
  {Bellazzini}, \citenamefont {Csaki}, \citenamefont {Hubisz}, \citenamefont
  {Serra},\ and\ \citenamefont {Terning}}]{Bellazzini:2012vz}%
  \BibitemOpen
  \bibfield  {author} {\bibinfo {author} {\bibnamefont {Bellazzini}
  \bibfnamefont {Brando}}, \bibinfo {author} {\bibfnamefont {Csaba}\
  \bibnamefont {Csaki}}, \bibinfo {author} {\bibfnamefont {Jay}\ \bibnamefont
  {Hubisz}}, \bibinfo {author} {\bibfnamefont {Javi}\ \bibnamefont {Serra}}, \
  and\ \bibinfo {author} {\bibfnamefont {John}\ \bibnamefont {Terning}}}
  (\bibinfo {year} {2013}),\ \bibfield  {title} {\enquote {\bibinfo {title} {{A
  Higgslike Dilaton}},}\ }\Doi {10.1140/epjc/s10052-013-2333-x} {\bibfield
  {journal} {\bibinfo  {journal} {Eur.~Phys.~J.},\ }\textbf {\bibinfo {volume}
  {C73}},\ \bibinfo {pages} {2333}},\ \Eprint {http://arxiv.org/abs/1209.3299}
  {arXiv:1209.3299 [hep-ph]} \BibitemShut {NoStop}%
%%CITATION = ARXIV:1209.3299;%%
\bibitem [{\citenamefont {Bellazzini}\ \emph {et~al.}(2014)\citenamefont
  {Bellazzini}, \citenamefont {Csáki},\ and\ \citenamefont
  {Serra}}]{Bellazzini:2014yua}%
  \BibitemOpen
  \bibfield  {author} {\bibinfo {author} {\bibnamefont {Bellazzini}
  \bibfnamefont {Brando}}, \bibinfo {author} {\bibfnamefont {Csaba}\
  \bibnamefont {Csáki}}, \ and\ \bibinfo {author} {\bibfnamefont {Javi}\
  \bibnamefont {Serra}}} (\bibinfo {year} {2014}),\ \bibfield  {title}
  {\enquote {\bibinfo {title} {{Composite Higgses}},}\ }\Doi
  {10.1140/epjc/s10052-014-2766-x} {\bibfield  {journal} {\bibinfo  {journal}
  {Eur.~Phys.~J.},\ }\textbf {\bibinfo {volume} {C74}},\ \bibinfo {pages}
  {2766}},\ \Eprint {http://arxiv.org/abs/1401.2457} {arXiv:1401.2457 [hep-ph]}
  \BibitemShut {NoStop}%
%%CITATION = ARXIV:1401.2457;%%
\bibitem [{\citenamefont {Bergner}\ \emph {et~al.}(2013)\citenamefont
  {Bergner}, \citenamefont {Montvay}, \citenamefont {M{\"u}nster},
  \citenamefont {{\"O}zugurel},\ and\ \citenamefont
  {Sandbrink}}]{Bergner:2013nwa}%
  \BibitemOpen
  \bibfield  {author} {\bibinfo {author} {\bibnamefont {Bergner} \bibfnamefont
  {Georg}}, \bibinfo {author} {\bibfnamefont {Istvan}\ \bibnamefont {Montvay}},
  \bibinfo {author} {\bibfnamefont {Gernot}\ \bibnamefont {M{\"u}nster}},
  \bibinfo {author} {\bibfnamefont {Umut~D.}\ \bibnamefont {{\"O}zugurel}}, \
  and\ \bibinfo {author} {\bibfnamefont {Dirk}\ \bibnamefont {Sandbrink}}}
  (\bibinfo {year} {2013}),\ \bibfield  {title} {\enquote {\bibinfo {title}
  {{Towards the spectrum of low-lying particles in supersymmetric Yang--Mills
  theory}},}\ }\Doi {10.1007/JHEP11(2013)061} {\bibfield  {journal} {\bibinfo
  {journal} {JHEP},\ }\textbf {\bibinfo {volume} {1311}},\ \bibinfo {pages}
  {061}},\ \Eprint {http://arxiv.org/abs/1304.2168} {arXiv:1304.2168}
  \BibitemShut {NoStop}%
\bibitem [{\citenamefont {Blairon}\ \emph {et~al.}(1981)\citenamefont
  {Blairon}, \citenamefont {Brout}, \citenamefont {Englert},\ and\
  \citenamefont {Greensite}}]{Blairon:1980pk}%
  \BibitemOpen
  \bibfield  {author} {\bibinfo {author} {\bibnamefont {Blairon} \bibfnamefont
  {JM}}, \bibinfo {author} {\bibfnamefont {R.}~\bibnamefont {Brout}}, \bibinfo
  {author} {\bibfnamefont {F.}~\bibnamefont {Englert}}, \ and\ \bibinfo
  {author} {\bibfnamefont {J.}~\bibnamefont {Greensite}}} (\bibinfo {year}
  {1981}),\ \bibfield  {title} {\enquote {\bibinfo {title} {{Chiral Symmetry
  Breaking in the Action Formulation of Lattice Gauge Theory}},}\ }\Doi
  {10.1016/0550-3213(81)90061-4} {\bibfield  {journal} {\bibinfo  {journal}
  {Nucl.~Phys.},\ }\textbf {\bibinfo {volume} {B180}},\ \bibinfo {pages}
  {439}}\BibitemShut {NoStop}%
%%CITATION = NUPHA,B180,439;%%
\bibitem [{\citenamefont {Bowler}\ \emph {et~al.}(1985)\citenamefont {Bowler},
  \citenamefont {Kenway}, \citenamefont {Pawley}, \citenamefont {Wallace},
  \citenamefont {Hasenfratz} \emph {et~al.}}]{Bowler:1984hv}%
  \BibitemOpen
  \bibfield  {author} {\bibinfo {author} {\bibnamefont {Bowler} \bibfnamefont
  {KC}}, \bibinfo {author} {\bibfnamefont {R.D.}\ \bibnamefont {Kenway}},
  \bibinfo {author} {\bibfnamefont {G.S.}\ \bibnamefont {Pawley}}, \bibinfo
  {author} {\bibfnamefont {D.J.}\ \bibnamefont {Wallace}}, \bibinfo {author}
  {\bibfnamefont {A.}~\bibnamefont {Hasenfratz}},  \emph {et~al.}} (\bibinfo
  {year} {1985}),\ \bibfield  {title} {\enquote {\bibinfo {title} {{Monte Carlo
  Renormalization Group Studies of SU(3) Lattice Gauge Theory}},}\ }\Doi
  {10.1016/0550-3213(85)90341-4} {\bibfield  {journal} {\bibinfo  {journal}
  {Nucl.~Phys.},\ }\textbf {\bibinfo {volume} {B257}},\ \bibinfo {pages}
  {155--172}}\BibitemShut {NoStop}%
%%CITATION = NUPHA,B257,155;%%
\bibitem [{\citenamefont {Boyle}\ \emph {et~al.}(2010)\citenamefont {Boyle},
  \citenamefont {Del~Debbio}, \citenamefont {Wennekers},\ and\ \citenamefont
  {Zanotti}}]{Boyle:2009xi}%
  \BibitemOpen
  \bibfield  {author} {\bibinfo {author} {\bibnamefont {Boyle} \bibfnamefont
  {Peter~A}}, \bibinfo {author} {\bibfnamefont {Luigi}\ \bibnamefont
  {Del~Debbio}}, \bibinfo {author} {\bibfnamefont {Jan}\ \bibnamefont
  {Wennekers}}, \ and\ \bibinfo {author} {\bibfnamefont {James~M.}\
  \bibnamefont {Zanotti}} (\bibinfo {collaboration} {RBC Collaborations, UKQCD
  Collaborations})} (\bibinfo {year} {2010}),\ \bibfield  {title} {\enquote
  {\bibinfo {title} {{The S Parameter in QCD from Domain Wall Fermions}},}\
  }\Doi {10.1103/PhysRevD.81.014504} {\bibfield  {journal} {\bibinfo  {journal}
  {Phys.~Rev.},\ }\textbf {\bibinfo {volume} {D81}},\ \bibinfo {pages}
  {014504}},\ \Eprint {http://arxiv.org/abs/0909.4931} {arXiv:0909.4931
  [hep-lat]} \BibitemShut {NoStop}%
%%CITATION = ARXIV:0909.4931;%%
\bibitem [{\citenamefont {Brower}\ \emph {et~al.}(2014)\citenamefont {Brower},
  \citenamefont {Hasenfratz}, \citenamefont {Rebbi}, \citenamefont {Weinberg},\
  and\ \citenamefont {Witzel}}]{Brower:2014ita}%
  \BibitemOpen
  \bibfield  {author} {\bibinfo {author} {\bibnamefont {Brower} \bibfnamefont
  {Rich}}, \bibinfo {author} {\bibfnamefont {Anna}\ \bibnamefont {Hasenfratz}},
  \bibinfo {author} {\bibfnamefont {Claudio}\ \bibnamefont {Rebbi}}, \bibinfo
  {author} {\bibfnamefont {Evan}\ \bibnamefont {Weinberg}}, \ and\ \bibinfo
  {author} {\bibfnamefont {Oliver}\ \bibnamefont {Witzel}}} (\bibinfo {year}
  {2014}),\ \bibfield  {title} {\enquote {\bibinfo {title} {{Targeting the
  conformal window with 4+8 flavors}},}\ }\bibfield  {booktitle} {\emph
  {\bibinfo {booktitle} {{Proceedings, 32nd International Symposium on Lattice
  Field Theory (Lattice 2014)}}},\ }\href@noop {} {\bibfield  {journal}
  {\bibinfo  {journal} {PoS},\ }\textbf {\bibinfo {volume} {LATTICE2014}},\
  \bibinfo {pages} {254}},\ \Eprint {http://arxiv.org/abs/1411.3243}
  {arXiv:1411.3243 [hep-lat]} \BibitemShut {NoStop}%
%%CITATION = ARXIV:1411.3243;%%
\bibitem [{\citenamefont {Buchalla}\ \emph {et~al.}(2015)\citenamefont
  {Buchalla}, \citenamefont {Cata},\ and\ \citenamefont
  {Krause}}]{Buchalla:2014eca}%
  \BibitemOpen
  \bibfield  {author} {\bibinfo {author} {\bibnamefont {Buchalla} \bibfnamefont
  {Gerhard}}, \bibinfo {author} {\bibfnamefont {Oscar}\ \bibnamefont {Cata}}, \
  and\ \bibinfo {author} {\bibfnamefont {Claudius}\ \bibnamefont {Krause}}}
  (\bibinfo {year} {2015}),\ \bibfield  {title} {\enquote {\bibinfo {title} {{A
  Systematic Approach to the SILH Lagrangian}},}\ }\Doi
  {10.1016/j.nuclphysb.2015.03.024} {\bibfield  {journal} {\bibinfo  {journal}
  {Nucl. Phys.},\ }\textbf {\bibinfo {volume} {B894}},\ \bibinfo {pages}
  {602--620}},\ \Eprint {http://arxiv.org/abs/1412.6356} {arXiv:1412.6356
  [hep-ph]} \BibitemShut {NoStop}%
%%CITATION = ARXIV:1412.6356;%%
\bibitem [{\citenamefont {Buchmuller}\ and\ \citenamefont
  {Wyler}(1986)}]{Buchmuller:1985jz}%
  \BibitemOpen
  \bibfield  {author} {\bibinfo {author} {\bibnamefont {Buchmuller}
  \bibfnamefont {W}}, \ and\ \bibinfo {author} {\bibfnamefont {D.}~\bibnamefont
  {Wyler}}} (\bibinfo {year} {1986}),\ \bibfield  {title} {\enquote {\bibinfo
  {title} {{Effective Lagrangian Analysis of New Interactions and Flavor
  Conservation}},}\ }\Doi {10.1016/0550-3213(86)90262-2} {\bibfield  {journal}
  {\bibinfo  {journal} {Nucl.~Phys.},\ }\textbf {\bibinfo {volume} {B268}},\
  \bibinfo {pages} {621--653}}\BibitemShut {NoStop}%
%%CITATION = NUPHA,B268,621;%%
\bibitem [{\citenamefont {Bulava}\ \emph {et~al.}(2013)\citenamefont {Bulava},
  \citenamefont {Jansen},\ and\ \citenamefont {Nagy}}]{Bulava:2013ep}%
  \BibitemOpen
  \bibfield  {author} {\bibinfo {author} {\bibnamefont {Bulava} \bibfnamefont
  {J}}, \bibinfo {author} {\bibfnamefont {K.}~\bibnamefont {Jansen}}, \ and\
  \bibinfo {author} {\bibfnamefont {A.}~\bibnamefont {Nagy}}} (\bibinfo {year}
  {2013}),\ \bibfield  {title} {\enquote {\bibinfo {title} {{Constraining a
  fourth generation of quarks: non-perturbative Higgs boson mass bounds}},}\
  }\Doi {10.1016/j.physletb.2013.04.041} {\bibfield  {journal} {\bibinfo
  {journal} {Phys.~Lett.},\ }\textbf {\bibinfo {volume} {B723}},\ \bibinfo
  {pages} {95--99}},\ \Eprint {http://arxiv.org/abs/1301.3416}
  {arXiv:1301.3416} \BibitemShut {NoStop}%
%%CITATION = ARXIV:1301.3416;%%
\bibitem [{\citenamefont {Bursa}\ \emph {et~al.}(2011a)\citenamefont {Bursa},
  \citenamefont {Del~Debbio}, \citenamefont {Henty}, \citenamefont {Kerrane},
  \citenamefont {Lucini}, \citenamefont {Patella}, \citenamefont {Pica},
  \citenamefont {Pickup},\ and\ \citenamefont {Rago}}]{Bursa:2011ru}%
  \BibitemOpen
  \bibfield  {author} {\bibinfo {author} {\bibnamefont {Bursa} \bibfnamefont
  {Francis}}, \bibinfo {author} {\bibfnamefont {Luigi}\ \bibnamefont
  {Del~Debbio}}, \bibinfo {author} {\bibfnamefont {David}\ \bibnamefont
  {Henty}}, \bibinfo {author} {\bibfnamefont {Eoin}\ \bibnamefont {Kerrane}},
  \bibinfo {author} {\bibfnamefont {Biagio}\ \bibnamefont {Lucini}}, \bibinfo
  {author} {\bibfnamefont {Agostino}\ \bibnamefont {Patella}}, \bibinfo
  {author} {\bibfnamefont {Claudio}\ \bibnamefont {Pica}}, \bibinfo {author}
  {\bibfnamefont {Thomas}\ \bibnamefont {Pickup}}, \ and\ \bibinfo {author}
  {\bibfnamefont {Antonio}\ \bibnamefont {Rago}}} (\bibinfo {year} {2011a}),\
  \bibfield  {title} {\enquote {\bibinfo {title} {{Improved Lattice
  Spectroscopy of Minimal Walking Technicolor}},}\ }\Doi
  {10.1103/PhysRevD.84.034506} {\bibfield  {journal} {\bibinfo  {journal}
  {Phys. Rev.},\ }\textbf {\bibinfo {volume} {D84}},\ \bibinfo {pages}
  {034506}},\ \Eprint {http://arxiv.org/abs/1104.4301} {arXiv:1104.4301
  [hep-lat]} \BibitemShut {NoStop}%
%%CITATION = ARXIV:1104.4301;%%
\bibitem [{\citenamefont {Bursa}\ \emph {et~al.}(2010)\citenamefont {Bursa},
  \citenamefont {Del~Debbio}, \citenamefont {Keegan}, \citenamefont {Pica},\
  and\ \citenamefont {Pickup}}]{Bursa:2009we}%
  \BibitemOpen
  \bibfield  {author} {\bibinfo {author} {\bibnamefont {Bursa} \bibfnamefont
  {Francis}}, \bibinfo {author} {\bibfnamefont {Luigi}\ \bibnamefont
  {Del~Debbio}}, \bibinfo {author} {\bibfnamefont {Liam}\ \bibnamefont
  {Keegan}}, \bibinfo {author} {\bibfnamefont {Claudio}\ \bibnamefont {Pica}},
  \ and\ \bibinfo {author} {\bibfnamefont {Thomas}\ \bibnamefont {Pickup}}}
  (\bibinfo {year} {2010}),\ \bibfield  {title} {\enquote {\bibinfo {title}
  {{Mass anomalous dimension in SU(2) with two adjoint fermions}},}\ }\Doi
  {10.1103/PhysRevD.81.014505} {\bibfield  {journal} {\bibinfo  {journal}
  {Phys.~Rev.},\ }\textbf {\bibinfo {volume} {D81}},\ \bibinfo {pages}
  {014505}},\ \Eprint {http://arxiv.org/abs/0910.4535} {arXiv:0910.4535
  [hep-ph]} \BibitemShut {NoStop}%
%%CITATION = ARXIV:0910.4535;%%
\bibitem [{\citenamefont {Bursa}\ \emph {et~al.}(2011b)\citenamefont {Bursa},
  \citenamefont {Del~Debbio}, \citenamefont {Keegan}, \citenamefont {Pica},\
  and\ \citenamefont {Pickup}}]{Bursa:2010xn}%
  \BibitemOpen
  \bibfield  {author} {\bibinfo {author} {\bibnamefont {Bursa} \bibfnamefont
  {Francis}}, \bibinfo {author} {\bibfnamefont {Luigi}\ \bibnamefont
  {Del~Debbio}}, \bibinfo {author} {\bibfnamefont {Liam}\ \bibnamefont
  {Keegan}}, \bibinfo {author} {\bibfnamefont {Claudio}\ \bibnamefont {Pica}},
  \ and\ \bibinfo {author} {\bibfnamefont {Thomas}\ \bibnamefont {Pickup}}}
  (\bibinfo {year} {2011b}),\ \bibfield  {title} {\enquote {\bibinfo {title}
  {{Mass anomalous dimension in SU(2) with six fundamental fermions}},}\ }\Doi
  {10.1016/j.physletb.2010.12.050} {\bibfield  {journal} {\bibinfo  {journal}
  {Phys.~Lett.},\ }\textbf {\bibinfo {volume} {B696}},\ \bibinfo {pages}
  {374--379}},\ \Eprint {http://arxiv.org/abs/1007.3067} {arXiv:1007.3067
  [hep-ph]} \BibitemShut {NoStop}%
%%CITATION = ARXIV:1007.3067;%%
\bibitem [{\citenamefont {Capitani}\ \emph {et~al.}(1999)\citenamefont
  {Capitani}, \citenamefont {Luscher}, \citenamefont {Sommer},\ and\
  \citenamefont {Wittig}}]{Capitani:1998mq}%
  \BibitemOpen
  \bibfield  {author} {\bibinfo {author} {\bibnamefont {Capitani} \bibfnamefont
  {Stefano}}, \bibinfo {author} {\bibfnamefont {Martin}\ \bibnamefont
  {Luscher}}, \bibinfo {author} {\bibfnamefont {Rainer}\ \bibnamefont
  {Sommer}}, \ and\ \bibinfo {author} {\bibfnamefont {Hartmut}\ \bibnamefont
  {Wittig}} (\bibinfo {collaboration} {ALPHA Collaboration})} (\bibinfo {year}
  {1999}),\ \bibfield  {title} {\enquote {\bibinfo {title} {{Nonperturbative
  quark mass renormalization in quenched lattice QCD}},}\ }\Doi
  {10.1016/S0550-3213(98)00857-8} {\bibfield  {journal} {\bibinfo  {journal}
  {Nucl.~Phys.},\ }\textbf {\bibinfo {volume} {B544}},\ \bibinfo {pages}
  {669--698}},\ \Eprint {http://arxiv.org/abs/hep-lat/9810063}
  {arXiv:hep-lat/9810063 [hep-lat]} \BibitemShut {NoStop}%
%%CITATION = HEP-LAT/9810063;%%
\bibitem [{\citenamefont {Cardy}(1996)}]{Cardy:1996xt}%
  \BibitemOpen
  \bibfield  {author} {\bibinfo {author} {\bibnamefont {Cardy} \bibfnamefont
  {John~L}}} (\bibinfo {year} {1996}),\ \href@noop {} {\emph {\bibinfo {title}
  {{Scaling and renormalization in statistical physics}}}}\ (\bibinfo
  {publisher} {Cambridge})\BibitemShut {NoStop}%
%%CITATION = INSPIRE-429658;%%
\bibitem [{\citenamefont {Caswell}(1974)}]{Caswell:1974gg}%
  \BibitemOpen
  \bibfield  {author} {\bibinfo {author} {\bibnamefont {Caswell} \bibfnamefont
  {William~E}}} (\bibinfo {year} {1974}),\ \bibfield  {title} {\enquote
  {\bibinfo {title} {{Asymptotic Behavior of Nonabelian Gauge Theories to Two
  Loop Order}},}\ }\Doi {10.1103/PhysRevLett.33.244} {\bibfield  {journal}
  {\bibinfo  {journal} {Phys.~Rev.Lett.~},\ }\textbf {\bibinfo {volume} {33}},\
  \bibinfo {pages} {244}}\BibitemShut {NoStop}%
%%CITATION = PRLTA,33,244;%%
\bibitem [{\citenamefont {Catterall}\ \emph {et~al.}(2009b)\citenamefont
  {Catterall}, \citenamefont {Giedt}, \citenamefont {Sannino},\ and\
  \citenamefont {Schneible}}]{Catterall:2009sb}%
  \BibitemOpen
  \bibfield  {author} {\bibinfo {author} {\bibnamefont {Catterall}
  \bibfnamefont {Simon}}, \bibinfo {author} {\bibfnamefont {Joel}\ \bibnamefont
  {Giedt}}, \bibinfo {author} {\bibfnamefont {Francesco}\ \bibnamefont
  {Sannino}}, \ and\ \bibinfo {author} {\bibfnamefont {Joe}\ \bibnamefont
  {Schneible}}} (\bibinfo {year} {2009b}),\ \bibfield  {title} {\enquote
  {\bibinfo {title} {{Probes of nearly conformal behavior in lattice
  simulations of minimal walking technicolor}},}\ }\href@noop {} {}\Eprint
  {http://arxiv.org/abs/0910.4387} {arXiv:0910.4387 [hep-lat]} \BibitemShut
  {NoStop}%
%%CITATION = ARXIV:0910.4387;%%
\bibitem [{\citenamefont {Catterall}\ \emph {et~al.}(2009a)\citenamefont
  {Catterall}, \citenamefont {Kaplan},\ and\ \citenamefont
  {Unsal}}]{Catterall:2009it}%
  \BibitemOpen
  \bibfield  {author} {\bibinfo {author} {\bibnamefont {Catterall}
  \bibfnamefont {Simon}}, \bibinfo {author} {\bibfnamefont {David~B.}\
  \bibnamefont {Kaplan}}, \ and\ \bibinfo {author} {\bibfnamefont {Mithat}\
  \bibnamefont {Unsal}}} (\bibinfo {year} {2009a}),\ \bibfield  {title}
  {\enquote {\bibinfo {title} {{Exact lattice supersymmetry}},}\ }\Doi
  {10.1016/j.physrep.2009.09.001} {\bibfield  {journal} {\bibinfo  {journal}
  {Phys.~Rept.},\ }\textbf {\bibinfo {volume} {484}},\ \bibinfo {pages}
  {71--130}},\ \Eprint {http://arxiv.org/abs/0903.4881} {arXiv:0903.4881
  [hep-lat]} \BibitemShut {NoStop}%
%%CITATION = ARXIV:0903.4881;%%
\bibitem [{\citenamefont {Catterall}\ and\ \citenamefont
  {Sannino}(2007)}]{Catterall:2007yx}%
  \BibitemOpen
  \bibfield  {author} {\bibinfo {author} {\bibnamefont {Catterall}
  \bibfnamefont {Simon}}, \ and\ \bibinfo {author} {\bibfnamefont {Francesco}\
  \bibnamefont {Sannino}}} (\bibinfo {year} {2007}),\ \bibfield  {title}
  {\enquote {\bibinfo {title} {{Minimal walking on the lattice}},}\ }\Doi
  {10.1103/PhysRevD.76.034504} {\bibfield  {journal} {\bibinfo  {journal}
  {Phys.~Rev.},\ }\textbf {\bibinfo {volume} {D76}},\ \bibinfo {pages}
  {034504}},\ \Eprint {http://arxiv.org/abs/0705.1664} {arXiv:0705.1664
  [hep-lat]} \BibitemShut {NoStop}%
%%CITATION = ARXIV:0705.1664;%%
\bibitem [{\citenamefont {Catterall}\ \emph {et~al.}(2014)\citenamefont
  {Catterall}, \citenamefont {Schaich}, \citenamefont {Damgaard}, \citenamefont
  {DeGrand},\ and\ \citenamefont {Giedt}}]{Catterall:2014vka}%
  \BibitemOpen
  \bibfield  {author} {\bibinfo {author} {\bibnamefont {Catterall}
  \bibfnamefont {Simon}}, \bibinfo {author} {\bibfnamefont {David}\
  \bibnamefont {Schaich}}, \bibinfo {author} {\bibfnamefont {Poul~H.}\
  \bibnamefont {Damgaard}}, \bibinfo {author} {\bibfnamefont {Thomas}\
  \bibnamefont {DeGrand}}, \ and\ \bibinfo {author} {\bibfnamefont {Joel}\
  \bibnamefont {Giedt}}} (\bibinfo {year} {2014}),\ \bibfield  {title}
  {\enquote {\bibinfo {title} {{N=4 Supersymmetry on a Space-Time Lattice}},}\
  }\Doi {10.1103/PhysRevD.90.065013} {\bibfield  {journal} {\bibinfo  {journal}
  {Phys.~Rev.},\ }\textbf {\bibinfo {volume} {D90}},\ \bibinfo {pages}
  {065013}},\ \Eprint {http://arxiv.org/abs/1405.0644} {arXiv:1405.0644
  [hep-lat]} \BibitemShut {NoStop}%
%%CITATION = ARXIV:1405.0644;%%
\bibitem [{\citenamefont {Chatrchyan}\ \emph {et~al.}(2012)\citenamefont
  {Chatrchyan} \emph {et~al.}}]{Chatrchyan:2012ufa}%
  \BibitemOpen
  \bibfield  {author} {\bibinfo {author} {\bibnamefont {Chatrchyan}
  \bibfnamefont {Serguei}},  \emph {et~al.} (\bibinfo {collaboration} {CMS
  Collaboration})} (\bibinfo {year} {2012}),\ \bibfield  {title} {\enquote
  {\bibinfo {title} {{Observation of a new boson at a mass of 125 GeV with the
  CMS experiment at the LHC}},}\ }\Doi {10.1016/j.physletb.2012.08.021}
  {\bibfield  {journal} {\bibinfo  {journal} {Phys.~Lett.},\ }\textbf {\bibinfo
  {volume} {B716}},\ \bibinfo {pages} {30--61}},\ \Eprint
  {http://arxiv.org/abs/1207.7235} {arXiv:1207.7235 [hep-ex]} \BibitemShut
  {NoStop}%
%%CITATION = ARXIV:1207.7235;%%
\bibitem [{\citenamefont {Cheng}\ \emph {et~al.}(2014a)\citenamefont {Cheng},
  \citenamefont {Hasenfratz}, \citenamefont {Liu}, \citenamefont
  {Petropoulos},\ and\ \citenamefont {Schaich}}]{Cheng:2013xha}%
  \BibitemOpen
  \bibfield  {author} {\bibinfo {author} {\bibnamefont {Cheng} \bibfnamefont
  {Anqi}}, \bibinfo {author} {\bibfnamefont {Anna}\ \bibnamefont {Hasenfratz}},
  \bibinfo {author} {\bibfnamefont {Yuzhi}\ \bibnamefont {Liu}}, \bibinfo
  {author} {\bibfnamefont {Gregory}\ \bibnamefont {Petropoulos}}, \ and\
  \bibinfo {author} {\bibfnamefont {David}\ \bibnamefont {Schaich}}} (\bibinfo
  {year} {2014a}),\ \bibfield  {title} {\enquote {\bibinfo {title} {{Finite
  size scaling of conformal theories in the presence of a near-marginal
  operator}},}\ }\Doi {10.1103/PhysRevD.90.014509} {\bibfield  {journal}
  {\bibinfo  {journal} {Phys.~Rev.},\ }\textbf {\bibinfo {volume} {D90}},\
  \bibinfo {pages} {014509}},\ \Eprint {http://arxiv.org/abs/1401.0195}
  {arXiv:1401.0195 [hep-lat]} \BibitemShut {NoStop}%
%%CITATION = ARXIV:1401.0195;%%
\bibitem [{\citenamefont {Cheng}\ \emph {et~al.}(2014b)\citenamefont {Cheng},
  \citenamefont {Hasenfratz}, \citenamefont {Liu}, \citenamefont
  {Petropoulos},\ and\ \citenamefont {Schaich}}]{Cheng:2014jba}%
  \BibitemOpen
  \bibfield  {author} {\bibinfo {author} {\bibnamefont {Cheng} \bibfnamefont
  {Anqi}}, \bibinfo {author} {\bibfnamefont {Anna}\ \bibnamefont {Hasenfratz}},
  \bibinfo {author} {\bibfnamefont {Yuzhi}\ \bibnamefont {Liu}}, \bibinfo
  {author} {\bibfnamefont {Gregory}\ \bibnamefont {Petropoulos}}, \ and\
  \bibinfo {author} {\bibfnamefont {David}\ \bibnamefont {Schaich}}} (\bibinfo
  {year} {2014b}),\ \bibfield  {title} {\enquote {\bibinfo {title} {{Improving
  the continuum limit of gradient flow step scaling}},}\ }\Doi
  {10.1007/JHEP05(2014)137} {\bibfield  {journal} {\bibinfo  {journal} {JHEP},\
  }\textbf {\bibinfo {volume} {1405}},\ \bibinfo {pages} {137}},\ \Eprint
  {http://arxiv.org/abs/1404.0984} {arXiv:1404.0984 [hep-lat]} \BibitemShut
  {NoStop}%
%%CITATION = ARXIV:1404.0984;%%
\bibitem [{\citenamefont {Cheng}\ \emph {et~al.}(2013)\citenamefont {Cheng},
  \citenamefont {Hasenfratz}, \citenamefont {Petropoulos},\ and\ \citenamefont
  {Schaich}}]{Cheng:2013eu}%
  \BibitemOpen
  \bibfield  {author} {\bibinfo {author} {\bibnamefont {Cheng} \bibfnamefont
  {Anqi}}, \bibinfo {author} {\bibfnamefont {Anna}\ \bibnamefont {Hasenfratz}},
  \bibinfo {author} {\bibfnamefont {Gregory}\ \bibnamefont {Petropoulos}}, \
  and\ \bibinfo {author} {\bibfnamefont {David}\ \bibnamefont {Schaich}}}
  (\bibinfo {year} {2013}),\ \bibfield  {title} {\enquote {\bibinfo {title}
  {{Scale-dependent mass anomalous dimension from Dirac eigenmodes}},}\ }\Doi
  {10.1007/JHEP07(2013)061} {\bibfield  {journal} {\bibinfo  {journal} {JHEP},\
  }\textbf {\bibinfo {volume} {1307}},\ \bibinfo {pages} {061}},\ \Eprint
  {http://arxiv.org/abs/1301.1355} {arXiv:1301.1355 [hep-lat]} \BibitemShut
  {NoStop}%
%%CITATION = ARXIV:1301.1355;%%
\bibitem [{\citenamefont {Cheng}\ \emph {et~al.}(2012)\citenamefont {Cheng},
  \citenamefont {Hasenfratz},\ and\ \citenamefont {Schaich}}]{Cheng:2011ic}%
  \BibitemOpen
  \bibfield  {author} {\bibinfo {author} {\bibnamefont {Cheng} \bibfnamefont
  {Anqi}}, \bibinfo {author} {\bibfnamefont {Anna}\ \bibnamefont {Hasenfratz}},
  \ and\ \bibinfo {author} {\bibfnamefont {David}\ \bibnamefont {Schaich}}}
  (\bibinfo {year} {2012}),\ \bibfield  {title} {\enquote {\bibinfo {title}
  {{Novel phase in SU(3) lattice gauge theory with 12 light fermions}},}\ }\Doi
  {10.1103/PhysRevD.85.094509} {\bibfield  {journal} {\bibinfo  {journal}
  {Phys.~Rev.},\ }\textbf {\bibinfo {volume} {D85}},\ \bibinfo {pages}
  {094509}},\ \Eprint {http://arxiv.org/abs/1111.2317} {arXiv:1111.2317
  [hep-lat]} \BibitemShut {NoStop}%
%%CITATION = ARXIV:1111.2317;%%
\bibitem [{\citenamefont {Cheng}\ \emph {et~al.}(2011)\citenamefont {Cheng},
  \citenamefont {Datta}, \citenamefont {Francis}, \citenamefont {van~der
  Heide}, \citenamefont {Jung} \emph {et~al.}}]{Cheng:2010fe}%
  \BibitemOpen
  \bibfield  {author} {\bibinfo {author} {\bibnamefont {Cheng} \bibfnamefont
  {M}}, \bibinfo {author} {\bibfnamefont {S.}~\bibnamefont {Datta}}, \bibinfo
  {author} {\bibfnamefont {A.}~\bibnamefont {Francis}}, \bibinfo {author}
  {\bibfnamefont {J.}~\bibnamefont {van~der Heide}}, \bibinfo {author}
  {\bibfnamefont {C.}~\bibnamefont {Jung}},  \emph {et~al.}} (\bibinfo {year}
  {2011}),\ \bibfield  {title} {\enquote {\bibinfo {title} {{Meson screening
  masses from lattice QCD with two light and the strange quark}},}\ }\Doi
  {10.1140/epjc/s10052-011-1564-y} {\bibfield  {journal} {\bibinfo  {journal}
  {Eur.~Phys.~J.},\ }\textbf {\bibinfo {volume} {C71}},\ \bibinfo {pages}
  {1564}},\ \Eprint {http://arxiv.org/abs/1010.1216} {arXiv:1010.1216
  [hep-lat]} \BibitemShut {NoStop}%
%%CITATION = ARXIV:1010.1216;%%
\bibitem [{\citenamefont {Ciuchini}\ \emph {et~al.}(2013)\citenamefont
  {Ciuchini}, \citenamefont {Franco}, \citenamefont {Mishima},\ and\
  \citenamefont {Silvestrini}}]{Ciuchini:2013pca}%
  \BibitemOpen
  \bibfield  {author} {\bibinfo {author} {\bibnamefont {Ciuchini} \bibfnamefont
  {Marco}}, \bibinfo {author} {\bibfnamefont {Enrico}\ \bibnamefont {Franco}},
  \bibinfo {author} {\bibfnamefont {Satoshi}\ \bibnamefont {Mishima}}, \ and\
  \bibinfo {author} {\bibfnamefont {Luca}\ \bibnamefont {Silvestrini}}}
  (\bibinfo {year} {2013}),\ \bibfield  {title} {\enquote {\bibinfo {title}
  {{Electroweak Precision Observables, New Physics and the Nature of a 126 GeV
  Higgs Boson}},}\ }\Doi {10.1007/JHEP08(2013)106} {\bibfield  {journal}
  {\bibinfo  {journal} {JHEP},\ }\textbf {\bibinfo {volume} {1308}},\ \bibinfo
  {pages} {106}},\ \Eprint {http://arxiv.org/abs/1306.4644} {arXiv:1306.4644
  [hep-ph]} \BibitemShut {NoStop}%
%%CITATION = ARXIV:1306.4644;%%
\bibitem [{\citenamefont {Contino}(2010)}]{Contino:2010rs}%
  \BibitemOpen
  \bibfield  {author} {\bibinfo {author} {\bibnamefont {Contino} \bibfnamefont
  {Roberto}}} (\bibinfo {year} {2010}),\ \bibfield  {title} {\enquote {\bibinfo
  {title} {{The Higgs as a Composite Nambu-Goldstone Boson}},}\ }\href@noop {}
  {}\Eprint {http://arxiv.org/abs/1005.4269} {arXiv:1005.4269 [hep-ph]}
  \BibitemShut {NoStop}%
%%CITATION = ARXIV:1005.4269;%%
\bibitem [{\citenamefont {Contino}\ \emph {et~al.}(2013)\citenamefont
  {Contino}, \citenamefont {Ghezzi}, \citenamefont {Grojean}, \citenamefont
  {Muhlleitner},\ and\ \citenamefont {Spira}}]{Contino:2013kra}%
  \BibitemOpen
  \bibfield  {author} {\bibinfo {author} {\bibnamefont {Contino} \bibfnamefont
  {Roberto}}, \bibinfo {author} {\bibfnamefont {Margherita}\ \bibnamefont
  {Ghezzi}}, \bibinfo {author} {\bibfnamefont {Christophe}\ \bibnamefont
  {Grojean}}, \bibinfo {author} {\bibfnamefont {Margarete}\ \bibnamefont
  {Muhlleitner}}, \ and\ \bibinfo {author} {\bibfnamefont {Michael}\
  \bibnamefont {Spira}}} (\bibinfo {year} {2013}),\ \bibfield  {title}
  {\enquote {\bibinfo {title} {{Effective Lagrangian for a light Higgs-like
  scalar}},}\ }\Doi {10.1007/JHEP07(2013)035} {\bibfield  {journal} {\bibinfo
  {journal} {JHEP},\ }\textbf {\bibinfo {volume} {1307}},\ \bibinfo {pages}
  {035}},\ \Eprint {http://arxiv.org/abs/1303.3876} {arXiv:1303.3876 [hep-ph]}
  \BibitemShut {NoStop}%
%%CITATION = ARXIV:1303.3876;%%
\bibitem [{\citenamefont {Contino}\ \emph {et~al.}(2010)\citenamefont
  {Contino}, \citenamefont {Grojean}, \citenamefont {Moretti}, \citenamefont
  {Piccinini},\ and\ \citenamefont {Rattazzi}}]{Contino:2010mh}%
  \BibitemOpen
  \bibfield  {author} {\bibinfo {author} {\bibnamefont {Contino} \bibfnamefont
  {Roberto}}, \bibinfo {author} {\bibfnamefont {Christophe}\ \bibnamefont
  {Grojean}}, \bibinfo {author} {\bibfnamefont {Mauro}\ \bibnamefont
  {Moretti}}, \bibinfo {author} {\bibfnamefont {Fulvio}\ \bibnamefont
  {Piccinini}}, \ and\ \bibinfo {author} {\bibfnamefont {Riccardo}\
  \bibnamefont {Rattazzi}}} (\bibinfo {year} {2010}),\ \bibfield  {title}
  {\enquote {\bibinfo {title} {{Strong Double Higgs Production at the LHC}},}\
  }\Doi {10.1007/JHEP05(2010)089} {\bibfield  {journal} {\bibinfo  {journal}
  {JHEP},\ }\textbf {\bibinfo {volume} {1005}},\ \bibinfo {pages} {089}},\
  \Eprint {http://arxiv.org/abs/1002.1011} {arXiv:1002.1011 [hep-ph]}
  \BibitemShut {NoStop}%
%%CITATION = ARXIV:1002.1011;%%
\bibitem [{\citenamefont {Creutz}(1980a)}]{Creutz:1980zw}%
  \BibitemOpen
  \bibfield  {author} {\bibinfo {author} {\bibnamefont {Creutz} \bibfnamefont
  {M}}} (\bibinfo {year} {1980a}),\ \bibfield  {title} {\enquote {\bibinfo
  {title} {{Monte Carlo Study of Quantized SU(2) Gauge Theory}},}\ }\Doi
  {10.1103/PhysRevD.21.2308} {\bibfield  {journal} {\bibinfo  {journal}
  {Phys.~Rev.},\ }\textbf {\bibinfo {volume} {D21}},\ \bibinfo {pages}
  {2308--2315}}\BibitemShut {NoStop}%
%%CITATION = PHRVA,D21,2308;%%
\bibitem [{\citenamefont {Creutz}(1980b)}]{Creutz:1980wj}%
  \BibitemOpen
  \bibfield  {author} {\bibinfo {author} {\bibnamefont {Creutz} \bibfnamefont
  {Michael}}} (\bibinfo {year} {1980b}),\ \bibfield  {title} {\enquote
  {\bibinfo {title} {{Asymptotic Freedom Scales}},}\ }\Doi
  {10.1103/PhysRevLett.45.313} {\bibfield  {journal} {\bibinfo  {journal}
  {Phys.~Rev.~Lett.},\ }\textbf {\bibinfo {volume} {45}},\ \bibinfo {pages}
  {313}}\BibitemShut {NoStop}%
%%CITATION = PRLTA,45,313;%%
\bibitem [{\citenamefont {Damgaard}\ \emph {et~al.}(1997)\citenamefont
  {Damgaard}, \citenamefont {Heller}, \citenamefont {Krasnitz},\ and\
  \citenamefont {Olesen}}]{Damgaard:1997ut}%
  \BibitemOpen
  \bibfield  {author} {\bibinfo {author} {\bibnamefont {Damgaard} \bibfnamefont
  {PH}}, \bibinfo {author} {\bibfnamefont {Urs~M.}\ \bibnamefont {Heller}},
  \bibinfo {author} {\bibfnamefont {A.}~\bibnamefont {Krasnitz}}, \ and\
  \bibinfo {author} {\bibfnamefont {P.}~\bibnamefont {Olesen}}} (\bibinfo
  {year} {1997}),\ \bibfield  {title} {\enquote {\bibinfo {title} {{On lattice
  QCD with many flavors}},}\ }\Doi {10.1016/S0370-2693(97)00355-9} {\bibfield
  {journal} {\bibinfo  {journal} {Phys.~Lett.},\ }\textbf {\bibinfo {volume}
  {B400}},\ \bibinfo {pages} {169--175}},\ \Eprint
  {http://arxiv.org/abs/hep-lat/9701008} {arXiv:hep-lat/9701008 [hep-lat]}
  \BibitemShut {NoStop}%
%%CITATION = HEP-LAT/9701008;%%
\bibitem [{\citenamefont {Damgaard}\ \emph {et~al.}(2002)\citenamefont
  {Damgaard}, \citenamefont {Heller}, \citenamefont {Niclasen},\ and\
  \citenamefont {Svetitsky}}]{Damgaard:2001fg}%
  \BibitemOpen
  \bibfield  {author} {\bibinfo {author} {\bibnamefont {Damgaard} \bibfnamefont
  {PH}}, \bibinfo {author} {\bibfnamefont {Urs~M.}\ \bibnamefont {Heller}},
  \bibinfo {author} {\bibfnamefont {R.}~\bibnamefont {Niclasen}}, \ and\
  \bibinfo {author} {\bibfnamefont {B.}~\bibnamefont {Svetitsky}}} (\bibinfo
  {year} {2002}),\ \bibfield  {title} {\enquote {\bibinfo {title} {{Patterns of
  spontaneous chiral symmetry breaking in vector - like gauge theories}},}\
  }\Doi {10.1016/S0550-3213(02)00252-3} {\bibfield  {journal} {\bibinfo
  {journal} {Nucl.~Phys.},\ }\textbf {\bibinfo {volume} {B633}},\ \bibinfo
  {pages} {97--113}},\ \Eprint {http://arxiv.org/abs/hep-lat/0110028}
  {arXiv:hep-lat/0110028 [hep-lat]} \BibitemShut {NoStop}%
%%CITATION = HEP-LAT/0110028;%%
\bibitem [{\citenamefont {Damgaard}\ \emph {et~al.}(1999)\citenamefont
  {Damgaard}, \citenamefont {Osborn}, \citenamefont {Toublan},\ and\
  \citenamefont {Verbaarschot}}]{Damgaard:1998xy}%
  \BibitemOpen
  \bibfield  {author} {\bibinfo {author} {\bibnamefont {Damgaard} \bibfnamefont
  {PH}}, \bibinfo {author} {\bibfnamefont {J.C.}\ \bibnamefont {Osborn}},
  \bibinfo {author} {\bibfnamefont {D.}~\bibnamefont {Toublan}}, \ and\
  \bibinfo {author} {\bibfnamefont {J.J.M.}\ \bibnamefont {Verbaarschot}}}
  (\bibinfo {year} {1999}),\ \bibfield  {title} {\enquote {\bibinfo {title}
  {{The microscopic spectral density of the QCD Dirac operator}},}\ }\Doi
  {10.1016/S0550-3213(99)00094-2} {\bibfield  {journal} {\bibinfo  {journal}
  {Nucl.~Phys.},\ }\textbf {\bibinfo {volume} {B547}},\ \bibinfo {pages}
  {305--328}},\ \Eprint {http://arxiv.org/abs/hep-th/9811212}
  {arXiv:hep-th/9811212 [hep-th]} \BibitemShut {NoStop}%
%%CITATION = HEP-TH/9811212;%%
\bibitem [{\citenamefont {Das}\ \emph {et~al.}(1967)\citenamefont {Das},
  \citenamefont {Guralnik}, \citenamefont {Mathur}, \citenamefont {Low},\ and\
  \citenamefont {Young}}]{Das:1967it}%
  \BibitemOpen
  \bibfield  {author} {\bibinfo {author} {\bibnamefont {Das} \bibfnamefont
  {T}}, \bibinfo {author} {\bibfnamefont {G.S.}\ \bibnamefont {Guralnik}},
  \bibinfo {author} {\bibfnamefont {V.S.}\ \bibnamefont {Mathur}}, \bibinfo
  {author} {\bibfnamefont {F.E.}\ \bibnamefont {Low}}, \ and\ \bibinfo {author}
  {\bibfnamefont {J.E.}\ \bibnamefont {Young}}} (\bibinfo {year} {1967}),\
  \bibfield  {title} {\enquote {\bibinfo {title} {{Electromagnetic mass
  difference of pions}},}\ }\Doi {10.1103/PhysRevLett.18.759} {\bibfield
  {journal} {\bibinfo  {journal} {Phys.~Rev.~Lett.},\ }\textbf {\bibinfo
  {volume} {18}},\ \bibinfo {pages} {759--761}}\BibitemShut {NoStop}%
%%CITATION = PRLTA,18,759;%%
\bibitem [{\citenamefont {DeGrand}(2009)}]{DeGrand:2009hu}%
  \BibitemOpen
  \bibfield  {author} {\bibinfo {author} {\bibnamefont {DeGrand} \bibfnamefont
  {Thomas}}} (\bibinfo {year} {2009}),\ \bibfield  {title} {\enquote {\bibinfo
  {title} {{Finite-size scaling tests for SU(3) lattice gauge theory with color
  sextet fermions}},}\ }\Doi {10.1103/PhysRevD.80.114507} {\bibfield  {journal}
  {\bibinfo  {journal} {Phys.~Rev.},\ }\textbf {\bibinfo {volume} {D80}},\
  \bibinfo {pages} {114507}},\ \Eprint {http://arxiv.org/abs/0910.3072}
  {arXiv:0910.3072 [hep-lat]} \BibitemShut {NoStop}%
%%CITATION = ARXIV:0910.3072;%%
\bibitem [{\citenamefont {DeGrand}(2011)}]{DeGrand:2011cu}%
  \BibitemOpen
  \bibfield  {author} {\bibinfo {author} {\bibnamefont {DeGrand} \bibfnamefont
  {Thomas}}} (\bibinfo {year} {2011}),\ \bibfield  {title} {\enquote {\bibinfo
  {title} {{Finite-size scaling tests for spectra in SU(3) lattice gauge theory
  coupled to 12 fundamental flavor fermions}},}\ }\Doi
  {10.1103/PhysRevD.84.116901} {\bibfield  {journal} {\bibinfo  {journal}
  {Phys.~Rev.},\ }\textbf {\bibinfo {volume} {D84}},\ \bibinfo {pages}
  {116901}},\ \Eprint {http://arxiv.org/abs/1109.1237} {arXiv:1109.1237
  [hep-lat]} \BibitemShut {NoStop}%
%%CITATION = ARXIV:1109.1237;%%
\bibitem [{\citenamefont {DeGrand}\ and\ \citenamefont
  {DeTar}(2006)}]{DeGrand:2006zz}%
  \BibitemOpen
  \bibfield  {author} {\bibinfo {author} {\bibnamefont {DeGrand} \bibfnamefont
  {Thomas}}, \ and\ \bibinfo {author} {\bibfnamefont {Carleton~E.}\
  \bibnamefont {DeTar}}} (\bibinfo {year} {2006}),\ \href@noop {} {\emph
  {\bibinfo {title} {{Lattice methods for quantum chromodynamics}}}}\ (\bibinfo
   {publisher} {World Scientific})\BibitemShut {NoStop}%
%%CITATION = INSPIRE-739893;%%
\bibitem [{\citenamefont {DeGrand}\ and\ \citenamefont
  {Hasenfratz}(2009)}]{DeGrand:2009mt}%
  \BibitemOpen
  \bibfield  {author} {\bibinfo {author} {\bibnamefont {DeGrand} \bibfnamefont
  {Thomas}}, \ and\ \bibinfo {author} {\bibfnamefont {Anna}\ \bibnamefont
  {Hasenfratz}}} (\bibinfo {year} {2009}),\ \bibfield  {title} {\enquote
  {\bibinfo {title} {{Remarks on lattice gauge theories with
  infrared-attractive fixed points}},}\ }\Doi {10.1103/PhysRevD.80.034506}
  {\bibfield  {journal} {\bibinfo  {journal} {Phys.~Rev.},\ }\textbf {\bibinfo
  {volume} {D80}},\ \bibinfo {pages} {034506}},\ \Eprint
  {http://arxiv.org/abs/0906.1976} {arXiv:0906.1976 [hep-lat]} \BibitemShut
  {NoStop}%
%%CITATION = ARXIV:0906.1976;%%
\bibitem [{\citenamefont {DeGrand}\ \emph {et~al.}(2015)\citenamefont
  {DeGrand}, \citenamefont {Liu}, \citenamefont {Neil}, \citenamefont
  {Shamir},\ and\ \citenamefont {Svetitsky}}]{DeGrand:2015lna}%
  \BibitemOpen
  \bibfield  {author} {\bibinfo {author} {\bibnamefont {DeGrand} \bibfnamefont
  {Thomas}}, \bibinfo {author} {\bibfnamefont {Yuzhi}\ \bibnamefont {Liu}},
  \bibinfo {author} {\bibfnamefont {Ethan~T.}\ \bibnamefont {Neil}}, \bibinfo
  {author} {\bibfnamefont {Yigal}\ \bibnamefont {Shamir}}, \ and\ \bibinfo
  {author} {\bibfnamefont {Benjamin}\ \bibnamefont {Svetitsky}}} (\bibinfo
  {year} {2015}),\ \bibfield  {title} {\enquote {\bibinfo {title}
  {{Spectroscopy of SU(4) gauge theory with two flavors of sextet fermions}},}\
  }\Doi {10.1103/PhysRevD.91.114502} {\bibfield  {journal} {\bibinfo  {journal}
  {Phys. Rev.},\ }\textbf {\bibinfo {volume} {D91}},\ \bibinfo {pages}
  {114502}},\ \Eprint {http://arxiv.org/abs/1501.05665} {arXiv:1501.05665
  [hep-lat]} \BibitemShut {NoStop}%
%%CITATION = ARXIV:1501.05665;%%
\bibitem [{\citenamefont {DeGrand}\ \emph {et~al.}(2009)\citenamefont
  {DeGrand}, \citenamefont {Shamir},\ and\ \citenamefont
  {Svetitsky}}]{DeGrand:2008kx}%
  \BibitemOpen
  \bibfield  {author} {\bibinfo {author} {\bibnamefont {DeGrand} \bibfnamefont
  {Thomas}}, \bibinfo {author} {\bibfnamefont {Yigal}\ \bibnamefont {Shamir}},
  \ and\ \bibinfo {author} {\bibfnamefont {Benjamin}\ \bibnamefont
  {Svetitsky}}} (\bibinfo {year} {2009}),\ \bibfield  {title} {\enquote
  {\bibinfo {title} {{Phase structure of SU(3) gauge theory with two flavors of
  symmetric-representation fermions}},}\ }\Doi {10.1103/PhysRevD.79.034501}
  {\bibfield  {journal} {\bibinfo  {journal} {Phys.~Rev.},\ }\textbf {\bibinfo
  {volume} {D79}},\ \bibinfo {pages} {034501}},\ \Eprint
  {http://arxiv.org/abs/0812.1427} {arXiv:0812.1427 [hep-lat]} \BibitemShut
  {NoStop}%
%%CITATION = ARXIV:0812.1427;%%
\bibitem [{\citenamefont {DeGrand}\ \emph {et~al.}(2010)\citenamefont
  {DeGrand}, \citenamefont {Shamir},\ and\ \citenamefont
  {Svetitsky}}]{DeGrand:2010na}%
  \BibitemOpen
  \bibfield  {author} {\bibinfo {author} {\bibnamefont {DeGrand} \bibfnamefont
  {Thomas}}, \bibinfo {author} {\bibfnamefont {Yigal}\ \bibnamefont {Shamir}},
  \ and\ \bibinfo {author} {\bibfnamefont {Benjamin}\ \bibnamefont
  {Svetitsky}}} (\bibinfo {year} {2010}),\ \bibfield  {title} {\enquote
  {\bibinfo {title} {{Running coupling and mass anomalous dimension of SU(3)
  gauge theory with two flavors of symmetric-representation fermions}},}\ }\Doi
  {10.1103/PhysRevD.82.054503} {\bibfield  {journal} {\bibinfo  {journal}
  {Phys.~Rev.},\ }\textbf {\bibinfo {volume} {D82}},\ \bibinfo {pages}
  {054503}},\ \Eprint {http://arxiv.org/abs/1006.0707} {arXiv:1006.0707
  [hep-lat]} \BibitemShut {NoStop}%
%%CITATION = ARXIV:1006.0707;%%
\bibitem [{\citenamefont {DeGrand}\ \emph {et~al.}(2011)\citenamefont
  {DeGrand}, \citenamefont {Shamir},\ and\ \citenamefont
  {Svetitsky}}]{DeGrand:2011qd}%
  \BibitemOpen
  \bibfield  {author} {\bibinfo {author} {\bibnamefont {DeGrand} \bibfnamefont
  {Thomas}}, \bibinfo {author} {\bibfnamefont {Yigal}\ \bibnamefont {Shamir}},
  \ and\ \bibinfo {author} {\bibfnamefont {Benjamin}\ \bibnamefont
  {Svetitsky}}} (\bibinfo {year} {2011}),\ \bibfield  {title} {\enquote
  {\bibinfo {title} {{Infrared fixed point in SU(2) gauge theory with adjoint
  fermions}},}\ }\Doi {10.1103/PhysRevD.83.074507} {\bibfield  {journal}
  {\bibinfo  {journal} {Phys.~Rev.},\ }\textbf {\bibinfo {volume} {D83}},\
  \bibinfo {pages} {074507}},\ \Eprint {http://arxiv.org/abs/1102.2843}
  {arXiv:1102.2843 [hep-lat]} \BibitemShut {NoStop}%
%%CITATION = ARXIV:1102.2843;%%
\bibitem [{\citenamefont {DeGrand}\ \emph {et~al.}(2012)\citenamefont
  {DeGrand}, \citenamefont {Shamir},\ and\ \citenamefont
  {Svetitsky}}]{DeGrand:2012qa}%
  \BibitemOpen
  \bibfield  {author} {\bibinfo {author} {\bibnamefont {DeGrand} \bibfnamefont
  {Thomas}}, \bibinfo {author} {\bibfnamefont {Yigal}\ \bibnamefont {Shamir}},
  \ and\ \bibinfo {author} {\bibfnamefont {Benjamin}\ \bibnamefont
  {Svetitsky}}} (\bibinfo {year} {2012}),\ \bibfield  {title} {\enquote
  {\bibinfo {title} {{SU(4) lattice gauge theory with decuplet fermions:
  Schrodinger functional analysis}},}\ }\Doi {10.1103/PhysRevD.85.074506}
  {\bibfield  {journal} {\bibinfo  {journal} {Phys.~Rev.},\ }\textbf {\bibinfo
  {volume} {D85}},\ \bibinfo {pages} {074506}},\ \Eprint
  {http://arxiv.org/abs/1202.2675} {arXiv:1202.2675 [hep-lat]} \BibitemShut
  {NoStop}%
%%CITATION = ARXIV:1202.2675;%%
\bibitem [{\citenamefont {DeGrand}\ \emph {et~al.}(2013a)\citenamefont
  {DeGrand}, \citenamefont {Shamir},\ and\ \citenamefont
  {Svetitsky}}]{DeGrand:2012yq}%
  \BibitemOpen
  \bibfield  {author} {\bibinfo {author} {\bibnamefont {DeGrand} \bibfnamefont
  {Thomas}}, \bibinfo {author} {\bibfnamefont {Yigal}\ \bibnamefont {Shamir}},
  \ and\ \bibinfo {author} {\bibfnamefont {Benjamin}\ \bibnamefont
  {Svetitsky}}} (\bibinfo {year} {2013a}),\ \bibfield  {title} {\enquote
  {\bibinfo {title} {{Mass anomalous dimension in sextet QCD}},}\ }\Doi
  {10.1103/PhysRevD.87.074507} {\bibfield  {journal} {\bibinfo  {journal}
  {Phys.~Rev.},\ }\textbf {\bibinfo {volume} {D87}},\ \bibinfo {pages}
  {074507}},\ \Eprint {http://arxiv.org/abs/1201.0935} {arXiv:1201.0935
  [hep-lat]} \BibitemShut {NoStop}%
%%CITATION = ARXIV:1201.0935;%%
\bibitem [{\citenamefont {DeGrand}\ \emph {et~al.}(2013b)\citenamefont
  {DeGrand}, \citenamefont {Shamir},\ and\ \citenamefont
  {Svetitsky}}]{DeGrand:2013uha}%
  \BibitemOpen
  \bibfield  {author} {\bibinfo {author} {\bibnamefont {DeGrand} \bibfnamefont
  {Thomas}}, \bibinfo {author} {\bibfnamefont {Yigal}\ \bibnamefont {Shamir}},
  \ and\ \bibinfo {author} {\bibfnamefont {Benjamin}\ \bibnamefont
  {Svetitsky}}} (\bibinfo {year} {2013b}),\ \bibfield  {title} {\enquote
  {\bibinfo {title} {{Near the sill of the conformal window: gauge theories
  with fermions in two-index representations}},}\ }\Doi
  {10.1103/PhysRevD.88.054505} {\bibfield  {journal} {\bibinfo  {journal}
  {Phys.~Rev.},\ }\textbf {\bibinfo {volume} {D88}},\ \bibinfo {pages}
  {054505}},\ \Eprint {http://arxiv.org/abs/1307.2425} {arXiv:1307.2425}
  \BibitemShut {NoStop}%
%%CITATION = ARXIV:1307.2425;%%
\bibitem [{\citenamefont {DeGrand}(2004)}]{DeGrand:2003in}%
  \BibitemOpen
  \bibfield  {author} {\bibinfo {author} {\bibnamefont {DeGrand} \bibfnamefont
  {Thomas~A}} (\bibinfo {collaboration} {MILC Collaboration})} (\bibinfo {year}
  {2004}),\ \bibfield  {title} {\enquote {\bibinfo {title} {{Kaon B parameter
  in quenched QCD}},}\ }\Doi {10.1103/PhysRevD.69.014504} {\bibfield  {journal}
  {\bibinfo  {journal} {Phys.~Rev.},\ }\textbf {\bibinfo {volume} {D69}},\
  \bibinfo {pages} {014504}},\ \Eprint {http://arxiv.org/abs/hep-lat/0309026}
  {arXiv:hep-lat/0309026 [hep-lat]} \BibitemShut {NoStop}%
%%CITATION = HEP-LAT/0309026;%%
\bibitem [{\citenamefont {Del~Debbio}\ \emph {et~al.}(2010)\citenamefont
  {Del~Debbio}, \citenamefont {Lucini}, \citenamefont {Patella}, \citenamefont
  {Pica},\ and\ \citenamefont {Rago}}]{DelDebbio:2010hx}%
  \BibitemOpen
  \bibfield  {author} {\bibinfo {author} {\bibnamefont {Del~Debbio}
  \bibfnamefont {Luigi}}, \bibinfo {author} {\bibfnamefont {Biagio}\
  \bibnamefont {Lucini}}, \bibinfo {author} {\bibfnamefont {Agostino}\
  \bibnamefont {Patella}}, \bibinfo {author} {\bibfnamefont {Claudio}\
  \bibnamefont {Pica}}, \ and\ \bibinfo {author} {\bibfnamefont {Antonio}\
  \bibnamefont {Rago}}} (\bibinfo {year} {2010}),\ \bibfield  {title} {\enquote
  {\bibinfo {title} {{The infrared dynamics of Minimal Walking Technicolor}},}\
  }\Doi {10.1103/PhysRevD.82.014510} {\bibfield  {journal} {\bibinfo  {journal}
  {Phys.~Rev.},\ }\textbf {\bibinfo {volume} {D82}},\ \bibinfo {pages}
  {014510}},\ \Eprint {http://arxiv.org/abs/1004.3206} {arXiv:1004.3206
  [hep-lat]} \BibitemShut {NoStop}%
%%CITATION = ARXIV:1004.3206;%%
\bibitem [{\citenamefont {Del~Debbio}\ \emph {et~al.}(2013)\citenamefont
  {Del~Debbio}, \citenamefont {Lucini}, \citenamefont {Pica}, \citenamefont
  {Patella}, \citenamefont {Rago} \emph {et~al.}}]{DelDebbio:2013hha}%
  \BibitemOpen
  \bibfield  {author} {\bibinfo {author} {\bibnamefont {Del~Debbio}
  \bibfnamefont {Luigi}}, \bibinfo {author} {\bibfnamefont {Biagio}\
  \bibnamefont {Lucini}}, \bibinfo {author} {\bibfnamefont {Claudio}\
  \bibnamefont {Pica}}, \bibinfo {author} {\bibfnamefont {Agostino}\
  \bibnamefont {Patella}}, \bibinfo {author} {\bibfnamefont {Antonio}\
  \bibnamefont {Rago}},  \emph {et~al.}} (\bibinfo {year} {2013}),\ \bibfield
  {title} {\enquote {\bibinfo {title} {{Large-volume results in SU(2) with
  adjoint fermions}},}\ }\href@noop {} {}\Eprint
  {http://arxiv.org/abs/1311.5597} {arXiv:1311.5597 [hep-lat]} \BibitemShut
  {NoStop}%
%%CITATION = ARXIV:1311.5597;%%
\bibitem [{\citenamefont {Del~Debbio}\ and\ \citenamefont
  {Zwicky}(2010)}]{DelDebbio:2010ze}%
  \BibitemOpen
  \bibfield  {author} {\bibinfo {author} {\bibnamefont {Del~Debbio}
  \bibfnamefont {Luigi}}, \ and\ \bibinfo {author} {\bibfnamefont {Roman}\
  \bibnamefont {Zwicky}}} (\bibinfo {year} {2010}),\ \bibfield  {title}
  {\enquote {\bibinfo {title} {{Hyperscaling relations in mass-deformed
  conformal gauge theories}},}\ }\Doi {10.1103/PhysRevD.82.014502} {\bibfield
  {journal} {\bibinfo  {journal} {Phys.~Rev.},\ }\textbf {\bibinfo {volume}
  {D82}},\ \bibinfo {pages} {014502}},\ \Eprint
  {http://arxiv.org/abs/1005.2371} {arXiv:1005.2371 [hep-ph]} \BibitemShut
  {NoStop}%
%%CITATION = ARXIV:1005.2371;%%
\bibitem [{\citenamefont {Del~Debbio}\ and\ \citenamefont
  {Zwicky}(2011)}]{DelDebbio:2010jy}%
  \BibitemOpen
  \bibfield  {author} {\bibinfo {author} {\bibnamefont {Del~Debbio}
  \bibfnamefont {Luigi}}, \ and\ \bibinfo {author} {\bibfnamefont {Roman}\
  \bibnamefont {Zwicky}}} (\bibinfo {year} {2011}),\ \bibfield  {title}
  {\enquote {\bibinfo {title} {{Scaling relations for the entire spectrum in
  mass-deformed conformal gauge theories}},}\ }\Doi
  {10.1016/j.physletb.2011.04.059} {\bibfield  {journal} {\bibinfo  {journal}
  {Phys.~Lett.},\ }\textbf {\bibinfo {volume} {B700}},\ \bibinfo {pages}
  {217--220}},\ \Eprint {http://arxiv.org/abs/1009.2894} {arXiv:1009.2894
  [hep-ph]} \BibitemShut {NoStop}%
%%CITATION = ARXIV:1009.2894;%%
\bibitem [{\citenamefont {Del~Debbio}\ and\ \citenamefont
  {Zwicky}(2014)}]{DelDebbio:2013qta}%
  \BibitemOpen
  \bibfield  {author} {\bibinfo {author} {\bibnamefont {Del~Debbio}
  \bibfnamefont {Luigi}}, \ and\ \bibinfo {author} {\bibfnamefont {Roman}\
  \bibnamefont {Zwicky}}} (\bibinfo {year} {2014}),\ \bibfield  {title}
  {\enquote {\bibinfo {title} {{Conformal scaling and the size of
  $m$-hadrons}},}\ }\Doi {10.1103/PhysRevD.89.014503} {\bibfield  {journal}
  {\bibinfo  {journal} {Phys.~Rev.},\ }\textbf {\bibinfo {volume} {D89}},\
  \bibinfo {pages} {014503}},\ \Eprint {http://arxiv.org/abs/1306.4038}
  {arXiv:1306.4038 [hep-ph]} \BibitemShut {NoStop}%
%%CITATION = ARXIV:1306.4038;%%
\bibitem [{\citenamefont {Della~Morte}\ \emph {et~al.}(2005a)\citenamefont
  {Della~Morte} \emph {et~al.}}]{DellaMorte:2004bc}%
  \BibitemOpen
  \bibfield  {author} {\bibinfo {author} {\bibnamefont {Della~Morte}
  \bibfnamefont {Michele}},  \emph {et~al.} (\bibinfo {collaboration} {ALPHA
  Collaboration})} (\bibinfo {year} {2005a}),\ \bibfield  {title} {\enquote
  {\bibinfo {title} {{Computation of the strong coupling in QCD with two
  dynamical flavors}},}\ }\Doi {10.1016/j.nuclphysb.2005.02.013} {\bibfield
  {journal} {\bibinfo  {journal} {Nucl.~Phys.},\ }\textbf {\bibinfo {volume}
  {B713}},\ \bibinfo {pages} {378--406}},\ \Eprint
  {http://arxiv.org/abs/hep-lat/0411025} {arXiv:hep-lat/0411025 [hep-lat]}
  \BibitemShut {NoStop}%
%%CITATION = HEP-LAT/0411025;%%
\bibitem [{\citenamefont {Della~Morte}\ \emph {et~al.}(2005b)\citenamefont
  {Della~Morte} \emph {et~al.}}]{DellaMorte:2005kg}%
  \BibitemOpen
  \bibfield  {author} {\bibinfo {author} {\bibnamefont {Della~Morte}
  \bibfnamefont {Michele}},  \emph {et~al.} (\bibinfo {collaboration} {ALPHA
  Collaboration})} (\bibinfo {year} {2005b}),\ \bibfield  {title} {\enquote
  {\bibinfo {title} {{Non-perturbative quark mass renormalization in two-flavor
  QCD}},}\ }\Doi {10.1016/j.nuclphysb.2005.09.028} {\bibfield  {journal}
  {\bibinfo  {journal} {Nucl.~Phys.},\ }\textbf {\bibinfo {volume} {B729}},\
  \bibinfo {pages} {117--134}},\ \Eprint {http://arxiv.org/abs/hep-lat/0507035}
  {arXiv:hep-lat/0507035 [hep-lat]} \BibitemShut {NoStop}%
%%CITATION = HEP-LAT/0507035;%%
\bibitem [{\citenamefont {Detmold}\ \emph {et~al.}(2014a)\citenamefont
  {Detmold}, \citenamefont {McCullough},\ and\ \citenamefont
  {Pochinsky}}]{Detmold:2014qqa}%
  \BibitemOpen
  \bibfield  {author} {\bibinfo {author} {\bibnamefont {Detmold} \bibfnamefont
  {William}}, \bibinfo {author} {\bibfnamefont {Matthew}\ \bibnamefont
  {McCullough}}, \ and\ \bibinfo {author} {\bibfnamefont {Andrew}\ \bibnamefont
  {Pochinsky}}} (\bibinfo {year} {2014a}),\ \bibfield  {title} {\enquote
  {\bibinfo {title} {{Dark Nuclei I: Cosmology and Indirect Detection}},}\
  }\Doi {10.1103/PhysRevD.90.115013} {\bibfield  {journal} {\bibinfo  {journal}
  {Phys.~Rev.},\ }\textbf {\bibinfo {volume} {D90}},\ \bibinfo {pages}
  {115013}},\ \Eprint {http://arxiv.org/abs/1406.2276} {arXiv:1406.2276
  [hep-ph]} \BibitemShut {NoStop}%
%%CITATION = ARXIV:1406.2276;%%
\bibitem [{\citenamefont {Detmold}\ \emph {et~al.}(2014b)\citenamefont
  {Detmold}, \citenamefont {McCullough},\ and\ \citenamefont
  {Pochinsky}}]{Detmold:2014kba}%
  \BibitemOpen
  \bibfield  {author} {\bibinfo {author} {\bibnamefont {Detmold} \bibfnamefont
  {William}}, \bibinfo {author} {\bibfnamefont {Matthew}\ \bibnamefont
  {McCullough}}, \ and\ \bibinfo {author} {\bibfnamefont {Andrew}\ \bibnamefont
  {Pochinsky}}} (\bibinfo {year} {2014b}),\ \bibfield  {title} {\enquote
  {\bibinfo {title} {{Dark Nuclei II: Nuclear Spectroscopy in Two-Colour
  QCD}},}\ }\Doi {10.1103/PhysRevD.90.114506} {\bibfield  {journal} {\bibinfo
  {journal} {Phys.~Rev.},\ }\textbf {\bibinfo {volume} {D90}},\ \bibinfo
  {pages} {114506}},\ \Eprint {http://arxiv.org/abs/1406.4116} {arXiv:1406.4116
  [hep-lat]} \BibitemShut {NoStop}%
%%CITATION = ARXIV:1406.4116;%%
\bibitem [{\citenamefont {Deuzeman}\ \emph {et~al.}(2013)\citenamefont
  {Deuzeman}, \citenamefont {Lombardo}, \citenamefont {Nunes Da~Silva},\ and\
  \citenamefont {Pallante}}]{Deuzeman:2012ee}%
  \BibitemOpen
  \bibfield  {author} {\bibinfo {author} {\bibnamefont {Deuzeman} \bibfnamefont
  {Albert}}, \bibinfo {author} {\bibfnamefont {Maria~Paola}\ \bibnamefont
  {Lombardo}}, \bibinfo {author} {\bibfnamefont {Tiago}\ \bibnamefont {Nunes
  Da~Silva}}, \ and\ \bibinfo {author} {\bibfnamefont {Elisabetta}\
  \bibnamefont {Pallante}}} (\bibinfo {year} {2013}),\ \bibfield  {title}
  {\enquote {\bibinfo {title} {{The bulk transition of QCD with twelve flavors
  and the role of improvement}},}\ }\Doi {10.1016/j.physletb.2013.02.030}
  {\bibfield  {journal} {\bibinfo  {journal} {Phys.~Lett.},\ }\textbf {\bibinfo
  {volume} {B720}},\ \bibinfo {pages} {358--365}},\ \Eprint
  {http://arxiv.org/abs/1209.5720} {arXiv:1209.5720 [hep-lat]} \BibitemShut
  {NoStop}%
%%CITATION = ARXIV:1209.5720;%%
\bibitem [{\citenamefont {Deuzeman}\ \emph {et~al.}(2008)\citenamefont
  {Deuzeman}, \citenamefont {Lombardo},\ and\ \citenamefont
  {Pallante}}]{Deuzeman:2008sc}%
  \BibitemOpen
  \bibfield  {author} {\bibinfo {author} {\bibnamefont {Deuzeman} \bibfnamefont
  {Albert}}, \bibinfo {author} {\bibfnamefont {Maria~Paola}\ \bibnamefont
  {Lombardo}}, \ and\ \bibinfo {author} {\bibfnamefont {Elisabetta}\
  \bibnamefont {Pallante}}} (\bibinfo {year} {2008}),\ \bibfield  {title}
  {\enquote {\bibinfo {title} {{The Physics of eight flavours}},}\ }\Doi
  {10.1016/j.physletb.2008.10.039} {\bibfield  {journal} {\bibinfo  {journal}
  {Phys.~Lett.},\ }\textbf {\bibinfo {volume} {B670}},\ \bibinfo {pages}
  {41--48}},\ \Eprint {http://arxiv.org/abs/0804.2905} {arXiv:0804.2905
  [hep-lat]} \BibitemShut {NoStop}%
%%CITATION = ARXIV:0804.2905;%%
\bibitem [{\citenamefont {Dietrich}\ and\ \citenamefont
  {Sannino}(2007)}]{Dietrich:2006cm}%
  \BibitemOpen
  \bibfield  {author} {\bibinfo {author} {\bibnamefont {Dietrich} \bibfnamefont
  {Dennis~D}}, \ and\ \bibinfo {author} {\bibfnamefont {Francesco}\
  \bibnamefont {Sannino}}} (\bibinfo {year} {2007}),\ \bibfield  {title}
  {\enquote {\bibinfo {title} {{Conformal window of SU(N) gauge theories with
  fermions in higher dimensional representations}},}\ }\Doi
  {10.1103/PhysRevD.75.085018} {\bibfield  {journal} {\bibinfo  {journal}
  {Phys.~Rev.},\ }\textbf {\bibinfo {volume} {D75}},\ \bibinfo {pages}
  {085018}},\ \Eprint {http://arxiv.org/abs/hep-ph/0611341}
  {arXiv:hep-ph/0611341 [hep-ph]} \BibitemShut {NoStop}%
%%CITATION = HEP-PH/0611341;%%
\bibitem [{\citenamefont {Dimopoulos}\ and\ \citenamefont
  {Susskind}(1979)}]{Dimopoulos:1979es}%
  \BibitemOpen
  \bibfield  {author} {\bibinfo {author} {\bibnamefont {Dimopoulos}
  \bibfnamefont {Savas}}, \ and\ \bibinfo {author} {\bibfnamefont {Leonard}\
  \bibnamefont {Susskind}}} (\bibinfo {year} {1979}),\ \bibfield  {title}
  {\enquote {\bibinfo {title} {{Mass Without Scalars}},}\ }\Doi
  {10.1016/0550-3213(79)90364-X} {\bibfield  {journal} {\bibinfo  {journal}
  {Nucl.~Phys.},\ }\textbf {\bibinfo {volume} {B155}},\ \bibinfo {pages}
  {237--252}}\BibitemShut {NoStop}%
%%CITATION = NUPHA,B155,237;%%
\bibitem [{\citenamefont {de~Divitiis}\ \emph {et~al.}(1994)\citenamefont
  {de~Divitiis}, \citenamefont {Frezzotti}, \citenamefont {Guagnelli},\ and\
  \citenamefont {Petronzio}}]{deDivitiis:1993hj}%
  \BibitemOpen
  \bibfield  {author} {\bibinfo {author} {\bibnamefont {de~Divitiis}
  \bibfnamefont {GM}}, \bibinfo {author} {\bibfnamefont {R.}~\bibnamefont
  {Frezzotti}}, \bibinfo {author} {\bibfnamefont {M.}~\bibnamefont
  {Guagnelli}}, \ and\ \bibinfo {author} {\bibfnamefont {R.}~\bibnamefont
  {Petronzio}}} (\bibinfo {year} {1994}),\ \bibfield  {title} {\enquote
  {\bibinfo {title} {{A Definition of the running coupling constant in a
  twisted SU(2) lattice gauge theory}},}\ }\Doi {10.1016/0550-3213(94)00126-X}
  {\bibfield  {journal} {\bibinfo  {journal} {Nucl.~Phys.},\ }\textbf {\bibinfo
  {volume} {B422}},\ \bibinfo {pages} {382--396}},\ \Eprint
  {http://arxiv.org/abs/hep-lat/9312085} {arXiv:hep-lat/9312085 [hep-lat]}
  \BibitemShut {NoStop}%
%%CITATION = HEP-LAT/9312085;%%
\bibitem [{\citenamefont {D'Onofrio}\ \emph {et~al.}(2014)\citenamefont
  {D'Onofrio}, \citenamefont {Rummukainen},\ and\ \citenamefont
  {Tranberg}}]{D'Onofrio:2014kta}%
  \BibitemOpen
  \bibfield  {author} {\bibinfo {author} {\bibnamefont {D'Onofrio}
  \bibfnamefont {Michela}}, \bibinfo {author} {\bibfnamefont {Kari}\
  \bibnamefont {Rummukainen}}, \ and\ \bibinfo {author} {\bibfnamefont
  {Anders}\ \bibnamefont {Tranberg}}} (\bibinfo {year} {2014}),\ \bibfield
  {title} {\enquote {\bibinfo {title} {{The Sphaleron Rate in the Minimal
  Standard Model}},}\ }\Doi {10.1103/PhysRevLett.113.141602} {\bibfield
  {journal} {\bibinfo  {journal} {Phys.~Rev.~Lett.},\ }\textbf {\bibinfo
  {volume} {113}},\ \bibinfo {pages} {141602}},\ \Eprint
  {http://arxiv.org/abs/1404.3565} {arXiv:1404.3565 [hep-ph]} \BibitemShut
  {NoStop}%
%%CITATION = ARXIV:1404.3565;%%
\bibitem [{\citenamefont {Dugan}\ \emph {et~al.}(1985)\citenamefont {Dugan},
  \citenamefont {Georgi},\ and\ \citenamefont {Kaplan}}]{Dugan:1984hq}%
  \BibitemOpen
  \bibfield  {author} {\bibinfo {author} {\bibnamefont {Dugan} \bibfnamefont
  {Michael~J}}, \bibinfo {author} {\bibfnamefont {Howard}\ \bibnamefont
  {Georgi}}, \ and\ \bibinfo {author} {\bibfnamefont {David~B.}\ \bibnamefont
  {Kaplan}}} (\bibinfo {year} {1985}),\ \bibfield  {title} {\enquote {\bibinfo
  {title} {{Anatomy of a Composite Higgs Model}},}\ }\Doi
  {10.1016/0550-3213(85)90221-4} {\bibfield  {journal} {\bibinfo  {journal}
  {Nucl.~Phys.},\ }\textbf {\bibinfo {volume} {B254}},\ \bibinfo {pages}
  {299}}\BibitemShut {NoStop}%
%%CITATION = NUPHA,B254,299;%%
\bibitem [{\citenamefont {Eichten}\ and\ \citenamefont
  {Lane}(1980)}]{Eichten:1979ah}%
  \BibitemOpen
  \bibfield  {author} {\bibinfo {author} {\bibnamefont {Eichten} \bibfnamefont
  {Estia}}, \ and\ \bibinfo {author} {\bibfnamefont {Kenneth~D.}\ \bibnamefont
  {Lane}}} (\bibinfo {year} {1980}),\ \bibfield  {title} {\enquote {\bibinfo
  {title} {{Dynamical Breaking of Weak Interaction Symmetries}},}\ }\Doi
  {10.1016/0370-2693(80)90065-9} {\bibfield  {journal} {\bibinfo  {journal}
  {Phys.~Lett.},\ }\textbf {\bibinfo {volume} {B90}},\ \bibinfo {pages}
  {125--130}}\BibitemShut {NoStop}%
%%CITATION = PHLTA,B90,125;%%
\bibitem [{\citenamefont {Endres}(2009)}]{Endres:2009yp}%
  \BibitemOpen
  \bibfield  {author} {\bibinfo {author} {\bibnamefont {Endres} \bibfnamefont
  {Michael~G}}} (\bibinfo {year} {2009}),\ \bibfield  {title} {\enquote
  {\bibinfo {title} {{Dynamical simulation of N=1 supersymmetric Yang-Mills
  theory with domain wall fermions}},}\ }\Doi {10.1103/PhysRevD.79.094503}
  {\bibfield  {journal} {\bibinfo  {journal} {Phys.~Rev.},\ }\textbf {\bibinfo
  {volume} {D79}},\ \bibinfo {pages} {094503}},\ \Eprint
  {http://arxiv.org/abs/0902.4267} {arXiv:0902.4267 [hep-lat]} \BibitemShut
  {NoStop}%
%%CITATION = ARXIV:0902.4267;%%
\bibitem [{\citenamefont {Engels}\ and\ \citenamefont
  {Karsch}(2012)}]{Engels:2011km}%
  \BibitemOpen
  \bibfield  {author} {\bibinfo {author} {\bibnamefont {Engels} \bibfnamefont
  {J}}, \ and\ \bibinfo {author} {\bibfnamefont {F.}~\bibnamefont {Karsch}}}
  (\bibinfo {year} {2012}),\ \bibfield  {title} {\enquote {\bibinfo {title}
  {{The scaling functions of the free energy density and its derivatives for
  the 3d O(4) model}},}\ }\Doi {10.1103/PhysRevD.85.094506} {\bibfield
  {journal} {\bibinfo  {journal} {Phys.~Rev.},\ }\textbf {\bibinfo {volume}
  {D85}},\ \bibinfo {pages} {094506}},\ \Eprint
  {http://arxiv.org/abs/1105.0584} {arXiv:1105.0584 [hep-lat]} \BibitemShut
  {NoStop}%
%%CITATION = ARXIV:1105.0584;%%
\bibitem [{\citenamefont {Ferretti}(2014)}]{Ferretti:2014qta}%
  \BibitemOpen
  \bibfield  {author} {\bibinfo {author} {\bibnamefont {Ferretti} \bibfnamefont
  {Gabriele}}} (\bibinfo {year} {2014}),\ \bibfield  {title} {\enquote
  {\bibinfo {title} {{UV Completions of Partial Compositeness: The Case for a
  SU(4) Gauge Group}},}\ }\Doi {10.1007/JHEP06(2014)142} {\bibfield  {journal}
  {\bibinfo  {journal} {JHEP},\ }\textbf {\bibinfo {volume} {1406}},\ \bibinfo
  {pages} {142}},\ \Eprint {http://arxiv.org/abs/1404.7137} {arXiv:1404.7137
  [hep-ph]} \BibitemShut {NoStop}%
%%CITATION = ARXIV:1404.7137;%%
\bibitem [{\citenamefont {Ferretti}\ and\ \citenamefont
  {Karateev}(2014)}]{Ferretti:2013kya}%
  \BibitemOpen
  \bibfield  {author} {\bibinfo {author} {\bibnamefont {Ferretti} \bibfnamefont
  {Gabriele}}, \ and\ \bibinfo {author} {\bibfnamefont {Denis}\ \bibnamefont
  {Karateev}}} (\bibinfo {year} {2014}),\ \bibfield  {title} {\enquote
  {\bibinfo {title} {{Fermionic UV completions of Composite Higgs models}},}\
  }\Doi {10.1007/JHEP03(2014)077} {\bibfield  {journal} {\bibinfo  {journal}
  {JHEP},\ }\textbf {\bibinfo {volume} {1403}},\ \bibinfo {pages} {077}},\
  \Eprint {http://arxiv.org/abs/1312.5330} {arXiv:1312.5330 [hep-ph]}
  \BibitemShut {NoStop}%
%%CITATION = ARXIV:1312.5330;%%
\bibitem [{\citenamefont {Fleming}(2008)}]{Fleming:2008gy}%
  \BibitemOpen
  \bibfield  {author} {\bibinfo {author} {\bibnamefont {Fleming} \bibfnamefont
  {George~T}}} (\bibinfo {year} {2008}),\ \bibfield  {title} {\enquote
  {\bibinfo {title} {{Strong Interactions for the LHC}},}\ }\href@noop {}
  {\bibfield  {journal} {\bibinfo  {journal} {PoS},\ }\textbf {\bibinfo
  {volume} {LATTICE2008}},\ \bibinfo {pages} {021}},\ \Eprint
  {http://arxiv.org/abs/0812.2035} {arXiv:0812.2035 [hep-lat]} \BibitemShut
  {NoStop}%
%%CITATION = ARXIV:0812.2035;%%
\bibitem [{\citenamefont {Fleming}\ \emph {et~al.}(2001)\citenamefont
  {Fleming}, \citenamefont {Kogut},\ and\ \citenamefont
  {Vranas}}]{Fleming:2000fa}%
  \BibitemOpen
  \bibfield  {author} {\bibinfo {author} {\bibnamefont {Fleming} \bibfnamefont
  {George~Tamminga}}, \bibinfo {author} {\bibfnamefont {John~B.}\ \bibnamefont
  {Kogut}}, \ and\ \bibinfo {author} {\bibfnamefont {Pavlos~M.}\ \bibnamefont
  {Vranas}}} (\bibinfo {year} {2001}),\ \bibfield  {title} {\enquote {\bibinfo
  {title} {{SuperYang-Mills on the lattice with domain wall fermions}},}\ }\Doi
  {10.1103/PhysRevD.64.034510} {\bibfield  {journal} {\bibinfo  {journal}
  {Phys.~Rev.},\ }\textbf {\bibinfo {volume} {D64}},\ \bibinfo {pages}
  {034510}},\ \Eprint {http://arxiv.org/abs/hep-lat/0008009}
  {arXiv:hep-lat/0008009 [hep-lat]} \BibitemShut {NoStop}%
%%CITATION = HEP-LAT/0008009;%%
\bibitem [{\citenamefont {Fodor}\ \emph {et~al.}(2015a)\citenamefont {Fodor},
  \citenamefont {Holland}, \citenamefont {Kuti}, \citenamefont {Mondal},
  \citenamefont {Nogradi},\ and\ \citenamefont {Wong}}]{Fodor:2015baa}%
  \BibitemOpen
  \bibfield  {author} {\bibinfo {author} {\bibnamefont {Fodor} \bibfnamefont
  {Zoltan}}, \bibinfo {author} {\bibfnamefont {Kieran}\ \bibnamefont
  {Holland}}, \bibinfo {author} {\bibfnamefont {Julius}\ \bibnamefont {Kuti}},
  \bibinfo {author} {\bibfnamefont {Santanu}\ \bibnamefont {Mondal}}, \bibinfo
  {author} {\bibfnamefont {Daniel}\ \bibnamefont {Nogradi}}, \ and\ \bibinfo
  {author} {\bibfnamefont {Chik~Him}\ \bibnamefont {Wong}}} (\bibinfo {year}
  {2015a}),\ \bibfield  {title} {\enquote {\bibinfo {title} {{The running
  coupling of 8 flavors and 3 colors}},}\ }\Doi {10.1007/JHEP06(2015)019}
  {\bibfield  {journal} {\bibinfo  {journal} {JHEP},\ }\textbf {\bibinfo
  {volume} {06}},\ \bibinfo {pages} {019}},\ \Eprint
  {http://arxiv.org/abs/1503.01132} {arXiv:1503.01132 [hep-lat]} \BibitemShut
  {NoStop}%
%%CITATION = ARXIV:1503.01132;%%
\bibitem [{\citenamefont {Fodor}\ \emph {et~al.}(2015b)\citenamefont {Fodor},
  \citenamefont {Holland}, \citenamefont {Kuti}, \citenamefont {Mondal},
  \citenamefont {Nogradi},\ and\ \citenamefont {Wong}}]{Fodor:2015vwa}%
  \BibitemOpen
  \bibfield  {author} {\bibinfo {author} {\bibnamefont {Fodor} \bibfnamefont
  {Zoltan}}, \bibinfo {author} {\bibfnamefont {Kieran}\ \bibnamefont
  {Holland}}, \bibinfo {author} {\bibfnamefont {Julius}\ \bibnamefont {Kuti}},
  \bibinfo {author} {\bibfnamefont {Santanu}\ \bibnamefont {Mondal}}, \bibinfo
  {author} {\bibfnamefont {Daniel}\ \bibnamefont {Nogradi}}, \ and\ \bibinfo
  {author} {\bibfnamefont {Chik~Him}\ \bibnamefont {Wong}}} (\bibinfo {year}
  {2015b}),\ \bibfield  {title} {\enquote {\bibinfo {title} {{Toward the
  minimal realization of a light composite Higgs}},}\ }\bibfield  {booktitle}
  {\emph {\bibinfo {booktitle} {{Proceedings, 32nd International Symposium on
  Lattice Field Theory (Lattice 2014)}}},\ }\href@noop {} {\bibfield  {journal}
  {\bibinfo  {journal} {PoS},\ }\textbf {\bibinfo {volume} {LATTICE2014}},\
  \bibinfo {pages} {244}},\ \Eprint {http://arxiv.org/abs/1502.00028}
  {arXiv:1502.00028 [hep-lat]} \BibitemShut {NoStop}%
%%CITATION = ARXIV:1502.00028;%%
\bibitem [{\citenamefont {Fodor}\ \emph {et~al.}(2014b)\citenamefont {Fodor},
  \citenamefont {Holland}, \citenamefont {Kuti}, \citenamefont {Mondal},
  \citenamefont {Nogradi} \emph {et~al.}}]{Fodor:2014cpa}%
  \BibitemOpen
  \bibfield  {author} {\bibinfo {author} {\bibnamefont {Fodor} \bibfnamefont
  {Zoltan}}, \bibinfo {author} {\bibfnamefont {Kieran}\ \bibnamefont
  {Holland}}, \bibinfo {author} {\bibfnamefont {Julius}\ \bibnamefont {Kuti}},
  \bibinfo {author} {\bibfnamefont {Santanu}\ \bibnamefont {Mondal}}, \bibinfo
  {author} {\bibfnamefont {Daniel}\ \bibnamefont {Nogradi}},  \emph {et~al.}}
  (\bibinfo {year} {2014b}),\ \bibfield  {title} {\enquote {\bibinfo {title}
  {{The lattice gradient flow at tree-level and its improvement}},}\ }\Doi
  {10.1007/JHEP09(2014)018} {\bibfield  {journal} {\bibinfo  {journal} {JHEP},\
  }\textbf {\bibinfo {volume} {1409}},\ \bibinfo {pages} {018}},\ \Eprint
  {http://arxiv.org/abs/1406.0827} {arXiv:1406.0827 [hep-lat]} \BibitemShut
  {NoStop}%
%%CITATION = ARXIV:1406.0827;%%
\bibitem [{\citenamefont {Fodor}\ \emph {et~al.}(2007)\citenamefont {Fodor},
  \citenamefont {Holland}, \citenamefont {Kuti}, \citenamefont {Nogradi},\ and\
  \citenamefont {Schroeder}}]{Fodor:2007fn}%
  \BibitemOpen
  \bibfield  {author} {\bibinfo {author} {\bibnamefont {Fodor} \bibfnamefont
  {Zoltan}}, \bibinfo {author} {\bibfnamefont {Kieran}\ \bibnamefont
  {Holland}}, \bibinfo {author} {\bibfnamefont {Julius}\ \bibnamefont {Kuti}},
  \bibinfo {author} {\bibfnamefont {Daniel}\ \bibnamefont {Nogradi}}, \ and\
  \bibinfo {author} {\bibfnamefont {Chris}\ \bibnamefont {Schroeder}}}
  (\bibinfo {year} {2007}),\ \bibfield  {title} {\enquote {\bibinfo {title}
  {{New Higgs physics from the lattice}},}\ }\href@noop {} {\bibfield
  {journal} {\bibinfo  {journal} {PoS},\ }\textbf {\bibinfo {volume}
  {LAT2007}},\ \bibinfo {pages} {056}},\ \Eprint
  {http://arxiv.org/abs/0710.3151} {arXiv:0710.3151 [hep-lat]} \BibitemShut
  {NoStop}%
%%CITATION = ARXIV:0710.3151;%%
\bibitem [{\citenamefont {Fodor}\ \emph {et~al.}(2009)\citenamefont {Fodor},
  \citenamefont {Holland}, \citenamefont {Kuti}, \citenamefont {Nogradi},\ and\
  \citenamefont {Schroeder}}]{Fodor:2009wk}%
  \BibitemOpen
  \bibfield  {author} {\bibinfo {author} {\bibnamefont {Fodor} \bibfnamefont
  {Zoltan}}, \bibinfo {author} {\bibfnamefont {Kieran}\ \bibnamefont
  {Holland}}, \bibinfo {author} {\bibfnamefont {Julius}\ \bibnamefont {Kuti}},
  \bibinfo {author} {\bibfnamefont {Daniel}\ \bibnamefont {Nogradi}}, \ and\
  \bibinfo {author} {\bibfnamefont {Chris}\ \bibnamefont {Schroeder}}}
  (\bibinfo {year} {2009}),\ \bibfield  {title} {\enquote {\bibinfo {title}
  {{Nearly conformal gauge theories in finite volume}},}\ }\Doi
  {10.1016/j.physletb.2009.10.040} {\bibfield  {journal} {\bibinfo  {journal}
  {Phys.~Lett.},\ }\textbf {\bibinfo {volume} {B681}},\ \bibinfo {pages}
  {353--361}},\ \Eprint {http://arxiv.org/abs/0907.4562} {arXiv:0907.4562
  [hep-lat]} \BibitemShut {NoStop}%
%%CITATION = ARXIV:0907.4562;%%
\bibitem [{\citenamefont {Fodor}\ \emph {et~al.}(2011a)\citenamefont {Fodor},
  \citenamefont {Holland}, \citenamefont {Kuti}, \citenamefont {Nogradi},
  \citenamefont {Schroeder} \emph {et~al.}}]{Fodor:2011tu}%
  \BibitemOpen
  \bibfield  {author} {\bibinfo {author} {\bibnamefont {Fodor} \bibfnamefont
  {Zoltan}}, \bibinfo {author} {\bibfnamefont {Kieran}\ \bibnamefont
  {Holland}}, \bibinfo {author} {\bibfnamefont {Julius}\ \bibnamefont {Kuti}},
  \bibinfo {author} {\bibfnamefont {Daniel}\ \bibnamefont {Nogradi}}, \bibinfo
  {author} {\bibfnamefont {Chris}\ \bibnamefont {Schroeder}},  \emph {et~al.}}
  (\bibinfo {year} {2011a}),\ \bibfield  {title} {\enquote {\bibinfo {title}
  {{Twelve massless flavors and three colors below the conformal window}},}\
  }\Doi {10.1016/j.physletb.2011.07.037} {\bibfield  {journal} {\bibinfo
  {journal} {Phys.~Lett.},\ }\textbf {\bibinfo {volume} {B703}},\ \bibinfo
  {pages} {348--358}},\ \Eprint {http://arxiv.org/abs/1104.3124}
  {arXiv:1104.3124 [hep-lat]} \BibitemShut {NoStop}%
%%CITATION = ARXIV:1104.3124;%%
\bibitem [{\citenamefont {Fodor}\ \emph {et~al.}(2011b)\citenamefont {Fodor},
  \citenamefont {Holland}, \citenamefont {Kuti}, \citenamefont {Nogradi},
  \citenamefont {Schroeder} \emph {et~al.}}]{Fodor:2012uu}%
  \BibitemOpen
  \bibfield  {author} {\bibinfo {author} {\bibnamefont {Fodor} \bibfnamefont
  {Zoltan}}, \bibinfo {author} {\bibfnamefont {Kieran}\ \bibnamefont
  {Holland}}, \bibinfo {author} {\bibfnamefont {Julius}\ \bibnamefont {Kuti}},
  \bibinfo {author} {\bibfnamefont {Daniel}\ \bibnamefont {Nogradi}}, \bibinfo
  {author} {\bibfnamefont {Chris}\ \bibnamefont {Schroeder}},  \emph {et~al.}}
  (\bibinfo {year} {2011b}),\ \bibfield  {title} {\enquote {\bibinfo {title}
  {{Twelve fundamental and two sextet fermion flavors}},}\ }\href@noop {}
  {\bibfield  {journal} {\bibinfo  {journal} {PoS},\ }\textbf {\bibinfo
  {volume} {Lattice2011}},\ \bibinfo {pages} {073}},\ \Eprint
  {http://arxiv.org/abs/1205.1878} {arXiv:1205.1878 [hep-lat]} \BibitemShut
  {NoStop}%
%%CITATION = ARXIV:1205.1878;%%
\bibitem [{\citenamefont {Fodor}\ \emph {et~al.}(2012a)\citenamefont {Fodor},
  \citenamefont {Holland}, \citenamefont {Kuti}, \citenamefont {Nogradi},
  \citenamefont {Schroeder} \emph {et~al.}}]{Fodor:2012ty}%
  \BibitemOpen
  \bibfield  {author} {\bibinfo {author} {\bibnamefont {Fodor} \bibfnamefont
  {Zoltan}}, \bibinfo {author} {\bibfnamefont {Kieran}\ \bibnamefont
  {Holland}}, \bibinfo {author} {\bibfnamefont {Julius}\ \bibnamefont {Kuti}},
  \bibinfo {author} {\bibfnamefont {Daniel}\ \bibnamefont {Nogradi}}, \bibinfo
  {author} {\bibfnamefont {Chris}\ \bibnamefont {Schroeder}},  \emph {et~al.}}
  (\bibinfo {year} {2012a}),\ \bibfield  {title} {\enquote {\bibinfo {title}
  {{Can the nearly conformal sextet gauge model hide the Higgs impostor?}}}\
  }\Doi {10.1016/j.physletb.2012.10.079} {\bibfield  {journal} {\bibinfo
  {journal} {Phys.~Lett.},\ }\textbf {\bibinfo {volume} {B718}},\ \bibinfo
  {pages} {657--666}},\ \Eprint {http://arxiv.org/abs/1209.0391}
  {arXiv:1209.0391 [hep-lat]} \BibitemShut {NoStop}%
%%CITATION = ARXIV:1209.0391;%%
\bibitem [{\citenamefont {Fodor}\ \emph {et~al.}(2012b)\citenamefont {Fodor},
  \citenamefont {Holland}, \citenamefont {Kuti}, \citenamefont {Nogradi},
  \citenamefont {Schroeder} \emph {et~al.}}]{Fodor:2012uw}%
  \BibitemOpen
  \bibfield  {author} {\bibinfo {author} {\bibnamefont {Fodor} \bibfnamefont
  {Zoltan}}, \bibinfo {author} {\bibfnamefont {Kieran}\ \bibnamefont
  {Holland}}, \bibinfo {author} {\bibfnamefont {Julius}\ \bibnamefont {Kuti}},
  \bibinfo {author} {\bibfnamefont {Daniel}\ \bibnamefont {Nogradi}}, \bibinfo
  {author} {\bibfnamefont {Chris}\ \bibnamefont {Schroeder}},  \emph {et~al.}}
  (\bibinfo {year} {2012b}),\ \bibfield  {title} {\enquote {\bibinfo {title}
  {{Confining force and running coupling with twelve fundamental and two sextet
  fermions}},}\ }\href@noop {} {\bibfield  {journal} {\bibinfo  {journal}
  {PoS},\ }\textbf {\bibinfo {volume} {LATTICE2012}},\ \bibinfo {pages}
  {025}},\ \Eprint {http://arxiv.org/abs/1211.3548} {arXiv:1211.3548 [hep-lat]}
  \BibitemShut {NoStop}%
%%CITATION = ARXIV:1211.3548;%%
\bibitem [{\citenamefont {Fodor}\ \emph {et~al.}(2012)\citenamefont {Fodor},
  \citenamefont {Holland}, \citenamefont {Kuti}, \citenamefont {Nogradi},\ and\
  \citenamefont {Wong}}]{Fodor:2012td}%
  \BibitemOpen
  \bibfield  {author} {\bibinfo {author} {\bibnamefont {Fodor} \bibfnamefont
  {Zoltan}}, \bibinfo {author} {\bibfnamefont {Kieran}\ \bibnamefont
  {Holland}}, \bibinfo {author} {\bibfnamefont {Julius}\ \bibnamefont {Kuti}},
  \bibinfo {author} {\bibfnamefont {Daniel}\ \bibnamefont {Nogradi}}, \ and\
  \bibinfo {author} {\bibfnamefont {Chik~Him}\ \bibnamefont {Wong}}} (\bibinfo
  {year} {2012}),\ \bibfield  {title} {\enquote {\bibinfo {title} {{The
  Yang-Mills gradient flow in finite volume}},}\ }\Doi
  {10.1007/JHEP11(2012)007} {\bibfield  {journal} {\bibinfo  {journal} {JHEP},\
  }\textbf {\bibinfo {volume} {1211}},\ \bibinfo {pages} {007}},\ \Eprint
  {http://arxiv.org/abs/1208.1051} {arXiv:1208.1051 [hep-lat]} \BibitemShut
  {NoStop}%
%%CITATION = ARXIV:1208.1051;%%
\bibitem [{\citenamefont {Fodor}\ \emph {et~al.}(2014a)\citenamefont {Fodor},
  \citenamefont {Holland}, \citenamefont {Kuti}, \citenamefont {Nogradi},\ and\
  \citenamefont {Wong}}]{Fodor:2014pqa}%
  \BibitemOpen
  \bibfield  {author} {\bibinfo {author} {\bibnamefont {Fodor} \bibfnamefont
  {Zoltan}}, \bibinfo {author} {\bibfnamefont {Kieran}\ \bibnamefont
  {Holland}}, \bibinfo {author} {\bibfnamefont {Julius}\ \bibnamefont {Kuti}},
  \bibinfo {author} {\bibfnamefont {Daniel}\ \bibnamefont {Nogradi}}, \ and\
  \bibinfo {author} {\bibfnamefont {Chik~Him}\ \bibnamefont {Wong}}} (\bibinfo
  {year} {2014a}),\ \bibfield  {title} {\enquote {\bibinfo {title} {{Can a
  light Higgs impostor hide in composite gauge models?}}}\ }\href@noop {}
  {\bibfield  {journal} {\bibinfo  {journal} {PoS},\ }\textbf {\bibinfo
  {volume} {LATTICE2013}},\ \bibinfo {pages} {062}},\ \Eprint
  {http://arxiv.org/abs/1401.2176} {arXiv:1401.2176 [hep-lat]} \BibitemShut
  {NoStop}%
%%CITATION = ARXIV:1401.2176;%%
\bibitem [{\citenamefont {de~Forcrand}\ \emph {et~al.}(2010)\citenamefont
  {de~Forcrand}, \citenamefont {Kurkela},\ and\ \citenamefont
  {Panero}}]{deForcrand:2010be}%
  \BibitemOpen
  \bibfield  {author} {\bibinfo {author} {\bibnamefont {de~Forcrand}
  \bibfnamefont {Philippe}}, \bibinfo {author} {\bibfnamefont {Aleksi}\
  \bibnamefont {Kurkela}}, \ and\ \bibinfo {author} {\bibfnamefont {Marco}\
  \bibnamefont {Panero}}} (\bibinfo {year} {2010}),\ \bibfield  {title}
  {\enquote {\bibinfo {title} {{The phase diagram of Yang-Mills theory with a
  compact extra dimension}},}\ }\Doi {10.1007/JHEP06(2010)050} {\bibfield
  {journal} {\bibinfo  {journal} {JHEP},\ }\textbf {\bibinfo {volume} {1006}},\
  \bibinfo {pages} {050}},\ \Eprint {http://arxiv.org/abs/1003.4643}
  {arXiv:1003.4643 [hep-lat]} \BibitemShut {NoStop}%
%%CITATION = ARXIV:1003.4643;%%
\bibitem [{\citenamefont {Fritzsch}\ and\ \citenamefont
  {Ramos}(2013)}]{Fritzsch:2013je}%
  \BibitemOpen
  \bibfield  {author} {\bibinfo {author} {\bibnamefont {Fritzsch} \bibfnamefont
  {Patrick}}, \ and\ \bibinfo {author} {\bibfnamefont {Alberto}\ \bibnamefont
  {Ramos}}} (\bibinfo {year} {2013}),\ \bibfield  {title} {\enquote {\bibinfo
  {title} {{The gradient flow coupling in the Schrödinger Functional}},}\ }\Doi
  {10.1007/JHEP10(2013)008} {\bibfield  {journal} {\bibinfo  {journal} {JHEP},\
  }\textbf {\bibinfo {volume} {1310}},\ \bibinfo {pages} {008}},\ \Eprint
  {http://arxiv.org/abs/1301.4388} {arXiv:1301.4388 [hep-lat]} \BibitemShut
  {NoStop}%
%%CITATION = ARXIV:1301.4388;%%
\bibitem [{\citenamefont {Gattringer}\ and\ \citenamefont
  {Lang}(2010)}]{Gattringer:2010zz}%
  \BibitemOpen
  \bibfield  {author} {\bibinfo {author} {\bibnamefont {Gattringer}
  \bibfnamefont {Christof}}, \ and\ \bibinfo {author} {\bibfnamefont
  {Christian~B.}\ \bibnamefont {Lang}}} (\bibinfo {year} {2010}),\ \Doi
  {10.1007/978-3-642-01850-3} {\emph {\bibinfo {title} {{Quantum chromodynamics
  on the lattice}}}},\ Vol.\ \bibinfo {volume} {788}\ (\bibinfo  {publisher}
  {Springer})\BibitemShut {NoStop}%
%%CITATION = LNPHA,788,1;%%
\bibitem [{\citenamefont {Georgi}(1986)}]{Georgi:1985hf}%
  \BibitemOpen
  \bibfield  {author} {\bibinfo {author} {\bibnamefont {Georgi} \bibfnamefont
  {Howard}}} (\bibinfo {year} {1986}),\ \bibfield  {title} {\enquote {\bibinfo
  {title} {{A Tool Kit for Builders of Composite Models}},}\ }\Doi
  {10.1016/0550-3213(86)90092-1} {\bibfield  {journal} {\bibinfo  {journal}
  {Nucl.~Phys.},\ }\textbf {\bibinfo {volume} {B266}},\ \bibinfo {pages}
  {274}}\BibitemShut {NoStop}%
%%CITATION = NUPHA,B266,274;%%
\bibitem [{\citenamefont {Georgi}\ and\ \citenamefont
  {Kaplan}(1984)}]{Georgi:1984af}%
  \BibitemOpen
  \bibfield  {author} {\bibinfo {author} {\bibnamefont {Georgi} \bibfnamefont
  {Howard}}, \ and\ \bibinfo {author} {\bibfnamefont {David~B.}\ \bibnamefont
  {Kaplan}}} (\bibinfo {year} {1984}),\ \bibfield  {title} {\enquote {\bibinfo
  {title} {{Composite Higgs and Custodial SU(2)}},}\ }\Doi
  {10.1016/0370-2693(84)90341-1} {\bibfield  {journal} {\bibinfo  {journal}
  {Phys.~Lett.},\ }\textbf {\bibinfo {volume} {B145}},\ \bibinfo {pages}
  {216}}\BibitemShut {NoStop}%
%%CITATION = PHLTA,B145,216;%%
\bibitem [{\citenamefont {Georgi}\ \emph {et~al.}(1984)\citenamefont {Georgi},
  \citenamefont {Kaplan},\ and\ \citenamefont {Galison}}]{Georgi:1984ef}%
  \BibitemOpen
  \bibfield  {author} {\bibinfo {author} {\bibnamefont {Georgi} \bibfnamefont
  {Howard}}, \bibinfo {author} {\bibfnamefont {David~B.}\ \bibnamefont
  {Kaplan}}, \ and\ \bibinfo {author} {\bibfnamefont {Peter}\ \bibnamefont
  {Galison}}} (\bibinfo {year} {1984}),\ \bibfield  {title} {\enquote {\bibinfo
  {title} {{Calculation of the Composite Higgs Mass}},}\ }\Doi
  {10.1016/0370-2693(84)90823-2} {\bibfield  {journal} {\bibinfo  {journal}
  {Phys.~Lett.},\ }\textbf {\bibinfo {volume} {B143}},\ \bibinfo {pages}
  {152}}\BibitemShut {NoStop}%
%%CITATION = PHLTA,B143,152;%%
\bibitem [{\citenamefont {Gerhold}\ and\ \citenamefont
  {Jansen}(2010)}]{Gerhold:2010bh}%
  \BibitemOpen
  \bibfield  {author} {\bibinfo {author} {\bibnamefont {Gerhold} \bibfnamefont
  {P}}, \ and\ \bibinfo {author} {\bibfnamefont {K.}~\bibnamefont {Jansen}}}
  (\bibinfo {year} {2010}),\ \bibfield  {title} {\enquote {\bibinfo {title}
  {{Upper Higgs boson mass bounds from a chirally invariant lattice
  Higgs-Yukawa model}},}\ }\Doi {10.1007/s13130-010-0464-1} {\bibfield
  {journal} {\bibinfo  {journal} {JHEP},\ }\textbf {\bibinfo {volume} {1004}},\
  \bibinfo {pages} {094}},\ \Eprint {http://arxiv.org/abs/1002.4336}
  {arXiv:1002.4336 [hep-lat]} \BibitemShut {NoStop}%
%%CITATION = ARXIV:1002.4336;%%
\bibitem [{\citenamefont {Giedt}\ \emph {et~al.}(2009)\citenamefont {Giedt},
  \citenamefont {Brower}, \citenamefont {Catterall}, \citenamefont {Fleming},\
  and\ \citenamefont {Vranas}}]{Giedt:2008xm}%
  \BibitemOpen
  \bibfield  {author} {\bibinfo {author} {\bibnamefont {Giedt} \bibfnamefont
  {Joel}}, \bibinfo {author} {\bibfnamefont {Richard}\ \bibnamefont {Brower}},
  \bibinfo {author} {\bibfnamefont {Simon}\ \bibnamefont {Catterall}}, \bibinfo
  {author} {\bibfnamefont {George~T.}\ \bibnamefont {Fleming}}, \ and\ \bibinfo
  {author} {\bibfnamefont {Pavlos}\ \bibnamefont {Vranas}}} (\bibinfo {year}
  {2009}),\ \bibfield  {title} {\enquote {\bibinfo {title} {{Lattice
  super-Yang-Mills using domain wall fermions in the chiral limit}},}\ }\Doi
  {10.1103/PhysRevD.79.025015} {\bibfield  {journal} {\bibinfo  {journal}
  {Phys.~Rev.},\ }\textbf {\bibinfo {volume} {D79}},\ \bibinfo {pages}
  {025015}},\ \Eprint {http://arxiv.org/abs/0810.5746} {arXiv:0810.5746
  [hep-lat]} \BibitemShut {NoStop}%
%%CITATION = ARXIV:0810.5746;%%
\bibitem [{\citenamefont {Giedt}\ and\ \citenamefont
  {Weinberg}(2012)}]{Giedt:2012rj}%
  \BibitemOpen
  \bibfield  {author} {\bibinfo {author} {\bibnamefont {Giedt} \bibfnamefont
  {Joel}}, \ and\ \bibinfo {author} {\bibfnamefont {Evan}\ \bibnamefont
  {Weinberg}}} (\bibinfo {year} {2012}),\ \bibfield  {title} {\enquote
  {\bibinfo {title} {{Finite size scaling in minimal walking technicolor}},}\
  }\Doi {10.1103/PhysRevD.85.097503} {\bibfield  {journal} {\bibinfo  {journal}
  {Phys.~Rev.},\ }\textbf {\bibinfo {volume} {D85}},\ \bibinfo {pages}
  {097503}},\ \Eprint {http://arxiv.org/abs/1201.6262} {arXiv:1201.6262
  [hep-lat]} \BibitemShut {NoStop}%
%%CITATION = ARXIV:1201.6262;%%
\bibitem [{\citenamefont {Gildener}\ and\ \citenamefont
  {Weinberg}(1976)}]{Gildener:1976ih}%
  \BibitemOpen
  \bibfield  {author} {\bibinfo {author} {\bibnamefont {Gildener} \bibfnamefont
  {Eldad}}, \ and\ \bibinfo {author} {\bibfnamefont {Steven}\ \bibnamefont
  {Weinberg}}} (\bibinfo {year} {1976}),\ \bibfield  {title} {\enquote
  {\bibinfo {title} {{Symmetry Breaking and Scalar Bosons}},}\ }\Doi
  {10.1103/PhysRevD.13.3333} {\bibfield  {journal} {\bibinfo  {journal}
  {Phys.~Rev.},\ }\textbf {\bibinfo {volume} {D13}},\ \bibinfo {pages}
  {3333}}\BibitemShut {NoStop}%
%%CITATION = PHRVA,D13,3333;%%
\bibitem [{\citenamefont {Giudice}\ \emph {et~al.}(2007)\citenamefont
  {Giudice}, \citenamefont {Grojean}, \citenamefont {Pomarol},\ and\
  \citenamefont {Rattazzi}}]{Giudice:2007fh}%
  \BibitemOpen
  \bibfield  {author} {\bibinfo {author} {\bibnamefont {Giudice} \bibfnamefont
  {GF}}, \bibinfo {author} {\bibfnamefont {C.}~\bibnamefont {Grojean}},
  \bibinfo {author} {\bibfnamefont {A.}~\bibnamefont {Pomarol}}, \ and\
  \bibinfo {author} {\bibfnamefont {R.}~\bibnamefont {Rattazzi}}} (\bibinfo
  {year} {2007}),\ \bibfield  {title} {\enquote {\bibinfo {title} {{The
  Strongly-Interacting Light Higgs}},}\ }\Doi {10.1088/1126-6708/2007/06/045}
  {\bibfield  {journal} {\bibinfo  {journal} {JHEP},\ }\textbf {\bibinfo
  {volume} {0706}},\ \bibinfo {pages} {045}},\ \Eprint
  {http://arxiv.org/abs/hep-ph/0703164} {arXiv:hep-ph/0703164 [hep-ph]}
  \BibitemShut {NoStop}%
%%CITATION = HEP-PH/0703164;%%
\bibitem [{\citenamefont {Giusti}\ and\ \citenamefont
  {Luscher}(2009)}]{Giusti:2008vb}%
  \BibitemOpen
  \bibfield  {author} {\bibinfo {author} {\bibnamefont {Giusti} \bibfnamefont
  {Leonardo}}, \ and\ \bibinfo {author} {\bibfnamefont {Martin}\ \bibnamefont
  {Luscher}}} (\bibinfo {year} {2009}),\ \bibfield  {title} {\enquote {\bibinfo
  {title} {{Chiral symmetry breaking and the Banks-Casher relation in lattice
  QCD with Wilson quarks}},}\ }\Doi {10.1088/1126-6708/2009/03/013} {\bibfield
  {journal} {\bibinfo  {journal} {JHEP},\ }\textbf {\bibinfo {volume} {0903}},\
  \bibinfo {pages} {013}},\ \Eprint {http://arxiv.org/abs/0812.3638}
  {arXiv:0812.3638 [hep-lat]} \BibitemShut {NoStop}%
%%CITATION = ARXIV:0812.3638;%%
\bibitem [{\citenamefont {Golterman}(2001)}]{Golterman:2000hr}%
  \BibitemOpen
  \bibfield  {author} {\bibinfo {author} {\bibnamefont {Golterman}
  \bibfnamefont {Maarten}}} (\bibinfo {year} {2001}),\ \bibfield  {title}
  {\enquote {\bibinfo {title} {{Lattice chiral gauge theories}},}\ }\Doi
  {10.1016/S0920-5632(01)00953-7} {\bibfield  {journal} {\bibinfo  {journal}
  {Nucl.Phys.Proc.Suppl.},\ }\textbf {\bibinfo {volume} {94}},\ \bibinfo
  {pages} {189--203}},\ \Eprint {http://arxiv.org/abs/hep-lat/0011027}
  {arXiv:hep-lat/0011027 [hep-lat]} \BibitemShut {NoStop}%
%%CITATION = HEP-LAT/0011027;%%
\bibitem [{\citenamefont {Golterman}\ and\ \citenamefont
  {Shamir}(2004)}]{Golterman:2004qv}%
  \BibitemOpen
  \bibfield  {author} {\bibinfo {author} {\bibnamefont {Golterman}
  \bibfnamefont {Maarten}}, \ and\ \bibinfo {author} {\bibfnamefont {Yigal}\
  \bibnamefont {Shamir}}} (\bibinfo {year} {2004}),\ \bibfield  {title}
  {\enquote {\bibinfo {title} {{SU(N) chiral gauge theories on the lattice}},}\
  }\Doi {10.1103/PhysRevD.70.094506} {\bibfield  {journal} {\bibinfo  {journal}
  {Phys.Rev.},\ }\textbf {\bibinfo {volume} {D70}},\ \bibinfo {pages}
  {094506}},\ \Eprint {http://arxiv.org/abs/hep-lat/0404011}
  {arXiv:hep-lat/0404011 [hep-lat]} \BibitemShut {NoStop}%
%%CITATION = HEP-LAT/0404011;%%
\bibitem [{\citenamefont {Golterman}\ and\ \citenamefont
  {Shamir}(2014a)}]{Golterman:2014yha}%
  \BibitemOpen
  \bibfield  {author} {\bibinfo {author} {\bibnamefont {Golterman}
  \bibfnamefont {Maarten}}, \ and\ \bibinfo {author} {\bibfnamefont {Yigal}\
  \bibnamefont {Shamir}}} (\bibinfo {year} {2014a}),\ \bibfield  {title}
  {\enquote {\bibinfo {title} {{Vacuum alignment and lattice artifacts: Wilson
  fermions}},}\ }\Doi {10.1103/PhysRevD.89.054501} {\bibfield  {journal}
  {\bibinfo  {journal} {Phys.~Rev.},\ }\textbf {\bibinfo {volume} {D89}},\
  \bibinfo {pages} {054501}},\ \Eprint {http://arxiv.org/abs/1401.0356}
  {arXiv:1401.0356 [hep-lat]} \BibitemShut {NoStop}%
%%CITATION = ARXIV:1401.0356;%%
\bibitem [{\citenamefont {Golterman}\ and\ \citenamefont
  {Shamir}(2014b)}]{Golterman:2014lea}%
  \BibitemOpen
  \bibfield  {author} {\bibinfo {author} {\bibnamefont {Golterman}
  \bibfnamefont {Maarten}}, \ and\ \bibinfo {author} {\bibfnamefont {Yigal}\
  \bibnamefont {Shamir}}} (\bibinfo {year} {2014b}),\ \bibfield  {title}
  {\enquote {\bibinfo {title} {{Vacuum alignment and lattice artifacts:
  staggered fermions}},}\ }\Doi {10.1103/PhysRevD.89.074502} {\bibfield
  {journal} {\bibinfo  {journal} {Phys.~Rev.},\ }\textbf {\bibinfo {volume}
  {D89}},\ \bibinfo {pages} {074502}},\ \Eprint
  {http://arxiv.org/abs/1401.3151} {arXiv:1401.3151 [hep-lat]} \BibitemShut
  {NoStop}%
%%CITATION = ARXIV:1401.3151;%%
\bibitem [{\citenamefont {Golterman}\ and\ \citenamefont
  {Shamir}(2015)}]{Golterman:2015zwa}%
  \BibitemOpen
  \bibfield  {author} {\bibinfo {author} {\bibnamefont {Golterman}
  \bibfnamefont {Maarten}}, \ and\ \bibinfo {author} {\bibfnamefont {Yigal}\
  \bibnamefont {Shamir}}} (\bibinfo {year} {2015}),\ \bibfield  {title}
  {\enquote {\bibinfo {title} {{Top quark induced effective potential in a
  composite Higgs model}},}\ }\Doi {10.1103/PhysRevD.91.094506} {\bibfield
  {journal} {\bibinfo  {journal} {Phys. Rev.},\ }\textbf {\bibinfo {volume}
  {D91}}~(\bibinfo {number} {9}),\ \bibinfo {pages} {094506}},\ \Eprint
  {http://arxiv.org/abs/1502.00390} {arXiv:1502.00390 [hep-ph]} \BibitemShut
  {NoStop}%
%%CITATION = ARXIV:1502.00390;%%
\bibitem [{\citenamefont {Greensite}\ and\ \citenamefont
  {Primack}(1981)}]{Greensite:1980hy}%
  \BibitemOpen
  \bibfield  {author} {\bibinfo {author} {\bibnamefont {Greensite}
  \bibfnamefont {Jeff}}, \ and\ \bibinfo {author} {\bibfnamefont {Joel}\
  \bibnamefont {Primack}}} (\bibinfo {year} {1981}),\ \bibfield  {title}
  {\enquote {\bibinfo {title} {{Pions as Spin Waves: Chiral Symmetry Breaking
  in Lattice Gauge Theory}},}\ }\Doi {10.1016/0550-3213(81)90160-7} {\bibfield
  {journal} {\bibinfo  {journal} {Nucl.~Phys.},\ }\textbf {\bibinfo {volume}
  {B180}},\ \bibinfo {pages} {170}}\BibitemShut {NoStop}%
%%CITATION = NUPHA,B180,170;%%
\bibitem [{\citenamefont {Halkiadakis}\ \emph {et~al.}(2014)\citenamefont
  {Halkiadakis}, \citenamefont {Redlinger},\ and\ \citenamefont
  {Shih}}]{Halkiadakis:2014qda}%
  \BibitemOpen
  \bibfield  {author} {\bibinfo {author} {\bibnamefont {Halkiadakis}
  \bibfnamefont {Eva}}, \bibinfo {author} {\bibfnamefont {George}\ \bibnamefont
  {Redlinger}}, \ and\ \bibinfo {author} {\bibfnamefont {David}\ \bibnamefont
  {Shih}}} (\bibinfo {year} {2014}),\ \bibfield  {title} {\enquote {\bibinfo
  {title} {{Status and Implications of Beyond-the-Standard-Model Searches at
  the LHC}},}\ }\Doi {10.1146/annurev-nucl-102313-025632} {\bibfield  {journal}
  {\bibinfo  {journal} {Ann.Rev.Nucl.Part.Sci.},\ }\textbf {\bibinfo {volume}
  {64}},\ \bibinfo {pages} {319--342}},\ \Eprint
  {http://arxiv.org/abs/1411.1427} {arXiv:1411.1427 [hep-ex]} \BibitemShut
  {NoStop}%
%%CITATION = ARXIV:1411.1427;%%
\bibitem [{\citenamefont {Han}\ and\ \citenamefont {Skiba}(2005)}]{Han:2004az}%
  \BibitemOpen
  \bibfield  {author} {\bibinfo {author} {\bibnamefont {Han} \bibfnamefont
  {Zhenyu}}, \ and\ \bibinfo {author} {\bibfnamefont {Witold}\ \bibnamefont
  {Skiba}}} (\bibinfo {year} {2005}),\ \bibfield  {title} {\enquote {\bibinfo
  {title} {{Effective theory analysis of precision electroweak data}},}\ }\Doi
  {10.1103/PhysRevD.71.075009} {\bibfield  {journal} {\bibinfo  {journal}
  {Phys.~Rev.},\ }\textbf {\bibinfo {volume} {D71}},\ \bibinfo {pages}
  {075009}},\ \Eprint {http://arxiv.org/abs/hep-ph/0412166}
  {arXiv:hep-ph/0412166 [hep-ph]} \BibitemShut {NoStop}%
%%CITATION = HEP-PH/0412166;%%
\bibitem [{\citenamefont {Hanada}\ \emph {et~al.}(2014)\citenamefont {Hanada},
  \citenamefont {Hyakutake}, \citenamefont {Ishiki},\ and\ \citenamefont
  {Nishimura}}]{Hanada:2013rga}%
  \BibitemOpen
  \bibfield  {author} {\bibinfo {author} {\bibnamefont {Hanada} \bibfnamefont
  {Masanori}}, \bibinfo {author} {\bibfnamefont {Yoshifumi}\ \bibnamefont
  {Hyakutake}}, \bibinfo {author} {\bibfnamefont {Goro}\ \bibnamefont
  {Ishiki}}, \ and\ \bibinfo {author} {\bibfnamefont {Jun}\ \bibnamefont
  {Nishimura}}} (\bibinfo {year} {2014}),\ \bibfield  {title} {\enquote
  {\bibinfo {title} {{Holographic description of quantum black hole on a
  computer}},}\ }\Doi {10.1126/science.1250122} {\bibfield  {journal} {\bibinfo
   {journal} {Science},\ }\textbf {\bibinfo {volume} {344}},\ \bibinfo {pages}
  {882--885}},\ \Eprint {http://arxiv.org/abs/1311.5607} {arXiv:1311.5607
  [hep-th]} \BibitemShut {NoStop}%
%%CITATION = ARXIV:1311.5607;%%
\bibitem [{\citenamefont {Hasenfratz}\ \emph {et~al.}(1984a)\citenamefont
  {Hasenfratz}, \citenamefont {Hasenfratz}, \citenamefont {Heller},\ and\
  \citenamefont {Karsch}}]{Hasenfratz:1984bx}%
  \BibitemOpen
  \bibfield  {author} {\bibinfo {author} {\bibnamefont {Hasenfratz}
  \bibfnamefont {A}}, \bibinfo {author} {\bibfnamefont {P.}~\bibnamefont
  {Hasenfratz}}, \bibinfo {author} {\bibfnamefont {Urs~M.}\ \bibnamefont
  {Heller}}, \ and\ \bibinfo {author} {\bibfnamefont {F.}~\bibnamefont
  {Karsch}}} (\bibinfo {year} {1984a}),\ \bibfield  {title} {\enquote {\bibinfo
  {title} {{The Beta Function of the SU(3) Wilson Action}},}\ }\Doi
  {10.1016/0370-2693(84)90833-5} {\bibfield  {journal} {\bibinfo  {journal}
  {Phys.~Lett.},\ }\textbf {\bibinfo {volume} {B143}},\ \bibinfo {pages}
  {193}}\BibitemShut {NoStop}%
%%CITATION = PHLTA,B143,193;%%
\bibitem [{\citenamefont {Hasenfratz}\ \emph {et~al.}(1984b)\citenamefont
  {Hasenfratz}, \citenamefont {Hasenfratz}, \citenamefont {Heller},\ and\
  \citenamefont {Karsch}}]{Hasenfratz:1984hx}%
  \BibitemOpen
  \bibfield  {author} {\bibinfo {author} {\bibnamefont {Hasenfratz}
  \bibfnamefont {A}}, \bibinfo {author} {\bibfnamefont {P.}~\bibnamefont
  {Hasenfratz}}, \bibinfo {author} {\bibfnamefont {Urs~M.}\ \bibnamefont
  {Heller}}, \ and\ \bibinfo {author} {\bibfnamefont {F.}~\bibnamefont
  {Karsch}}} (\bibinfo {year} {1984b}),\ \bibfield  {title} {\enquote {\bibinfo
  {title} {{Improved Monte Carlo Renormalization Group Methods}},}\ }\Doi
  {10.1016/0370-2693(84)91051-7} {\bibfield  {journal} {\bibinfo  {journal}
  {Phys.~Lett.},\ }\textbf {\bibinfo {volume} {B140}},\ \bibinfo {pages}
  {76}}\BibitemShut {NoStop}%
%%CITATION = PHLTA,B140,76;%%
\bibitem [{\citenamefont {Hasenfratz}(2009)}]{Hasenfratz:2009ea}%
  \BibitemOpen
  \bibfield  {author} {\bibinfo {author} {\bibnamefont {Hasenfratz}
  \bibfnamefont {Anna}}} (\bibinfo {year} {2009}),\ \bibfield  {title}
  {\enquote {\bibinfo {title} {{Investigating the critical properties of
  beyond-QCD theories using Monte Carlo Renormalization Group matching}},}\
  }\Doi {10.1103/PhysRevD.80.034505} {\bibfield  {journal} {\bibinfo  {journal}
  {Phys.~Rev.},\ }\textbf {\bibinfo {volume} {D80}},\ \bibinfo {pages}
  {034505}},\ \Eprint {http://arxiv.org/abs/0907.0919} {arXiv:0907.0919
  [hep-lat]} \BibitemShut {NoStop}%
%%CITATION = ARXIV:0907.0919;%%
\bibitem [{\citenamefont {Hasenfratz}(2010)}]{Hasenfratz:2010fi}%
  \BibitemOpen
  \bibfield  {author} {\bibinfo {author} {\bibnamefont {Hasenfratz}
  \bibfnamefont {Anna}}} (\bibinfo {year} {2010}),\ \bibfield  {title}
  {\enquote {\bibinfo {title} {{Conformal or Walking? Monte Carlo
  renormalization group studies of SU(3) gauge models with fundamental
  fermions}},}\ }\Doi {10.1103/PhysRevD.82.014506} {\bibfield  {journal}
  {\bibinfo  {journal} {Phys.~Rev.},\ }\textbf {\bibinfo {volume} {D82}},\
  \bibinfo {pages} {014506}},\ \Eprint {http://arxiv.org/abs/1004.1004}
  {arXiv:1004.1004 [hep-lat]} \BibitemShut {NoStop}%
%%CITATION = ARXIV:1004.1004;%%
\bibitem [{\citenamefont {Hasenfratz}(2012)}]{Hasenfratz:2011xn}%
  \BibitemOpen
  \bibfield  {author} {\bibinfo {author} {\bibnamefont {Hasenfratz}
  \bibfnamefont {Anna}}} (\bibinfo {year} {2012}),\ \bibfield  {title}
  {\enquote {\bibinfo {title} {{Infrared fixed point of the 12-fermion SU(3)
  gauge model based on 2-lattice MCRG matching}},}\ }\Doi
  {10.1103/PhysRevLett.108.061601} {\bibfield  {journal} {\bibinfo  {journal}
  {Phys.~Rev.~Lett.},\ }\textbf {\bibinfo {volume} {108}},\ \bibinfo {pages}
  {061601}},\ \Eprint {http://arxiv.org/abs/1106.5293} {arXiv:1106.5293
  [hep-lat]} \BibitemShut {NoStop}%
%%CITATION = ARXIV:1106.5293;%%
\bibitem [{\citenamefont {Hasenfratz}\ \emph {et~al.}(2015)\citenamefont
  {Hasenfratz}, \citenamefont {Schaich},\ and\ \citenamefont
  {Veernala}}]{Hasenfratz:2014rna}%
  \BibitemOpen
  \bibfield  {author} {\bibinfo {author} {\bibnamefont {Hasenfratz}
  \bibfnamefont {Anna}}, \bibinfo {author} {\bibfnamefont {David}\ \bibnamefont
  {Schaich}}, \ and\ \bibinfo {author} {\bibfnamefont {Aarti}\ \bibnamefont
  {Veernala}}} (\bibinfo {year} {2015}),\ \bibfield  {title} {\enquote
  {\bibinfo {title} {{Nonperturbative $\beta$ function of eight-flavor SU(3)
  gauge theory}},}\ }\Doi {10.1007/JHEP06(2015)143} {\bibfield  {journal}
  {\bibinfo  {journal} {JHEP},\ }\textbf {\bibinfo {volume} {06}},\ \bibinfo
  {pages} {143}},\ \Eprint {http://arxiv.org/abs/1410.5886} {arXiv:1410.5886
  [hep-lat]} \BibitemShut {NoStop}%
%%CITATION = ARXIV:1410.5886;%%
\bibitem [{\citenamefont {Hayakawa}\ \emph {et~al.}(2011)\citenamefont
  {Hayakawa}, \citenamefont {Ishikawa}, \citenamefont {Osaki}, \citenamefont
  {Takeda}, \citenamefont {Uno} \emph {et~al.}}]{Hayakawa:2010yn}%
  \BibitemOpen
  \bibfield  {author} {\bibinfo {author} {\bibnamefont {Hayakawa} \bibfnamefont
  {M}}, \bibinfo {author} {\bibfnamefont {K.-I.}\ \bibnamefont {Ishikawa}},
  \bibinfo {author} {\bibfnamefont {Y.}~\bibnamefont {Osaki}}, \bibinfo
  {author} {\bibfnamefont {S.}~\bibnamefont {Takeda}}, \bibinfo {author}
  {\bibfnamefont {S.}~\bibnamefont {Uno}},  \emph {et~al.}} (\bibinfo {year}
  {2011}),\ \bibfield  {title} {\enquote {\bibinfo {title} {{Running coupling
  constant of ten-flavor QCD with the Schr\'odinger functional method}},}\
  }\Doi {10.1103/PhysRevD.83.074509} {\bibfield  {journal} {\bibinfo  {journal}
  {Phys.~Rev.},\ }\textbf {\bibinfo {volume} {D83}},\ \bibinfo {pages}
  {074509}},\ \Eprint {http://arxiv.org/abs/1011.2577} {arXiv:1011.2577
  [hep-lat]} \BibitemShut {NoStop}%
%%CITATION = ARXIV:1011.2577;%%
\bibitem [{\citenamefont {Hayakawa}\ \emph {et~al.}(2013)\citenamefont
  {Hayakawa}, \citenamefont {Ishikawa}, \citenamefont {Takeda},\ and\
  \citenamefont {Yamada}}]{Hayakawa:2013yfa}%
  \BibitemOpen
  \bibfield  {author} {\bibinfo {author} {\bibnamefont {Hayakawa} \bibfnamefont
  {M}}, \bibinfo {author} {\bibfnamefont {K.-I.}\ \bibnamefont {Ishikawa}},
  \bibinfo {author} {\bibfnamefont {S.}~\bibnamefont {Takeda}}, \ and\ \bibinfo
  {author} {\bibfnamefont {N.}~\bibnamefont {Yamada}}} (\bibinfo {year}
  {2013}),\ \bibfield  {title} {\enquote {\bibinfo {title} {{Running coupling
  constant and mass anomalous dimension of six-flavor SU(2) gauge theory}},}\
  }\Doi {10.1103/PhysRevD.88.094504} {\bibfield  {journal} {\bibinfo  {journal}
  {Phys.~Rev.},\ }\textbf {\bibinfo {volume} {D88}}~(\bibinfo {number} {9}),\
  \bibinfo {pages} {094504}},\ \Eprint {http://arxiv.org/abs/1307.6997}
  {arXiv:1307.6997 [hep-lat]} \BibitemShut {NoStop}%
%%CITATION = ARXIV:1307.6997;%%
\bibitem [{\citenamefont {Heller}(1998)}]{Heller:1997vh}%
  \BibitemOpen
  \bibfield  {author} {\bibinfo {author} {\bibnamefont {Heller} \bibfnamefont
  {Urs~M}}} (\bibinfo {year} {1998}),\ \bibfield  {title} {\enquote {\bibinfo
  {title} {{The Schrodinger functional running coupling with staggered fermions
  and its application to many flavor QCD}},}\ }\Doi
  {10.1016/S0920-5632(97)00735-4} {\bibfield  {journal} {\bibinfo  {journal}
  {Nucl.~Phys.~Proc.~Suppl.},\ }\textbf {\bibinfo {volume} {63}},\ \bibinfo
  {pages} {248--250}},\ \Eprint {http://arxiv.org/abs/hep-lat/9709159}
  {arXiv:hep-lat/9709159 [hep-lat]} \BibitemShut {NoStop}%
%%CITATION = HEP-LAT/9709159;%%
\bibitem [{\citenamefont {Hietanen}\ \emph {et~al.}(2014a)\citenamefont
  {Hietanen}, \citenamefont {Lewis}, \citenamefont {Pica},\ and\ \citenamefont
  {Sannino}}]{Hietanen:2014xca}%
  \BibitemOpen
  \bibfield  {author} {\bibinfo {author} {\bibnamefont {Hietanen} \bibfnamefont
  {Ari}}, \bibinfo {author} {\bibfnamefont {Randy}\ \bibnamefont {Lewis}},
  \bibinfo {author} {\bibfnamefont {Claudio}\ \bibnamefont {Pica}}, \ and\
  \bibinfo {author} {\bibfnamefont {Francesco}\ \bibnamefont {Sannino}}}
  (\bibinfo {year} {2014a}),\ \bibfield  {title} {\enquote {\bibinfo {title}
  {{Fundamental Composite Higgs Dynamics on the Lattice: SU(2) with Two
  Flavors}},}\ }\Doi {10.1007/JHEP07(2014)116} {\bibfield  {journal} {\bibinfo
  {journal} {JHEP},\ }\textbf {\bibinfo {volume} {1407}},\ \bibinfo {pages}
  {116}},\ \Eprint {http://arxiv.org/abs/1404.2794} {arXiv:1404.2794 [hep-lat]}
  \BibitemShut {NoStop}%
%%CITATION = ARXIV:1404.2794;%%
\bibitem [{\citenamefont {Hietanen}\ \emph {et~al.}(2014b)\citenamefont
  {Hietanen}, \citenamefont {Lewis}, \citenamefont {Pica},\ and\ \citenamefont
  {Sannino}}]{Hietanen:2013fya}%
  \BibitemOpen
  \bibfield  {author} {\bibinfo {author} {\bibnamefont {Hietanen} \bibfnamefont
  {Ari}}, \bibinfo {author} {\bibfnamefont {Randy}\ \bibnamefont {Lewis}},
  \bibinfo {author} {\bibfnamefont {Claudio}\ \bibnamefont {Pica}}, \ and\
  \bibinfo {author} {\bibfnamefont {Francesco}\ \bibnamefont {Sannino}}}
  (\bibinfo {year} {2014b}),\ \bibfield  {title} {\enquote {\bibinfo {title}
  {{Composite Goldstone Dark Matter: Experimental Predictions from the
  Lattice}},}\ }\Doi {10.1007/JHEP12(2014)130} {\bibfield  {journal} {\bibinfo
  {journal} {JHEP},\ }\textbf {\bibinfo {volume} {1412}},\ \bibinfo {pages}
  {130}},\ \Eprint {http://arxiv.org/abs/1308.4130} {arXiv:1308.4130 [hep-ph]}
  \BibitemShut {NoStop}%
%%CITATION = ARXIV:1308.4130;%%
\bibitem [{\citenamefont {Hietanen}\ \emph {et~al.}(2009a)\citenamefont
  {Hietanen}, \citenamefont {Rantaharju}, \citenamefont {Rummukainen},\ and\
  \citenamefont {Tuominen}}]{Hietanen:2008mr}%
  \BibitemOpen
  \bibfield  {author} {\bibinfo {author} {\bibnamefont {Hietanen} \bibfnamefont
  {Ari~J}}, \bibinfo {author} {\bibfnamefont {Jarno}\ \bibnamefont
  {Rantaharju}}, \bibinfo {author} {\bibfnamefont {Kari}\ \bibnamefont
  {Rummukainen}}, \ and\ \bibinfo {author} {\bibfnamefont {Kimmo}\ \bibnamefont
  {Tuominen}}} (\bibinfo {year} {2009a}),\ \bibfield  {title} {\enquote
  {\bibinfo {title} {{Spectrum of SU(2) lattice gauge theory with two adjoint
  Dirac flavours}},}\ }\Doi {10.1088/1126-6708/2009/05/025} {\bibfield
  {journal} {\bibinfo  {journal} {JHEP},\ }\textbf {\bibinfo {volume} {0905}},\
  \bibinfo {pages} {025}},\ \Eprint {http://arxiv.org/abs/0812.1467}
  {arXiv:0812.1467 [hep-lat]} \BibitemShut {NoStop}%
%%CITATION = ARXIV:0812.1467;%%
\bibitem [{\citenamefont {Hietanen}\ \emph {et~al.}(2009b)\citenamefont
  {Hietanen}, \citenamefont {Rummukainen},\ and\ \citenamefont
  {Tuominen}}]{Hietanen:2009az}%
  \BibitemOpen
  \bibfield  {author} {\bibinfo {author} {\bibnamefont {Hietanen} \bibfnamefont
  {Ari~J}}, \bibinfo {author} {\bibfnamefont {Kari}\ \bibnamefont
  {Rummukainen}}, \ and\ \bibinfo {author} {\bibfnamefont {Kimmo}\ \bibnamefont
  {Tuominen}}} (\bibinfo {year} {2009b}),\ \bibfield  {title} {\enquote
  {\bibinfo {title} {{Evolution of the coupling constant in SU(2) lattice gauge
  theory with two adjoint fermions}},}\ }\Doi {10.1103/PhysRevD.80.094504}
  {\bibfield  {journal} {\bibinfo  {journal} {Phys.~Rev.},\ }\textbf {\bibinfo
  {volume} {D80}},\ \bibinfo {pages} {094504}},\ \Eprint
  {http://arxiv.org/abs/0904.0864} {arXiv:0904.0864 [hep-lat]} \BibitemShut
  {NoStop}%
%%CITATION = ARXIV:0904.0864;%%
\bibitem [{\citenamefont {Hill}\ and\ \citenamefont
  {Simmons}(2003)}]{Hill:2002ap}%
  \BibitemOpen
  \bibfield  {author} {\bibinfo {author} {\bibnamefont {Hill} \bibfnamefont
  {Christopher~T}}, \ and\ \bibinfo {author} {\bibfnamefont {Elizabeth~H.}\
  \bibnamefont {Simmons}}} (\bibinfo {year} {2003}),\ \bibfield  {title}
  {\enquote {\bibinfo {title} {{Strong dynamics and electroweak symmetry
  breaking}},}\ }\Doi {10.1016/S0370-1573(03)00140-6} {\bibfield  {journal}
  {\bibinfo  {journal} {Phys.~Rept.},\ }\textbf {\bibinfo {volume} {381}},\
  \bibinfo {pages} {235--402}},\ \Eprint {http://arxiv.org/abs/hep-ph/0203079}
  {arXiv:hep-ph/0203079 [hep-ph]} \BibitemShut {NoStop}%
%%CITATION = HEP-PH/0203079;%%
\bibitem [{\citenamefont {Holdom}(1981)}]{Holdom:1981rm}%
  \BibitemOpen
  \bibfield  {author} {\bibinfo {author} {\bibnamefont {Holdom} \bibfnamefont
  {Bob}}} (\bibinfo {year} {1981}),\ \bibfield  {title} {\enquote {\bibinfo
  {title} {{Raising the Sideways Scale}},}\ }\Doi {10.1103/PhysRevD.24.1441}
  {\bibfield  {journal} {\bibinfo  {journal} {Phys.~Rev.},\ }\textbf {\bibinfo
  {volume} {D24}},\ \bibinfo {pages} {1441}}\BibitemShut {NoStop}%
%%CITATION = PHRVA,D24,1441;%%
\bibitem [{\citenamefont {Holdom}(1985)}]{Holdom:1984sk}%
  \BibitemOpen
  \bibfield  {author} {\bibinfo {author} {\bibnamefont {Holdom} \bibfnamefont
  {Bob}}} (\bibinfo {year} {1985}),\ \bibfield  {title} {\enquote {\bibinfo
  {title} {{Techniodor}},}\ }\Doi {10.1016/0370-2693(85)91015-9} {\bibfield
  {journal} {\bibinfo  {journal} {Phys.~Lett.},\ }\textbf {\bibinfo {volume}
  {B150}},\ \bibinfo {pages} {301}}\BibitemShut {NoStop}%
%%CITATION = PHLTA,B150,301;%%
\bibitem [{\citenamefont {Honda}\ \emph {et~al.}(2013)\citenamefont {Honda},
  \citenamefont {Ishiki}, \citenamefont {Kim}, \citenamefont {Nishimura},\ and\
  \citenamefont {Tsuchiya}}]{Honda:2013nfa}%
  \BibitemOpen
  \bibfield  {author} {\bibinfo {author} {\bibnamefont {Honda} \bibfnamefont
  {Masazumi}}, \bibinfo {author} {\bibfnamefont {Goro}\ \bibnamefont {Ishiki}},
  \bibinfo {author} {\bibfnamefont {Sang-Woo}\ \bibnamefont {Kim}}, \bibinfo
  {author} {\bibfnamefont {Jun}\ \bibnamefont {Nishimura}}, \ and\ \bibinfo
  {author} {\bibfnamefont {Asato}\ \bibnamefont {Tsuchiya}}} (\bibinfo {year}
  {2013}),\ \bibfield  {title} {\enquote {\bibinfo {title} {{Direct test of the
  AdS/CFT correspondence by Monte Carlo studies of N=4 super Yang-Mills
  theory}},}\ }\Doi {10.1007/JHEP11(2013)200} {\bibfield  {journal} {\bibinfo
  {journal} {JHEP},\ }\textbf {\bibinfo {volume} {1311}},\ \bibinfo {pages}
  {200}},\ \Eprint {http://arxiv.org/abs/1308.3525} {arXiv:1308.3525}
  \BibitemShut {NoStop}%
\bibitem [{\citenamefont {Honda}\ \emph {et~al.}(2011)\citenamefont {Honda},
  \citenamefont {Ishiki}, \citenamefont {Nishimura},\ and\ \citenamefont
  {Tsuchiya}}]{Honda:2011qk}%
  \BibitemOpen
  \bibfield  {author} {\bibinfo {author} {\bibnamefont {Honda} \bibfnamefont
  {Masazumi}}, \bibinfo {author} {\bibfnamefont {Goro}\ \bibnamefont {Ishiki}},
  \bibinfo {author} {\bibfnamefont {Jun}\ \bibnamefont {Nishimura}}, \ and\
  \bibinfo {author} {\bibfnamefont {Asato}\ \bibnamefont {Tsuchiya}}} (\bibinfo
  {year} {2011}),\ \bibfield  {title} {\enquote {\bibinfo {title} {{Testing the
  AdS/CFT correspondence by Monte Carlo calculation of BPS and non-BPS Wilson
  loops in 4d N=4 super-Yang-Mills theory}},}\ }\href@noop {} {\bibfield
  {journal} {\bibinfo  {journal} {PoS},\ }\textbf {\bibinfo {volume}
  {LATTICE2011}},\ \bibinfo {pages} {244}},\ \Eprint
  {http://arxiv.org/abs/1112.4274} {arXiv:1112.4274 [hep-lat]} \BibitemShut
  {NoStop}%
%%CITATION = ARXIV:1112.4274;%%
\bibitem [{\citenamefont {'t~Hooft}(1974)}]{'tHooft:1973jz}%
  \BibitemOpen
  \bibfield  {author} {\bibinfo {author} {\bibnamefont {'t~Hooft} \bibfnamefont
  {Gerard}}} (\bibinfo {year} {1974}),\ \bibfield  {title} {\enquote {\bibinfo
  {title} {{A Planar Diagram Theory for Strong Interactions}},}\ }\Doi
  {10.1016/0550-3213(74)90154-0} {\bibfield  {journal} {\bibinfo  {journal}
  {Nucl.Phys.},\ }\textbf {\bibinfo {volume} {B72}},\ \bibinfo {pages}
  {461}}\BibitemShut {NoStop}%
%%CITATION = NUPHA,B72,461;%%
\bibitem [{\citenamefont {Irges}\ and\ \citenamefont
  {Knechtli}(2007)}]{Irges:2006hg}%
  \BibitemOpen
  \bibfield  {author} {\bibinfo {author} {\bibnamefont {Irges} \bibfnamefont
  {Nikos}}, \ and\ \bibinfo {author} {\bibfnamefont {Francesco}\ \bibnamefont
  {Knechtli}}} (\bibinfo {year} {2007}),\ \bibfield  {title} {\enquote
  {\bibinfo {title} {{Lattice gauge theory approach to spontaneous symmetry
  breaking from an extra dimension}},}\ }\Doi {10.1016/j.nuclphysb.2007.01.023}
  {\bibfield  {journal} {\bibinfo  {journal} {Nucl.~Phys.},\ }\textbf {\bibinfo
  {volume} {B775}},\ \bibinfo {pages} {283--311}},\ \Eprint
  {http://arxiv.org/abs/hep-lat/0609045} {arXiv:hep-lat/0609045 [hep-lat]}
  \BibitemShut {NoStop}%
%%CITATION = HEP-LAT/0609045;%%
\bibitem [{\citenamefont {Irges}\ and\ \citenamefont
  {Knechtli}(2014)}]{Irges:2013rya}%
  \BibitemOpen
  \bibfield  {author} {\bibinfo {author} {\bibnamefont {Irges} \bibfnamefont
  {Nikos}}, \ and\ \bibinfo {author} {\bibfnamefont {Francesco}\ \bibnamefont
  {Knechtli}}} (\bibinfo {year} {2014}),\ \bibfield  {title} {\enquote
  {\bibinfo {title} {{Non-perturbative Gauge-Higgs Unification: Symmetries and
  Order Parameters}},}\ }\Doi {10.1007/JHEP06(2014)070} {\bibfield  {journal}
  {\bibinfo  {journal} {JHEP},\ }\textbf {\bibinfo {volume} {1406}},\ \bibinfo
  {pages} {070}},\ \Eprint {http://arxiv.org/abs/1312.3142} {arXiv:1312.3142
  [hep-lat]} \BibitemShut {NoStop}%
%%CITATION = ARXIV:1312.3142;%%
\bibitem [{\citenamefont {Irges}\ \emph {et~al.}(2013)\citenamefont {Irges},
  \citenamefont {Knechtli},\ and\ \citenamefont {Yoneyama}}]{Irges:2012mp}%
  \BibitemOpen
  \bibfield  {author} {\bibinfo {author} {\bibnamefont {Irges} \bibfnamefont
  {Nikos}}, \bibinfo {author} {\bibfnamefont {Francesco}\ \bibnamefont
  {Knechtli}}, \ and\ \bibinfo {author} {\bibfnamefont {Kyoko}\ \bibnamefont
  {Yoneyama}}} (\bibinfo {year} {2013}),\ \bibfield  {title} {\enquote
  {\bibinfo {title} {{Higgs mechanism near the 5d bulk phase transition}},}\
  }\Doi {10.1016/j.physletb.2013.04.032} {\bibfield  {journal} {\bibinfo
  {journal} {Phys.~Lett.},\ }\textbf {\bibinfo {volume} {B722}},\ \bibinfo
  {pages} {378--383}},\ \Eprint {http://arxiv.org/abs/1212.5514}
  {arXiv:1212.5514} \BibitemShut {NoStop}%
%%CITATION = ARXIV:1212.5514;%%
\bibitem [{\citenamefont {Ishii}\ \emph {et~al.}(2008)\citenamefont {Ishii},
  \citenamefont {Ishiki}, \citenamefont {Shimasaki},\ and\ \citenamefont
  {Tsuchiya}}]{Ishii:2008ib}%
  \BibitemOpen
  \bibfield  {author} {\bibinfo {author} {\bibnamefont {Ishii} \bibfnamefont
  {Takaaki}}, \bibinfo {author} {\bibfnamefont {Goro}\ \bibnamefont {Ishiki}},
  \bibinfo {author} {\bibfnamefont {Shinji}\ \bibnamefont {Shimasaki}}, \ and\
  \bibinfo {author} {\bibfnamefont {Asato}\ \bibnamefont {Tsuchiya}}} (\bibinfo
  {year} {2008}),\ \bibfield  {title} {\enquote {\bibinfo {title} {{N=4 Super
  Yang-Mills from the Plane Wave Matrix Model}},}\ }\Doi
  {10.1103/PhysRevD.78.106001} {\bibfield  {journal} {\bibinfo  {journal}
  {Phys.~Rev.},\ }\textbf {\bibinfo {volume} {D78}},\ \bibinfo {pages}
  {106001}},\ \Eprint {http://arxiv.org/abs/0807.2352} {arXiv:0807.2352
  [hep-th]} \BibitemShut {NoStop}%
%%CITATION = ARXIV:0807.2352;%%
\bibitem [{\citenamefont {Ishiki}\ \emph {et~al.}(2009a)\citenamefont {Ishiki},
  \citenamefont {Kim}, \citenamefont {Nishimura},\ and\ \citenamefont
  {Tsuchiya}}]{Ishiki:2009sg}%
  \BibitemOpen
  \bibfield  {author} {\bibinfo {author} {\bibnamefont {Ishiki} \bibfnamefont
  {Goro}}, \bibinfo {author} {\bibfnamefont {Sang-Woo}\ \bibnamefont {Kim}},
  \bibinfo {author} {\bibfnamefont {Jun}\ \bibnamefont {Nishimura}}, \ and\
  \bibinfo {author} {\bibfnamefont {Asato}\ \bibnamefont {Tsuchiya}}} (\bibinfo
  {year} {2009a}),\ \bibfield  {title} {\enquote {\bibinfo {title} {{Testing a
  novel large-N reduction for N=4 super Yang-Mills theory on R x S**3}},}\
  }\Doi {10.1088/1126-6708/2009/09/029} {\bibfield  {journal} {\bibinfo
  {journal} {JHEP},\ }\textbf {\bibinfo {volume} {0909}},\ \bibinfo {pages}
  {029}},\ \Eprint {http://arxiv.org/abs/0907.1488} {arXiv:0907.1488 [hep-th]}
  \BibitemShut {NoStop}%
%%CITATION = ARXIV:0907.1488;%%
\bibitem [{\citenamefont {Ishiki}\ \emph {et~al.}(2009b)\citenamefont {Ishiki},
  \citenamefont {Kim}, \citenamefont {Nishimura},\ and\ \citenamefont
  {Tsuchiya}}]{Ishiki:2008te}%
  \BibitemOpen
  \bibfield  {author} {\bibinfo {author} {\bibnamefont {Ishiki} \bibfnamefont
  {Goro}}, \bibinfo {author} {\bibfnamefont {Sang-Woo}\ \bibnamefont {Kim}},
  \bibinfo {author} {\bibfnamefont {Jun}\ \bibnamefont {Nishimura}}, \ and\
  \bibinfo {author} {\bibfnamefont {Asato}\ \bibnamefont {Tsuchiya}}} (\bibinfo
  {year} {2009b}),\ \bibfield  {title} {\enquote {\bibinfo {title}
  {{Deconfinement phase transition in N=4 super Yang-Mills theory on R x S**3
  from supersymmetric matrix quantum mechanics}},}\ }\Doi
  {10.1103/PhysRevLett.102.111601} {\bibfield  {journal} {\bibinfo  {journal}
  {Phys.~Rev.~Lett.},\ }\textbf {\bibinfo {volume} {102}},\ \bibinfo {pages}
  {111601}},\ \Eprint {http://arxiv.org/abs/0810.2884} {arXiv:0810.2884
  [hep-th]} \BibitemShut {NoStop}%
%%CITATION = ARXIV:0810.2884;%%
\bibitem [{\citenamefont {Iwasaki}\ \emph {et~al.}(2004)\citenamefont
  {Iwasaki}, \citenamefont {Kanaya}, \citenamefont {Kaya}, \citenamefont
  {Sakai},\ and\ \citenamefont {Yoshie}}]{Iwasaki:2003de}%
  \BibitemOpen
  \bibfield  {author} {\bibinfo {author} {\bibnamefont {Iwasaki} \bibfnamefont
  {Y}}, \bibinfo {author} {\bibfnamefont {K.}~\bibnamefont {Kanaya}}, \bibinfo
  {author} {\bibfnamefont {S.}~\bibnamefont {Kaya}}, \bibinfo {author}
  {\bibfnamefont {S.}~\bibnamefont {Sakai}}, \ and\ \bibinfo {author}
  {\bibfnamefont {T.}~\bibnamefont {Yoshie}}} (\bibinfo {year} {2004}),\
  \bibfield  {title} {\enquote {\bibinfo {title} {{Phase structure of lattice
  QCD for general number of flavors}},}\ }\Doi {10.1103/PhysRevD.69.014507}
  {\bibfield  {journal} {\bibinfo  {journal} {Phys.~Rev.},\ }\textbf {\bibinfo
  {volume} {D69}},\ \bibinfo {pages} {014507}},\ \Eprint
  {http://arxiv.org/abs/hep-lat/0309159} {arXiv:hep-lat/0309159 [hep-lat]}
  \BibitemShut {NoStop}%
%%CITATION = HEP-LAT/0309159;%%
\bibitem [{\citenamefont {Iwasaki}\ \emph {et~al.}(1992)\citenamefont
  {Iwasaki}, \citenamefont {Kanaya}, \citenamefont {Sakai},\ and\ \citenamefont
  {Yoshie}}]{Iwasaki:1991mr}%
  \BibitemOpen
  \bibfield  {author} {\bibinfo {author} {\bibnamefont {Iwasaki} \bibfnamefont
  {Y}}, \bibinfo {author} {\bibfnamefont {K.}~\bibnamefont {Kanaya}}, \bibinfo
  {author} {\bibfnamefont {S.}~\bibnamefont {Sakai}}, \ and\ \bibinfo {author}
  {\bibfnamefont {T.}~\bibnamefont {Yoshie}}} (\bibinfo {year} {1992}),\
  \bibfield  {title} {\enquote {\bibinfo {title} {{Quark confinement and number
  of flavors in strong coupling lattice QCD}},}\ }\Doi
  {10.1103/PhysRevLett.69.21} {\bibfield  {journal} {\bibinfo  {journal}
  {Phys.~Rev.~Lett.},\ }\textbf {\bibinfo {volume} {69}},\ \bibinfo {pages}
  {21--24}}\BibitemShut {NoStop}%
%%CITATION = PRLTA,69,21;%%
\bibitem [{\citenamefont {Jansen}\ and\ \citenamefont
  {Sommer}(1998)}]{Jansen:1998mx}%
  \BibitemOpen
  \bibfield  {author} {\bibinfo {author} {\bibnamefont {Jansen} \bibfnamefont
  {Karl}}, \ and\ \bibinfo {author} {\bibfnamefont {Rainer}\ \bibnamefont
  {Sommer}} (\bibinfo {collaboration} {ALPHA collaboration})} (\bibinfo {year}
  {1998}),\ \bibfield  {title} {\enquote {\bibinfo {title} {{O(alpha)
  improvement of lattice QCD with two flavors of Wilson quarks}},}\ }\Doi
  {10.1016/S0550-3213(98)00396-4} {\bibfield  {journal} {\bibinfo  {journal}
  {Nucl.~Phys.},\ }\textbf {\bibinfo {volume} {B530}},\ \bibinfo {pages}
  {185--203}},\ \Eprint {http://arxiv.org/abs/hep-lat/9803017}
  {arXiv:hep-lat/9803017 [hep-lat]} \BibitemShut {NoStop}%
%%CITATION = HEP-LAT/9803017;%%
\bibitem [{\citenamefont {Jin}\ and\ \citenamefont
  {Mawhinney}(2013)}]{Jin:2013hpa}%
  \BibitemOpen
  \bibfield  {author} {\bibinfo {author} {\bibnamefont {Jin} \bibfnamefont
  {Xiao-Yong}}, \ and\ \bibinfo {author} {\bibfnamefont {Robert~D.}\
  \bibnamefont {Mawhinney}}} (\bibinfo {year} {2013}),\ \bibfield  {title}
  {\enquote {\bibinfo {title} {{Lattice QCD with 12 Quark Flavors: A Careful
  Scrutiny}},}\ }\href@noop {} {}\Eprint {http://arxiv.org/abs/1304.0312}
  {arXiv:1304.0312 [hep-lat]} \BibitemShut {NoStop}%
%%CITATION = ARXIV:1304.0312;%%
\bibitem [{\citenamefont {Jones}(1974)}]{Jones:1974mm}%
  \BibitemOpen
  \bibfield  {author} {\bibinfo {author} {\bibnamefont {Jones} \bibfnamefont
  {DRT}}} (\bibinfo {year} {1974}),\ \bibfield  {title} {\enquote {\bibinfo
  {title} {{Two Loop Diagrams in Yang-Mills Theory}},}\ }\Doi
  {10.1016/0550-3213(74)90093-5} {\bibfield  {journal} {\bibinfo  {journal}
  {Nucl.~Phys.},\ }\textbf {\bibinfo {volume} {B75}},\ \bibinfo {pages}
  {531}}\BibitemShut {NoStop}%
%%CITATION = NUPHA,B75,531;%%
\bibitem [{\citenamefont {Kaplan}(1991)}]{Kaplan:1991dc}%
  \BibitemOpen
  \bibfield  {author} {\bibinfo {author} {\bibnamefont {Kaplan} \bibfnamefont
  {David~B}}} (\bibinfo {year} {1991}),\ \bibfield  {title} {\enquote {\bibinfo
  {title} {{Flavor at SSC energies: A New mechanism for dynamically generated
  fermion masses}},}\ }\Doi {10.1016/S0550-3213(05)80021-5} {\bibfield
  {journal} {\bibinfo  {journal} {Nucl.~Phys.},\ }\textbf {\bibinfo {volume}
  {B365}},\ \bibinfo {pages} {259--278}}\BibitemShut {NoStop}%
%%CITATION = NUPHA,B365,259;%%
\bibitem [{\citenamefont {Kaplan}\ and\ \citenamefont
  {Georgi}(1984)}]{Kaplan:1983fs}%
  \BibitemOpen
  \bibfield  {author} {\bibinfo {author} {\bibnamefont {Kaplan} \bibfnamefont
  {David~B}}, \ and\ \bibinfo {author} {\bibfnamefont {Howard}\ \bibnamefont
  {Georgi}}} (\bibinfo {year} {1984}),\ \bibfield  {title} {\enquote {\bibinfo
  {title} {{SU(2) x U(1) Breaking by Vacuum Misalignment}},}\ }\Doi
  {10.1016/0370-2693(84)91177-8} {\bibfield  {journal} {\bibinfo  {journal}
  {Phys.~Lett.},\ }\textbf {\bibinfo {volume} {B136}},\ \bibinfo {pages}
  {183}}\BibitemShut {NoStop}%
%%CITATION = PHLTA,B136,183;%%
\bibitem [{\citenamefont {Kaplan}\ \emph {et~al.}(1984)\citenamefont {Kaplan},
  \citenamefont {Georgi},\ and\ \citenamefont {Dimopoulos}}]{Kaplan:1983sm}%
  \BibitemOpen
  \bibfield  {author} {\bibinfo {author} {\bibnamefont {Kaplan} \bibfnamefont
  {David~B}}, \bibinfo {author} {\bibfnamefont {Howard}\ \bibnamefont
  {Georgi}}, \ and\ \bibinfo {author} {\bibfnamefont {Savas}\ \bibnamefont
  {Dimopoulos}}} (\bibinfo {year} {1984}),\ \bibfield  {title} {\enquote
  {\bibinfo {title} {{Composite Higgs Scalars}},}\ }\Doi
  {10.1016/0370-2693(84)91178-X} {\bibfield  {journal} {\bibinfo  {journal}
  {Phys.~Lett.},\ }\textbf {\bibinfo {volume} {B136}},\ \bibinfo {pages}
  {187}}\BibitemShut {NoStop}%
%%CITATION = PHLTA,B136,187;%%
\bibitem [{\citenamefont {Kaplan}\ \emph {et~al.}(2009)\citenamefont {Kaplan},
  \citenamefont {Lee}, \citenamefont {Son},\ and\ \citenamefont
  {Stephanov}}]{Kaplan:2009kr}%
  \BibitemOpen
  \bibfield  {author} {\bibinfo {author} {\bibnamefont {Kaplan} \bibfnamefont
  {David~B}}, \bibinfo {author} {\bibfnamefont {Jong-Wan}\ \bibnamefont {Lee}},
  \bibinfo {author} {\bibfnamefont {Dam~T.}\ \bibnamefont {Son}}, \ and\
  \bibinfo {author} {\bibfnamefont {Mikhail~A.}\ \bibnamefont {Stephanov}}}
  (\bibinfo {year} {2009}),\ \bibfield  {title} {\enquote {\bibinfo {title}
  {{Conformality Lost}},}\ }\Doi {10.1103/PhysRevD.80.125005} {\bibfield
  {journal} {\bibinfo  {journal} {Phys.~Rev.},\ }\textbf {\bibinfo {volume}
  {D80}},\ \bibinfo {pages} {125005}},\ \Eprint
  {http://arxiv.org/abs/0905.4752} {arXiv:0905.4752 [hep-th]} \BibitemShut
  {NoStop}%
%%CITATION = ARXIV:0905.4752;%%
\bibitem [{\citenamefont {Karavirta}\ \emph {et~al.}(2012a)\citenamefont
  {Karavirta}, \citenamefont {Rantaharju}, \citenamefont {Rummukainen},\ and\
  \citenamefont {Tuominen}}]{Karavirta:2011zg}%
  \BibitemOpen
  \bibfield  {author} {\bibinfo {author} {\bibnamefont {Karavirta}
  \bibfnamefont {Tuomas}}, \bibinfo {author} {\bibfnamefont {Jarno}\
  \bibnamefont {Rantaharju}}, \bibinfo {author} {\bibfnamefont {Kari}\
  \bibnamefont {Rummukainen}}, \ and\ \bibinfo {author} {\bibfnamefont {Kimmo}\
  \bibnamefont {Tuominen}}} (\bibinfo {year} {2012a}),\ \bibfield  {title}
  {\enquote {\bibinfo {title} {{Determining the conformal window: SU(2) gauge
  theory with $N_f$ = 4, 6 and 10 fermion flavours}},}\ }\Doi
  {10.1007/JHEP05(2012)003} {\bibfield  {journal} {\bibinfo  {journal} {JHEP},\
  }\textbf {\bibinfo {volume} {1205}},\ \bibinfo {pages} {003}},\ \Eprint
  {http://arxiv.org/abs/1111.4104} {arXiv:1111.4104 [hep-lat]} \BibitemShut
  {NoStop}%
%%CITATION = ARXIV:1111.4104;%%
\bibitem [{\citenamefont {Karavirta}\ \emph {et~al.}(2012b)\citenamefont
  {Karavirta}, \citenamefont {Tuominen},\ and\ \citenamefont
  {Rummukainen}}]{Karavirta:2012qd}%
  \BibitemOpen
  \bibfield  {author} {\bibinfo {author} {\bibnamefont {Karavirta}
  \bibfnamefont {Tuomas}}, \bibinfo {author} {\bibfnamefont {Kimmo}\
  \bibnamefont {Tuominen}}, \ and\ \bibinfo {author} {\bibfnamefont {Kari}\
  \bibnamefont {Rummukainen}}} (\bibinfo {year} {2012b}),\ \bibfield  {title}
  {\enquote {\bibinfo {title} {{Perturbative Improvement of the Schrodinger
  Functional for Lattice Strong Dynamics}},}\ }\Doi
  {10.1103/PhysRevD.85.054506} {\bibfield  {journal} {\bibinfo  {journal}
  {Phys.~Rev.},\ }\textbf {\bibinfo {volume} {D85}},\ \bibinfo {pages}
  {054506}},\ \Eprint {http://arxiv.org/abs/1201.1883} {arXiv:1201.1883
  [hep-lat]} \BibitemShut {NoStop}%
%%CITATION = ARXIV:1201.1883;%%
\bibitem [{\citenamefont {Kim}\ \emph {et~al.}(2011)\citenamefont {Kim},
  \citenamefont {Fukaya}, \citenamefont {Hashimoto}, \citenamefont {Matsufuru},
  \citenamefont {Nishimura},\ and\ \citenamefont {Onogi}}]{Kim:2011fw}%
  \BibitemOpen
  \bibfield  {author} {\bibinfo {author} {\bibnamefont {Kim} \bibfnamefont
  {S~W}}, \bibinfo {author} {\bibfnamefont {H.}~\bibnamefont {Fukaya}},
  \bibinfo {author} {\bibfnamefont {S.}~\bibnamefont {Hashimoto}}, \bibinfo
  {author} {\bibfnamefont {H.}~\bibnamefont {Matsufuru}}, \bibinfo {author}
  {\bibfnamefont {J.}~\bibnamefont {Nishimura}}, \ and\ \bibinfo {author}
  {\bibfnamefont {T.}~\bibnamefont {Onogi}} (\bibinfo {collaboration} {JLQCD})}
  (\bibinfo {year} {2011}),\ \bibfield  {title} {\enquote {\bibinfo {title}
  {{Lattice study of 4d $\mathcal N = 1$ super Yang--Mills theory with
  dynamical overlap gluino}},}\ }\href@noop {} {\bibfield  {journal} {\bibinfo
  {journal} {PoS},\ }\textbf {\bibinfo {volume} {Lattice 2011}},\ \bibinfo
  {pages} {069}},\ \Eprint {http://arxiv.org/abs/1111.2180} {arXiv:1111.2180}
  \BibitemShut {NoStop}%
\bibitem [{\citenamefont {Kluberg-Stern}\ \emph {et~al.}(1982)\citenamefont
  {Kluberg-Stern}, \citenamefont {Morel},\ and\ \citenamefont
  {Petersson}}]{KlubergStern:1982eh}%
  \BibitemOpen
  \bibfield  {author} {\bibinfo {author} {\bibnamefont {Kluberg-Stern}
  \bibfnamefont {H}}, \bibinfo {author} {\bibfnamefont {A.}~\bibnamefont
  {Morel}}, \ and\ \bibinfo {author} {\bibfnamefont {B.}~\bibnamefont
  {Petersson}}} (\bibinfo {year} {1982}),\ \bibfield  {title} {\enquote
  {\bibinfo {title} {{The strong coupling limit of gauge theories with fermions
  on a lattice}},}\ }\Doi {10.1016/0370-2693(82)90135-6} {\bibfield  {journal}
  {\bibinfo  {journal} {Phys.~Lett.},\ }\textbf {\bibinfo {volume} {B114}},\
  \bibinfo {pages} {152--156}}\BibitemShut {NoStop}%
%%CITATION = PHLTA,B114,152;%%
\bibitem [{\citenamefont {Knechtli}\ \emph {et~al.}(2014)\citenamefont
  {Knechtli}, \citenamefont {Yoneyama}, \citenamefont {Dziennik},\ and\
  \citenamefont {Irges}}]{Knechtli:2014ioa}%
  \BibitemOpen
  \bibfield  {author} {\bibinfo {author} {\bibnamefont {Knechtli} \bibfnamefont
  {Francesco}}, \bibinfo {author} {\bibfnamefont {Kyoko}\ \bibnamefont
  {Yoneyama}}, \bibinfo {author} {\bibfnamefont {Peter}\ \bibnamefont
  {Dziennik}}, \ and\ \bibinfo {author} {\bibfnamefont {Nikos}\ \bibnamefont
  {Irges}}} (\bibinfo {year} {2014}),\ \bibfield  {title} {\enquote {\bibinfo
  {title} {{Progress in Gauge-Higgs Unification on the Lattice}},}\ }\href@noop
  {} {\bibfield  {journal} {\bibinfo  {journal} {PoS},\ }\textbf {\bibinfo
  {volume} {LATTICE2013}},\ \bibinfo {pages} {061}},\ \Eprint
  {http://arxiv.org/abs/1402.3491} {arXiv:1402.3491 [hep-lat]} \BibitemShut
  {NoStop}%
%%CITATION = ARXIV:1402.3491;%%
\bibitem [{\citenamefont {Kogut}\ and\ \citenamefont
  {Sinclair}(2010)}]{Kogut:2010cz}%
  \BibitemOpen
  \bibfield  {author} {\bibinfo {author} {\bibnamefont {Kogut} \bibfnamefont
  {JB}}, \ and\ \bibinfo {author} {\bibfnamefont {D.K.}\ \bibnamefont
  {Sinclair}}} (\bibinfo {year} {2010}),\ \bibfield  {title} {\enquote
  {\bibinfo {title} {{Thermodynamics of lattice QCD with 2 flavours of
  colour-sextet quarks: A model of walking/conformal Technicolor}},}\ }\Doi
  {10.1103/PhysRevD.81.114507} {\bibfield  {journal} {\bibinfo  {journal}
  {Phys.~Rev.},\ }\textbf {\bibinfo {volume} {D81}},\ \bibinfo {pages}
  {114507}},\ \Eprint {http://arxiv.org/abs/1002.2988} {arXiv:1002.2988
  [hep-lat]} \BibitemShut {NoStop}%
%%CITATION = ARXIV:1002.2988;%%
\bibitem [{\citenamefont {Kogut}\ and\ \citenamefont
  {Sinclair}(2011)}]{Kogut:2011ty}%
  \BibitemOpen
  \bibfield  {author} {\bibinfo {author} {\bibnamefont {Kogut} \bibfnamefont
  {JB}}, \ and\ \bibinfo {author} {\bibfnamefont {D.K.}\ \bibnamefont
  {Sinclair}}} (\bibinfo {year} {2011}),\ \bibfield  {title} {\enquote
  {\bibinfo {title} {{Thermodynamics of lattice QCD with 2 sextet quarks on
  $N_t$=8 lattices}},}\ }\Doi {10.1103/PhysRevD.84.074504} {\bibfield
  {journal} {\bibinfo  {journal} {Phys.~Rev.},\ }\textbf {\bibinfo {volume}
  {D84}},\ \bibinfo {pages} {074504}},\ \Eprint
  {http://arxiv.org/abs/1105.3749} {arXiv:1105.3749 [hep-lat]} \BibitemShut
  {NoStop}%
%%CITATION = ARXIV:1105.3749;%%
\bibitem [{\citenamefont {Kogut}\ and\ \citenamefont
  {Sinclair}(2014)}]{Kogut:2014kla}%
  \BibitemOpen
  \bibfield  {author} {\bibinfo {author} {\bibnamefont {Kogut} \bibfnamefont
  {JB}}, \ and\ \bibinfo {author} {\bibfnamefont {D.K.}\ \bibnamefont
  {Sinclair}}} (\bibinfo {year} {2014}),\ \bibfield  {title} {\enquote
  {\bibinfo {title} {{Thermodynamics of lattice QCD with 3 flavours of
  colour-sextet quarks II: $N_t=6$ and $N_t=8$}},}\ }\Doi
  {10.1103/PhysRevD.90.014506} {\bibfield  {journal} {\bibinfo  {journal}
  {Phys.~Rev.},\ }\textbf {\bibinfo {volume} {D90}},\ \bibinfo {pages}
  {014506}},\ \Eprint {http://arxiv.org/abs/1406.1524} {arXiv:1406.1524
  [hep-lat]} \BibitemShut {NoStop}%
%%CITATION = ARXIV:1406.1524;%%
\bibitem [{\citenamefont {Kosower}(1984)}]{Kosower:1984aw}%
  \BibitemOpen
  \bibfield  {author} {\bibinfo {author} {\bibnamefont {Kosower} \bibfnamefont
  {D~A}}} (\bibinfo {year} {1984}),\ \bibfield  {title} {\enquote {\bibinfo
  {title} {{Symmetry Breaking Patterns in Pseudoreal and Real Gauge
  Theories}},}\ }\Doi {10.1016/0370-2693(84)91806-9} {\bibfield  {journal}
  {\bibinfo  {journal} {Phys. Lett.},\ }\textbf {\bibinfo {volume} {B144}},\
  \bibinfo {pages} {215--216}}\BibitemShut {NoStop}%
%%CITATION = PHLTA,B144,215;%%
\bibitem [{\citenamefont {Kuzmin}\ \emph {et~al.}(1985)\citenamefont {Kuzmin},
  \citenamefont {Rubakov},\ and\ \citenamefont {Shaposhnikov}}]{Kuzmin:1985mm}%
  \BibitemOpen
  \bibfield  {author} {\bibinfo {author} {\bibnamefont {Kuzmin} \bibfnamefont
  {VA}}, \bibinfo {author} {\bibfnamefont {V.A.}\ \bibnamefont {Rubakov}}, \
  and\ \bibinfo {author} {\bibfnamefont {M.E.}\ \bibnamefont {Shaposhnikov}}}
  (\bibinfo {year} {1985}),\ \bibfield  {title} {\enquote {\bibinfo {title}
  {{On the Anomalous Electroweak Baryon Number Nonconservation in the Early
  Universe}},}\ }\Doi {10.1016/0370-2693(85)91028-7} {\bibfield  {journal}
  {\bibinfo  {journal} {Phys.~Lett.},\ }\textbf {\bibinfo {volume} {B155}},\
  \bibinfo {pages} {36}}\BibitemShut {NoStop}%
%%CITATION = PHLTA,B155,36;%%
\bibitem [{\citenamefont {Laine}\ \emph {et~al.}(2013)\citenamefont {Laine},
  \citenamefont {Nardini},\ and\ \citenamefont {Rummukainen}}]{Laine:2013raa}%
  \BibitemOpen
  \bibfield  {author} {\bibinfo {author} {\bibnamefont {Laine} \bibfnamefont
  {M}}, \bibinfo {author} {\bibfnamefont {G.}~\bibnamefont {Nardini}}, \ and\
  \bibinfo {author} {\bibfnamefont {K.}~\bibnamefont {Rummukainen}}} (\bibinfo
  {year} {2013}),\ \bibfield  {title} {\enquote {\bibinfo {title} {{First order
  thermal phase transition with 126 GeV Higgs mass}},}\ }\href@noop {}
  {\bibfield  {journal} {\bibinfo  {journal} {PoS},\ }\textbf {\bibinfo
  {volume} {LATTICE2013}},\ \bibinfo {pages} {104}},\ \Eprint
  {http://arxiv.org/abs/1311.4424} {arXiv:1311.4424 [hep-lat]} \BibitemShut
  {NoStop}%
%%CITATION = ARXIV:1311.4424;%%
\bibitem [{\citenamefont {Lepage}\ \emph {et~al.}(2014)\citenamefont {Lepage},
  \citenamefont {Mackenzie},\ and\ \citenamefont {Peskin}}]{Lepage:2014fla}%
  \BibitemOpen
  \bibfield  {author} {\bibinfo {author} {\bibnamefont {Lepage} \bibfnamefont
  {G~Peter}}, \bibinfo {author} {\bibfnamefont {Paul~B.}\ \bibnamefont
  {Mackenzie}}, \ and\ \bibinfo {author} {\bibfnamefont {Michael~E.}\
  \bibnamefont {Peskin}}} (\bibinfo {year} {2014}),\ \bibfield  {title}
  {\enquote {\bibinfo {title} {{Expected Precision of Higgs Boson Partial
  Widths within the Standard Model}},}\ }\href@noop {} {}\Eprint
  {http://arxiv.org/abs/1404.0319} {arXiv:1404.0319 [hep-ph]} \BibitemShut
  {NoStop}%
%%CITATION = ARXIV:1404.0319;%%
\bibitem [{\citenamefont {Lewis}\ \emph {et~al.}(2012)\citenamefont {Lewis},
  \citenamefont {Pica},\ and\ \citenamefont {Sannino}}]{Lewis:2011zb}%
  \BibitemOpen
  \bibfield  {author} {\bibinfo {author} {\bibnamefont {Lewis} \bibfnamefont
  {Randy}}, \bibinfo {author} {\bibfnamefont {Claudio}\ \bibnamefont {Pica}}, \
  and\ \bibinfo {author} {\bibfnamefont {Francesco}\ \bibnamefont {Sannino}}}
  (\bibinfo {year} {2012}),\ \bibfield  {title} {\enquote {\bibinfo {title}
  {{Light Asymmetric Dark Matter on the Lattice: SU(2) Technicolor with Two
  Fundamental Flavors}},}\ }\Doi {10.1103/PhysRevD.85.014504} {\bibfield
  {journal} {\bibinfo  {journal} {Phys.~Rev.},\ }\textbf {\bibinfo {volume}
  {D85}},\ \bibinfo {pages} {014504}},\ \Eprint
  {http://arxiv.org/abs/1109.3513} {arXiv:1109.3513 [hep-ph]} \BibitemShut
  {NoStop}%
%%CITATION = ARXIV:1109.3513;%%
\bibitem [{\citenamefont {Lin}\ \emph {et~al.}(2012)\citenamefont {Lin},
  \citenamefont {Ogawa}, \citenamefont {Ohki},\ and\ \citenamefont
  {Shintani}}]{Lin:2012iw}%
  \BibitemOpen
  \bibfield  {author} {\bibinfo {author} {\bibnamefont {Lin} \bibfnamefont
  {C-J~David}}, \bibinfo {author} {\bibfnamefont {Kenji}\ \bibnamefont
  {Ogawa}}, \bibinfo {author} {\bibfnamefont {Hiroshi}\ \bibnamefont {Ohki}}, \
  and\ \bibinfo {author} {\bibfnamefont {Eigo}\ \bibnamefont {Shintani}}}
  (\bibinfo {year} {2012}),\ \bibfield  {title} {\enquote {\bibinfo {title}
  {{Lattice study of infrared behaviour in SU(3) gauge theory with twelve
  massless flavours}},}\ }\Doi {10.1007/JHEP08(2012)096} {\bibfield  {journal}
  {\bibinfo  {journal} {JHEP},\ }\textbf {\bibinfo {volume} {1208}},\ \bibinfo
  {pages} {096}},\ \Eprint {http://arxiv.org/abs/1205.6076} {arXiv:1205.6076
  [hep-lat]} \BibitemShut {NoStop}%
%%CITATION = ARXIV:1205.6076;%%
\bibitem [{\citenamefont {Logan}(2014)}]{Logan:2014jla}%
  \BibitemOpen
  \bibfield  {author} {\bibinfo {author} {\bibnamefont {Logan} \bibfnamefont
  {Heather~E}}} (\bibinfo {year} {2014}),\ \bibfield  {title} {\enquote
  {\bibinfo {title} {{TASI 2013 lectures on Higgs physics within and beyond the
  Standard Model}},}\ }\href@noop {} {}\Eprint {http://arxiv.org/abs/1406.1786}
  {arXiv:1406.1786 [hep-ph]} \BibitemShut {NoStop}%
%%CITATION = ARXIV:1406.1786;%%
\bibitem [{\citenamefont {Lombardo}\ \emph {et~al.}(2014)\citenamefont
  {Lombardo}, \citenamefont {Miura}, \citenamefont {da~Silva},\ and\
  \citenamefont {Pallante}}]{Lombardo:2014pda}%
  \BibitemOpen
  \bibfield  {author} {\bibinfo {author} {\bibnamefont {Lombardo} \bibfnamefont
  {MP}}, \bibinfo {author} {\bibfnamefont {K.}~\bibnamefont {Miura}}, \bibinfo
  {author} {\bibfnamefont {T.~J.~Nunes}\ \bibnamefont {da~Silva}}, \ and\
  \bibinfo {author} {\bibfnamefont {E.}~\bibnamefont {Pallante}}} (\bibinfo
  {year} {2014}),\ \bibfield  {title} {\enquote {\bibinfo {title} {{On the
  particle spectrum and the conformal window}},}\ }\Doi
  {10.1007/JHEP12(2014)183} {\bibfield  {journal} {\bibinfo  {journal} {JHEP},\
  }\textbf {\bibinfo {volume} {1412}},\ \bibinfo {pages} {183}},\ \Eprint
  {http://arxiv.org/abs/1410.0298} {arXiv:1410.0298 [hep-lat]} \BibitemShut
  {NoStop}%
%%CITATION = ARXIV:1410.0298;%%
\bibitem [{\citenamefont {Lucini}\ and\ \citenamefont
  {Panero}(2013)}]{Lucini:2012gg}%
  \BibitemOpen
  \bibfield  {author} {\bibinfo {author} {\bibnamefont {Lucini} \bibfnamefont
  {Biagio}}, \ and\ \bibinfo {author} {\bibfnamefont {Marco}\ \bibnamefont
  {Panero}}} (\bibinfo {year} {2013}),\ \bibfield  {title} {\enquote {\bibinfo
  {title} {{SU(N) gauge theories at large N}},}\ }\Doi
  {10.1016/j.physrep.2013.01.001} {\bibfield  {journal} {\bibinfo  {journal}
  {Phys.~Rept.},\ }\textbf {\bibinfo {volume} {526}},\ \bibinfo {pages}
  {93--163}},\ \Eprint {http://arxiv.org/abs/1210.4997} {arXiv:1210.4997
  [hep-th]} \BibitemShut {NoStop}%
%%CITATION = ARXIV:1210.4997;%%
\bibitem [{\citenamefont {Luscher}(2000)}]{Luscher:2000hn}%
  \BibitemOpen
  \bibfield  {author} {\bibinfo {author} {\bibnamefont {Luscher} \bibfnamefont
  {Martin}}} (\bibinfo {year} {2000}),\ \bibfield  {title} {\enquote {\bibinfo
  {title} {{Chiral gauge theories revisited}},}\ }\href@noop {} {,\ \bibinfo
  {pages} {41--89}}\Eprint {http://arxiv.org/abs/hep-th/0102028}
  {arXiv:hep-th/0102028 [hep-th]} \BibitemShut {NoStop}%
%%CITATION = HEP-TH/0102028;%%
\bibitem [{\citenamefont {Luscher}(2010)}]{Luscher:2010iy}%
  \BibitemOpen
  \bibfield  {author} {\bibinfo {author} {\bibnamefont {Luscher} \bibfnamefont
  {Martin}}} (\bibinfo {year} {2010}),\ \bibfield  {title} {\enquote {\bibinfo
  {title} {{Properties and uses of the Wilson flow in lattice QCD}},}\ }\Doi
  {10.1007/JHEP08(2010)071} {\bibfield  {journal} {\bibinfo  {journal} {JHEP},\
  }\textbf {\bibinfo {volume} {1008}},\ \bibinfo {pages} {071}},\ \Eprint
  {http://arxiv.org/abs/1006.4518} {arXiv:1006.4518 [hep-lat]} \BibitemShut
  {NoStop}%
%%CITATION = ARXIV:1006.4518;%%
\bibitem [{\citenamefont {Luscher}\ \emph {et~al.}(1992)\citenamefont
  {Luscher}, \citenamefont {Narayanan}, \citenamefont {Weisz},\ and\
  \citenamefont {Wolff}}]{Luscher:1992an}%
  \BibitemOpen
  \bibfield  {author} {\bibinfo {author} {\bibnamefont {Luscher} \bibfnamefont
  {Martin}}, \bibinfo {author} {\bibfnamefont {Rajamani}\ \bibnamefont
  {Narayanan}}, \bibinfo {author} {\bibfnamefont {Peter}\ \bibnamefont
  {Weisz}}, \ and\ \bibinfo {author} {\bibfnamefont {Ulli}\ \bibnamefont
  {Wolff}}} (\bibinfo {year} {1992}),\ \bibfield  {title} {\enquote {\bibinfo
  {title} {{The Schrodinger functional: A Renormalizable probe for nonAbelian
  gauge theories}},}\ }\Doi {10.1016/0550-3213(92)90466-O} {\bibfield
  {journal} {\bibinfo  {journal} {Nucl.~Phys.},\ }\textbf {\bibinfo {volume}
  {B384}},\ \bibinfo {pages} {168--228}},\ \Eprint
  {http://arxiv.org/abs/hep-lat/9207009} {arXiv:hep-lat/9207009 [hep-lat]}
  \BibitemShut {NoStop}%
%%CITATION = HEP-LAT/9207009;%%
\bibitem [{\citenamefont {Luscher}\ \emph {et~al.}(1994)\citenamefont
  {Luscher}, \citenamefont {Sommer}, \citenamefont {Weisz},\ and\ \citenamefont
  {Wolff}}]{Luscher:1993gh}%
  \BibitemOpen
  \bibfield  {author} {\bibinfo {author} {\bibnamefont {Luscher} \bibfnamefont
  {Martin}}, \bibinfo {author} {\bibfnamefont {Rainer}\ \bibnamefont {Sommer}},
  \bibinfo {author} {\bibfnamefont {Peter}\ \bibnamefont {Weisz}}, \ and\
  \bibinfo {author} {\bibfnamefont {Ulli}\ \bibnamefont {Wolff}}} (\bibinfo
  {year} {1994}),\ \bibfield  {title} {\enquote {\bibinfo {title} {{A Precise
  determination of the running coupling in the SU(3) Yang-Mills theory}},}\
  }\Doi {10.1016/0550-3213(94)90629-7} {\bibfield  {journal} {\bibinfo
  {journal} {Nucl.~Phys.},\ }\textbf {\bibinfo {volume} {B413}},\ \bibinfo
  {pages} {481--502}},\ \Eprint {http://arxiv.org/abs/hep-lat/9309005}
  {arXiv:hep-lat/9309005 [hep-lat]} \BibitemShut {NoStop}%
%%CITATION = HEP-LAT/9309005;%%
\bibitem [{\citenamefont {Maas}(2013)}]{Maas:2012tj}%
  \BibitemOpen
  \bibfield  {author} {\bibinfo {author} {\bibnamefont {Maas} \bibfnamefont
  {Axel}}} (\bibinfo {year} {2013}),\ \bibfield  {title} {\enquote {\bibinfo
  {title} {{Bound-state/elementary-particle duality in the Higgs sector and the
  case for an excited 'Higgs' within the standard model}},}\ }\Doi
  {10.1142/S0217732313501034} {\bibfield  {journal} {\bibinfo  {journal}
  {Mod.~Phys.~Lett.},\ }\textbf {\bibinfo {volume} {A28}},\ \bibinfo {pages}
  {1350103}},\ \Eprint {http://arxiv.org/abs/1205.6625} {arXiv:1205.6625
  [hep-lat]} \BibitemShut {NoStop}%
%%CITATION = ARXIV:1205.6625;%%
\bibitem [{\citenamefont {Maas}\ and\ \citenamefont
  {Mufti}(2014)}]{Maas:2013aia}%
  \BibitemOpen
  \bibfield  {author} {\bibinfo {author} {\bibnamefont {Maas} \bibfnamefont
  {Axel}}, \ and\ \bibinfo {author} {\bibfnamefont {Tajdar}\ \bibnamefont
  {Mufti}}} (\bibinfo {year} {2014}),\ \bibfield  {title} {\enquote {\bibinfo
  {title} {{Two- and three-point functions in Landau gauge Yang-Mills-Higgs
  theory}},}\ }\Doi {10.1007/JHEP04(2014)006} {\bibfield  {journal} {\bibinfo
  {journal} {JHEP},\ }\textbf {\bibinfo {volume} {1404}},\ \bibinfo {pages}
  {006}},\ \Eprint {http://arxiv.org/abs/1312.4873} {arXiv:1312.4873 [hep-lat]}
  \BibitemShut {NoStop}%
%%CITATION = ARXIV:1312.4873;%%
\bibitem [{\citenamefont {Maas}\ and\ \citenamefont
  {Mufti}(2015)}]{Maas:2014pba}%
  \BibitemOpen
  \bibfield  {author} {\bibinfo {author} {\bibnamefont {Maas} \bibfnamefont
  {Axel}}, \ and\ \bibinfo {author} {\bibfnamefont {Tajdar}\ \bibnamefont
  {Mufti}}} (\bibinfo {year} {2015}),\ \bibfield  {title} {\enquote {\bibinfo
  {title} {{Spectroscopic analysis of the phase diagram of Yang-Mills-Higgs
  theory}},}\ }\Doi {10.1103/PhysRevD.91.113011} {\bibfield  {journal}
  {\bibinfo  {journal} {Phys. Rev.},\ }\textbf {\bibinfo {volume}
  {D91}}~(\bibinfo {number} {11}),\ \bibinfo {pages} {113011}},\ \Eprint
  {http://arxiv.org/abs/1412.6440} {arXiv:1412.6440 [hep-lat]} \BibitemShut
  {NoStop}%
%%CITATION = ARXIV:1412.6440;%%
\bibitem [{\citenamefont {Marques~Tavares}\ \emph {et~al.}(2014)\citenamefont
  {Marques~Tavares}, \citenamefont {Schmaltz},\ and\ \citenamefont
  {Skiba}}]{Tavares:2013dga}%
  \BibitemOpen
  \bibfield  {author} {\bibinfo {author} {\bibnamefont {Marques~Tavares}
  \bibfnamefont {Gustavo}}, \bibinfo {author} {\bibfnamefont {Martin}\
  \bibnamefont {Schmaltz}}, \ and\ \bibinfo {author} {\bibfnamefont {Witold}\
  \bibnamefont {Skiba}}} (\bibinfo {year} {2014}),\ \bibfield  {title}
  {\enquote {\bibinfo {title} {{Higgs mass naturalness and scale invariance in
  the UV}},}\ }\Doi {10.1103/PhysRevD.89.015009} {\bibfield  {journal}
  {\bibinfo  {journal} {Phys.~Rev.},\ }\textbf {\bibinfo {volume} {D89}},\
  \bibinfo {pages} {015009}},\ \Eprint {http://arxiv.org/abs/1308.0025}
  {arXiv:1308.0025 [hep-ph]} \BibitemShut {NoStop}%
%%CITATION = ARXIV:1308.0025;%%
\bibitem [{\citenamefont {Miransky}\ and\ \citenamefont
  {Yamawaki}(1997)}]{Miransky:1996pd}%
  \BibitemOpen
  \bibfield  {author} {\bibinfo {author} {\bibnamefont {Miransky} \bibfnamefont
  {VA}}, \ and\ \bibinfo {author} {\bibfnamefont {Koichi}\ \bibnamefont
  {Yamawaki}}} (\bibinfo {year} {1997}),\ \bibfield  {title} {\enquote
  {\bibinfo {title} {{Conformal phase transition in gauge theories}},}\ }\Doi
  {10.1103/PhysRevD.56.3768, 10.1103/PhysRevD.55.5051} {\bibfield  {journal}
  {\bibinfo  {journal} {Phys.~Rev.},\ }\textbf {\bibinfo {volume} {D55}},\
  \bibinfo {pages} {5051--5066}},\ \Eprint
  {http://arxiv.org/abs/hep-th/9611142} {arXiv:hep-th/9611142 [hep-th]}
  \BibitemShut {NoStop}%
%%CITATION = HEP-TH/9611142;%%
\bibitem [{\citenamefont {Miura}\ \emph {et~al.}(2012)\citenamefont {Miura},
  \citenamefont {Lombardo},\ and\ \citenamefont {Pallante}}]{Miura:2011mc}%
  \BibitemOpen
  \bibfield  {author} {\bibinfo {author} {\bibnamefont {Miura} \bibfnamefont
  {Kohtaroh}}, \bibinfo {author} {\bibfnamefont {Maria~Paola}\ \bibnamefont
  {Lombardo}}, \ and\ \bibinfo {author} {\bibfnamefont {Elisabetta}\
  \bibnamefont {Pallante}}} (\bibinfo {year} {2012}),\ \bibfield  {title}
  {\enquote {\bibinfo {title} {{Chiral phase transition at finite temperature
  and conformal dynamics in large Nf QCD}},}\ }\Doi
  {10.1016/j.physletb.2012.03.017} {\bibfield  {journal} {\bibinfo  {journal}
  {Phys.~Lett.},\ }\textbf {\bibinfo {volume} {B710}},\ \bibinfo {pages}
  {676--682}},\ \Eprint {http://arxiv.org/abs/1110.3152} {arXiv:1110.3152
  [hep-lat]} \BibitemShut {NoStop}%
%%CITATION = ARXIV:1110.3152;%%
\bibitem [{\citenamefont {Nagai}\ \emph {et~al.}(2009)\citenamefont {Nagai},
  \citenamefont {Carrillo-Ruiz}, \citenamefont {Koleva},\ and\ \citenamefont
  {Lewis}}]{Nagai:2009ip}%
  \BibitemOpen
  \bibfield  {author} {\bibinfo {author} {\bibnamefont {Nagai} \bibfnamefont
  {Kei-ichi}}, \bibinfo {author} {\bibfnamefont {Georgina}\ \bibnamefont
  {Carrillo-Ruiz}}, \bibinfo {author} {\bibfnamefont {Gergana}\ \bibnamefont
  {Koleva}}, \ and\ \bibinfo {author} {\bibfnamefont {Randy}\ \bibnamefont
  {Lewis}}} (\bibinfo {year} {2009}),\ \bibfield  {title} {\enquote {\bibinfo
  {title} {{Exploration of SU(N(c)) gauge theory with many Wilson fermions at
  strong coupling}},}\ }\Doi {10.1103/PhysRevD.80.074508} {\bibfield  {journal}
  {\bibinfo  {journal} {Phys.~Rev.},\ }\textbf {\bibinfo {volume} {D80}},\
  \bibinfo {pages} {074508}},\ \Eprint {http://arxiv.org/abs/0908.0166}
  {arXiv:0908.0166 [hep-lat]} \BibitemShut {NoStop}%
%%CITATION = ARXIV:0908.0166;%%
\bibitem [{\citenamefont {Nogradi}(2012)}]{Nogradi:2012dj}%
  \BibitemOpen
  \bibfield  {author} {\bibinfo {author} {\bibnamefont {Nogradi} \bibfnamefont
  {Daniel}}} (\bibinfo {year} {2012}),\ \bibfield  {title} {\enquote {\bibinfo
  {title} {{An ideal toy model for confining, walking and conformal gauge
  theories: the O(3) sigma model with theta-term}},}\ }\Doi
  {10.1007/JHEP05(2012)089} {\bibfield  {journal} {\bibinfo  {journal} {JHEP},\
  }\textbf {\bibinfo {volume} {1205}},\ \bibinfo {pages} {089}},\ \Eprint
  {http://arxiv.org/abs/1202.4616} {arXiv:1202.4616 [hep-lat]} \BibitemShut
  {NoStop}%
%%CITATION = ARXIV:1202.4616;%%
\bibitem [{\citenamefont {Nussinov}(1985)}]{Nussinov:1985xr}%
  \BibitemOpen
  \bibfield  {author} {\bibinfo {author} {\bibnamefont {Nussinov} \bibfnamefont
  {S}}} (\bibinfo {year} {1985}),\ \bibfield  {title} {\enquote {\bibinfo
  {title} {{Technocosmology: could a technibaryon excess provide a natural
  missing mass candidate?}}}\ }\Doi {10.1016/0370-2693(85)90689-6} {\bibfield
  {journal} {\bibinfo  {journal} {Phys.~Lett.},\ }\textbf {\bibinfo {volume}
  {B165}},\ \bibinfo {pages} {55}}\BibitemShut {NoStop}%
%%CITATION = PHLTA,B165,55;%%
\bibitem [{\citenamefont {Ohki}\ \emph {et~al.}(2010)\citenamefont {Ohki},
  \citenamefont {Aoyama}, \citenamefont {Itou}, \citenamefont {Kurachi},
  \citenamefont {Lin} \emph {et~al.}}]{Ohki:2010sr}%
  \BibitemOpen
  \bibfield  {author} {\bibinfo {author} {\bibnamefont {Ohki} \bibfnamefont
  {Hiroshi}}, \bibinfo {author} {\bibfnamefont {Tatsumi}\ \bibnamefont
  {Aoyama}}, \bibinfo {author} {\bibfnamefont {Etsuko}\ \bibnamefont {Itou}},
  \bibinfo {author} {\bibfnamefont {Masafumi}\ \bibnamefont {Kurachi}},
  \bibinfo {author} {\bibfnamefont {C.-J.~David}\ \bibnamefont {Lin}},  \emph
  {et~al.}} (\bibinfo {year} {2010}),\ \bibfield  {title} {\enquote {\bibinfo
  {title} {{Study of the scaling properties in SU(2) gauge theory with eight
  flavors}},}\ }\href@noop {} {\bibfield  {journal} {\bibinfo  {journal}
  {PoS},\ }\textbf {\bibinfo {volume} {LATTICE2010}},\ \bibinfo {pages}
  {066}},\ \Eprint {http://arxiv.org/abs/1011.0373} {arXiv:1011.0373 [hep-lat]}
  \BibitemShut {NoStop}%
%%CITATION = ARXIV:1011.0373;%%
\bibitem [{\citenamefont {Olive}\ \emph {et~al.}(2014)\citenamefont {Olive}
  \emph {et~al.}}]{Agashe:2014kda}%
  \BibitemOpen
  \bibfield  {author} {\bibinfo {author} {\bibnamefont {Olive} \bibfnamefont
  {KA}},  \emph {et~al.} (\bibinfo {collaboration} {Particle Data Group})}
  (\bibinfo {year} {2014}),\ \bibfield  {title} {\enquote {\bibinfo {title}
  {{Review of Particle Physics}},}\ }\Doi {10.1088/1674-1137/38/9/090001}
  {\bibfield  {journal} {\bibinfo  {journal} {Chin.~Phys.},\ }\textbf {\bibinfo
  {volume} {C38}},\ \bibinfo {pages} {090001}}\BibitemShut {NoStop}%
%%CITATION = CHPHD,C38,090001;%%
\bibitem [{\citenamefont {Osborn}\ \emph {et~al.}(1999)\citenamefont {Osborn},
  \citenamefont {Toublan},\ and\ \citenamefont {Verbaarschot}}]{Osborn:1998qb}%
  \BibitemOpen
  \bibfield  {author} {\bibinfo {author} {\bibnamefont {Osborn} \bibfnamefont
  {JC}}, \bibinfo {author} {\bibfnamefont {D.}~\bibnamefont {Toublan}}, \ and\
  \bibinfo {author} {\bibfnamefont {J.J.M.}\ \bibnamefont {Verbaarschot}}}
  (\bibinfo {year} {1999}),\ \bibfield  {title} {\enquote {\bibinfo {title}
  {{From chiral random matrix theory to chiral perturbation theory}},}\ }\Doi
  {10.1016/S0550-3213(98)00716-0} {\bibfield  {journal} {\bibinfo  {journal}
  {Nucl.~Phys.},\ }\textbf {\bibinfo {volume} {B540}},\ \bibinfo {pages}
  {317--344}},\ \Eprint {http://arxiv.org/abs/hep-th/9806110}
  {arXiv:hep-th/9806110 [hep-th]} \BibitemShut {NoStop}%
%%CITATION = HEP-TH/9806110;%%
\bibitem [{\citenamefont {Patella}(2012)}]{Patella:2012da}%
  \BibitemOpen
  \bibfield  {author} {\bibinfo {author} {\bibnamefont {Patella} \bibfnamefont
  {Agostino}}} (\bibinfo {year} {2012}),\ \bibfield  {title} {\enquote
  {\bibinfo {title} {{A precise determination of the psibar-psi anomalous
  dimension in conformal gauge theories}},}\ }\Doi {10.1103/PhysRevD.86.025006}
  {\bibfield  {journal} {\bibinfo  {journal} {Phys.~Rev.},\ }\textbf {\bibinfo
  {volume} {D86}},\ \bibinfo {pages} {025006}},\ \Eprint
  {http://arxiv.org/abs/1204.4432} {arXiv:1204.4432 [hep-lat]} \BibitemShut
  {NoStop}%
%%CITATION = ARXIV:1204.4432;%%
\bibitem [{\citenamefont {Perelstein}(2007)}]{Perelstein:2005ka}%
  \BibitemOpen
  \bibfield  {author} {\bibinfo {author} {\bibnamefont {Perelstein}
  \bibfnamefont {Maxim}}} (\bibinfo {year} {2007}),\ \bibfield  {title}
  {\enquote {\bibinfo {title} {{Little Higgs models and their
  phenomenology}},}\ }\Doi {10.1016/j.ppnp.2006.04.001} {\bibfield  {journal}
  {\bibinfo  {journal} {Prog.~Part.~Nucl.~Phys.},\ }\textbf {\bibinfo {volume}
  {58}},\ \bibinfo {pages} {247--291}},\ \Eprint
  {http://arxiv.org/abs/hep-ph/0512128} {arXiv:hep-ph/0512128 [hep-ph]}
  \BibitemShut {NoStop}%
%%CITATION = HEP-PH/0512128;%%
\bibitem [{\citenamefont {Peskin}(1980)}]{Peskin:1980gc}%
  \BibitemOpen
  \bibfield  {author} {\bibinfo {author} {\bibnamefont {Peskin} \bibfnamefont
  {Michael~E}}} (\bibinfo {year} {1980}),\ \bibfield  {title} {\enquote
  {\bibinfo {title} {{The Alignment of the Vacuum in Theories of
  Technicolor}},}\ }\Doi {10.1016/0550-3213(80)90051-6} {\bibfield  {journal}
  {\bibinfo  {journal} {Nucl.~Phys.},\ }\textbf {\bibinfo {volume} {B175}},\
  \bibinfo {pages} {197--233}}\BibitemShut {NoStop}%
%%CITATION = NUPHA,B175,197;%%
\bibitem [{\citenamefont {Peskin}\ and\ \citenamefont
  {Takeuchi}(1990)}]{Peskin:1990zt}%
  \BibitemOpen
  \bibfield  {author} {\bibinfo {author} {\bibnamefont {Peskin} \bibfnamefont
  {Michael~E}}, \ and\ \bibinfo {author} {\bibfnamefont {Tatsu}\ \bibnamefont
  {Takeuchi}}} (\bibinfo {year} {1990}),\ \bibfield  {title} {\enquote
  {\bibinfo {title} {{A New constraint on a strongly interacting Higgs
  sector}},}\ }\Doi {10.1103/PhysRevLett.65.964} {\bibfield  {journal}
  {\bibinfo  {journal} {Phys.~Rev.~Lett.},\ }\textbf {\bibinfo {volume} {65}},\
  \bibinfo {pages} {964--967}}\BibitemShut {NoStop}%
%%CITATION = PRLTA,65,964;%%
\bibitem [{\citenamefont {Peskin}\ and\ \citenamefont
  {Takeuchi}(1992)}]{Peskin:1991sw}%
  \BibitemOpen
  \bibfield  {author} {\bibinfo {author} {\bibnamefont {Peskin} \bibfnamefont
  {Michael~E}}, \ and\ \bibinfo {author} {\bibfnamefont {Tatsu}\ \bibnamefont
  {Takeuchi}}} (\bibinfo {year} {1992}),\ \bibfield  {title} {\enquote
  {\bibinfo {title} {{Estimation of oblique electroweak corrections}},}\ }\Doi
  {10.1103/PhysRevD.46.381} {\bibfield  {journal} {\bibinfo  {journal}
  {Phys.~Rev.},\ }\textbf {\bibinfo {volume} {D46}},\ \bibinfo {pages}
  {381--409}}\BibitemShut {NoStop}%
%%CITATION = PHRVA,D46,381;%%
\bibitem [{\citenamefont {Pica}\ and\ \citenamefont
  {Sannino}(2011)}]{Pica:2010xq}%
  \BibitemOpen
  \bibfield  {author} {\bibinfo {author} {\bibnamefont {Pica} \bibfnamefont
  {Claudio}}, \ and\ \bibinfo {author} {\bibfnamefont {Francesco}\ \bibnamefont
  {Sannino}}} (\bibinfo {year} {2011}),\ \bibfield  {title} {\enquote {\bibinfo
  {title} {{UV and IR Zeros of Gauge Theories at The Four Loop Order and
  Beyond}},}\ }\Doi {10.1103/PhysRevD.83.035013} {\bibfield  {journal}
  {\bibinfo  {journal} {Phys.~Rev.},\ }\textbf {\bibinfo {volume} {D83}},\
  \bibinfo {pages} {035013}},\ \Eprint {http://arxiv.org/abs/1011.5917}
  {arXiv:1011.5917 [hep-ph]} \BibitemShut {NoStop}%
%%CITATION = ARXIV:1011.5917;%%
\bibitem [{\citenamefont {Preskill}(1981)}]{Preskill:1980mz}%
  \BibitemOpen
  \bibfield  {author} {\bibinfo {author} {\bibnamefont {Preskill} \bibfnamefont
  {John}}} (\bibinfo {year} {1981}),\ \bibfield  {title} {\enquote {\bibinfo
  {title} {{Subgroup Alignment in Hypercolor Theories}},}\ }\Doi
  {10.1016/0550-3213(81)90265-0} {\bibfield  {journal} {\bibinfo  {journal}
  {Nucl. Phys.},\ }\textbf {\bibinfo {volume} {B177}},\ \bibinfo {pages}
  {21--59}}\BibitemShut {NoStop}%
%%CITATION = NUPHA,B177,21;%%
\bibitem [{\citenamefont {Rantaharju}(2014)}]{Rantaharju:2013bva}%
  \BibitemOpen
  \bibfield  {author} {\bibinfo {author} {\bibnamefont {Rantaharju}
  \bibfnamefont {Jarno}}} (\bibinfo {year} {2014}),\ \bibfield  {title}
  {\enquote {\bibinfo {title} {{The Gradient Flow Coupling in Minimal Walking
  Technicolor}},}\ }\href@noop {} {\bibfield  {journal} {\bibinfo  {journal}
  {PoS},\ }\textbf {\bibinfo {volume} {Lattice2013}},\ \bibinfo {pages}
  {084}},\ \Eprint {http://arxiv.org/abs/1311.3719} {arXiv:1311.3719 [hep-lat]}
  \BibitemShut {NoStop}%
%%CITATION = ARXIV:1311.3719;%%
\bibitem [{\citenamefont {Rantaharju}\ \emph {et~al.}(2014)\citenamefont
  {Rantaharju}, \citenamefont {Karavirta}, \citenamefont {Leino}, \citenamefont
  {Rantalaiho}, \citenamefont {Rummukainen},\ and\ \citenamefont
  {Tuominen}}]{Rantaharju:2014ila}%
  \BibitemOpen
  \bibfield  {author} {\bibinfo {author} {\bibnamefont {Rantaharju}
  \bibfnamefont {Jarno}}, \bibinfo {author} {\bibfnamefont {Tuomas}\
  \bibnamefont {Karavirta}}, \bibinfo {author} {\bibfnamefont {Viljami}\
  \bibnamefont {Leino}}, \bibinfo {author} {\bibfnamefont {Teemu}\ \bibnamefont
  {Rantalaiho}}, \bibinfo {author} {\bibfnamefont {Kari}\ \bibnamefont
  {Rummukainen}}, \ and\ \bibinfo {author} {\bibfnamefont {Kimmo}\ \bibnamefont
  {Tuominen}}} (\bibinfo {year} {2014}),\ \bibfield  {title} {\enquote
  {\bibinfo {title} {{The gradient flow running coupling in SU2 with 8
  flavors}},}\ }\bibfield  {booktitle} {\emph {\bibinfo {booktitle}
  {{Proceedings, 32nd International Symposium on Lattice Field Theory (Lattice
  2014)}}},\ }\href@noop {} {\bibfield  {journal} {\bibinfo  {journal} {PoS},\
  }\textbf {\bibinfo {volume} {LATTICE2014}},\ \bibinfo {pages} {258}},\
  \Eprint {http://arxiv.org/abs/1411.4879} {arXiv:1411.4879 [hep-lat]}
  \BibitemShut {NoStop}%
%%CITATION = ARXIV:1411.4879;%%
\bibitem [{\citenamefont {Rodrigo}\ and\ \citenamefont
  {Santamaria}(1993)}]{Rodrigo:1993hc}%
  \BibitemOpen
  \bibfield  {author} {\bibinfo {author} {\bibnamefont {Rodrigo} \bibfnamefont
  {German}}, \ and\ \bibinfo {author} {\bibfnamefont {Arcadi}\ \bibnamefont
  {Santamaria}}} (\bibinfo {year} {1993}),\ \bibfield  {title} {\enquote
  {\bibinfo {title} {{QCD matching conditions at thresholds}},}\ }\Doi
  {10.1016/0370-2693(93)90016-B} {\bibfield  {journal} {\bibinfo  {journal}
  {Phys.~Lett.},\ }\textbf {\bibinfo {volume} {B313}},\ \bibinfo {pages}
  {441--446}},\ \Eprint {http://arxiv.org/abs/hep-ph/9305305}
  {arXiv:hep-ph/9305305 [hep-ph]} \BibitemShut {NoStop}%
%%CITATION = HEP-PH/9305305;%%
\bibitem [{\citenamefont {Ryttov}\ and\ \citenamefont
  {Shrock}(2011)}]{Ryttov:2010iz}%
  \BibitemOpen
  \bibfield  {author} {\bibinfo {author} {\bibnamefont {Ryttov} \bibfnamefont
  {Thomas~A}}, \ and\ \bibinfo {author} {\bibfnamefont {Robert}\ \bibnamefont
  {Shrock}}} (\bibinfo {year} {2011}),\ \bibfield  {title} {\enquote {\bibinfo
  {title} {{Higher-Loop Corrections to the Infrared Evolution of a Gauge Theory
  with Fermions}},}\ }\Doi {10.1103/PhysRevD.83.056011} {\bibfield  {journal}
  {\bibinfo  {journal} {Phys.~Rev.},\ }\textbf {\bibinfo {volume} {D83}},\
  \bibinfo {pages} {056011}},\ \Eprint {http://arxiv.org/abs/1011.4542}
  {arXiv:1011.4542 [hep-ph]} \BibitemShut {NoStop}%
%%CITATION = ARXIV:1011.4542;%%
\bibitem [{\citenamefont {Sannino}\ and\ \citenamefont
  {Tuominen}(2005)}]{Sannino:2004qp}%
  \BibitemOpen
  \bibfield  {author} {\bibinfo {author} {\bibnamefont {Sannino} \bibfnamefont
  {Francesco}}, \ and\ \bibinfo {author} {\bibfnamefont {Kimmo}\ \bibnamefont
  {Tuominen}}} (\bibinfo {year} {2005}),\ \bibfield  {title} {\enquote
  {\bibinfo {title} {{Orientifold theory dynamics and symmetry breaking}},}\
  }\Doi {10.1103/PhysRevD.71.051901} {\bibfield  {journal} {\bibinfo  {journal}
  {Phys.~Rev.},\ }\textbf {\bibinfo {volume} {D71}},\ \bibinfo {pages}
  {051901}},\ \Eprint {http://arxiv.org/abs/hep-ph/0405209}
  {arXiv:hep-ph/0405209 [hep-ph]} \BibitemShut {NoStop}%
%%CITATION = HEP-PH/0405209;%%
\bibitem [{\citenamefont {Sharpe}(2006)}]{Sharpe:2006re}%
  \BibitemOpen
  \bibfield  {author} {\bibinfo {author} {\bibnamefont {Sharpe} \bibfnamefont
  {Stephen~R}}} (\bibinfo {year} {2006}),\ \bibfield  {title} {\enquote
  {\bibinfo {title} {{Rooted staggered fermions: Good, bad or ugly?}}}\
  }\href@noop {} {\bibfield  {journal} {\bibinfo  {journal} {PoS},\ }\textbf
  {\bibinfo {volume} {LAT2006}},\ \bibinfo {pages} {022}},\ \Eprint
  {http://arxiv.org/abs/hep-lat/0610094} {arXiv:hep-lat/0610094 [hep-lat]}
  \BibitemShut {NoStop}%
%%CITATION = HEP-LAT/0610094;%%
\bibitem [{\citenamefont {Shintani}\ \emph {et~al.}(2008)\citenamefont
  {Shintani} \emph {et~al.}}]{Shintani:2008qe}%
  \BibitemOpen
  \bibfield  {author} {\bibinfo {author} {\bibnamefont {Shintani} \bibfnamefont
  {E}},  \emph {et~al.} (\bibinfo {collaboration} {JLQCD Collaboration})}
  (\bibinfo {year} {2008}),\ \bibfield  {title} {\enquote {\bibinfo {title}
  {{S-parameter and pseudo-Nambu-Goldstone boson mass from lattice QCD}},}\
  }\Doi {10.1103/PhysRevLett.101.242001} {\bibfield  {journal} {\bibinfo
  {journal} {Phys.~Rev.~Lett.},\ }\textbf {\bibinfo {volume} {101}},\ \bibinfo
  {pages} {242001}},\ \Eprint {http://arxiv.org/abs/0806.4222} {arXiv:0806.4222
  [hep-lat]} \BibitemShut {NoStop}%
%%CITATION = ARXIV:0806.4222;%%
\bibitem [{\citenamefont {Sint}\ and\ \citenamefont
  {Ramos}(2015)}]{Ramos:2014kka}%
  \BibitemOpen
  \bibfield  {author} {\bibinfo {author} {\bibnamefont {Sint} \bibfnamefont
  {Stefan}}, \ and\ \bibinfo {author} {\bibfnamefont {Alberto}\ \bibnamefont
  {Ramos}}} (\bibinfo {year} {2015}),\ \bibfield  {title} {\enquote {\bibinfo
  {title} {{On O($a^2$) effects in gradient flow observables}},}\ }\bibfield
  {booktitle} {\emph {\bibinfo {booktitle} {{Proceedings, 32nd International
  Symposium on Lattice Field Theory (Lattice 2014)}}},\ }\href@noop {}
  {\bibfield  {journal} {\bibinfo  {journal} {PoS},\ }\textbf {\bibinfo
  {volume} {LATTICE2014}},\ \bibinfo {pages} {329}},\ \Eprint
  {http://arxiv.org/abs/1411.6706} {arXiv:1411.6706 [hep-lat]} \BibitemShut
  {NoStop}%
%%CITATION = ARXIV:1411.6706;%%
\bibitem [{\citenamefont {Sint}\ and\ \citenamefont
  {Sommer}(1996)}]{Sint:1995ch}%
  \BibitemOpen
  \bibfield  {author} {\bibinfo {author} {\bibnamefont {Sint} \bibfnamefont
  {Stefan}}, \ and\ \bibinfo {author} {\bibfnamefont {Rainer}\ \bibnamefont
  {Sommer}}} (\bibinfo {year} {1996}),\ \bibfield  {title} {\enquote {\bibinfo
  {title} {{The Running coupling from the QCD Schrodinger functional: A One
  loop analysis}},}\ }\Doi {10.1016/0550-3213(96)00020-X} {\bibfield  {journal}
  {\bibinfo  {journal} {Nucl.~Phys.},\ }\textbf {\bibinfo {volume} {B465}},\
  \bibinfo {pages} {71--98}},\ \Eprint {http://arxiv.org/abs/hep-lat/9508012}
  {arXiv:hep-lat/9508012 [hep-lat]} \BibitemShut {NoStop}%
%%CITATION = HEP-LAT/9508012;%%
\bibitem [{\citenamefont {Sint}\ and\ \citenamefont
  {Vilaseca}(2011)}]{Sint:2011gv}%
  \BibitemOpen
  \bibfield  {author} {\bibinfo {author} {\bibnamefont {Sint} \bibfnamefont
  {Stefan}}, \ and\ \bibinfo {author} {\bibfnamefont {Pol}\ \bibnamefont
  {Vilaseca}}} (\bibinfo {year} {2011}),\ \bibfield  {title} {\enquote
  {\bibinfo {title} {{Perturbative lattice artefacts in the SF coupling for
  technicolor-inspired models}},}\ }\href@noop {} {\bibfield  {journal}
  {\bibinfo  {journal} {PoS},\ }\textbf {\bibinfo {volume} {LATTICE2011}},\
  \bibinfo {pages} {091}},\ \Eprint {http://arxiv.org/abs/1111.2227}
  {arXiv:1111.2227 [hep-lat]} \BibitemShut {NoStop}%
%%CITATION = ARXIV:1111.2227;%%
\bibitem [{\citenamefont {Sint}\ and\ \citenamefont
  {Vilaseca}(2012)}]{Sint:2012ae}%
  \BibitemOpen
  \bibfield  {author} {\bibinfo {author} {\bibnamefont {Sint} \bibfnamefont
  {Stefan}}, \ and\ \bibinfo {author} {\bibfnamefont {Pol}\ \bibnamefont
  {Vilaseca}}} (\bibinfo {year} {2012}),\ \bibfield  {title} {\enquote
  {\bibinfo {title} {{Lattice artefacts in the Schr\'odinger Functional
  coupling for strongly interacting theories}},}\ }\href@noop {} {\bibfield
  {journal} {\bibinfo  {journal} {PoS},\ }\textbf {\bibinfo {volume}
  {LATTICE2012}},\ \bibinfo {pages} {031}},\ \Eprint
  {http://arxiv.org/abs/1211.0411} {arXiv:1211.0411 [hep-lat]} \BibitemShut
  {NoStop}%
%%CITATION = ARXIV:1211.0411;%%
\bibitem [{\citenamefont {Sint}\ and\ \citenamefont
  {Weisz}(1999)}]{Sint:1998iq}%
  \BibitemOpen
  \bibfield  {author} {\bibinfo {author} {\bibnamefont {Sint} \bibfnamefont
  {Stefan}}, \ and\ \bibinfo {author} {\bibfnamefont {Peter}\ \bibnamefont
  {Weisz}} (\bibinfo {collaboration} {ALPHA collaboration})} (\bibinfo {year}
  {1999}),\ \bibfield  {title} {\enquote {\bibinfo {title} {{The Running quark
  mass in the SF scheme and its two loop anomalous dimension}},}\ }\Doi
  {10.1016/S0550-3213(98)00874-8} {\bibfield  {journal} {\bibinfo  {journal}
  {Nucl.~Phys.},\ }\textbf {\bibinfo {volume} {B545}},\ \bibinfo {pages}
  {529--542}},\ \Eprint {http://arxiv.org/abs/hep-lat/9808013}
  {arXiv:hep-lat/9808013 [hep-lat]} \BibitemShut {NoStop}%
%%CITATION = HEP-LAT/9808013;%%
\bibitem [{\citenamefont {Susskind}(1979)}]{Susskind:1978ms}%
  \BibitemOpen
  \bibfield  {author} {\bibinfo {author} {\bibnamefont {Susskind} \bibfnamefont
  {Leonard}}} (\bibinfo {year} {1979}),\ \bibfield  {title} {\enquote {\bibinfo
  {title} {{Dynamics of Spontaneous Symmetry Breaking in the Weinberg-Salam
  Theory}},}\ }\Doi {10.1103/PhysRevD.20.2619} {\bibfield  {journal} {\bibinfo
  {journal} {Phys.~Rev.},\ }\textbf {\bibinfo {volume} {D20}},\ \bibinfo
  {pages} {2619--2625}}\BibitemShut {NoStop}%
%%CITATION = PHRVA,D20,2619;%%
\bibitem [{\citenamefont {Svetitsky}\ \emph {et~al.}(1980)\citenamefont
  {Svetitsky}, \citenamefont {Drell}, \citenamefont {Quinn},\ and\
  \citenamefont {Weinstein}}]{Svetitsky:1980pa}%
  \BibitemOpen
  \bibfield  {author} {\bibinfo {author} {\bibnamefont {Svetitsky}
  \bibfnamefont {Benjamin}}, \bibinfo {author} {\bibfnamefont {S.D.}\
  \bibnamefont {Drell}}, \bibinfo {author} {\bibfnamefont {Helen~R.}\
  \bibnamefont {Quinn}}, \ and\ \bibinfo {author} {\bibfnamefont {Marvin}\
  \bibnamefont {Weinstein}}} (\bibinfo {year} {1980}),\ \bibfield  {title}
  {\enquote {\bibinfo {title} {{Dynamical Breaking of Chiral Symmetry in
  Lattice Gauge Theories}},}\ }\Doi {10.1103/PhysRevD.22.490} {\bibfield
  {journal} {\bibinfo  {journal} {Phys.~Rev.},\ }\textbf {\bibinfo {volume}
  {D22}},\ \bibinfo {pages} {490}}\BibitemShut {NoStop}%
%%CITATION = PHRVA,D22,490;%%
\bibitem [{\citenamefont {Swendsen}(1979)}]{Swendsen:1979gn}%
  \BibitemOpen
  \bibfield  {author} {\bibinfo {author} {\bibnamefont {Swendsen} \bibfnamefont
  {RH}}} (\bibinfo {year} {1979}),\ \bibfield  {title} {\enquote {\bibinfo
  {title} {{Monte Carlo Renormalization Group}},}\ }\Doi
  {10.1103/PhysRevLett.42.859} {\bibfield  {journal} {\bibinfo  {journal}
  {Phys.~Rev.~Lett.},\ }\textbf {\bibinfo {volume} {42}},\ \bibinfo {pages}
  {859--861}}\BibitemShut {NoStop}%
%%CITATION = PRLTA,42,859;%%
\bibitem [{\citenamefont {Swendsen}(1984)}]{Swendsen:1984vu}%
  \BibitemOpen
  \bibfield  {author} {\bibinfo {author} {\bibnamefont {Swendsen} \bibfnamefont
  {RH}}} (\bibinfo {year} {1984}),\ \bibfield  {title} {\enquote {\bibinfo
  {title} {{Optimization of real space renormalization group
  transformations}},}\ }\Doi {10.1103/PhysRevLett.52.2321} {\bibfield
  {journal} {\bibinfo  {journal} {Phys.~Rev.~Lett.},\ }\textbf {\bibinfo
  {volume} {52}},\ \bibinfo {pages} {2321--2323}}\BibitemShut {NoStop}%
%%CITATION = PRLTA,52,2321;%%
\bibitem [{\citenamefont {Vecchi}(2013)}]{Vecchi:2013bja}%
  \BibitemOpen
  \bibfield  {author} {\bibinfo {author} {\bibnamefont {Vecchi} \bibfnamefont
  {Luca}}} (\bibinfo {year} {2013}),\ \bibfield  {title} {\enquote {\bibinfo
  {title} {{The Natural Composite Higgs}},}\ }\href@noop {} {}\Eprint
  {http://arxiv.org/abs/1304.4579} {arXiv:1304.4579 [hep-ph]} \BibitemShut
  {NoStop}%
%%CITATION = ARXIV:1304.4579;%%
\bibitem [{\citenamefont {Weinberg}(1979)}]{Weinberg:1979bn}%
  \BibitemOpen
  \bibfield  {author} {\bibinfo {author} {\bibnamefont {Weinberg} \bibfnamefont
  {Steven}}} (\bibinfo {year} {1979}),\ \bibfield  {title} {\enquote {\bibinfo
  {title} {{Implications of Dynamical Symmetry Breaking: An Addendum}},}\ }\Doi
  {10.1103/PhysRevD.19.1277} {\bibfield  {journal} {\bibinfo  {journal}
  {Phys.~Rev.},\ }\textbf {\bibinfo {volume} {D19}},\ \bibinfo {pages}
  {1277--1280}}\BibitemShut {NoStop}%
%%CITATION = PHRVA,D19,1277;%%
\bibitem [{\citenamefont {Weinstein}\ \emph {et~al.}(1980)\citenamefont
  {Weinstein}, \citenamefont {Drell}, \citenamefont {Quinn},\ and\
  \citenamefont {Svetitsky}}]{Weinstein:1980jk}%
  \BibitemOpen
  \bibfield  {author} {\bibinfo {author} {\bibnamefont {Weinstein}
  \bibfnamefont {Marvin}}, \bibinfo {author} {\bibfnamefont {S.D.}\
  \bibnamefont {Drell}}, \bibinfo {author} {\bibfnamefont {Helen~R.}\
  \bibnamefont {Quinn}}, \ and\ \bibinfo {author} {\bibfnamefont {Benjamin}\
  \bibnamefont {Svetitsky}}} (\bibinfo {year} {1980}),\ \bibfield  {title}
  {\enquote {\bibinfo {title} {{Approximate Dynamical Symmetry in Lattice
  Quantum Chromodynamics}},}\ }\Doi {10.1103/PhysRevD.22.1190} {\bibfield
  {journal} {\bibinfo  {journal} {Phys.~Rev.},\ }\textbf {\bibinfo {volume}
  {D22}},\ \bibinfo {pages} {1190}}\BibitemShut {NoStop}%
%%CITATION = PHRVA,D22,1190;%%
\bibitem [{\citenamefont {Wilson}(1974)}]{Wilson:1974sk}%
  \BibitemOpen
  \bibfield  {author} {\bibinfo {author} {\bibnamefont {Wilson} \bibfnamefont
  {Kenneth~G}}} (\bibinfo {year} {1974}),\ \bibfield  {title} {\enquote
  {\bibinfo {title} {{Confinement of Quarks}},}\ }\Doi
  {10.1103/PhysRevD.10.2445} {\bibfield  {journal} {\bibinfo  {journal}
  {Phys.~Rev.},\ }\textbf {\bibinfo {volume} {D10}},\ \bibinfo {pages}
  {2445--2459}}\BibitemShut {NoStop}%
%%CITATION = PHRVA,D10,2445;%%
\bibitem [{\citenamefont {Witten}(1983)}]{Witten:1983ut}%
  \BibitemOpen
  \bibfield  {author} {\bibinfo {author} {\bibnamefont {Witten} \bibfnamefont
  {Edward}}} (\bibinfo {year} {1983}),\ \bibfield  {title} {\enquote {\bibinfo
  {title} {{Some Inequalities Among Hadron Masses}},}\ }\Doi
  {10.1103/PhysRevLett.51.2351} {\bibfield  {journal} {\bibinfo  {journal}
  {Phys.~Rev.~Lett.},\ }\textbf {\bibinfo {volume} {51}},\ \bibinfo {pages}
  {2351}}\BibitemShut {NoStop}%
%%CITATION = PRLTA,51,2351;%%
\bibitem [{\citenamefont {Yamawaki}\ \emph {et~al.}(1986)\citenamefont
  {Yamawaki}, \citenamefont {Bando},\ and\ \citenamefont
  {Matumoto}}]{Yamawaki:1985zg}%
  \BibitemOpen
  \bibfield  {author} {\bibinfo {author} {\bibnamefont {Yamawaki} \bibfnamefont
  {Koichi}}, \bibinfo {author} {\bibfnamefont {Masako}\ \bibnamefont {Bando}},
  \ and\ \bibinfo {author} {\bibfnamefont {Ken-iti}\ \bibnamefont {Matumoto}}}
  (\bibinfo {year} {1986}),\ \bibfield  {title} {\enquote {\bibinfo {title}
  {{Scale Invariant Technicolor Model and a Technidilaton}},}\ }\Doi
  {10.1103/PhysRevLett.56.1335} {\bibfield  {journal} {\bibinfo  {journal}
  {Phys.~Rev.~Lett.},\ }\textbf {\bibinfo {volume} {56}},\ \bibinfo {pages}
  {1335}}\BibitemShut {NoStop}%
%%CITATION = PRLTA,56,1335;%%
\end{thebibliography}%

\end{document}